\documentclass[pdflatex,sn-basic]{sn-jnl}% Basic Springer Nature Reference 

\usepackage{graphicx}%
\usepackage{multirow}%
\usepackage{amsmath,amssymb,amsfonts}%
\usepackage{amsthm}%
\usepackage{mathrsfs}%
\usepackage[title]{appendix}%
\usepackage{xcolor}%
\usepackage{textcomp}%
\usepackage{manyfoot}%
\usepackage{booktabs}%
\usepackage{algorithm}%
\usepackage{algorithmicx}%
\usepackage{algpseudocode}%
\usepackage{listings}%
\begin{document}

\title[A review of galaxy quenching --- Part I]{A review of galaxy quenching --- Part I: Defining the problem and observational results}


\author*[1]{\fnm{Asa F. L.} \sur{Bluck}}\email{abluck@fiu.edu}

\affil*[1]{\orgdiv{Stocker AstroScience Center, Dept. of Physics}, \orgname{Florida International University}, \orgaddress{\street{11200 SW 8th St.}, \city{Miami}, \postcode{33199}, \state{FL}, \country{USA}}}


\abstract{The goal of this review article series is to provide a comprehensive overview of galactic star formation and quenching from both an observational and theoretical perspective. Drawing on a vast quantity of literature, we attempt to answer a deceptively simple question: \textit{why do galaxies cease forming stars?} In Part I, we concentrate on observational results, especially from the past two decades, where there have been enormous advances made towards answering this fundamental question. In Part II we focus on theory and simulations, including discussion of direct observational tests thereof. Observationally, the advent of wide-field spectroscopic galaxy surveys, spatially resolved spectroscopy, high resolution sub-mm and radio observations, X-ray observations, the Hubble Space Telescope (HST), and most recently the James Webb Space Telescope (JWST) have all contributed significantly to our empirical knowledge of star formation and its demise in galaxy quenching. We discuss the fundamental problems which quenching aims to solve, various routes to identifying quiescent galaxies in observations, observational constraints on quenching from studies of galaxy populations, and the spatially resolved view of quenching. We end this part of the review with a detailed discussion of the latest results on galaxy quenching at the high redshift frontier from JWST observations.}




\maketitle

%%\newpage

\setcounter{tocdepth}{3} % TOC subsubsections
\tableofcontents

%%%%%%%%%%%%%%%%%%
%                %
%  INTRODUCTION  %
%                %
%%%%%%%%%%%%%%%%%%

\section{Introduction}\label{s1}

\subsection{Invitation -- Why do galaxies cease forming stars?}\label{s11}

Ultimately, the entirety of this review series is concerned with a single straightforward question: \textit{why do galaxies cease forming stars?} In an ideal world, we would answer this question beyond reasonable doubt here. However, in reality this turns out to be a particularly complex and hotly debated question, which has garnered intense speculation and research over the past several decades. According to NASA ADS, over 8000 research articles have referred to quenching in their abstracts since the year 2000 (with over 1500 articles mentioning quenching in their titles)\footnote{It may interest some readers to know that these articles have collectively received over 250,000 citations since the turn of the millennium. Source: NASA ADS search on 15 June 2026.}. Needless to say, these publications are not in general agreement with each other. As such, the goal of this review series is to critically assess some of this extensive body of research in an attempt to bring order and clarity to the various, often conflicting, perspectives.

\begin{figure}[ht]
\centering
\includegraphics[width=0.9\textwidth]{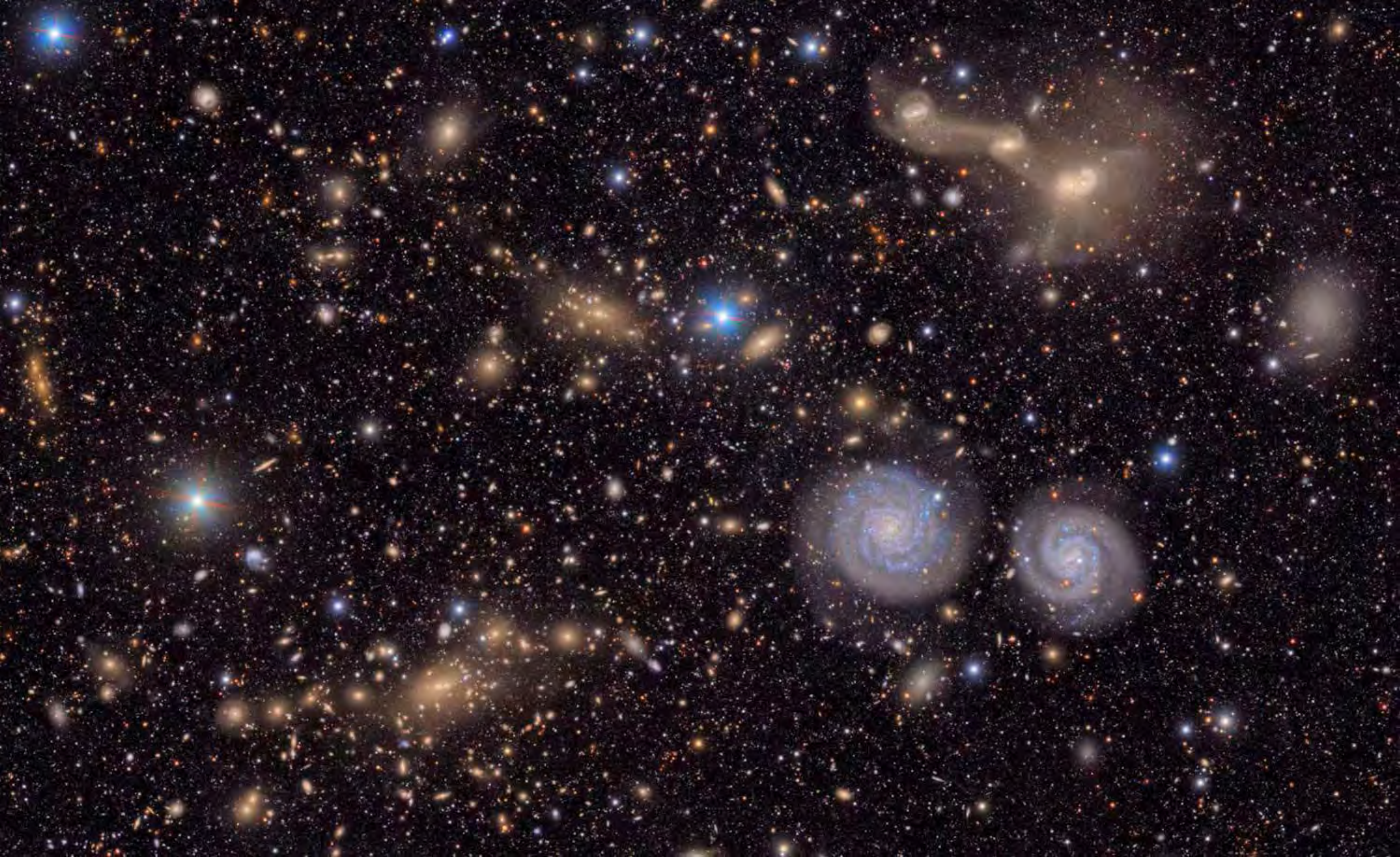}
\caption{A detail from the first publicly released multi-waveband image form the LSST (the legacy survey of space and time, from the Vera C. Rubin Observatory; \citealt{Iversic2008, Iversic2019}). Almost every source of light in this image is a galaxy (with a few interloper stars also present). From visual inspection alone, it is clear that there is remarkable diversity within the galaxy population. Much of this variation is caused by the great range in distances to these objects from Earth. However, even after accounting for the cosmological distance effects (on luminosity, color, angular size, etc.), a great diversity in galaxy types persists. Of particular importance for this review is the existence of distinctly red and blue galaxies, even when accounting for redshift and dust attenuation. This further implies that there is a profound diversity in galactic stellar population ages (discussed in Sect.~\ref{s2}), indicative of some systems which have not formed stars in a very long time. These are the `quenched' galaxies which form the primary subject of this review article. Image Credit: Vera C. Rubin Observatory.}\label{f1} 
\end{figure}

To begin, in Fig.~\ref{f1} we show a detail from the first publicly available image released from the Vera C. Rubin observatory. Almost every source of light in this image is a galaxy and, hence, it becomes immediately clear that there is a great diversity in the size, shape, and color of galaxies within the Universe. Much of this apparent diversity is reduced when accounting for the varying distances to galaxies in this image (spanning $\sim$20\,Gpc), but crucially not all of it. In fact, even when restricting galaxy samples to the same epoch and stellar mass range, one still finds a great diversity in the color of galaxies, revealing a deeper diversity in the ages of their stellar populations. 

Low-mass stars peak in emission at red wavelengths and are long-lived, whereas high-mass stars peak in emission at blue wavelengths and are short-lived (e.g., \citealt{Bruzual2003, Thomas2005, Maraston2005}). Hence, the color of a galaxy (essentially just the integral over its stellar populations) reveals its level of star formation. Galaxies without young stellar populations have, obviously enough, not formed stars in a long time. Hence, the existence of red galaxies is clear evidence for the existence of `quenched' (i.e., non-star forming) galaxies, provided one first rules out the cosmological redshift and dust attenuation as contributing factors.

The simplest possible solution for why some galaxies cease forming stars, and hence become red in optical colors, is that they have run out of the fuel (primarily Hydrogen gas) required for star formation. However, this is emphatically not the case. As we will see, the vast majority of baryons are never converted into stars in the Universe, despite the clear potential for electromagnetic cooling and gravitational collapse (e.g., \citealt{Fukugita2004, Shull2012}). Even within individual dark matter haloes, the vast majority of baryons never form stars, with as little as $\sim$5 - 10\,\% found in stars by the present epoch in the highest mass haloes (e.g., \citealt{Fabian2006, McNamara2007, Fabian2012}).

Hence, we have a simple sounding problem, which obviously deserves an answer --- how come some galaxies have ceased forming stars while others have not? Furthermore, we must reject the obvious possibility that this is simply because they have run out of fuel. Clearly, the problem is deeper than it first appears. Step by step, in this review we will carefully peel back the layers of complexity surrounding this issue. But first we give a brief historical overview.

A distinct population of quenched, early-type galaxies has been recognized observationally for nearly a century. Early evidence emerged from the morphological classification work of Edwin Hubble, who identified elliptical and lenticular galaxies as smooth, centrally concentrated systems lacking the prominent spiral structure associated with active star formation (\citealt{Hubble1926, Hubble1936}). Subsequent studies of galaxy clusters revealed that these morphologies were preferentially associated with dense environments. In particular, the morphology--density relation established by \cite{Dressler1980} demonstrated that early-type galaxies dominate the cores of rich clusters, while later-type, star-forming spirals are more common in low-density regions. At the same time, color studies of distant clusters by \cite{Butcher1978, Butcher1984} revealed an evolving population of blue star forming galaxies at intermediate redshift (the now well-known Butcher-Oemler effect), which implies that the quenched galaxy population must assemble progressively over cosmic time.

The modern quantitative picture of galaxy bimodality emerged with the advent of large wide-field galaxy surveys, particularly the Sloan Digital Sky Survey (\citealt{York2000}). Using large statistical samples, \cite{Strateva2001} demonstrated a clear bimodality in galaxy colors, separating red quiescent galaxies from blue star-forming systems. This result was later strengthened and extended by \cite{Baldry2004, Baldry2006, Peng2010}, who showed that the bimodality persists across luminosity and stellar mass, establishing the now-familiar division between the `red sequence' and `blue cloud' in color - magnitude and color - mass space. These observational discoveries provide the empirical foundation for modern studies of galaxy quenching and established that the cessation of star formation is one of the principal processes underpinning the evolution of galaxies.

In the first part of this review series, we focus on: (i)~setting up the problem of galaxy quenching; and (ii)~presenting key observational results. In the second part of this review series, we explore in detail the theory of galaxy quenching in contemporary models and simulations, and assess direct observational tests of various theoretically proposed quenching paradigms. Finally, we conclude this review series with tentative answers to the fundamental question of why galaxies cease forming stars.

\subsection{Three fundamental problems with $\Lambda$CDM}\label{s12}

\subsubsection{The cosmological problem: Why is star formation so inefficient?}\label{s121}

Perhaps the most basic thing one can hope to predict about the galaxy population from a galaxy formation model is the number density\,--\,stellar mass relation, i.e., the stellar mass function. This is essentially just the distribution of the stellar masses of galaxies in a given volume of space, within a given range of cosmic history. Nonetheless, it is important to highlight that this is a derived quantity, not a direct observable, which depends on an assumed initial mass function (IMF) for star formation, see detailed discussion in Sect.~\ref{s2}. 

\begin{figure}[ht]
\centering
\includegraphics[width=0.65\textwidth]{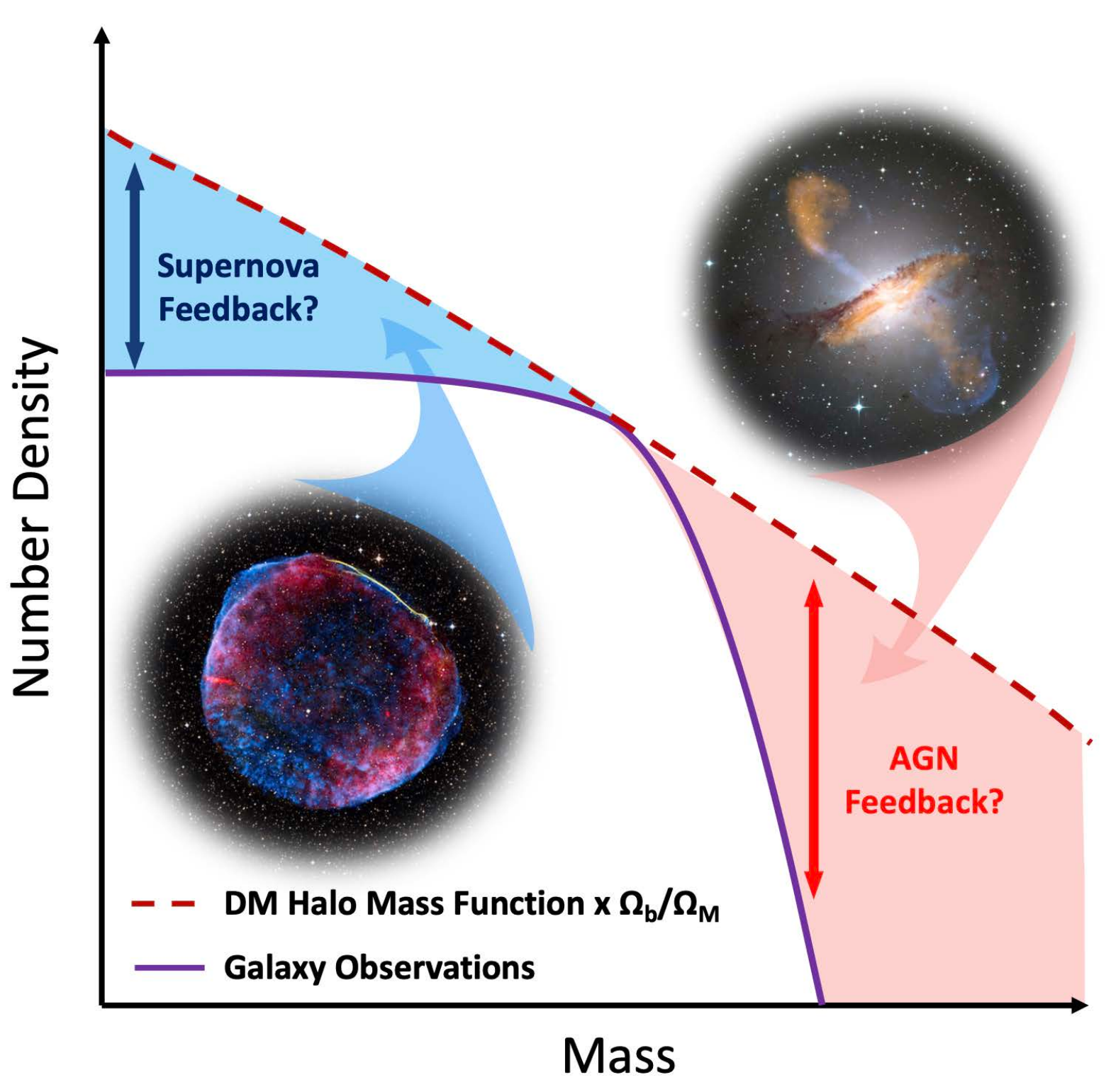}
\caption{A cartoon illustration of one of the main problems with $\Lambda$CDM as a theory of galaxy formation, in the absence of feedback processes. This figure compares the halo mass function scaled by the Universal baryon fraction (shown as a dashed red line) to the observed stellar mass function (shown as a solid purple line). Note that the scaled halo mass function indicates the maximum potential value of the stellar mass function, representing the case where all baryons are converted into stars. In the case of no feedback, the vast majority of baryons are expected to reside within stars by the present epoch (e.g., \citealt{Cole2000, Cole2001, Somerville1999, Somerville2015}). Yet, there are significant discrepancies between these mass functions at both the high- and low-mass ends of the diagram. In contemporary cosmological simulations, feedback from supernovae is utilized to resolve the low mass discrepancy and feedback from AGN is utilized to resolve the high mass discrepancy (as illustrated on the figure). A consequence of this is that there exists a peak efficiency of galaxy formation at intermediate halo masses, where the conversion of baryons into stars within a halo reaches a maximum. Image credit: J. M. Piotrowska (private communication).}\label{f2}
\end{figure}

In Fig.~\ref{f2} we present a cartoon representation of the observed stellar mass function in the local Universe (solid purple line). It has a very distinctive shape, demarcated by a gradual linearly declining phase at low masses, followed by an exponential cutoff at high masses (e.g., \citealt{Bell2003, Peng2010, Baldry2012, Muzzin2013, Thanjavur2016}). This is typically fitted with the theoretically motivated Schechter function, which combines a power law with an exponential function (e.g., \citealt{Bell2003, Marchesini2009}).

The simplest possible model of galaxy formation posits just three things: (i)~that baryons reside within dark matter haloes in proportion to the universal baryon fraction; (ii)~that baryons cool effectively through electromagnetic interactions (especially the bremsstrahlung free\,--\,free interaction); and (iii)~that high densities of baryons form stars, in accordance with some simple density threshold prescription (e.g., \citealt{Kennicutt1998}). Following reasonable empirical constraints on the details within these assumptions leads to the conclusion that the vast majority of baryons should reside within stars by the present epoch (e.g., \citealt{White1991, Kauffmann1993, Cole2000, Cole2001, Croton2006, Bower2006, Bower2008}). Moreover, simple models of this type predict that the stellar mass function of galaxies should be well represented by the halo mass function scaled by the universal baryon fraction (e.g., \citealt{Cole2001, Somerville1999, Bower2006, Bower2008, Somerville2015, Henriques2015, Naab2017}).

In Fig.~\ref{f2}, we compare the scaled halo mass function to the observed stellar mass function of galaxies. Clearly, the agreement is very poor. First, only $\sim$5-10\% of baryons reside in stars by the present epoch, which is an order of magnitude lower than that predicted by the simple galaxy formation model (outlined above). Worse than this, however, is the fact that the stellar mass function does not trace the scaled halo mass function. There is a distinct under-abundance of both low- and high-mass galaxies, relative to the naive prediction from $\Lambda$CDM. This structure cannot be accounted for with gravitational physics alone. Hence, one is forced to turn to the other forces of nature, which by assumption do not impact dark matter directly. Therefore, one must look for baryonic physical processes to explain the deviation in the distribution of galaxies compared to dark matter haloes.

This problem cannot be overstated. If one cannot rectify the shapes of the mass functions in Fig.~\ref{f2}, then (despite all of its successes) $\Lambda$CDM cannot be an accurate description of the Universe. That said, there is cause for hope. There are many things baryons can do which are not solely gravitational. Two examples of which are: (i)~baryons produce stars, some of which go supernova (SN), releasing vast quantities of energy into the galactic system; and (ii)~baryons release vast quantities of energy from accretion around supermassive black holes in active galactic nuclei (AGN), which is expected to impact both the interstellar medium (ISM) within galaxies and the circum-galactic medium (CGM) surrounding galaxies.

Due to the energy scales involved, it is widely believed by theorists today that SN feedback on galaxy evolution can rectify the low-mass end of the mass function disagreement (e.g., \citealt{Kauffmann1993, Cole2000, Bower2006, Bower2008, Henriques2013, Vogelsberger2013, Crain2015}). Briefly, this works by expelling gas from the system, which must then wait to re-accrete new material for future star formation, greatly slowing the rate of star formation in galaxies. However, at around the peak halo mass of star formation (the closest point of agreement in the mass functions), SN become ineffective at removing baryons from haloes (due to their deep potential wells) and actually exacerbate the problem thereafter at higher masses by rapidly building up gas reservoirs (e.g., \citealt{Dekel2009, Dekel2014, Henriques2019, Dekel2019}). 

To combat this even more challenging disagreement, many theorists have turned to the most effective mechanism for matter\,--\,energy conversion (short of annihilation with anti-matter), that of gravitational accretion onto a compact object (e.g., \citealt{Misner1973, Thorne1974}). Consequently, AGN feedback is now widely invoked in models as the key to resolving the high-mass discrepancy in the mass functions of Fig.~\ref{f2} (see, e.g., \citealt{Benson2003, Croton2006, Bower2006, Sijacki2007, Crain2015, Weinberger2017, Weinberger2018, Zinger2020, Piotrowska2022}).

The theory of feedback from SN and AGN will be discussed at length in Part~II of this review series, alongside detailed observational tests of this paradigm. For now, the main take-away is that, in the absence of strong baryonic feedback, $\Lambda$CDM is not an adequate theory of galaxy formation and evolution. Some extra baryonic process(es) must inhibit star formation at low masses and prevent it altogether at high masses, in order for theory and observations to agree.

\subsubsection{The clusters problem: Why is the hot gas halo stable?}\label{s122}

If the vast majority of baryons are not in stars, where are they? In the highest mass haloes, containing clusters and groups of galaxies, the answer is in a hot plasma filling the dark matter halo (see, e.g., \citealt{Fabian2006, McNamara2007, Fabian2012, HlavacekLarrondo2012, HlavacekLarrondo2015, HlavacekLarrondo2018}). These baryons are typically at the virial temperature, which is computed by setting the mean particle kinetic energy equal to the thermal energy. Explicitly, this is given by (e.g., \citealt{Fabian2012}):

\begin{equation}
T_\mathrm{vir} \sim \frac{\mu m_p V_c^2}{2K_B}\,; \quad \text{where} \quad V_c = \sqrt{\frac{G M_\mathrm{vir}}{R_\mathrm{vir}}} \quad \text{and} \quad R_\mathrm{vir} \propto \bigg(\frac{M_\mathrm{vir}}{H(z)^2}\bigg)^{1/3} ,
\label{e1}
\end{equation}

\noindent and where $T_\mathrm{vir}$ is the virial temperature, $V_c$ is the circular velocity, $M_\mathrm{vir}$ is the virial mass, and $R_\mathrm{vir}$ is the virial radius.

As a result of Eq.~\eqref{e1}, it is clear that higher mass haloes host higher temperature baryonic plasmas, ranging from $10^6 - 10^9$\,K across halo mass scales of $10^{11} - 10^{15} M_{\odot}$. Of course, this is embedded in the intergalactic medium (IGM), with a global CMB temperature of:

\begin{equation}
T_{\mathrm{CMB}}(z) = T_0 \, (1+z) \approx 2.73(1+z)\,{\mathrm K}.
\end{equation}

\noindent The temperature of the intergalactic medium (IGM) is actually much hotter than the CMB, due to photo-ionization from galaxies and AGN ($T_{\rm IGM} \sim 10^{4-5}$\,K, e.g., \citealt{Cen1994, Hui1997}). However, it remains much cooler than the virial temperature of high mass haloes. Ultimately, the problem arises from the second law of thermodynamics: hot objects must cool over time when placed in a cooler ambient medium in order for entropy to increase. 

The exact process of that cooling in the highest mass haloes is through the bremsstrahlung (free\,--\,free) interaction. Although it should be noted that in lower-mass haloes metal line cooling dominates. The cooling rate from bremsstrahlung is given explicitly by (e.g., \citealt{Fabian2006, Henriques2015}):

\begin{equation}
\frac{d \mathscr{E}}{dt} = n_e n_I \, \Lambda(T_{\mathrm{vir}}, Z) \propto \rho^2 \, T_\mathrm{vir}^{1/2} \, Z^2.
\end{equation}

\noindent where, $\mathscr{E}$ is the energy per unit volume, $n_e$ and $n_I$ represent the number densities of electrons and ions in the plasma, and $\Lambda$ is the cooling function, which depends on density ($\rho$), temperature ($T$) and the metallicity of the plasma ($Z$). Consequently, as density and temperature increase in the hot gas halo, the rate of cooling also increases. From the above expression, the cooling time may be computed as:

\begin{equation}
t_\mathrm{cool}(r) = \frac{\mathscr{E}(r)}{|d\mathscr{E}(r)/dt|} = \frac{3\mu m_p K_{B} T_\mathrm{vir}}{2 \rho(r) \Lambda(T_{\mathrm{vir}}, Z)},
\end{equation}

\noindent where we emphasize that this varies as a function of radius within the hot gas halo. The simplest density profile compatible with hydrostatic equilibrium is the isothermal sphere, which serves as a useful example here. The density profile of which is derived by setting a constant temperature in the ideal gas law and plugging into the equation for hydrostatic equilibrium, yielding:

\begin{equation}
\rho(r) = \frac{M_\mathrm{Hot}}{4\pi \, R_\mathrm{vir} \, r^2} \propto \frac{M_\mathrm{vir}^{2/3}}{r^2} \, ,
\end{equation}

\noindent where we assume the hot gas mass ($M_\mathrm{Hot}$) is proportional to the virial mass ($M_\mathrm{vir}$), and the virial radius is given by, $R_\mathrm{vir}$. Whilst this is only a very simple case, it does illuminate several general properties of the hot gas halo which turn out to be at least approximately true in more complex models (see, e.g., \citealt{Vogelsberger2013, Henriques2015, Somerville2015}). As the mass of the dark matter halo increases, so does the density at a given radius and the constant temperature throughout the halo. As a result of this, the cooling rate increases (and hence the cooling time decreases) progressively as a function of halo mass. Therefore, the most massive haloes are the least thermodynamically stable. This is known in the literature as the `cooling catastrophe' since it is in direct contradiction with observations of stable hot atmospheres in galaxy clusters, groups, and high mass isolated systems (see \citealt{Fabian1994, Fabian1999, Fabian2006}).

If these hot atmospheres were not thermodynamically stabilized, this would yield run-away accretion of baryons into galaxies, leading to star formation and, ultimately, the production of over-massive galaxies. Hence, the cluster problem is directly related to the cosmological problem discussed in the previous sub-section, given the empirically strong positive correlation between halo and stellar mass (as will be discussed in detail in Sect.~\ref{s3}). Halo stabilization directly impacts (and may even be the primary cause of) the offset in the stellar mass function compared to the scaled halo mass function at high masses. Nonetheless, these two issues in $\Lambda$CDM as a theory of galaxy formation are not usually considered as being one and the same.

The obvious solution to the cluster cooling problem is to evoke heating of some sort (e.g., \citealt{Fabian2006, Croton2006, Bower2006, Bower2008}). Possible heating sources include: (i) shock heating of the hot gas halo from inflowing cold gas streams (which may reach up to $\sqrt{2}$ times the virial velocity upon impact, see \citealt{Dekel2006}); (ii) heating and gas circulation via supernovae (e.g., \citealt{Cole2000, Guo2011, Henriques2013}); (iii) heating from radio jets launched from the accretion discs around supermassive black holes (e.g., \citealt{Sijacki2007, HlavacekLarrondo2012, Weinberger2017}); and (iv) heating and turbulence injection from galactic winds launched by either AGN or starbursts (e.g., \citealt{Springel2005c, DiMatteo2005, Hopkins2006, Hopkins2008}). Also potentially highly important is the formation of hot atmospheres through galaxy--galaxy mergers (e.g., \citealt{Hopkins2006, Hopkins2008, Moreno2015}). For now it suffices to appreciate that a heating source is required in high-mass haloes to achieve both hot gaseous halo stability and to prevent star formation within the central galaxy.

\subsubsection{The bimodality problem: Why are there two types of galaxies?}\label{s123}

The first publications to demonstrate bimodality in the galaxy population investigated the color - magnitude space for wide-field galaxy surveys (see, \citealt{Strateva2001, Baldry2004, Balogh2004, Bell2004}). These works highlight that at a fundamental level there are two types of galaxies in the local Universe - red and blue systems. In this sub-section we explore the bimodality in galaxy color as a fundamental observational constraint on galaxy evolution, evidencing the need for galaxy quenching. 

\begin{figure}[ht]
\centering
\includegraphics[width=0.9\textwidth]{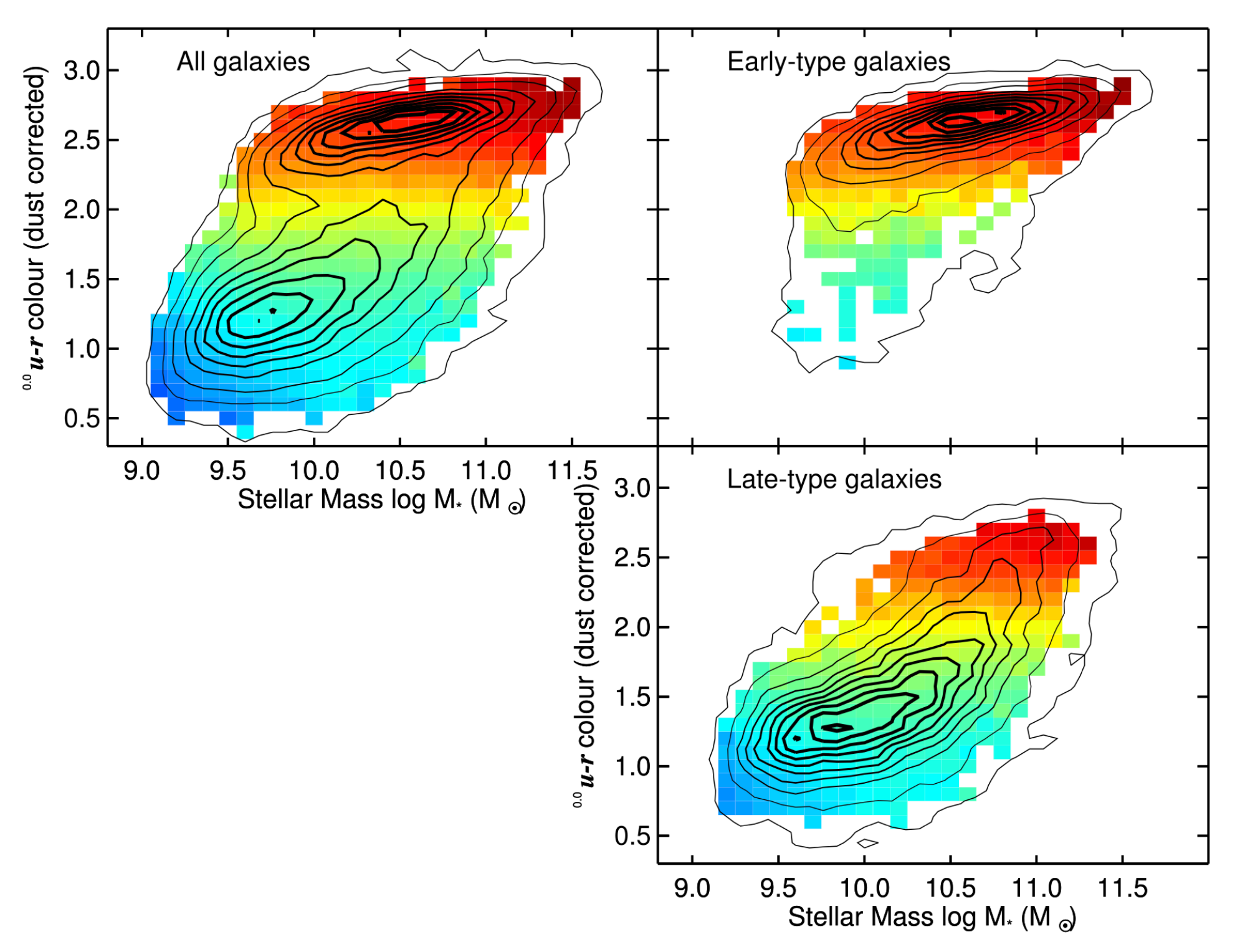}
\caption{Illustration of galaxy bimodality in observations from the SDSS, reproduced from \cite{Schawinski2014}. The top-left panel shows density distributions in the rest-frame, dust corrected optical color vs. stellar mass plane for all galaxies in the survey. Note that galaxies in the local Universe are segregated into two density peaks -- one representing red systems and one representing blue systems. Additionally, one can immediately see that the red galaxy population is skewed to higher stellar masses than the blue galaxy population. The color coding on all panels indicates the sSFR of the galaxies, clearly showing that the red systems are quenched (low sSFR, red colors) and the blue systems are actively star forming (high sSFR, blue colors). Interestingly, in the top-right panel it is clear that early-type galaxies (those presenting with elliptical morphologies) are primarily concentrated on the red sequence of quenched objects. However, in the bottom-right panel, late-type galaxies (those presenting with disc-like morphologies) are spread between the two density peaks, although the majority of these galaxies are centered in the blue/ star forming region. Any successful model of galaxy formation and evolution must explain this fundamental bimodality in galaxy properties with respect to color, sSFR, stellar mass, and morphology.}\label{f3}
\end{figure}

In Fig.~\ref{f3} we present the rest-frame, dust corrected color of local galaxies from the SDSS plot as a function of stellar mass, reproduced from \cite{Schawinski2014}. The top-left panel shows the results for all galaxies together. There is obvious structure in this plot, whereby galaxies are largely segregated into two high density regions -- a `blue cloud' (to the lower left) and a `red sequence' (to the upper right). Additionally, each region of the plot is color coded by the mean specific star formation rate (sSFR~=~SFR/$M_*$) within that region. Star formation rates were inferred from a hybrid emission line and spectral break technique in \cite{Brinchmann2004}\footnote{More details on this are provided in Section 2.}. It is clear that red galaxies have low sSFR and blue galaxies have high sSFR. This is as expected from the evolution of stellar populations. As noted before, the most massive stars are the hottest and hence appear blue in color, but these are also the shortest lived. Conversely, lower mass stars are cooler, appear redder in optical color, and live much longer (see, \citealt{Hertzsprung1909, Russell1914, Eddington1926}). Hence, in a galaxy without recent star formation, one anticipates only long-lived, low-mass stars to be present, and therefore the color will appear redder than in a galaxy with recent star formation (see, e.g., \citealt{Bruzual2003, Thomas2003, Maraston2005, Maraston2010}).

The problem of galaxy bimodality is simply -- why are there two fundamental classes of galaxies, instead of one or, say, seven? This is indeed an important question, which is certainly worth answering. Many extragalactic astronomers feel that this is the most pertinent observation we have of the local galaxy population and, hence, is deserving of paramount attention. Moreover, this clearly relates intimately to our primary purpose of understanding why galaxies cease forming stars (e.g., \citealt{Strateva2001, Brinchmann2004, Driver2006, Cameron2009a, Cameron2009b}). Indeed, some may consider galaxy bimodality as the obvious place to start in a review on galaxy quenching. 

Historically, quenching was first discussed in terms of galaxy bimodality and many extragalactic astronomers today still think primarily in terms of this problem (see \citealt{Strateva2001, Kauffmann2003c, Brinchmann2004, Baldry2006, Driver2006, Peng2010}). However, the first two problems outlined above in this section relate specifically to violations of clear expectations from the $\Lambda$CDM cosmology, in conjunction with the application of well-established physics. The bimodality problem is different. There is nothing about $\Lambda$CDM \textit{prima face} which suggests that galaxies having a bimodal distribution in color\,--\,stellar mass space is challenging. Nonetheless, this does in fact turn out to be the case (see, e.g., \citealt{Crain2015, Schaye2015, Vogelsberger2014a, Vogelsberger2014b, Nelson2018}).

As illustrated by the right-hand panels in Fig.~\ref{f3} (reproduced from \citealt{Schawinski2014}), the bimodality in color is closely linked to galaxy morphology. Early-type galaxies (those presenting with elliptical morphologies) are typically found on the red sequence. On the other hand, late-type galaxies (those presenting with disc-like morphologies) are found throughout the color\,--\,stellar mass diagram, but congregate mostly on the blue cloud. Any successful model of galaxy formation and evolution must explain this bimodality in color and sSFR, and its close (though far from perfect) relationship with morphology (see also, \citealt{Strateva2001, Baldry2004, Wuyts2011, Bluck2014, Lang2014, Omand2014}). Within this framework, quenching is the process, or processes, which drive transition from the `blue cloud' to the `red sequence'. 

This ordering may seem premature. However, upon a moments reflection one sees that in order to have an old stellar population (which appears red) one must have started with a young one (which appears blue). This directionality is further established in observations, which find that the fraction of blue/ star forming galaxies rises with redshift, at a fixed stellar mass (see Sects.~\ref{s2}~\&~\ref{s3}; and, e.g., \citealt{Lilly1996, Madau1996, Mortlock2013, Madau2014, Lang2014, Cheung2012}). This is not to rule out the possibility of rejuvenation (transition from the red sequence to the blue cloud) or multi-quenching (periodic changes in color) as possibilities, but rather just to emphasize that the predominant arrow of causality in this diagram must lead from blue objects to red objects as a function of cosmic time. Presumably, this also requires transition from lower-mass, disk-dominated systems to higher-mass, spheroidal-dominated systems as well (see, e.g., \citealt{Baldry2004, Baldry2006, Driver2006, Driver2011, Bluck2014, Lang2014}).

\subsection{A unified solution -- galaxy quenching}\label{s13}

For the sake of concreteness, for the duration of this review series we define quenching as follows: \\

\noindent \textit{Quenching is the process, or set of processes, which cause galaxies to reduce their star formation significantly for long cosmological time periods, relative to what is typical for actively star forming systems at the same epoch and mass.}\\

\noindent On the one hand, this is quite a broad definition. There is no assumption here that galaxies quench due to a single cause, or that all galaxies quench in the same way. Additionally, in principle, the definition can be further split into two sub-clauses: (i)~the process(es) which cause galaxies to reduce star formation initially, triggering quenching; and (ii)~the process(es) which keep star formation low for long cosmological time periods, relative to the Hubble time. As we will see, some quenching mechanisms can do one of these but not the other, while other quenching mechanisms can explain both aspects. On the other hand, this definition is somewhat narrow. It excludes the cosmological evolution in star formation rate density from consideration, and seeks to establish quenching as a mass-dependent deviation in star formation rate. 

As will become apparent in Sects.~\ref{s2} and \ref{s3}, there is a strong dependence of the star formation rate (SFR) on stellar mass ($M_*$), see \cite{Brinchmann2004}, and, hence, how many stars a galaxy forms in a given time period is not by itself evidence that a system has been quenched (expect in the obvious case of this being zero). Furthermore, at a fixed stellar mass, galaxies form more stars per unit time at earlier epochs than later ones (see, \citealt{Lilly1996, Madau1996, Madau2014}). Both of these features of galactic star formation are well understood theoretically (as will be discussed in later sections) and, hence, ought not to be confused with the outstanding problem of galaxy quenching.

The quenching definition given above is designed to exclude the cosmological evolution in star formation rate, and the close correlation with stellar mass for star forming systems within a given redshift range. There are good conceptual and theoretical reasons to make this distinction, which will be made explicit in Sects.~\ref{s2} and \ref{s3}. Moreover, this is standard practice in the literature (e.g., \citealt{Strateva2001, Baldry2004, Baldry2006, Peng2010, Peng2012, Bluck2014, Bluck2016}). Additionally, note that a return to `normal' levels of star formation following a star formation burst will also not be considered as quenching in this review. Rather, this is considered as temporary deviation and discussed in the context of main sequence oscillations.

How does the concept of quenching help with the three big problems of galaxy evolution (outlined in the previous sub-section)? Returning again to Fig.~\ref{f2}, if galaxies cease forming stars in a mass-dependent manner, this can explain the observed deviation in the stellar mass function from the underlying halo mass function. Indeed, if there is an approximate threshold in stellar mass at which no further star formation is possible, the exponential cutoff at the high-mass end may be fully explained (e.g., \citealt{Peng2010, Peng2012}). The issue at the low-mass end is more subtle, and may or may not be closely liked to quenching.

Furthermore, if one can find a process or set of processes which stabilize baryons in a hot atmosphere around high mass galaxies, groups, and clusters, this will inevitable lead the central galaxy to quench via starvation of future gas supply (e.g., \citealt{Croton2006, Bower2006, Bower2008, Somerville2015, Henriques2015}). This works in two stages: (i)~the hot statistic atmosphere, once stabilized to cooling, will no longer collapse onto the central galaxy replenishing cool gas reservoirs; and (ii)~the stabilized hot gas halo will act as a shield from cosmic gas inflows, shocking these to the virial temperature, and then keeping them hot via whatever process is posited to resolve the cooling catastrophe. Thereafter, assuming no other quenching mechanisms within the galaxy, the central is free to use up its extant cold gas reservoir within its ISM. Once this is fully depleted, the central galaxy will quench and, crucially, remain quenched. Hence, this type of quenching mechanism can effectively explain both aspects of the quenching definition -- why reduction in star formation begins, and why it persists over cosmological times. That said, there are other competing theoretical narratives, including via halo mass quenching (e.g., \cite{Dekel2006}). We will discuss these in detail in Part~II.

Finally, returning to Fig.~\ref{f3}, if one can understand how and why galaxies cease forming stars then one can explain the origin of red galaxies. The red sequence then emerges naturally as a maximally old stellar population (see, e.g., \citealt{Cameron2009a, Schawinski2014}). Moreover, a successful theory of quenching must also account for its strong correlation with both stellar mass and visual morphology. One example of which is linking quenching to galaxy mergers, which are well demonstrated to cause discs to transition into spheroids, as well as being an obvious route to stellar mass growth (see \citealt{Toomre1972, Hopkins2006, Hopkins2008, Hopkins2010}).

In conclusion, understanding how and why galaxies cease forming stars is of paramount importance to three of the biggest mysteries remaining in extragalactic astrophysics. As such, it deserves paramount attention from the extragalactic scientific community.

\subsection{What could cause galaxy quenching?}\label{s14}

This entire review series is concerned with this sub-section's titular question. Here we briefly introduce some possibilities to help the reader start thinking about quenching as a process occurring within and around galaxies.

\begin{figure}[ht]  
\centering
\includegraphics[width=0.95\textwidth]{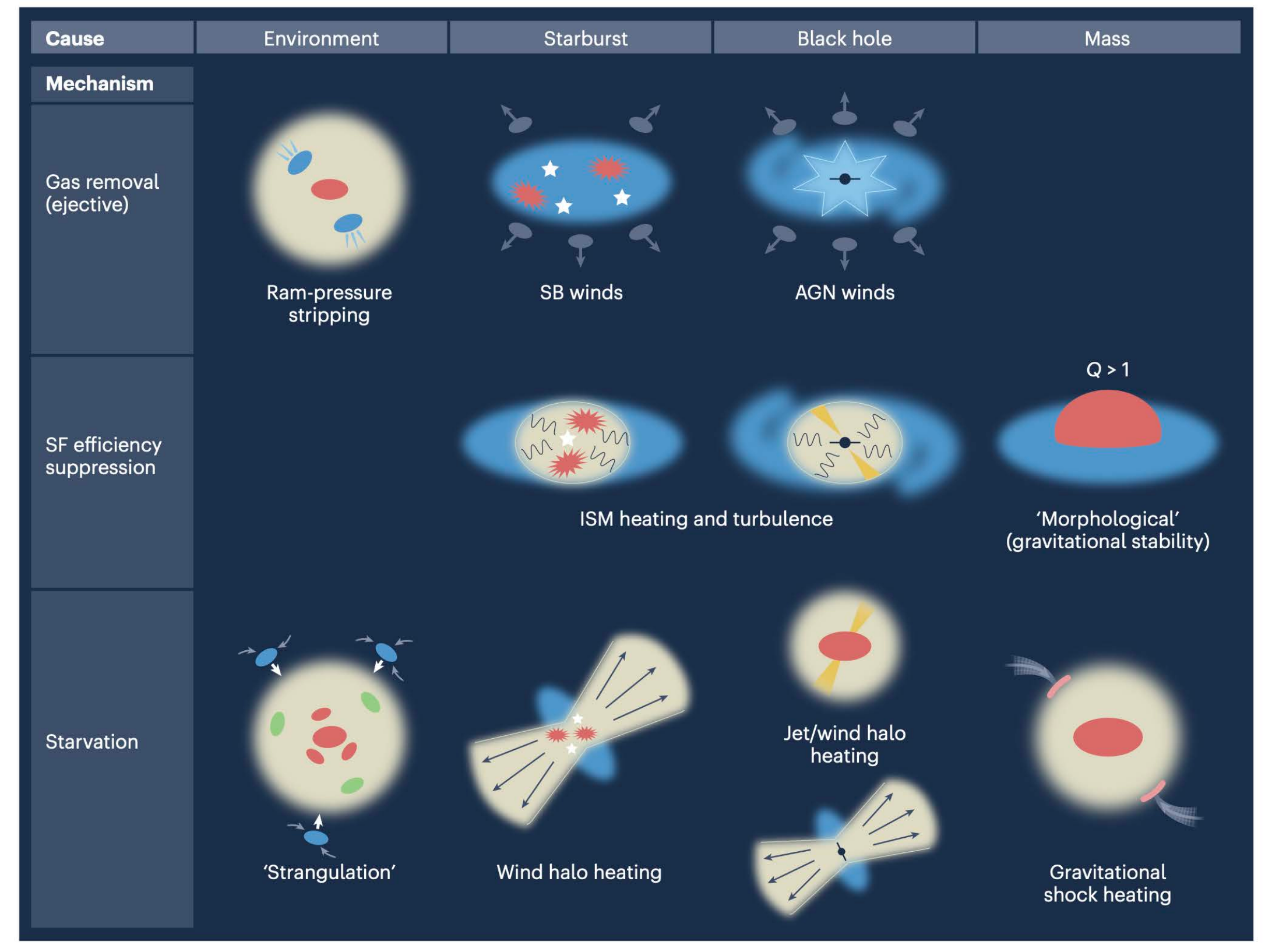}
\caption{A cartoon illustration of possible routes to quenching galaxies, reproduced from \cite{Curtis-Lake2023}. The underlying potential causes of quenching are listed as columns, including: environment, starbursts and supernovae, supermassive black hole regulated feedback, and gravitational physics (`mass'). The possible mechanisms of quenching are listed as rows, including: ejection, reduction in star forming efficiency, and starvation of gas supply. In the cartoon images, blue represents star forming gas, red represents gas stabilized against collapse, and yellow indicates thermalized/ hot gas. By way of examples, environmental quenching by gas removal can be achieved via ram-pressure stripping, and mass quenching by starvation may be caused by gravitational virial shock heating. This is intended as a fairly comprehensive list of possible quenching routes, although note that there are some more speculative ideas not represented here (e.g., magnetic field heating and heating from dark matter annihilation). All of these modes of quenching will be discussed in some detail in this review series.}\label{f4}
\end{figure}

In Fig.~\ref{f4}, we illustrate several possible quenching avenues, which are subdivided into `causes' (shown along columns) and `mechanisms' (shown along rows). For the purpose of this review, causes refer to the underlying physical origin of the quenching process, whereas mechanisms refer to the manner in which it is executed in practice. As such, the same cause can operate by various mechanisms, and the same mechanism can be achieved by various causes. It is only by isolating both the cause and the mechanism that one has established the quenching process in full.

It is clear that in order to quench a galaxy, either the gas needed as fuel for star formation must be removed (top row of Fig.~\ref{f4}) , or exhausted (bottom row of Fig.~\ref{f4}), or else the efficiency with which stars form from the gas reservoir must be reduced (middle row of Fig.~\ref{f4}). 

Removal of gas from the galaxy (particularly the ISM) can be caused in a number of ways and is commonly referred to as `ejection'. These modes typically act differently for central and satellite galaxies. The former is defined in this review as the most massive galaxy within a dark matter halo, with the latter referring to any other galaxy within the halo. For satellites, ram pressure stripping can lead to the ISM being extracted from galaxies (e.g., \citealt{Ebeling2014, Poggianti2017}). This is typically caused by high velocity relative motion through the intra-cluster medium (ICM, \citealt{vandenBosch2008, Peng2012, Goubert2024}). Alternatively, galactic winds driven by either SN or AGN feedback can lead to ejection of vast quantities of molecular, atomic, and ionized gas from the ISM (e.g., \citealt{Hopkins2006, Hopkins2008, Maiolino2012, Cicone2012, Cicone2014, Cicone2015, Fluetsch2019}). Clearly, these processes are not restricted to satellites, but may instead impact the whole galaxy population, including centrals.

The next possibility is that gas is simply exhausted in the ISM via star formation itself. In order to achieve this, replenishment of ISM gas from CGM cooling and cosmic cold gas streams must be eliminated. Additionally, mass return from stellar evolution must be stabilized from forming stars via continuous feedback (e.g., \citealt{Ciotti1991, Ciotti1997}). As such, this mode is typically referred to as `starvation'. There are many routes to starvation. For satellites, the same ram pressure stripping process (discussed above in relation to the ISM) may even more effectively remove the weakly bound CGM around the subhalo of the satellite, preventing its potential for replenishing the ISM through cooling. Additionally, as a satellite transitions from being a central in its own right to merely being in orbit of the central galaxy of the group or cluster (essentially the definition of a satellite), it is no longer efficiently fed by cold gas streams (e.g., \citealt{Dekel2014, Henriques2019}).

For central galaxies, the CGM may be heated by radio jets launched from the accretion discs around AGN (e.g., \citealt{Fabian1994, Fabian2006, McNamara2007, Fabian2012, HlavacekLarrondo2012, HlavacekLarrondo2015}). Additionally, the CGM may be heated through shocks from galactic scale winds, driven either by starbursts or AGN (e.g., \citealt{Nelson2018, Weinberger2017, Fluetsch2019, Fluetsch2021}). Alternatively, cold gas streams into high mass haloes likely shock upon entry, releasing energy into the system (e.g., \citealt{Dekel2006, Dekel2009, Woo2013}). All of these causes can lead to the mechanism of starvation-driven quenching, and clearly they all have a direct bearing upon the cluster problem (outlined in the previous sub-section).

Finally, it remains possible that gas is neither expelled nor exhausted from galaxies, but rather pools and stagnates within the ISM, unable to form stars due to some internal ISM physics. This is the reduction in efficiency mechanism. This may be achieved via heating or adding turbulence to the gas in the ISM from starbursts and/or AGN feedback (e.g., \citealt{Weinberger2017, Weinberger2018, Zinger2020, Piotrowska2022}). Alternatively, dynamical torques from a high central density (e.g., a massive galaxy bulge) may increase the stability of giant molecular clouds (GMCs, \citealt{Martig2009, Gensior2020}), preventing them collapsing to form stars.

It should be stressed at this point that these proposed quenching processes operate on very different scales to one another. In the case of starvation, typically the physics of relevance operates on halo-wide scales of 100\,kpc -- 1\,Mpc. Whereas, the ejective mode operates within each galaxy, typically on scales of 1 -- 10\,kpc. Alternatively, reduction in star forming efficiency is localized to $\leq$kpc scale regions within galaxies. Moreover, the entire backdrop to quenching is star formation, which is globally regulated by accretion onto haloes, which itself is regulated by the large-scale cosmology on $\sim$Gpc scales. At the other extreme, the physics of AGN, SN, and stars themselves are, of course, regulated on $\leq$pc scales. Hence, it is no exaggeration to say that quenching spans $\sim$10 orders of magnitude in physical scale. Consequently, the resolution of this problem must incorporate a highly diverse set of observations, simulation types, and underlying theoretical understanding. To do justice to this enormous complexity, we separate the review into two parts, concentrating on observations (Part I) and theory plus direct observational tests (Part~II).

\subsection{Observational galaxy surveys}

\begin{table*}
\centering
\caption{Major observational surveys relevant to galaxy quenching. All values are approximate and depend upon data release and quality control cuts.}\label{T1a}
\label{tab:surveys}
\scriptsize
\setlength{\tabcolsep}{4pt}
\renewcommand{\arraystretch}{1.15}

\begin{tabular}{lccccc}
\hline
Survey & Area & Sources & Redshift & Type & Ref. \\
 & [deg$^2$] &  & range &  &  \\
\hline

\\[-0.8ex]
\multicolumn{6}{l}{\textbf{Wide-field spectroscopic surveys}} \\
\\[-0.4ex]

SDSS & $\sim10\,000$ & $\sim10^6$ & $z\lesssim0.3$ & Spec. & \cite{York2000} \\
2dFGRS & $\sim1500$ & $\sim2.2\times10^5$ & $z\lesssim0.3$ & Spec. & \cite{Colless2001} \\
GAMA & $\sim286$ & $\sim2.4\times10^5$ & $z\lesssim0.5$ & Spec. & \cite{Driver2011} \\
zCOSMOS & $\sim1.7$ & $\sim2\times10^4$ & $0.1<z<1.2$ & Spec. & \cite{Lilly2007} \\
DEEP2 & $\sim3$ & $\sim3.8\times10^4$ & $0.7<z<1.4$ & Spec. & \cite{Newman2013} \\
VIPERS & $\sim23.5$ & $\sim9\times10^4$ & $0.5<z<1.2$ & Spec. & \cite{Guzzo2014} \\

\hline

\\[-0.8ex]
\multicolumn{6}{l}{\textbf{Deep imaging / photo-$z$ surveys}} \\
\\[-0.4ex]

CANDELS & $\sim0.22$ & $\sim2.5\times10^5$ & $1.5<z\lesssim6$ & HST img. & \cite{Koekemoer2011} \\
3D-HST & $\sim0.17$ & $\sim10^5$ & $0<z\lesssim4$ & Grism & \cite{Momcheva2016} \\
COSMOS2020 & $\sim2$ & $\sim1.7\times10^6$ & $0<z\lesssim6$ & Phot. & \cite{Weaver2022} \\
ZFOURGE & $\sim0.11$ & $\sim7\times10^4$ & $0<z\lesssim4$ & NIR phot. & \cite{Straatman2016} \\
UltraVISTA & $\sim1.5$ & $\sim2.6\times10^5$ & $0<z\lesssim4$ & NIR img. & \cite{Muzzin2013} \\

\hline

\\[-0.8ex]
\multicolumn{6}{l}{\textbf{Spatially resolved IFU surveys}} \\
\\[-0.4ex]

ATLAS$^{\rm 3D}$ & targeted & 260 & $z\lesssim0.01$ & IFU & \cite{Cappellari2011} \\
CALIFA & targeted & $\sim600$ & $0.005<z<0.03$ & IFU & \cite{Sanchez2012} \\
SAMI & targeted & $\sim3000$ & $0.004<z<0.095$ & IFU & \cite{Bryant2015} \\
MaNGA & targeted & $\sim10^4$ & $0.01<z<0.15$ & IFU & \cite{Bundy2015} \\

\hline

\\[-0.8ex]
\multicolumn{6}{l}{\textbf{Molecular gas surveys}} \\
\\[-0.4ex]

xCOLD GASS & targeted & 532 & $0.01<z<0.05$ & CO & \cite{Saintonge2017} \\
PHIBSS & targeted & $\sim10^2$ & $0.5<z<2.5$ & CO & \cite{Tacconi2013} \\
ALMaQUEST & targeted & 46 & $0.01<z<0.05$ & CO+IFU & \cite{Lin2020} \\
ASPECS & $1.3\times10^{-3}$ & tens & $0<z\lesssim4$ & CO/[CII] & \cite{Decarli2019} \\

\hline
\end{tabular}
\end{table*}

Large observational surveys have been central to establishing the empirical framework for galaxy quenching across cosmic time. Wide-field low-redshift spectroscopic surveys such as 2dF Galaxy Redshift Survey (\citealt{Colless2001}), Galaxy And Mass Assembly (\citealt{Driver2011, Baldry2018}), and Sloan Digital Sky Survey (\citealt{York2000, Abazajian2009}) enabled the first statistically robust characterization of the bimodality between star-forming and quiescent galaxies, and revealed strong correlations between quenching, stellar mass, environment, morphology, and central structure (e.g., \citealt{Baldry2004, Baldry2006, Kauffmann2003b, Peng2010}). We cover these topics in detail in Sect.~\ref{s4}

Extending these studies to earlier cosmic epochs, intermediate- and high-redshift surveys including COSMOS (\citealt{Scoville2007}), GOODS (\citealt{Giavalisco2004}), GOODS-NICMOS (\citealt{Conselice2011}), the DEEP2 Galaxy Redshift Survey (\citealt{Newman2013}), VIMOS Public Extragalactic Redshift Survey (\citealt{Guzzo2014}), zCOSMOS (\citealt{Lilly2007}), 3D-HST (\citealt{Brammer2012, Skelton2014}), COSMOS2020 (\citealt{Weaver2022}), UltraVISTA (\citealt{McCracken2012}), ZFOURGE (\citealt{Straatman2016}), and HST-CANDELS (\citealt{Grogin2011, Koekemoer2011}) traced the build-up of the quenched population out to $z \gtrsim 2 - 3$, demonstrating that massive galaxies quench rapidly and early, while lower-mass systems continue forming stars to later times (e.g., \citealt{Bell2004, Faber2007, Ilbert2013, Muzzin2013, Tomczak2014, Davidzon2017}). Collectively, these surveys established the evolving stellar mass function, the cosmic star formation history, and the environmental dependence of quenching over most of cosmic history. We cover these topics in detail in Sects.~\ref{s3} \& \ref{s5}.

More recently, spatially resolved IFU surveys such as ATLAS3D (\citealt{Cappellari2011}), CALIFA (\citealt{Sanchez2012}), SAMI (\citealt{Bryant2015}), and MaNGA (\citealt{Bundy2015}) transformed the field by enabling kpc-scale measurements of stellar populations, ionized gas, metallicities, and internal kinematics across large galaxy samples. These surveys revealed that quenching often proceeds inside-out in massive galaxies, linked closely to bulge growth, rising central velocity dispersions in the center of galaxies, and declining central gas fractions and star forming efficiencies (e.g., \citealt{Belfiore2017, Schaefer2017, Ellison2018, Medling2018, Bluck2020b}). These topics are covered in Sect.~\ref{s6}.

Complementing this, molecular gas surveys such as  COLD GASS (\citealt{Saintonge2016, Saintonge2017}), Phibbs (\citealt{Tacconi2013, Tacconi2018}), ALMaQUEST (\citealt{Lin2019, Ellison2020c}), and ALMA Spectroscopic Surveys in the Hubble Ultra Deep Field (e.g., \citealt{Walter2016, Decarli2019}) connected quenching directly to the depletion, stabilization, or removal of cold gas reservoirs over cosmic time. We cover these topics in Sect~\ref{s7}. 

Together, these complementary survey programs now provide a multi-dimensional observational framework for quenching studies, spanning global demographics, spatially resolved internal structure, gas content, and cosmic evolution. In Table~\ref{T1a} we provide a guide the the most important galaxy surveys for the field of quenching as a reference.

\subsection{Outline of Part I}\label{s15}

Throughout this review we assume a spacially flat $\Lambda$CDM cosmology and look for solutions to the outstanding issues with this theory as a model of galaxy formation and evolution through baryonic processes. In Part~I, where we focus on observations, this only impacts cosmography (i.e., the determination of distances, times, luminosities and so on). However, in Part~II, this will limit our discussion to frameworks for accounting for galactic star formation and quenching within the dominant cosmological paradigm.

In Sect.~\ref{s2}, we discuss methods to identify quenched galaxies throughout cosmic time, including via the star forming main sequence (SFMS), galaxy colors, and spectral indices. In Sect.~\ref{s3}, we present foundational observational results of relevance to galaxy quenching, particularly focusing on the evolution in the star formation rate density (SFRD) and the relationship between halo mass and stellar mass functions.

In Sect.~\ref{s4}, we consider a host of intrinsic observational correlators to quenching, including morphology, stellar mass, central density, and kinematics. In Sect.~\ref{s5}, we go on to  consider a host of environmental correlators to quenching, including local galaxy density, halo mass, and the location of satellite galaxies within their parent groups and clusters. 

In Sect.~\ref{s6}, we review results on quenching within galaxies on spatially resolved scales. Here we assess how quenching operates (i.e., `inside-out' vs. `outside-in'), whether galaxies quench as a whole or in part, and whether the global system dictates the local properties within galaxies, or vice versa.

In Sect.~\ref{s7a}, we consider the gas physics of quenching. In particular, we establish that the SFMS arises out of more fundamental gas scaling laws. We then investigate how these scaling laws evolve over cosmic time. Finally, we assess how quenching operates -- via reduction in {\it both} gas fraction and star forming efficiency -- on both global and resolved scales.

In Sect.~\ref{s7}, we review the exciting new evidence for quenched galaxies in the very early Universe with JWST observations, including assessing several ways to resolve the current apparent tensions with theory. Finally, in Sect.~\ref{s8} we summarize this part of the review series.

%%%%%%%%%%%%%%%%%%
%                %
%  Main Sequence %
%                %
%%%%%%%%%%%%%%%%%%

\section{Identifying quenched galaxies}\label{s2}

In this section we explore some of the most popular observational methods for identifying quenched galaxies and separating them from their actively star forming counterparts. First, we give a brief overview of the methods used in the literature to infer star formation rates (SFRs) and stellar masses ($M_*$), which are crucial for studying galaxy quenching. We go on to discuss the star forming main sequence (SFMS), offsets from the SFMS, specific star formation rates (${\rm sSFR} \equiv {\rm SFR}/M_*$), galaxy color-based approaches, the ages of stellar populations, and spectral index-based methods.

\subsection{Measuring Star Formation rates \& stellar masses}

\subsubsection{Star formation rates}\label{s21a}

There are a variety of methods commonly used in the literature to infer the star formation rates of galaxies. We summarize several common calibrations in Table~\ref{t2}. 

The first method is via UV luminosity (e.g., \citealt{Kennicutt1998}). This method traces the stellar continuum contributed by O- and B-type stars, which peaks in the UV. No other population of stars contribute significantly in these wavebands. Due to the very high masses of these stars, their lifetime is short (typically $\sim$10 Myr for O-type stars and $\sim$100 Myr for B-type stars; e.g., \citealt{Bruzual2003, Maraston2005}). This enables a calibration between star formation rate and UV luminosity to be developed via stellar population synthesis models (as discussed in \citealt{Kennicutt1998}). In the first row of Table~\ref{t2} the SFR - $L_\nu({\rm UV})$ calibration from \cite{Kennicutt1998} is shown, where the luminosity per unit frequency may be given anywhere in the region 1500 -- 2800\,\AA. This represents a direct observational constraint on current star formation, in the ideal case of no dust attenuation. One should also be aware that in the case of AGN, contribution in the UV is expected (unless this can be adequately resolved and modeled appropriately).

Unfortunately, UV photons are extremely susceptible to extinction via dust absorption, which is anticipated to impact essentially all star forming galaxies (although this is most severe for discs viewed edge-on). To combat this, a correction must be employed. A common approach to account for extinction is to use the UV slope, $\beta$ (see, e.g., \citealt{Kennicutt1998, Meurer1999, Hao2011}).

% TABLE 1: SFRs

\begin{table}[t]
\centering
\caption{A summary of methods to infer star formation rates}

\begin{tabular}{llcl}
\toprule
\textbf{Tracer} & \textbf{SFR Equation} & \textbf{Timescale} & \textbf{Notes} \\\\
\midrule
UV & $\text{SFR}_{\mathrm{UV}} = 1.4 \times 10^{-28} \, L_\nu(\mathrm{UV})$ & $\sim$100 Myr & Sensitive to extinction/ AGN \\\\
%%\hline
H$\alpha$ & $\text{SFR}_{\mathrm{H}\alpha} = 7.9 \times 10^{-42} \, L(\mathrm{H}\alpha)$ & $\sim$10 Myr & Requires (H$\alpha$ / H$\beta$) corr. \\\\
%%\hline
FIR & $\text{SFR}_{\mathrm{IR}} = 4.5 \times 10^{-44} \, L({\rm IR})$ & $\sim$100 Myr & Traces obscured SFR only \\\\
%%\hline
UV + FIR & $\text{SFR}_{\mathrm{UV+FIR}} = \text{SFR}_{\mathrm{UV,obs}} + \text{SFR}_{\mathrm{FIR}}$ & $\sim$100 Myr & Complete SFR \\\\
%%\hline
MIR (24 $\mu m$) & $\text{SFR}_{24 \mu m} = 1.27 \times 10^{-38} \, \big( L_\nu(24 \, \mu m\,)\big)^{0.885}$ & $\sim$100 Myr & Probes warm dust \\\\
%%\hline
1.4 GHz Radio & $\text{SFR}_{1.4\,\mathrm{GHz}} = 6.35 \times 10^{-29} \, L_\nu(1.4\,\mathrm{GHz})$ & $\sim$100 Myr & Unaffected by dust \\\\
%%\hline
$D_n4000$ Index & $\text{SFR}_{D_n4000} = M_* \times \text{sSFR}(D_n4000)$ & $100+ \, {\rm Myrs}$  & Empirical calibration \\\\
%%\hline
SED Fitting & $M_* = (1-R) \int \text{SFR}(t) \, dt $ \,\, (inversion) & Variable & See Sect.~\ref{s21b} \\\\
\bottomrule
\end{tabular}
\vspace{0.2cm}
{\small {\bf Notes:} Luminosities (indicated by $L$) are given in units of [erg/s] and specific luminosities (indicated by $L_{\nu})$ are given in units of [erg/s/Hz]. All star formation rates are given in units of [$M_{\odot}$/yr]. All equations assume a Salpeter IMF. Calibrations in the first three rows are reproduced from \cite{Kennicutt1998}. The combined UV + IR calibration is reproduced from \cite{Hao2011}, the 24\,$\mu m$ calibration is reproduced from \cite{Calzetti2007}, and the radio frequency tracer is reproduced from \cite{Murphy2011}. Finally, for the $D4000$ index see \cite{Brinchmann2004}, and for SED fitting see Sect.~\ref{s21b} (and, e.g., \citealt{Conroy2013}).}\label{t2}

\end{table}

Alternatively, star formation rates may be inferred from the luminosity of the H$\alpha$ emission line (i.e., $n = 3 \rightarrow 2$ atomic Hydrogen energy level transition; see second row in Table~\ref{t2}; \citealt{Kennicutt1998}). The essential idea here is that O-type stars ionize a region of the ISM gas around the star, from which the H$\alpha$ photons are produced via recombination in equilibrium (see \citealt{Kennicutt1998}). Only O-type stars emit at blue enough wavelengths to fully ionize Hydrogen. Consequently, this approach offers a shorter time window SFR constraint to other methods ($\sim$10\,Myr, i.e. the lifetime of O-type stars). However, AGN may also cause ionization (and, hence, recombination lines) as well. Therefore, this technique is not suitable for constraining star formation in galaxies with optical AGN, unless substantial additional modeling is performed. As with UV (though to a lesser degree), H$\alpha$ photons are also subject to dust extinction. The most common way to correct for this is to measure the Balmer decrement (H$\alpha$/H$\beta$ ratio) and compare to the theoretical expectation from Case B recombination ($\chi_\mathrm{rec}~=~2.86$), by assuming an extinction law (see, e.g., \citealt{Cardelli1989, Kennicutt1998, Calzetti2000, Brinchmann2004}).

Star formation rates may also be traced in the infrared (IR) via various calibrations (see third row in Table~\ref{t2}; \citealt{Kennicutt1998, Murphy2011}). This method traces re-emission from dust absorption at longer wavelengths. The IR flux is integrated from mid-to-far IR, i.e., 8 - $1000\,\mu m$ (\citealt{Kennicutt1998}). It is crucial to appreciate that this method is only sensitive to the obscured star formation rate. As such, this method is usually combined with the UV method (discussed above), without $\beta$-slope correction. This provides one method to get the full obscured + unobscured SFR (see fourth row in Table~\ref{t2}). Additionally, the full IR method may be simplified with an MIR calibration at $24\,\mu m$ (see fifth row in Table~\ref{t2}; e.g., \citealt{Calzetti2007, Rieke2009}). 

There also exist SFR calibrations with radio luminosity, particularly at 1.4 GHz (see sixth row in Table~\ref{t2}; e.g., \citealt{Condon1992, Murphy2011}). The physical rationale behind this relationship is that the radio traces synchrotron radiation from stellar remnants and, hence, is linked to the supernovae rate (and thus, star formation rate). This is usually calibrated against other (more direct) tracers. Again caution must be used when applying to systems with an AGN, which can also contribute flux in the radio.

SFRs may also be estimated via an empirical relationship between the strength of the 4000\,\AA \, break and the specific SFR (sSFR; see penultimate row in Table~\ref{t2}; e.g., \citealt{Brinchmann2004, Bluck2020a}). We discuss this spectral index further in the context of identifying quenched galaxies in Sect.~\ref{s22}. Finally, one may also constrain SFRs through spectral energy distribution (SED) fitting (see final row in Table~\ref{t2}; e.g., \citealt{Conroy2013}), which we discuss in detail in the next part of this sub-section. Although there are other methods to infer SFRs in observational data, the ones provided above are by far the most common.

\subsubsection{Stellar masses from SED fitting}\label{s21b}

Spectral energy distribution (SED) fitting is one of the principal methods used to infer the stellar properties of galaxies from their observed continuum spectra or multi-band photometry (see, e.g., \citealt{Conroy2013, Carnall2019, Leja2019}). The underlying idea is that the spectral shape of a galaxy from the near-UV to near-IR encodes information on its stellar populations, including the stellar mass ($M_*$), stellar population age, star formation history (SFH), metallicity, and level of dust attenuation. In cases where spectroscopy is unavailable, SED fitting may also provide a photometric redshift estimate ($z_{\rm phot}$). These quantities are critical for the study of galaxy evolution and quenching in particular.

The fundamental building blocks of stellar population synthesis (SPS) models are simple stellar populations (SSPs; e.g., \citealt{Bruzual2003, Maraston2005, Conroy2013}). An SSP represents a coeval population of stars with a single age and metallicity. An SSP model is constructed by combining an assumed stellar initial mass function (IMF: $\Phi_{\rm IMF}$) with stellar spectra and stellar evolutionary tracks. Formally, the luminosity of an SSP may be written as:

\begin{equation}
L_\mathrm{SSP}(\lambda, t',Z) = \int \Phi_\mathrm{IMF}(M_\star)\, L_{\star,\lambda}(M_\star , t', Z) \, dM_\star ,
\end{equation}

\noindent where $t'$ is the age of the stellar population, $Z$ is its metallicity, and $L_{\star , \lambda}$ is the luminosity of a given star at a given wavelength (which is dependent upon its mass, age, and metallicity). Note that the integral is performed over all masses of stars within the stellar population. Hence, the SSP forms a grid of model spectra spanning different ages and metallicities, which are fundamentally dependent upon an assumed IMF (see, e.g., \citealt{Salpeter1955, Larson1998, Kroupa2001, Chabrier2003}).

Real galaxies are not well described by an SSP because star formation may occur over extended cosmic time periods. Consequently, realistic galaxy spectra are modeled as composite stellar populations (CSPs), formed by summing many SSPs weighted by the star formation history (SFH). In addition, CSP models typically include metallicity evolution and dust attenuation. A general expression for a CSP may be given by:

\begin{equation}
L_\mathrm{CSP}(\lambda)\bigg|_{t_z} = \int_0^{t_z} {\rm SFH}(t_z-t')\, L_\mathrm{SSP}(\lambda,t',Z)\, e^{-\tau_d(\lambda)} \,dt' ,
\end{equation}

\noindent where ${\rm SFH}(t)$ is the star formation history (i.e., the SFR normalized by the final stellar mass of the system), $\tau_d$ is the dust optical depth, and $t_z$ is the age of the Universe at the observed redshift. More sophisticated models may additionally include metallicity distributions and dust re-emission in the far-IR (e.g., \citealt{DaCunha2008, Conroy2013, Boquien2019}).

A key ingredient in CSP modeling is the assumed star formation history. Parametric SFHs are widely used, including exponentially declining ($\tau$) models, delayed-$\tau$ models, and double power-law forms (e.g., \citealt{Lee2010, Carnall2019}). Respectively, these are given by:

\begin{equation}
{\rm SFH}(t) \propto e^{-t/\tau}, \quad t\,e^{-t/\tau}, \quad {\rm or} \quad
\bigg[ \bigg(\frac{t}{\tau}\bigg)^\alpha + \bigg(\frac{t}{\tau}\bigg)^{-\beta} \bigg]^{-1}.
\end{equation}

\noindent Increasingly, non-parametric SFHs are also used, in which the SFH is reconstructed from a series of constant star formation rates with discrete time bins (e.g., \citealt{Leja2019}). These approaches provide substantially greater flexibility, especially for galaxies with bursty or complex assembly histories.

To compare CSP models to observations, the model luminosity spectrum must be converted into an observed flux spectrum via:

\begin{equation}
f_\mathrm{CSP}(\lambda,z) = \bigg( \frac{1}{1+z} \bigg) \bigg( \frac{L_\mathrm{CSP}(\lambda/(1+z))} {4\pi D_L^2(z)} \bigg),
\end{equation}

\noindent where $D_L(z)$ is the luminosity distance. The extra factor of $(1+z)^{-1}$ accounts for cosmological wavelength stretching.

Traditionally, SED fitting has been performed via chi-square minimization (e.g., FAST, EAZY, HyperZ; see, e.g., \citealt{Kriek2009, Brammer2008, Bolzonella2000}), in which the optimal stellar mass and CSP parameters are determined by minimizing:

\begin{equation}
\chi^2 = \sum_\lambda \bigg[ \frac{f_\mathrm{obs}(\lambda) - M_* \, f_\mathrm{CSP}(\lambda)} {\sigma_\lambda} \bigg]^2 ,
\end{equation}

\noindent where $\sigma_\lambda$ is the uncertainty on the observed flux at wavelength, $\lambda$, and $M_*$ is the stellar mass. The latter is determined for each possible model CSP as the maximum likelihood value. The final predicted stellar mass is taken as the maximum likelihood value of the best fit CSP model.

More recently, Bayesian methods have become increasingly popular (e.g., PROSPECTOR, BAGPIPES, BEAGLE; see, e.g., \citealt{Chevallard2016, Leja2017, Carnall2018}). In this framework, one seeks the posterior probability distribution of the model parameters:

\begin{equation}
P({\bf \Theta}|{\rm data}) \propto \mathcal{L}({\rm data}|{\bf \Theta})\, P({\bf \Theta}),
\end{equation}

\noindent where $\mathcal{L}$ is the likelihood function and $P({\bf \Theta})$ is the prior distribution on the model parameters. In the above equation, ${\bf \Theta}$ indicates the set of all model parameters to be fit. Bayesian approaches provide a more robust treatment of parameter degeneracies and uncertainties, especially in the age--dust--metallicity parameter space, albeit at substantially greater computational expense.

Overall, SED fitting provides one of the most powerful methods for deriving stellar masses, star formation histories, and related stellar population properties for galaxies across cosmic time. However, the inferred quantities remain dependent on assumptions regarding the IMF, SFH parameterization, dust attenuation law, metallicity evolution, and the stellar population synthesis model employed.

\subsection{The star forming main sequence (SFMS)}\label{s21}

\begin{figure}[ht]  
\centering
\includegraphics[width=0.9\textwidth]{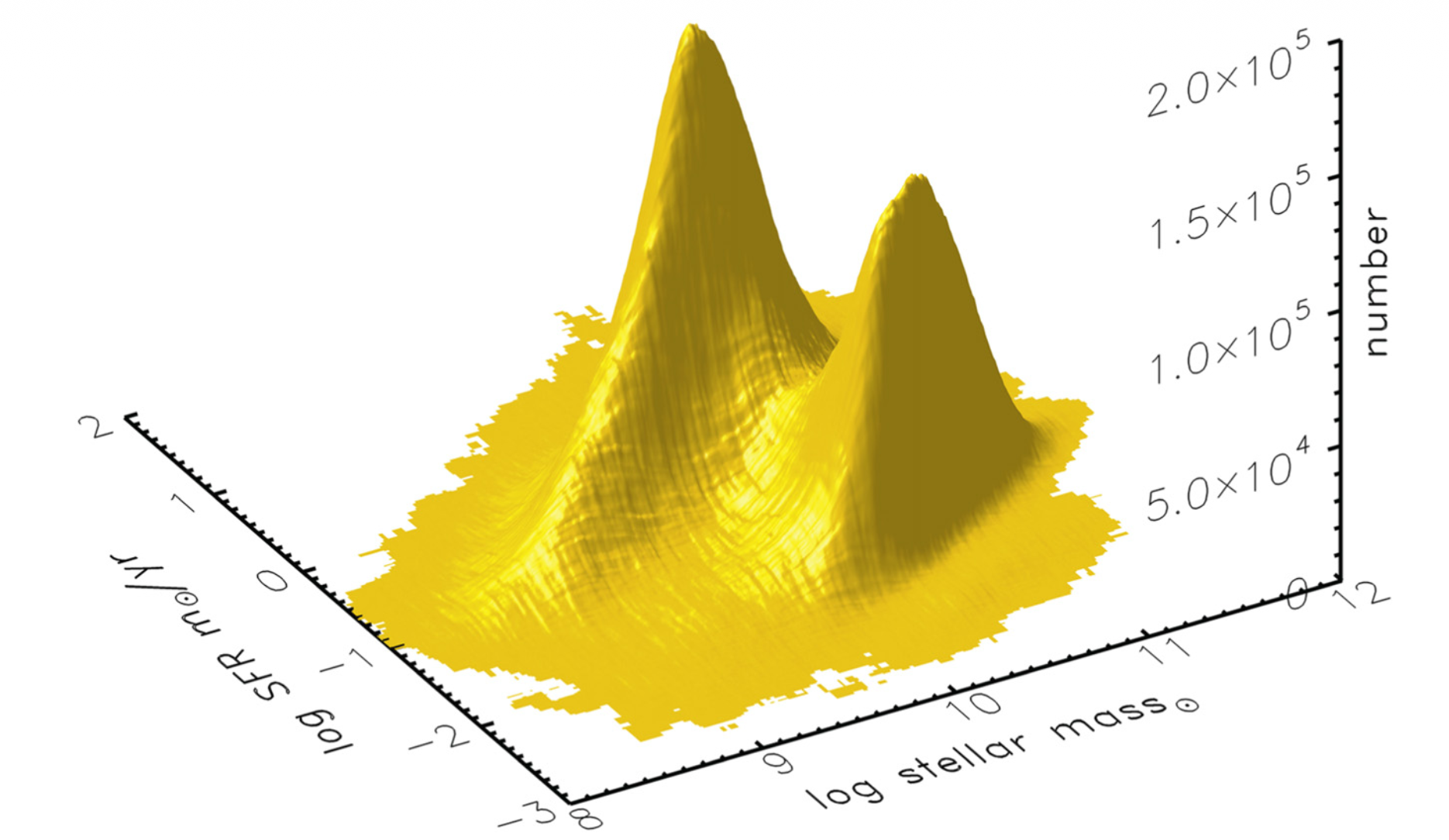}
\caption{A 3D rendering of the star forming main sequence (SFMS), i.e. the SFR - $M_*$ relationship for local ($z = 0.02 - 0.2$) SDSS galaxies, reproduced from \cite{Renzini2015}. The height of this figure indicates number of galaxies, with the 2D horizontal projection indicating the SFMS. It is clear that galaxies in the local Universe separate out into two distinct peaks in this diagram. The high-SFR peak represents actively star forming systems, with the low-SFR peak representing quenched galaxies.}\label{f5a}
\end{figure}

Having discussed methods to infer both star formation rates and stellar masses, we are now in a position to introduce the star forming main sequence (SFMS). The SFMS is a tight ($\sigma \sim 0.2-0.3\,$dex) relationship between star formation rate (SFR) and stellar mass ($M_*$) for actively star forming galaxies, which was first discussed in \cite{Brinchmann2004, Daddi2007, Noeske2007a}. 

In Fig.~\ref{f5a}, we present an example of a low-$z$ rendering of the SFMS for $\sim$500k galaxies observed from the SDSS DR7 (\citealt{York2000, Abazajian2009}), reproduced from \cite{Renzini2015}. Galaxies separate out into two density peaks (as clearly visible in this 3D rendering). The high-SFR peak represents star forming galaxies and the low-SFR peak represents quiescent (or quenched) galaxies.

This SFR\,--\,$M_*$ plane is often described as exhibiting significant bimodality. However, this is a contentious point because the SFRs of the lower peak are frequently just upper-limits. In reality they are expected to form a continuum to much lower SFRs than presented. Nonetheless, one thing is certain, that these `quenched' systems are forming stars at much lower rates than their actively star forming counterparts. 

The SFMS is often parameterized as a power-law in log-space, as follows:

\begin{equation}
\log_{10} \bigg( \text{SFR}_{\mathrm{SFMS}} \, \big[M_{\odot}/\mathrm{yr}\big] \bigg) = \alpha \times \log_{10}\bigg( M_* \, \big[M_{\odot}\big] \bigg) - \beta \, .
\end{equation}

\noindent  In \cite{Renzini2015}, the optimal fit for $z \sim 0$ galaxies was found to be: $\alpha = (0.76\pm0.01)$ and $\beta = (7.64\pm0.02)$. This leads to a number of methods to classify star forming and quenched systems. Two commonly used methods are illustrated below. First, via an offset from the SFMS ($\Delta$SFR; see, e.g., \citealt{Woo2013, Bluck2014, Ellison2018}):

\begin{equation}
\Delta \text{SFR} \equiv \log_{10}\bigg( \text{SFR}_i\bigg) -  \log_{10} \bigg( \text{SFR}_{\mathrm{SFMS}}(M_{*, \,i}) \bigg) < \text{lim.}
\end{equation}

\noindent and, second, via the specific SFR (sSFR; see, e.g., \citealt{Ilbert2010, Wuyts2011, Whitaker2012, Whitaker2014, Piotrowska2022}):

\begin{equation}
\log_{10} \bigg(s\text{SFR}_i \bigg) \equiv \log_{10} \bigg( \frac{\text{SFR}}{M_*} \bigg)  < \text{lim.}
\end{equation}

\noindent Note that these definitions are identical if the gradient of the SFMS is equal to unity. However, in practice, the gradient of the SFMS is found to be slightly sub-linear. Nonetheless, both methods have proved useful in segregating galaxy populations. At $z \sim 0$, a common choice for the limit in $\Delta \text{SFR}$ is -1\,dex (e.g., \citealt{Bluck2016, Teimoorinia2016}) and a common choice for the limit in sSFR is $10^{-11}\, \mathrm{[yr^{-1}]}$ (e.g., \citealt{Whitaker2012, Piotrowska2022}).

There is an important subtlety concerning the SFMS which is crucial to understand. For star forming systems, there is a strong positive relationship between SFR and stellar mass, such that more massive systems are forming stars at higher rates than their less massive counterparts. However, the fraction of galaxies which reside on the SFMS reduces as a function of increasing stellar mass, such that the fraction of quenched (non-star forming) galaxies also rises. At first exposure, this may appear almost oxymoronic: \textit{How can SFR both increase and decrease with stellar mass?} The answer, of course, is that it cannot. If one traces the median SFR as a function of stellar mass, one finds that SFR rises with stellar mass out to $M_* \sim 10^{10.5} \,M_{\odot}$, and then decreases rapidly, plateauing thereafter (e.g., \citealt{Baldry2006, Peng2010}). This suggests a strong correlation between quenching and stellar mass (which we discuss in Sect.~\ref{s42}). However, the SFMS does persist at high masses, it is just increasingly depleted of galaxies via quenching. 

Practically, there are three ways to quantify the SFMS in the high-mass regime, where it is no longer dominant: (i)~select `star forming' galaxies via another method (e.g., on the basis of emission lines, colors, or stellar population ages; see, e.g., \citealt{Brinchmann2004}); (ii)~extrapolate the low-mass SFMS to higher masses (to view what would be expected if quenching didn't occur; e.g., \citealt{Bluck2016}); or (iii)~weight the SFMS by star formation rate, to find the ridge-line (as in \citealt{Renzini2015}). Fortunately, all of these approaches tend to yield similar, though not identical, SFMS relations. Quenching is often \textit{defined} as the permanent (or long-term) departure of a galaxy from the SFMS, so it matters that we know what the SFMS is at all masses (and, indeed, all times; see next sub-section). However, for completeness, we note that the observational literature presents with many alternative definitions of quenching, including via optical/ NIR/ UV colors, lack of emission lines, and the strength of the Balmer break (all of which will be discussed further below; e.g., \citealt{Strateva2001, Baldry2006, Williams2009, Cameron2009b, Peng2010, Sanchez2019b}).

Some of the literature argues that the high-mass end of the SFMS exhibits curvature (e.g., \citealt{Whitaker2014, Lee2015, Tomczak2016, Popesso2019}), whilst other publications argue that this is solely a result of quiescent bulges forming within normal star forming discs (e.g., \citealt{Abramson2014, Guo2015, Leslie2020}). Ultimately, this seems to come down to whether or not one should remove intermediate star forming systems in the `green valley' before fitting the SFMS. Resolving this debate will not be critical for what follows in this review. With or without high-mass curvature, quenched galaxies tend to be very well separated from star forming systems on the SFMS plane.

\subsubsection{Evolution in the SFMS}

\begin{figure}[ht] 
\centering
\includegraphics[width=1.0\textwidth]{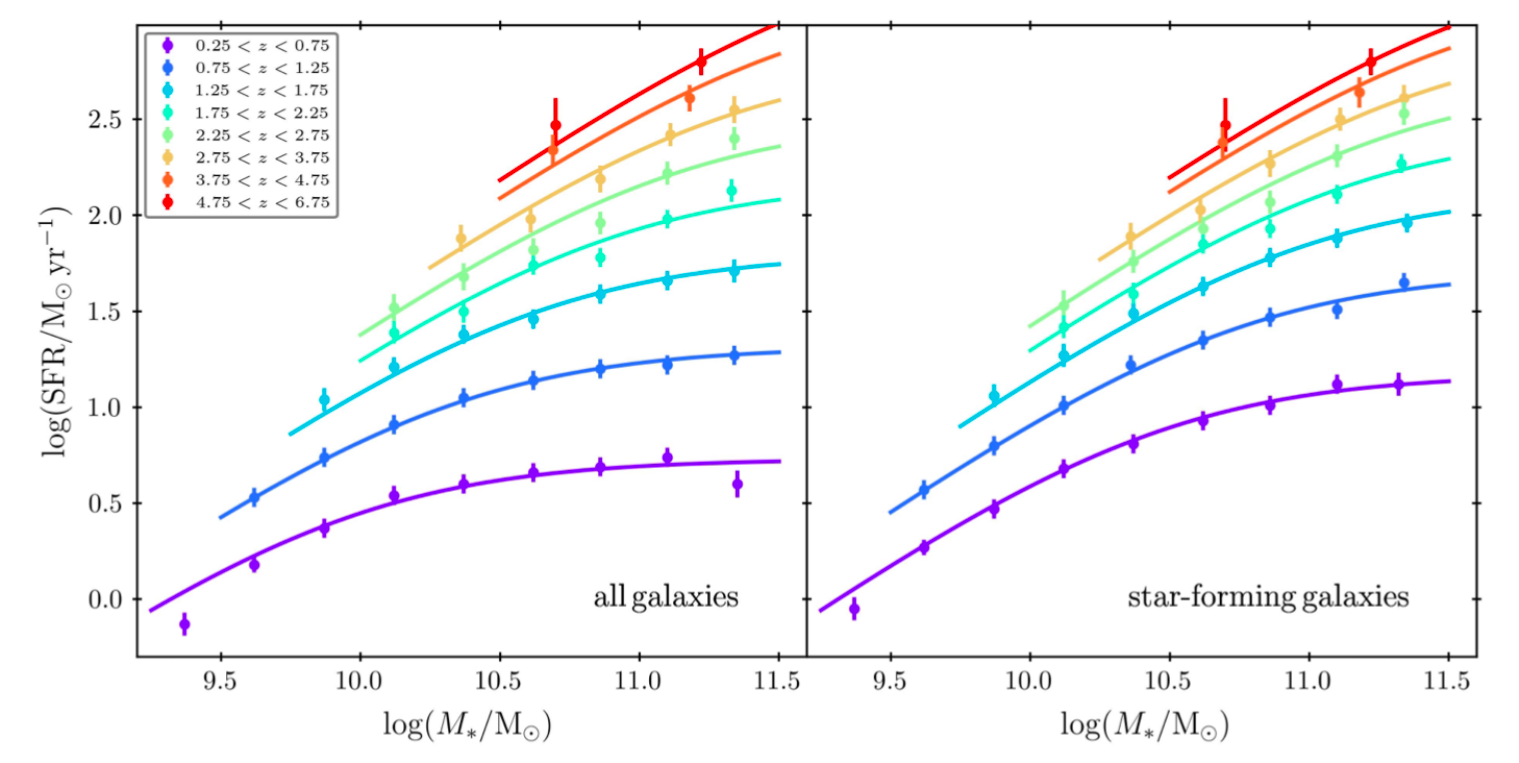}
\caption{Evolution in the star forming main sequence (SFR\,--\,$M_*$ relationship) for all galaxies (left panel) and actively star forming systems (right panel). Note that the typical SFR of galaxies at a fixed stellar mass rises as a strong function of redshift for both samples. This clearly indicates that, even in lieu of quenching, the rate of star formation is expected to reduce in galaxies at a fixed stellar mass as a function of time. Ultimately, this is a consequence of reducing accretion rates into haloes with cosmic time, which is regulated primarily via cosmic expansion. Quenching leads to even stronger evolution in these relationships at high masses and late cosmic times (compare the full galaxy sample to the star forming sample). This figure is reproduced from \cite{Koprowski2024}. }\label{f6}
\end{figure}

Many studies have found that the global SFMS exists up to high-redshifts (currently at least $z \sim 6$; see \citealt{Noeske2007a, Noeske2007b, Daddi2007, Elbaz2007, Speagle2014, Whitaker2012, Whitaker2014, Tomczak2016, Salmon2015, Koprowski2024}). 

In Fig.~\ref{f6} we show the evolution in the SFR\,--\,$M_*$ relationship for all galaxies (left-panel) and actively star forming galaxies (right panel), reproduced from \cite{Koprowski2024}. It is immediately clear that the SFR of a galaxy at a fixed stellar mass rises significantly (and monotonically) with increasing redshift. This implies that star formation was more prevalent at earlier cosmic times, for constant mass galaxies (which we discuss in more detail in Sect.~\ref{s3}). 

Consequently, the `universal' SFMS is often parameterized as a function of both stellar mass and redshift. Whilst there are an enormous array of fitting schemes (e.g., \citealt{Whitaker2014, Speagle2014, Salmon2015, Schreiber2015, Tomczak2016, Popesso2019, Koprowski2024}), one simple version (which makes contact with theory and achieves a reasonably good fit) is from \cite{Popesso2019}. Their parameterization is:

\begin{equation}
\bigg(\text{SFR} \,\mathrm{[M_\odot / yr]} \bigg)  = A \, \bigg(M_* \,[M_{\odot}] \bigg)^\gamma \cdot (1+z)^\delta 
\end{equation}

\noindent where, $A, \, \gamma, \,\, \text{and} \,\, \delta$ all represent positive constants to be fit to observations. In \cite{Popesso2019}, $\delta = 3.2\pm0.2$, with $\gamma$ broadly consistent with $z \sim 0$ works (see above). Although likely an over-simplification (since the gradient may evolve), this is a useful starting point for thinking about evolution in the SFMS. Alternatively, \cite{Tasca2015} find that the redshift dependence itself changes significantly with cosmic time, such that $\delta \sim 2.8$ at $z < 2.4$, and $\delta \sim 1.2$ at higher redshifts. However, note that this is computed for the evolving sSFR (rather than the SFMS directly). It is sufficient for now to appreciate that there remains much disagreement in the literature on how best to parameterize evolution in the SFMS (e.g., as a function of time vs. redshift, with or without an evolving SFMS gradient, accounting for high mass curvature or not, and fitting all epochs together vs. employing a multi-parameterization changing around $z \sim 1-2$). Nonetheless, that there is marked, monotonic evolution to higher SFMS normalizations at higher redshifts (and hence higher sSFRs) is very well established across the bulk of cosmic history.

As a result of the evolving SFMS, two important problems arise from the point of view of galaxy quenching. First, in order to identify a quenched galaxy, one must account for both its stellar mass and redshift, because what is `normal' for star forming systems varies as a strong function of both of these parameters. At $z > 0.5$, the evolution in sSFR appears to scale with the inverse of the Hubble time\footnote{But note that at later cosmic times this fails dramatically!}. As such, many authors utilize a quenching threshold of sSFR$ < \zeta/t_H(z)$, where $\zeta$ is a variable scaling factor, often in the range $\{0.1 - 1\}$ (see, e.g., \citealt{Franx2008, Williams2009, Tacchella2016, Pacifici2016, Tasca2015}). Second, at a deeper level, one must seek to understand the physical origin of the SFMS, and its evolution over cosmic time, in order to fully understand why some galaxies cease to form stars at the `expected' rate (which is, evidently, set by the evolving SFMS).

By reflecting on the evolution in the SFMS, an important question arises --- \emph{should the dramatic reduction in star formation as a function of cosmic time be considered as `quenching'?} We believe that the answer is, no. To understand this position, it is helpful to view again Fig.~\ref{f6}. Focusing first on the right-hand panel, there is marked evolution for \emph{star forming} galaxies, which few would argue to be quenched. These are just normal galaxies forming stars at a rate which is completely typical, given their mass and epoch. The only way around this would be to argue that all galaxies in the local Universe are quenched, which essentially no one does as this would only serve to limit our vocabulary in discussing systems which are forming stars much lower than expected (which are very numerous at low-$z$). To see this, viewing next the left-hand panel of Fig.~\ref{f6} (which shows all observed galaxies, regardless of their star forming state), one sees a similar evolution at low masses, but an increase in the shift to lower SFRs at high masses (relative to star forming systems). It is this change (clearly evident at a fixed epoch) which gets to the core of what quenching is really about. This cannot be explained as a function of accretion-limited star formation and, hence, requires a deeper explanation (likely involving feedback; see Part II of this review).

In principal, one could define quenching in an absolute sense, for example via a threshold in specific star formation rate (sSFR), or a reduction of some percentage in SFR, independent of epoch. However, such a definition would not account for the strong redshift evolution in the normalization of the star-forming main sequence, or the mass dependence thereof. Hence, in this work we adopt a relative definition, identifying quenched galaxies as systems which have SFRs significantly offset below the SFMS at a given mass and epoch (as is very common in the literature; see, e.g., \citealt{Barro2013, Lang2014, Tacchella2022, Bluck2024}). This approach more naturally isolates systems whose suppression of star formation cannot be explained by the global, accretion-driven decline in star formation activity.

\subsection{Galaxy colors, stellar ages and spectral indices}\label{s22}

In addition to leveraging measurements of star formation rates directly, there are a number of other important techniques used to identify quiescent galaxies. We briefly review some these approaches in this sub-section. These techniques are especially useful for photometric data, or for populations which either do not exhibit detectable emission lines, or else are contaminated by AGN. However, we note that technically these approaches tend to identify galaxies with old stellar populations, not necessarily wholly absent of ongoing star formation. Practically, however, these two classes (quiescent and old) tend to most often be at least similar.

First, there are many approaches which leverage bimodality in galaxy colors (e.g., \citealt{Strateva2001, Bell2004, Baldry2004, Williams2009, Peng2010, Whitaker2011}). Broadly speaking, color diagnostics come in four principal types: (i)~single color cuts; (ii)~color\,--\,stellar mass relations (like Fig.~\ref{f3}); (iii)~color\,--\,magnitude relations; and (iv)~color\,--\,color relations. Additionally, there are a host of different colors used, incorporating wavebands from UV to near-IR. 

Essentially all color based approaches for identifying quenched galaxies leverage correlations between observed galaxy colors and the luminosity-weighted age of stellar populations within galaxies, which in turn constrains the current star formation rate. The existence of a color--age connection is very well established (see, e.g., \citealt{Worthey1994, Kodama1997, Bruzual2003, Maraston2005, MacArthur2004, Gallazzi2005}). Essentially, this arises because of the color\,--\,temperature\,--\,mass\,--\,lifetime relationships in stars (e.g., \citealt{Bruzual2003, Maraston2005}). However, observed colors are also sensitive to dust extinction, viewing angle, contamination from AGN, and uncertainties in k-corrections (which can be hard to model directly). As a result, rest-frame, dust-corrected colors are most often used from models via SED fitting to observational data. Additionally, bi-color diagnostics may be used to further improve the resilience to these effects.

\begin{figure}[ht] 
\centering
\includegraphics[width=0.95\textwidth]{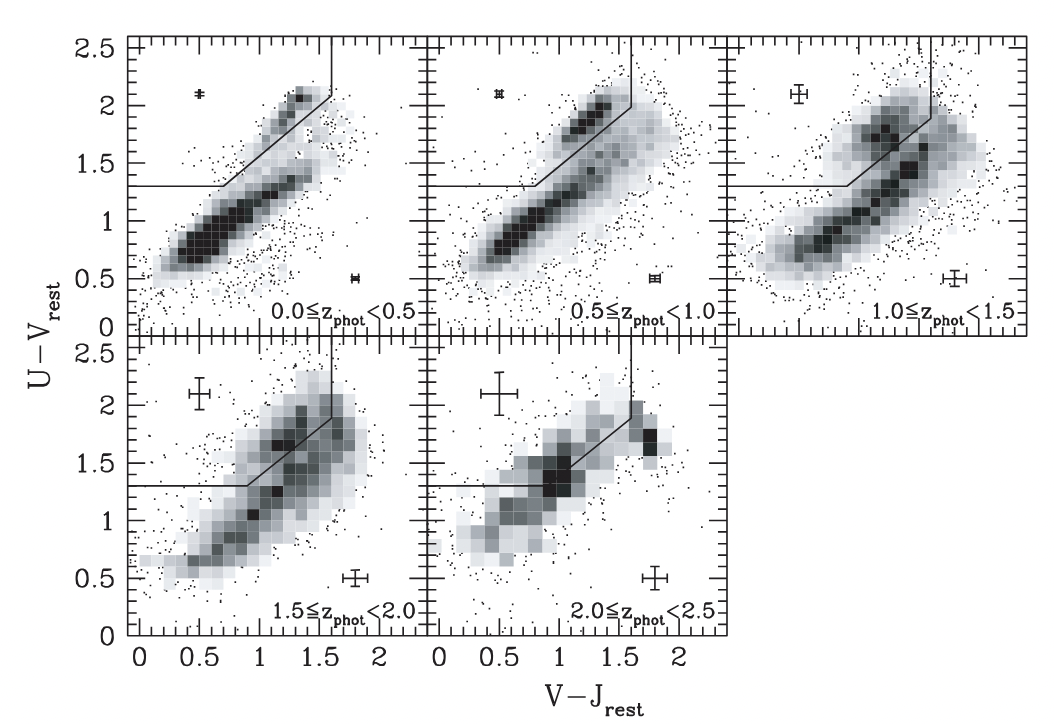}
\caption{Rest-frame UVJ diagram for UDS galaxies across five redshift bins, reproduced from \cite{Williams2009}. The grey-scale indicates the density of points within each region, with the typical uncertainty on individual colors given by a single error bar for each panel. Clearly, the galaxy population exhibits strong bimodality in this diagram, especially at lower redshifts. The red peak (upper left) maps onto systems with low sSFR, and the blue peak (lower region) maps onto systems with high sSFR. The polygon regions (indicated with a solid black line on each panel) indicate the location of UVJ selected quenched systems. Note that this evolves with redshift.}\label{f7}
\end{figure}

By way of an example, in Fig.~\ref{f7} we show the popular UVJ diagram (rest-frame (U-V) vs. (V-J) color space) for UDS data, reproduced from \cite{Williams2009}. The (U-V) color probes the 4000 \AA / Balmer break and (V-J) probes the level of  dust extinction. All of the rest-frame colors have to be constructed from SED fits (see back to the discussion at the start of Sect.~\ref{s2}). This typically requires 10+ observed bands over a range of UV - MIR (dependent upon redshift).

The UVJ relationship in Fig.~\ref{f7} is displayed for five redshift bins from the local Universe to cosmic noon (see labels on panels). There is clear bimodality evident in the UVJ diagram, especially at lower redshifts, as seen by the grey scale (which indicates the density of galaxies within each region). Generally speaking, quenched galaxies have redder colors in (U-V) than star forming systems. On the other hand, extremely red colors in (V-J) typically indicates significant dust extinction, rather than simply older stellar populations. Accounting for these effects leads to the polygon detection windows (shown with solid lines on each panel). 

As confirmed in \cite{Williams2009}, the quenched region (upper-left detection windows) correspond to relatively low sSFR values, compared to the star forming population at a given epoch. However, one should be aware that the SFRs are derived from photometry here and, hence, are less independent than those considered previously via spectroscopy, in \cite{Brinchmann2004}. Nonetheless, this consistency has been used to justify the use of the UVJ diagram as a method to form samples of potentially quenched systems in photometric data. The UVJ method is especially valuable at high-$z$, where spectroscopy is often lacking. It will be especially valuable to validate these color-based approaches against large statistical samples of galaxy spectra obtained at intermediate-to-high redshifts with the upcoming VLT-MOONRISE survey (e.g., \citealt{Cirasuolo2020, Maiolino2020}).

Alternatively, by performing SED fitting on observational data, one may extract photometric SFRs and, indeed, the full star formation history (SFH) for galaxies. However, the entire SFH tends to be less accurate than other output quantities (especially if the waveband coverage is limited). Consequently, many publications leverage the mean stellar ages within galaxies, as constrained via the best-fit composite stellar population model (or via marginalizing over the posteriors in the Bayesian case). Bimodality is clearly visible in the distribution of stellar population ages, especially when weighting by optical luminosity. Cuts to identify quenched galaxies are typically made at $\text{Age}_L \gtrsim 1 - 3$\,Gyr (e.g., \citealt{Gallazzi2005, Pacifici2016, Carnall2019, Bluck2020a}). 

Additionally, many authors have used spectral indices to separate star forming from quenched galaxies (see, e.g., \citealt{Balogh1999, Kauffmann2003b, Brinchmann2004, Gallazzi2005, Wild2009}). A particularly common approach is via the strength of the 4000\,\AA \, break, which may be defined as:

\begin{equation}
D_n(4000) \equiv \bigg( \int_{4000\,\text{\AA}}^{4100\,\text{\AA}} f_\lambda(\lambda) \, d\lambda \bigg) \,\, \bigg/ \,\,  \bigg( \int_{3900\,\text{\AA}}^{4000\,\text{\AA}} f_\lambda(\lambda) \, d\lambda \bigg)
\end{equation}

\noindent which is just the ratio of the integrated continuum spectrum either side of the break. The spectral discontinuity around 4000 \AA \, is commonly used as a stellar population age indicator because it strengthens as stellar populations age and become dominated by low-mass / cooler stars (e.g., \citealt{Bruzual1983, Balogh1999}). However, for post-starburst galaxies, the continuum is often dominated by A-stars, yielding strong Balmer absorption features and a prominent Balmer break at 3646 \AA. Indeed, at z $\gtrsim$ 2, many quiescent galaxy selections may be more sensitive to the Balmer break than the classical metal line 4000 \AA \, break, and the two features should not be conflated.

In contrast to the Balmer break, which is strongest in A-star dominated populations, the 4000\,\AA \, break depends primarily on stellar population age and metallicity in the local Universe. In the local Universe, the strength of the 4000\,\AA \, break depends primarily on stellar population age and metallicity. As the short-lived O- and B-type stars disappear, the integrated spectrum becomes increasingly dominated by cooler F-, G-, and K-type stars, causing the break to strengthen. Consequently, $D_n(4000)$ is widely used as a tracer of stellar population age for actively star forming and recently quenched systems. For older populations, $D_n(4000)$ continues to increase with age, but its interpretation becomes increasingly degenerate with metallicity, since the break is produced by the accumulation of numerous metal absorption features in stellar atmospheres. Nonetheless, the index remains a powerful discriminator between galaxies with and without recent star formation. Common thresholds used to identify quenched galaxies are $D_n(4000) \gtrsim 1.4 - 1.5$ (e.g., \citealt{Balogh1999, Brinchmann2004}).

In addition to the 4000\AA \,\,break, there are a few further spectral features which are often used to identify (or help confirm) quenching. First, an absence of emission lines itself is a good indicator of no recent star formation (provided there is not excessive dust extinction). Certainly, the presence of strong optical emission lines are unexpected in truly quenched galaxies (even if those emission lines are from AGN excitation), because this points to the existence of a large neutral gas reservoir which is being ionized. 

Additionally, the presence of strong H$\delta$ absorption features indicates the presence of A-types stars. In lineless systems, this combination suggests a high star formation rate within the past Gyr (or so) which has been rapidly quenched. These systems are often referred to as `post starbursts' or `E + A' systems, the latter standing for, elliptical-like continuum with strong A-type stellar population (see, e.g., \citealt{Dressler1983, Wild2009}).

Finally, in emission line systems, `retired' galaxies (those on the way to quenching but still forming some stars) may be identified by the H$\alpha$ equivalent width, which is defined as: 

\begin{equation}
\mathrm{EW}(\mathrm{H\alpha}) = \frac{ \displaystyle\int_{\lambda_1}^{\lambda_2} \left[ f_\lambda(\lambda) - f_{\lambda}^\mathrm{cont}(\lambda_\mathrm{H\alpha}) \right] \, d\lambda }{f_{\lambda}^\mathrm{cont}(\lambda_\mathrm{H\alpha}) } =  \frac{ F(\mathrm{H}\alpha-{\text{line}}) }{ f_{\lambda}^\mathrm{cont}(6563\,\text{\AA}) }
\end{equation}

\noindent where, the integral is performed around the H$\alpha$ emission line, and the spectral continuum at H$\alpha$ is subtracted from the measured flux ($f_\lambda(\lambda)$). The whole integral may be simply expressed as, $F(\mathrm{H}\alpha-{\text{line}})$, i.e., the total line flux. The above expression is normalized by the specific flux at the center of the H$\alpha$ line from the continuum ($f_{\lambda}^{\text{cont}}(\lambda_{\mathrm{H}\alpha})$), which is either estimated via extrapolation, or inferred via a continuum full-spectrum fit. Consequently, the equivalent width is given in units of \AA, and may be interpreted as a width (explicitly, the width needed for a box-shaped absorption line which reaches zero to have flux equal to the actual emission line). 

A commonly used definition for `retired' galaxies is EW$(\mathrm{H}\alpha) < 3$\,\AA \, (see, e.g., \citealt{Gavazzi2002, Kauffmann2003b, Salim2007, CidFernandes2011}). However, note that these systems are not fully quenched galaxies, which usually present with no emission lines at all (see \citealt{Bluck2020a} for a discussion). On the other hand, they may be interpreted as a type of `green valley' system. This nomenclature arises from the dearth of galaxies residing between the quenched (`red') and star forming (`blue') peaks in color diagnostic diagrams. Many have argued that these intermediate systems are currently undergoing quenching (e.g., \citealt{Strateva2001, Salim2007, Martin2007, Wyder2007}). However, others have warned that this may be an over-simplification (see, e.g., \citealt{Thomas2010, Schawinski2014}). This follows because some galaxies could be rejuvenating, and even amongst the quenching sample, different systems may be quenching at different rates, potentially due to different underlying causes.

\subsection{Identifying transitioning galaxies}

A key challenge in studies of galaxy quenching is the identification of systems currently transitioning between the star forming and quiescent populations. Since quenching is a time-dependent process, galaxies in transition are expected to be comparatively rare, making their selection particularly valuable for constraining quenching pathways and timescales. In short, via this approach, one ultimately seeks to `catch quenching in action'. This is particularly useful to constraining the triggers of galaxy quenching. On the other hand, this approach is much less constraining of the physical processes which conspire to maintain quiescence over long cosmological time periods.

The most common approach is to identify galaxies occupying the `green valley', located between the star forming, blue cloud and quenched, red sequence in color - magnitude, color - mass, sSFR - mass, and related parameter spaces (e.g., \citealt{Martin2007, Salim2014, Schawinski2014, Bremer2018}). However, the precise definition of the green valley varies considerably between studies, depending on the observables adopted and the evolving location of the star-forming sequence. Consequently, green-valley samples generally comprise a heterogeneous mixture of galaxies undergoing slow quenching, rapid quenching, rejuvenation, and systems affected by dust attenuation. Further modeling is usually required to establish the direction, and rate, of evolution through the green valley (e.g., \citealt{Schawinski2014, Smethurst2015}).

A complementary approach is provided by post-starburst galaxies (PSBs; also referred to in the literature as E+A or K+A galaxies\footnote{Where, E+A denotes elliptical like spectrum with A-stars present, and K+A indicates the prevalence of both K- and A-type stars in the galaxy spectrum.}). These systems exhibit strong Balmer absorption features from A-type stars together with weak or absent nebular emission lines (e.g., \citealt{Dressler1983, Couch1987, Wild2009, Wild2010, Wild2016, Rowlands2018, French2023}). These spectral signatures indicate a substantial decline in star formation within the previous $\sim 0.1 - 1$,Gyr. Hence, PSBs are amongst the clearest observational tracers of recent, rapid quenching. That said, some confusion remains as to the eventual fate of PSBs, particularly whether or not they will rejuvenate their star formation eventually. The abundance of PSBs evolves strongly with redshift and they appear increasingly important among recently quenched galaxies at earlier cosmic epochs (e.g., \citealt{Wild2009, Wild2016, Park2024}). Conversely, in the local Universe they make up only $\sim$1\% of systems (e.g., \citealt{Wild2016, French2021}). 

We return to the constraints on quenching triggers from observations of transitioning galaxies in Part II of this review (particularly in sect. 7). Briefly, many works have aimed to constrain the incidence rates of mergers, AGN feedback, AGN activity, and environmental location in the transitioning population (e.g., \citealt{Schawinski2014, Yesuf2017, Almaini2017, Pawlik2018, French2018, McNab2021, Li2023, Almaini2025}). We defer a detailed review of their findings to Part II, where we make specific connection with theoretical predictions.

%%%%%%%%%%%%%%%%%%
%                %
%  OBSERVATIONS  %
%                %
%%%%%%%%%%%%%%%%%%

\section{Foundational observational results}\label{s3}

In this section we explore two key observational results of direct relevance to the field of galaxy quenching. We begin by discussing the evolution in the star formation rate density (SFRD) with cosmic time. We then go on to discuss the peak efficiency of baryon-to-star conversion at a halo mass of $M_\mathrm{Halo} \sim 10^{12}\,M_{\odot}$.

\subsection{Cosmological evolution in the star formation rate density}\label{s31}

One of the most fundamental questions we can ask about star formation in the Universe is, \emph{when does it happen?} The other natural follow-up is, of course, \emph{where does it happen?} In this sub-section we will tackle the former, with the latter covered in the next sub-section. 

Early measurements from the Canada-France Redshift Survey demonstrated that the cosmic star formation rate density (SFRD) declines strongly from $z\sim1$ to the present day \citep{Lilly1996}. Subsequently, \citet{Madau1996} extended these measurements to higher redshifts using the Hubble Deep Field, revealing a characteristic rise in the SFRD at very high redshifts towards a peak at intermediate redshifts ($z \sim 2$). In the years since these pioneering observational works, a host of papers have expanded these observations across the bulk of the history of the Universe (see \citealt{Madau2014} for a dedicated review).

\begin{figure}[ht] 
\centering
\includegraphics[width=0.85\textwidth]{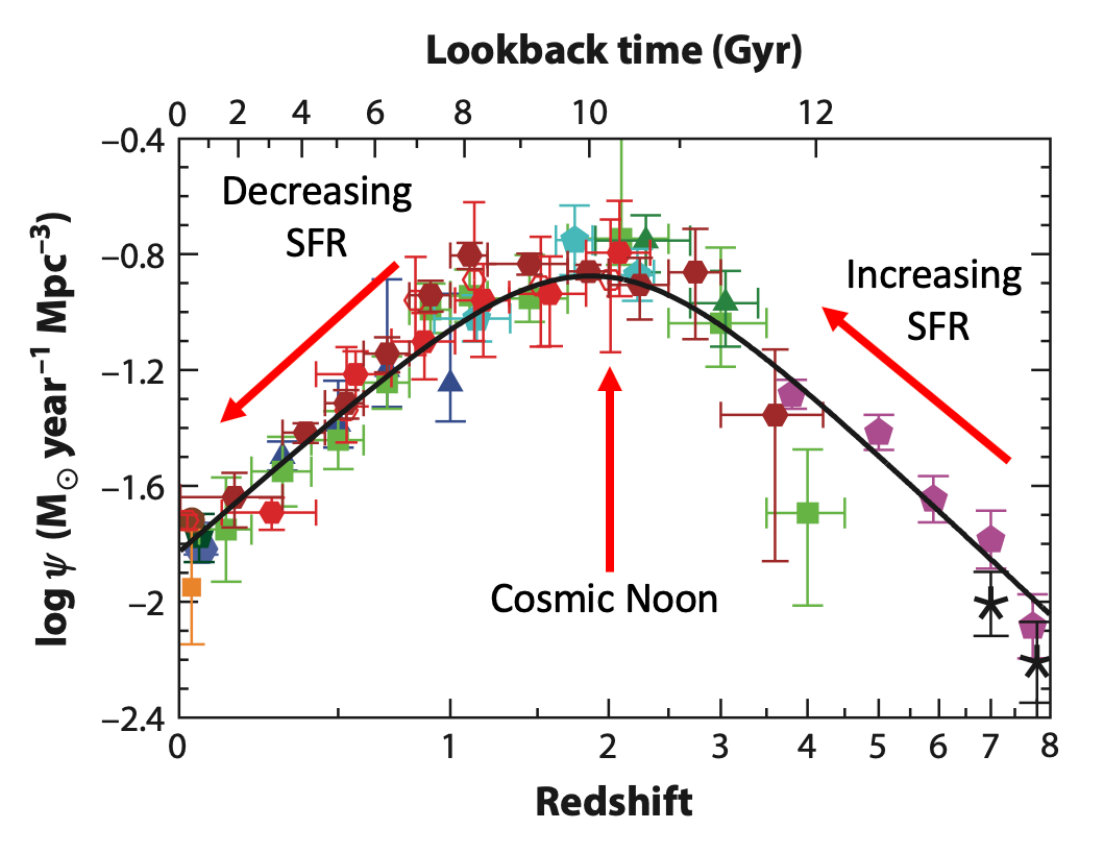}
\caption{Evolution in the comoving star formation rate density (SFRD: $\psi$) as a function of lookback time and redshift. This figure combines UV and FIR measurements of star formation rates from a wide variety of sources. Note that there is a rapid rise in the density of star formation in the early Universe ($\psi \sim (1+z)^{-2.9}$), leading to a peak at `cosmic noon' ($z \sim 2$). Thereafter, there is a slow decline in the star formation rate density over 10\,Gyrs to the present epoch ($\psi \sim (1+z)^{2.7}$). This figure is reproduced from \cite{Madau2014}, with additional annotation added here.}\label{f8}
\end{figure}

In Fig.~\ref{f8}, we show the SFRD (labeled as: $\psi$) as a function of redshift (and lookback time) from \cite{Madau2014}. Star formation rates are estimated via a range of methods, including dust-corrected UV measurements and far-infrared (FIR) calibrations. Uncertainties are relatively well constrained up to $z \sim 3$ through cross-validation between these tracers. However, at higher redshifts, constraints rely increasingly on UV luminosity functions, where uncertain dust corrections introduce significant systematic uncertainties.

The evolution in the SFRD is characterized by a rapid increase with time in the very early Universe, followed by a peak at cosmic noon ($z \sim 2$), and finally a long, slow decline over $\sim$10\,Gyr towards the present. This can be neatly packaged into the following empirical expression (see \citealt{Madau2014}):

\begin{equation}
\psi(t) = 0.015 \cdot \frac{(1 + z)^{2.7}}{1 + \left( \frac{1 + z}{2.9} \right)^{5.6}} \quad \left[ M_\odot\,\mathrm{yr}^{-1}\,\mathrm{Mpc}^{-3} \right].
\end{equation}

Consequently, there appears to be an optimal time for stars to form in our Universe (i.e., at $z \sim 2$), with marked reduction at both earlier and later times. This period in the Universe's history has become known as `cosmic noon', since it approximately represents the peak optical luminosity in the history of the Universe. This raises an interesting question: \textit{what was special about $z \sim 2$?}

To answer this question, it is helpful to factorize the SFRD into its constituent parts. Ultimately, there are three factors which contribute to the SFRD evolution -- (i)~the redshift evolving number density of dark matter haloes ($\phi_c(M,z)$); (ii)~the accretion rate of gas into the haloes, as a function of mass and redshift ($\dot{M}_\mathrm{acc}(M,z)$); and (iii)~the efficiency with which the accreted gas is converted into stars ($\mathcal{E}_\mathrm{SF}(M,z)$). Leveraging this insight, we may write down an ansatz for the evolution in star formation rate density as follows (see similar and complementary approaches in, e.g., \citealt{Sheth1999, Tinker2008, Behroozi2013, Lilly2013, Moster2013, Moster2018}):

\begin{equation}
\mathrm{SFRD: } \,\,\psi(z) = f_b \cdot \int_{M_\mathrm{min}}^{M_\mathrm{max}} \phi_c(M_H, z) \cdot \mathcal{E}_\mathrm{SF}(M_H,z) \cdot \dot{M}_\mathrm{acc}(M_H, z) \, dM_H.
\end{equation}

\noindent It is worth going through these terms in some detail as they reveal great insight on this problem. First, $f_b \equiv \Omega_b/\Omega_M$, which indicates the universal baryon fraction. This is a reasonable first-order approximation for the baryon content of haloes. However, note that simulations with feedback do show substantial variation in the baryonic content of haloes, beyond what is naively assumed from the cosmological baryon fraction (see, e.g., \citealt{Mitchell2022, Sorini2022}). Next,

\begin{equation}
\phi_c(M_H,z) \equiv \frac{d\,n_c(M_H,z)}{dM_H} \approx
\phi_c(z) \left( \frac{M_H}{M_c(z)} \right)^{\alpha}
\exp\left\{ -\left( \frac{M_H}{M_c(z)} \right)^{\beta} \right\},
\end{equation}

\noindent which indicates the comoving halo mass function as a function of redshift, which is often parameterized as a power law\,--\,exponential `Schechter-like' function (see \citealt{Sheth1999, Tinker2008}). The coefficients ($\alpha$ and $\beta$) are typically derived from fitting to dark matter simulations, or else via analytic extensions to linear growth theory. The critical mass of the exponential cutoff ($M_c(z)$) and the overall normalization ($\phi_c(z)$) both vary with redshift, such that more massive haloes may be formed at later cosmic times. Note that this term is governed by the hierarchical assembly of dark matter haloes, and is (to leading order) independent of baryonic physics.

The next term is:

\begin{equation}
\mathcal{E}_\mathrm{SF}(M,z) \equiv \frac{dM_*/dt}{dM_b/dt} \approx \frac{\mathrm{SFR}}{f_b\,\dot{M}_\mathrm{acc}}(M,z),
\end{equation}

\noindent which represents the efficiency of star formation at a fixed accretion rate into the halo. Here the efficiency describes the amount of star formation per unit baryonic accretion rate, rather than the amount of star formation per unit gas mass of the ISM (as in many other works; e.g., \citealt{Lin2019, Ellison2020}). This is the key term which baryonic physics impacts. It is also the term which most directly links to star formation and quenching. As such, we will discuss this at more length later on in this section.

Finally, we have:

\begin{equation}
\dot{M}_\mathrm{acc}(M, z) \equiv \frac{dM_H}{dt} \propto \big( M_H \big)^{a} \cdot (1 + z)^b \, ,
\end{equation}

\noindent which indicates the mass and redshift dependent accretion rate of matter into dark matter haloes. More massive haloes have higher accretion rates, but the accretion rate at a fixed halo mass is greater at earlier cosmic times. Hence, the coefficients ($a$ and $b$) indicate positive constants to be fit in simulations, or via analytical prescriptions, with typical values of $a \sim 1$ and $b \sim 2 - 2.5$ (e.g., \citealt{McBride2009, Dekel2009, Fakhouri2010, Lilly2013, Dekel2013, Dekel2019}).

The evolution in the number density of dark matter haloes ($\phi_c(z)$) and the accretion rate into haloes as a function of mass and redshift ($\dot{M}_\mathrm{acc}(M, z)$) are both set primarily by the underlying dark matter evolution, as traced by analytical extensions, N-body simulations, or cosmological hydrodynamical simulations. Only the efficiency term ($\mathcal{E}_\mathrm{SF}(M,z)$) is thought do depend on baryons to leading order. As we will see in the next subsection, a double power-law form (or, similarly, log-normal form) of the efficiency can bring theory and observations into accord.

However, before moving on, it is worth exploring precisely what this efficiency term really contains from a physical perspective. By construction, this variable encapsulates everything about star formation that is not governed by either accretion into haloes or the number densities of haloes of varying mass as a function of cosmic time. To look at this a bit more closely, from the definition, the halo efficiency may be written in terms of the standard star formation efficiency ($\mathrm{SFE} = \mathrm{SFR}/M_g = 1/\tau$), as follows:

\begin{equation}
\mathcal{E}_\mathrm{SF}(M,z,...) = \mathcal{R}_b \cdot   \mathcal{E}_\mathrm{p^+,\,e^- \rightarrow \, H} \cdot \mathcal{E}_\mathrm{H \rightarrow \, H_2} \cdot \mathrm{SFE} \,, 
\end{equation}

\noindent where each term is defined explicitly as:

\begin{equation}
\frac{\mathrm{SFR}}{\dot{M}_\mathrm{b, Halo}} = \bigg(\frac{M_\mathrm{b, Halo}}{\dot{M}_\mathrm{b, Halo}} \bigg) \cdot \bigg(\frac{M_\mathrm{H-neutral}}{M_\mathrm{b, Halo}}\bigg) \cdot \bigg(\frac{M_\mathrm{H-molecular}}{M_\mathrm{H-neutral}}\bigg) \cdot \bigg( \frac{\mathrm{SFR}}{M_\mathrm{H-molecular}} \bigg)
\label{e22}
\end{equation}

\noindent Every baryon within the halo has the potential to be processed into a star. Everything else follows as a trivial mathematical identity. This approach has value, however, because it enables us to separate out different physical processes operating on different scales within the halo. 

The first term on the right-hand side of Eq.~\eqref{e22} is used simply to convert from a mass dependence to a rate dependence, with the real physics contained in the following terms. The second term indicates the fraction of baryons that are in a neutral state, which requires cooling of the CGM (or else direct accretion from the IGM). The third term indicates the fraction of the neutral gas (largely contained within the ISM) which is dense enough to condense into a  molecular state. The final expression simply links the mass of molecular gas to the star formation rate, following \cite{Kennicutt1998}, which is the definition of SFE (see above and Sect.~\ref{s7}). 

It is important to appreciate that any of these stages might represent the `bottle neck' for star formation within galaxies, and that the specific limitation may vary with galaxy mass, environment, and epoch. Hence, we understand the cosmological history of SFRD in the broadest sense, but we lack (at least at this point in the review) a clear understanding of which specific stage in the gas cycle within haloes is the underlying cause of the mass-dependent inefficiency of star formation.

\subsection{The peak efficiency of galaxy formation}\label{s32}

\noindent Having addressed \emph{when} stars form (primarily around cosmic noon, with a fall off at both later and earlier times), we are now in a good position to ask \emph{where} they form. More specifically, we look in detail at which halo masses are most conducive for star formation. Many works have investigated the relationship between stellar mass and dark matter halo mass, utilizing a wide variety of methodologies including abundance matching, clustering statistics, and direct constraints from gravitational lensing (see, e.g., \citealt{Moster2010, Moster2013, Behroozi2010, Behroozi2013, Mandelbaum2006, Mandelbaum2016, Rodriguez-Puebla2017, Behroozi2019}). We provide a brief description of some of these methods at the start of Sect.~\ref{s5}, in relation to quantifying galaxy environments. For full details, see the above references. Generally, these works are in good agreement, finding a strong positive relationship between these mass components, but with notable sub-structure. 

\begin{figure}[ht]
\centering
\includegraphics[width=1\textwidth]{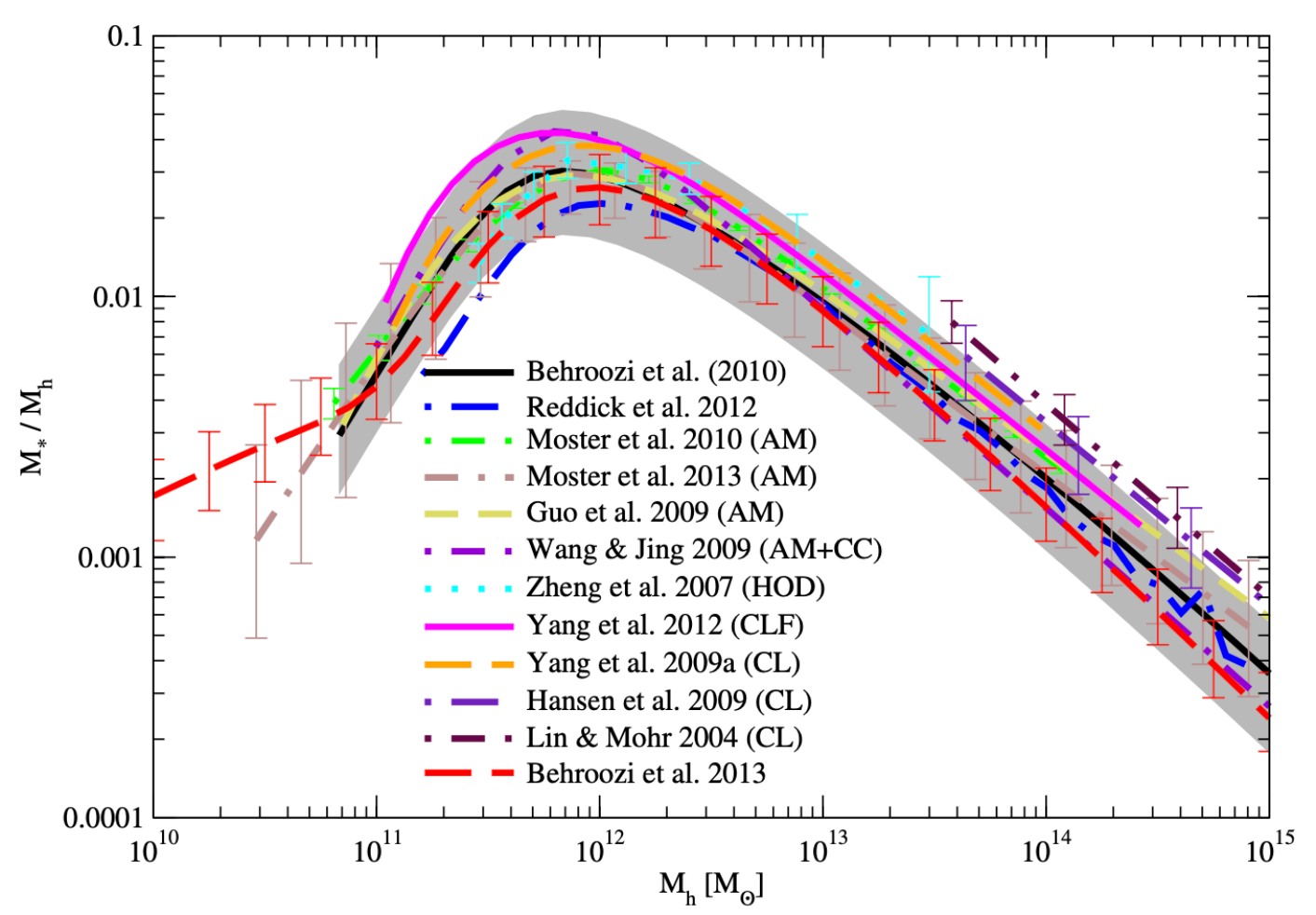}
\caption{The ratio of stellar mass to halo mass plot as a function of halo mass for a large variety of methods for inferring halo masses (see legend). In the legend, the following abbreviations are used: AM = abundance matching, CL = clustering statistics, HOD = halo occupation distributions, CC = conditional colors, CLF = conditional luminosity functions. This figure is reproduced from \cite{Behroozi2013}. It is clear that galaxy formation is most efficient at intermediate halo masses, where the highest fraction of baryons are converted into stars. Either side of this critical halo mass, dramatic suppression within both low- and high-mass haloes is evident. This figure may be compared to Fig.~\ref{f2} for a similar conclusions based on comparing the halo and stellar mass functions directly.}\label{f9}
\end{figure}

In Fig.~\ref{f9}, we present the ratio of stellar-to-halo mass as a function of halo mass from \cite{Behroozi2013}. It is clear that there is a peak halo mass for conversion of baryons into stars, at $M_{\rm Halo} \sim 10^{12}\,M_\odot$. At both higher and lower masses, the conversion of baryons into stars is suppressed, relative to simple (feedback-free) galaxy formation models.

As an example, in \cite{Moster2010} the $M_*$\,--\,$M_H$ relation is fit via a double power-law function, explicitly:

\begin{equation}
M_* = 2\mathcal{N}_p M_H \,\left[ \left( \frac{M_H}{M_c} \right)^{-\gamma} + \left( \frac{M_H}{M_c} \right)^{\delta} \, \right]^{-1}
\end{equation}

\noindent where, $M_c$ indicates the critical turn-over mass (fit to be at, $\log(M_c/M_\odot) = 11.93 \pm 0.03$); $\gamma = 1.00\pm0.03$, which is related to the the low-mass slope; $\delta = 0.56\pm0.01$, which is related to the high-mass slope; and $\mathcal{N}_p$ is a normalization constant (which is equal to the peak efficiency of star formation in haloes, fit to be: $(M_*/M_H)_\mathrm{peak} = 0.027\pm0.001$). Hence, at low masses, $M_* \sim (M_H )^2$, which indicates super-linear evolution. This is assumed to result from the efficiency of star formation rising with mass in this regime. At high masses, $M_* \sim (M_{H})^{0.44}$, which indicates sub-linear evolution.

\begin{figure}[ht]
\centering
\includegraphics[width=0.95\textwidth]{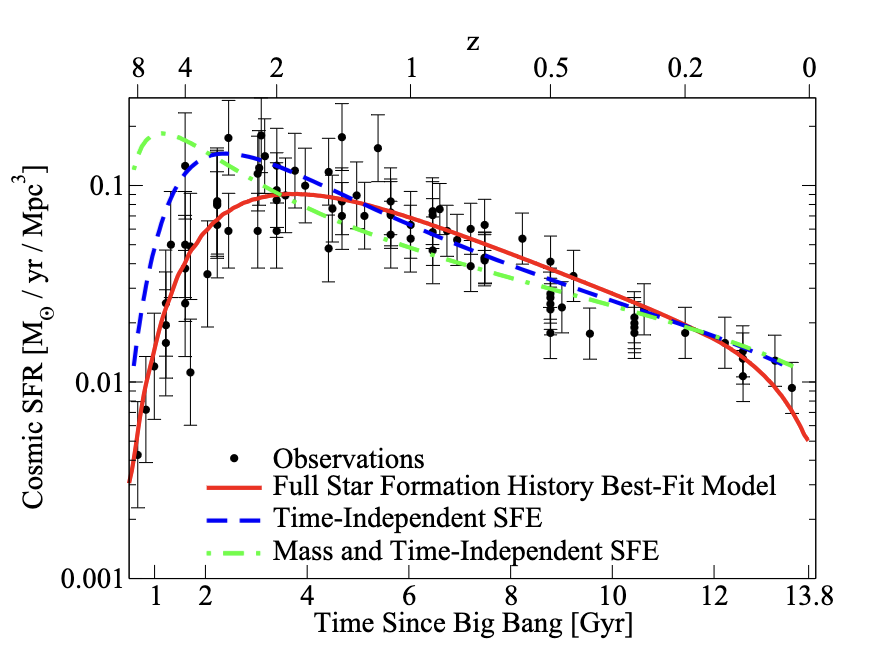}
\caption{Modeling the star formation rate density evolution. The cosmic SFR comoving density is plot against cosmic time and redshift for observational constraints (data points) and three best fit efficiency models (lines). In particular, the following models are shown: (i) constant efficiency in star formation with halo mass and redshift (green dashed line); (ii) variable efficiency in halo mass but constant in time (blue dashed line); and (iii) variable efficiency in halo mass and time (red solid line). It is clear that the evolving efficiency can reproduce the observations well. This figure is reproduced from \cite{Behroozi2013b}. }\label{f9b}
\end{figure}

There is now consensus in the observational literature for an approximately log-normal (or, similarly, double power-law) shape of the $M_*/M_H$\,--\,$M_H$ relationship. That is, star formation must be suppressed in haloes both above and below a critical mass ($M_\mathrm{crit} \sim 10^{12} M_{\odot}$). 

It is important to appreciate that there is nothing in the gravitational physics which can explain the key relationships in Fig.~\ref{f9}. At these masses and redshifts, the halo mass function is very well described by a single power-law form (e.g., \citealt{Press1974, Bond1991, Springel2006, Moster2010, Behroozi2013}). As such, these observations have forced theorists to incorporate non-gravitational baryonic feedback mechanisms into galaxy formation models and simulations, in order to get the basic properties of the galaxy stellar mass function correct (see, e.g., \citealt{Cole2000, Bower2006, Bower2008, Croton2006, Henriques2013, Vogelsberger2014a, Schaye2015, Weinberger2017, Nelson2018, Dave2019}). 

In Fig.~\ref{f9b} we present again the redshift evolution in the star formation rate density, but here overlaying various best-fit models for the efficiency of star formation. In particular three models are shown: (i) a constant efficiency (green dashed line); (ii) a variable efficiency in halo mass but constant in time (blue dashed line); and (iii) a variable efficiency in both halo mass and time. It is clear that the constant efficiency performs well at late cosmic times, indicating that the evolution in the SFRD is largely constrained by declining gas accretion into haloes throughout this epoch. However, to get the decline at high-$z$ correct with respect to observations, a variable efficiency in halo mass is essential (as independently confirmed from the halo mass - stellar mass relationship, discussed above). Additionally, some further mild evolution in the halo-dependent efficiency can yield even tighter accord with observations.

Ultimately, the evolution in the SFRD has two regimes: (i) an accretion limited regime at late cosmic times, where the reduction of baryonic accretion into haloes is governed primarily by cosmological expansion, and (ii) a halo mass limited regime, where the growth in dark matter haloes leads to increasing average accretion rates as a function of time in the very early Universe. The latter would almost always dominate, if it were not for the effect of the star formation efficiency dependence on halo mass (see green dot-dashed line in Fig.~\ref{f9b}). However, with a log-normal shape to the star formation efficiency (as in Fig.~\ref{f9}), at very high redshifts, the growth of dark matter haloes dominates over the reduction in specific accretion with redshift, causing the characteristic turn-over in the SFRD evolution (see blue dashed and red solid lines in Fig.~\ref{f9b}).

To make further progress it is essential to bring in cosmological hydrodynamical simulations into the discussion. These models are discussed at length in Part II of this review (see especially Sect.~2). Briefly, feedback from supernovae are employed to reduce the efficiency of star formation within low mass haloes, and feedback from AGN is employed to reduce the efficiency of star formation within high mass haloes (see, e.g., \citealt{Vogelsberger2013, Vogelsberger2014a, Vogelsberger2014b, Torrey2014, Schaye2015, Crain2015, Weinberger2017, Weinberger2018, Nelson2018, Pillepich2018, Dave2019}). Working together, these feedback channels lead to a log-normal shape of star formation efficiency as a function of halo mass, accounting for both the observed stellar mass - halo mass relationship and the evolution in the SFRD.

\section{Intrinsic correlators to quiescence}\label{s4}

In this section we discuss many of the observational correlators to quenching which are connected to the galaxy itself, i.e., intrinsic correlators to quenching. In the next section we will consider environmental correlators to quenching.

\subsection{Quantifying galaxy morphologies, structures \& kinematics}\label{41a}

Galaxy quenching is closely connected to galaxy morphology, structure, and kinematics. Historically, galaxies were classified visually via the Hubble sequence into ellipticals, lenticulars, spirals, and irregulars (\citealt{Hubble1926, Hubble1936}), with modern extensions including Galaxy Zoo citizen-science classifications (e.g., \citealt{Lintott2008, Willett2013}), and machine-learning approaches (e.g., \citealt{Huertas-Company2015, Dominguez-Sanchez2018}). Quantitative non-parametric methods such as Concentration--Asymmetry--Clumpiness (CAS; \citealt{Conselice2000, Conselice2003}) and Gini--$M_{20}$ (\citealt{Lotz2004, Lotz2008a}) are also widely used, particularly for identifying mergers and disturbed systems.

Modern structural analyses increasingly rely on parametric fitting of galaxy light profiles. Disc galaxies approximately follow exponential profiles (e.g., \citealt{Freeman1970}), while spheroids are often well described by the de Vaucouleurs law (e.g., \citealt{DeVaucouleurs1948}). These are unified through the S\'ersic profile (see, \citealt{Sersic1963}):

\begin{equation}
I_n(R) = I_e \exp \bigg\{-b_n \bigg[ \bigg( \frac{R}{R_e} \bigg)^{1/n} - 1 \bigg]  \bigg\},
\end{equation}

\noindent where $I_n(R)$ is the surface brightness at radius $R$, $I_e$ is the surface brightness at the effective radius, $R_e$ (i.e., the radius enclosing half the total light), $n$ is the S\'ersic index, and $b_n \approx 2n-1/3$ is chosen such that half the total luminosity lies within $R_e$. Typically, $n \sim 1$ describes exponential disks, while $n \gtrsim 4$ approximates spheroids. 

The elliptical radius ($R$), which the S\'ersic profile is dependent upon, is generally defined as:

\begin{equation}
R = \sqrt{X_\mathrm{gal}^2 + \bigg( \frac{Y_{\rm gal}^2}{q^2}\bigg)},
\end{equation}

\noindent where, $X_{\rm gal}$ and $Y_{\rm gal}$ are galaxy-frame coordinates and $q=b/a$ is the axial ratio (semi-minor axis divided by semi-major axis). To transform from sky coordinates to galaxy coordinates, one generally performs the following rotation and centering procedure:

\begin{equation}
\begin{pmatrix} X_\mathrm{gal} \\ Y_\mathrm{gal} \end{pmatrix}
=
\begin{pmatrix}
\cos(\mathrm{PA}) & \sin(\mathrm{PA}) \\
-\sin(\mathrm{PA}) & \cos(\mathrm{PA})
\end{pmatrix}
\begin{pmatrix}
X_\mathrm{sky} - X_c \\
Y_\mathrm{sky} - Y_c
\end{pmatrix},
\end{equation}

\noindent where, $(X_c,Y_c)$ are the galaxy center coordinates and PA is the position angle of the major axis relative to the X-axis in the sky-coordinate frame. 

In modern studies, galaxies are frequently decomposed into bulge and disk components via multi-component S\'ersic fitting (e.g., \citealt{Peng2002, Simard2002, Peng2010aa, Mendel2014, Dimauro2018}):

\begin{equation}
I_\mathrm{B+D}(R) = I_\mathrm{Bulge}(R)\big|_{n=n_b} + I_\mathrm{Disc}(R)\big|_{n=n_d},
\end{equation}

\noindent where, $n_b$ and $n_d$ are the S\'ersic indices of the bulge and disc, respectively. Model images are convolved with the point spread function (PSF) and optimized through $\chi^2$ minimization. Explicitly,:

\begin{equation}
\chi^2 = \sum_i \bigg( \frac{I_{\rm obs,i} - (I_{\rm mod} \ast {\rm PSF})_i}{\sigma_{{\rm obs},i}} \bigg)^2,
\end{equation}

\noindent where, $I_{\rm obs,i}$ is the observed surface brightness in pixel $i$, $I_{\rm mod}$ is the model image, $\sigma_{{\rm obs},i}$ is the observational uncertainty, and $\ast$ denotes convolution. Widely used fitting codes include GALFIT, GIM2D, ProFit, and StatMorph (\citealt{Peng2002, Simard2002, Robotham2017, Rodriguez-Gomez2019}). Multi-band fitting combined with SED modeling further enables stellar mass decompositions into bulge and disc components (see, e.g., \citealt{Gadotti2009, Mendel2014, Tacchella2015, Dimauro2018}).

While morphology constrains the stellar light distribution, kinematics probe the underlying dynamical mass distribution. Spatially resolved spectroscopy enables measurements of velocity and velocity dispersion maps, from which one may derive quantities such as the dimensionless spin parameter, $\lambda_s$, and stellar specific angular momentum, $j_\star$ (\citealt{Emsellem2007, Fall2018}):

\begin{equation}
\lambda_s = \frac{\sum_i f_i R_i |V_i|}{\sum_i f_i R_i \sqrt{V_i^2 + \sigma_i^2}}
\quad ; \quad
j_\star = \frac{\sum_i V_c(R_i)\Sigma_{\star,i}R_i}{\sum_i \Sigma_{\star,i}}.
\end{equation}

\noindent Here, $f_i$ is the flux in spaxel $i$, $R_i$ is the galacto-centric radius, $V_i$ is the line-of-sight velocity, $\sigma_i$ is the velocity dispersion, $\Sigma_{\star,i}$ is the stellar mass surface density, and $V_c(R_i)$ is the circular velocity. Rotating systems are commonly modeled with inclined disc (`tilted-ring') models (e.g., \citealt{Begeman1989, deBlok2008, Brownson2020}), where the line-of-sight velocity field is often parameterized as:

\begin{equation}
V_\mathrm{los}(R,\theta) = V_\mathrm{sys} + V_\mathrm{max}\tanh\bigg(\frac{R}{R_k}\bigg)\cos(\theta)\sin({\rm inc}),
\end{equation}

\noindent where, $V_{\rm los}$ is the observed line-of-sight velocity, $V_{\rm sys}$ is the systemic velocity, $V_{\rm max}$ is the maximum rotational velocity, $R_k$ is the kinematic scale radius, $\theta$ is the azimuthal angle in the galaxy plane (i.e., $\theta_i = \arctan (Y_{\rm sky,i}/X_{\rm sky,i})$), and inc is the inclination angle relative to the plane of the sky (which is given by inc $= \arccos(q)$ for a flat disk). More sophisticated analyses additionally model velocity dispersion profiles and correct for PSF/beam-smearing effects (e.g., \citealt{DiTeodoro2015, Brownson2022}).

Finally, pressure-supported and anisotropic systems are frequently analyzed via Jeans modeling (see. e.g., \citealt{Cappellari2006, Cappellari2008, Cappellari2013}), which solves the collisionless Jeans equations:

\begin{equation}
\frac{\partial (n_\star \overline{v_i v_j})}{\partial x_j} + n_\star \frac{\partial \Phi}{\partial x_i} = 0,
\end{equation}

\noindent where,

\begin{equation}
\overline{v_i v_j}(\vec{x}) =
\frac{1}{n_\star(\vec{x})}
\int v_i v_j f(\vec{x},\vec{v})\, d^3v
\quad ; \quad
n_\star(\vec{x}) =
\int f(\vec{x},\vec{v})\, d^3v.
\end{equation}

\noindent Here, $f(\vec{x},\vec{v})$ is the phase-space distribution function, $n_\star$ is the stellar number density, $\overline{v_i v_j}$ is the second velocity moment tensor, and $\Phi$ is the gravitational potential. These methods provide powerful constraints on galaxy dynamics, dark matter distributions, and the physical drivers of galaxy quenching (see \citealt{Cappellari2016} for a review).

\subsection{Morphology and structure}\label{s41}

The relationship between galaxy color and morphological type has a long history (e.g., \citealt{DeVaucouleurs1948, Visvanathan1977, Butcher1978, Dressler1980, Strateva2001, Baldry2006, Driver2006, Cameron2009a, Cameron2009b, Omand2014}). In more recent times, morphology has often been quantified via the S\'ersic index ($n$), which indicates the concentration of light within a galaxy (see, e.g., \citealt{Sersic1963, Peng2002, Simard2011}). Pure disc galaxies have $n = 1$, with pure spheroids having $n \geq 4$. Intermediate, bulge + disc, systems tend to exhibit with $n \sim 3$. Additionally, visual classifications via the Hubble sequence have also been used (e.g.,~\citealt{Lintott2008, Lintott2011, Schawinski2014}).

\begin{figure}[ht]  
\centering
\includegraphics[width=0.7\textwidth]{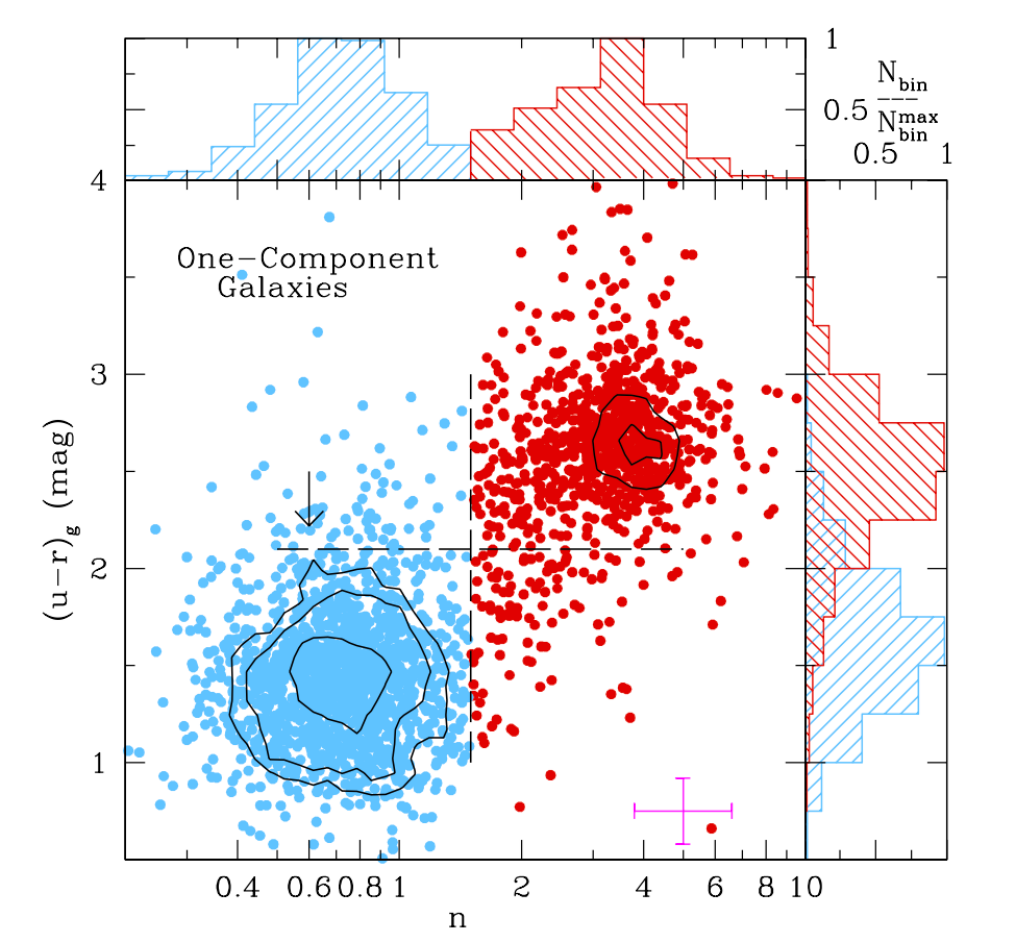}
\caption{Rest-frame, dust-corrected optical galaxy color plot as a function of S\'{e}rsic index ($n$, a parametric measurement of galactic light concentration). It is clear that galaxies form bimodal relationships in both color and structure. Furthermore, these two bimodalities are clearly related, such that there is a strong positive correlation between optical galaxy color and S\'{e}rsic index. Blue/ star forming systems tend to have low S\'{e}rsic indices, corresponding to disc-like light distributions. Conversely, red/ quiescent systems tend to have high S\'{e}rsic indices, corresponding to highly concentrated light distributions, consistent with spheroidal structure. This figure is reproduced from \cite{Cameron2009a}.  }\label{f10}
\end{figure}

In Fig.~\ref{f10}, we show the relationship between S\'ersic index and dust-corrected galaxy optical color from \cite{Cameron2009a}. There is a clear link here -- low-$n$ galaxies tend to be blue in optical color, and high-$n$ galaxies tend to be red in optical color. This further implies that more spheroidal galaxies tend to have older stellar populations, without evidence of recent star formation; whereas, disc galaxies have primarily younger stellar populations, evidencing recent star formation.

\begin{figure}[ht]  
\centering
\includegraphics[width=1.0\textwidth]{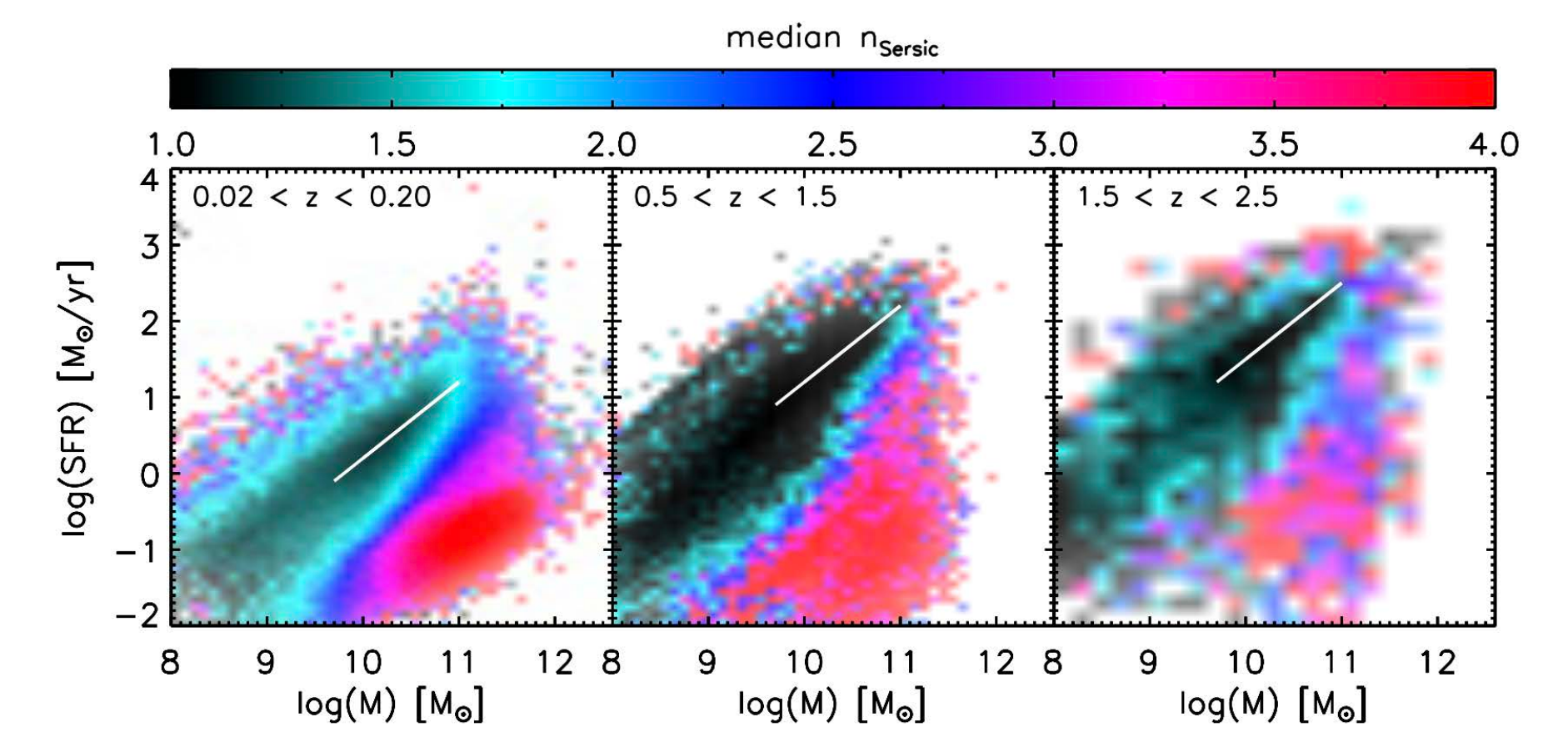}
\caption{The star forming main sequence (SFR\,--\,$M_*$ relationship) across a range of redshifts (as labeled on each panel). Each sub-region within the main sequence panel is color coded by the median S\'{e}rsic index ($n$) of galaxies at that location. It is clear that actively star forming systems have $n \sim 1$, indicating that they harbor a dominant exponential disc component. Conversely, quiescent galaxies tend to have $n > 3$, indicating that they harbor concentrated bulges in their cores, which dominate their light profiles. Interestingly, the relationship between structure and star formation quenching appears to be approximately stable from $z = 0 - 2.5$, even though the main sequence itself decreases in normalization substantially across this time frame. This figure is reproduced from \cite{Wuyts2011}.}\label{f11}
\end{figure}

These analyses have been extended to higher redshifts by \cite{Wuyts2011, Szomoru2011, Papovich2012, Mortlock2013}, among many others. Additionally, this result has been seen more directly as a function of star formation rate, rather than galaxy color (e.g., \citealt{Wuyts2011, Omand2014}). In Fig.~\ref{f11}, we show the star forming main sequence (SFR\,--\,$M_*$ relation; see back to Sect.~\ref{s21}) at various epochs from $z = 0 - 2.5$, color coded by median $n$, reproduced from \cite{Wuyts2011}. It is clear that quiescent galaxies (i.e., those with lower SFR values at a fixed stellar mass) are preferentially more spheroidal (higher $n$) than star forming systems. Moreover, these results hold up to cosmic noon.

This leads to an important question: \textit{what causes the relationship between galaxy morphology and star formation?} There are two possible answers here. First, this relationship may be caused by the distribution of star formation within galaxies. Due to conservation of angular momentum, accreting gas into a galaxy tends to settle into a disc structure. As such, it is natural to expect that new stars form in discs (e.g., \citealt{Swinbank2017, Peng2020}). If this is the correct explanation, it suggests that the link between lack of star formation and a lack of prominent disc component may be trivial (almost tautologous). Alternatively, the structure of a galaxy (or something that structure is correlated closely with) could directly cause a lack of star formation in galaxies. 

The degeneracy between possible causal explanations arises because the measurements of morphology in the works considered in this sub-section trace the distribution of optical light within galaxies, which is strongly correlated with ongoing star formation. Although the mass in stars formed in any given star forming period is small relative to the total stellar mass of the system, the luminosity\,--\,mass dependence of stars ensures that short-lived, massive stars dominate the rest-frame optical light budget of a stellar population (e.g., \citealt{Bruzual2003, Maraston2005}). To resolve this issue, one ideally needs a measurement of mass, not light. Additionally, one can observe in the rest-frame NIR, which is sensitive to the established (older) stellar populations and much less so to young/ new stars. As we will see in subsequent parts of this Section, both of these approaches lead to a confirmation of the close connection between morphology and quenching for central galaxies.

\subsection{Stellar mass}\label{s42}

\begin{figure}[ht]  
\centering
\includegraphics[width=0.6\textwidth]{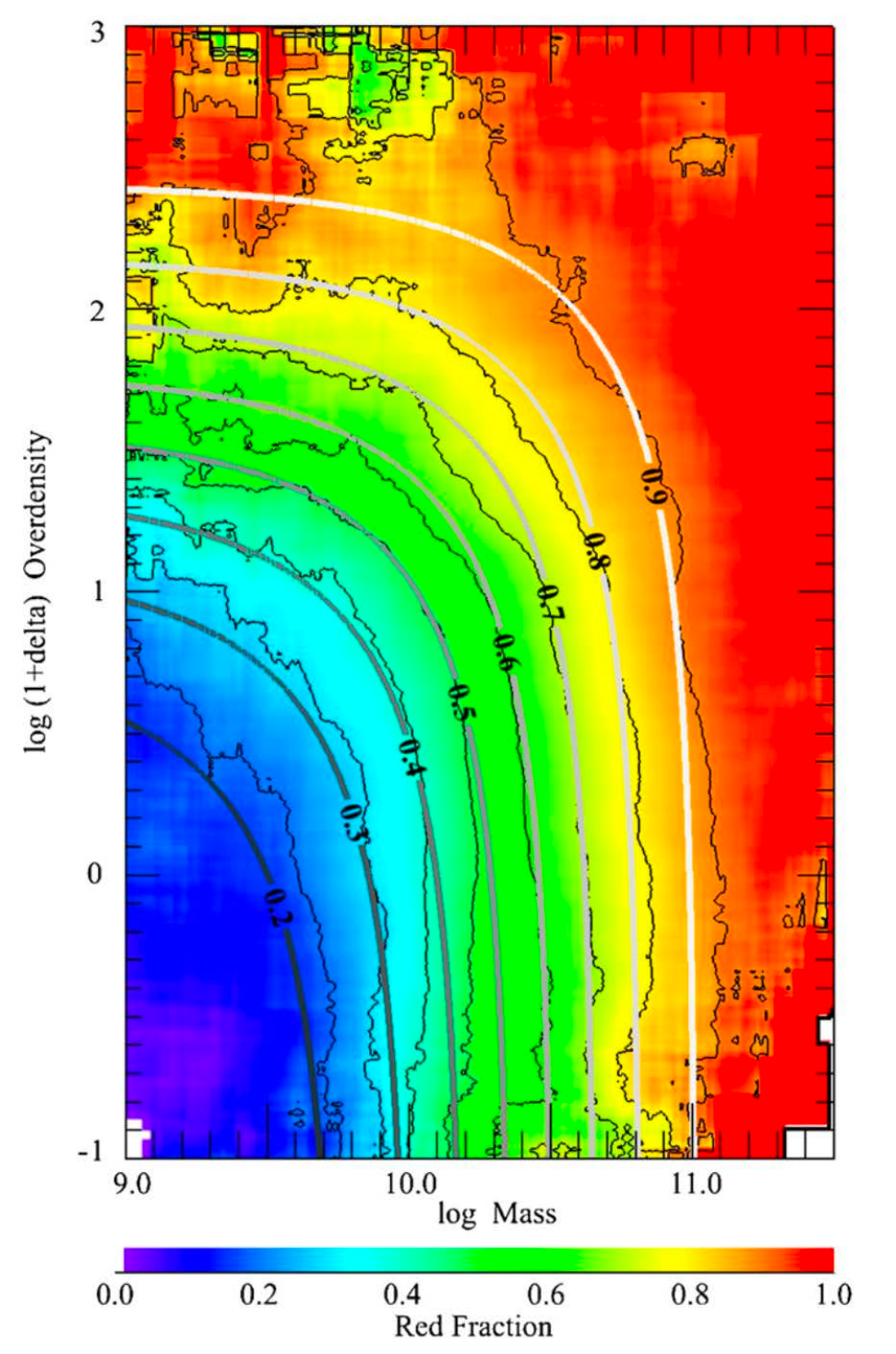}
\caption{Red (quenched) fraction map as a function of galaxy over-density ($Y$-axis) and stellar mass ($X$-axis). It is clear that star formation is most effective at lower masses and in lower density environments. Alternatively, quiescent systems are typically found at high stellar masses and within high density environments. Interestingly, increasing stellar mass at a fixed galaxy over-density increases the fraction of red/ quiescent systems, as does increasing galaxy over-density at a fixed stellar mass (for lower mass systems). This demonstrates that there are two, largely separable, routes to quenching galaxies from: (i) intrinsic (mass correlating) and (ii) extrinsic (environment correlating) mechanisms. This figure is reproduced from \cite{Peng2010}, but also see \cite{Baldry2006} for the first published result which establishes this duality in galaxy quenching.}\label{f12}
\end{figure}

Perhaps the most famous galaxy correlator to quenching is that of stellar mass (see \citealt{Baldry2006, Peng2010, Peng2012}). In Fig.~\ref{f12} we reproduce an important result from \cite{Peng2010}, which plots stellar mass against local galaxy over-density and color codes by the fraction of red (approximately `quenched') galaxies. It is evident that at essentially all environments, there is a strong trend whereby more massive galaxies are more frequently quenched. This is critical direct observational evidence for the high-mass end of the star formation efficiency relation discussed in Sect.~\ref{s3}. 

Moreover, this result also clearly demonstrates that there are two routes to quenching galaxies in the local Universe: (i) `mass quenching': which operates as a strong function of stellar mass, and is largely independent of environment; and (ii) `environment quenching': which operates as a strong function of local galaxy over-density, and is largely independent of stellar mass. We will discuss the latter at length in Sect.~\ref{s5}. For now we concentrate on mass-quenching.

Stellar mass is essentially just the integral of star formation over time in a system. Explicitly,

\begin{equation}
M_* = \int^{t_0}_{t_f} (1-R(t)) \, \cdot \, \mathrm{SFR}(t) \, dt,
\end{equation}

\noindent where $R(t)$ is the fraction of material returned to the ISM via stellar evolution (explicitly via stellar winds and supernovae), which may (at least in principle) be time dependent. Typical values of the return fraction are estimated to be in the range, $R \sim 0.25 - 0.5$; e.g., \citealt{Cole2000, Madau2014}). Note that stellar mass is sensitive to the total history of star formation and not just its instantaneous, contemporaneous value. Hence, a critical question emerges: \textit{why should the current state of star formation be constrained by it's history?}

There are many possibilities here. First, star formation could be `self-quenching', possibly via supernova feedback. Alternatively, gravitational effects within high mass galaxies could lead directly to suppression in star formation, perhaps via increased turbulence or ISM temperatures. Alternatively, stellar mass could merely correlate with some other more fundamental parameter, e.g. the mass of the dark matter halo or the mass of the central supermassive black hole. All of these potential explanations are discussed in detail in Part~II of this review. For now it suffices to appreciate that quenching is mass-dependent. This is required in order to bring the SFRD evolution (Sect.~\ref{s31}) and the halo mass--stellar mass relation (Sect.~\ref{s32}) into accord with observations, and (crucially) is seen directly in large populations of galaxies in the local Universe (i.e., Fig.~\ref{f12}).

\subsection{Bulge mass, central density \& central velocity dispersion}\label{s43}

\begin{figure}[ht] 
\centering
\includegraphics[width=0.95\textwidth]{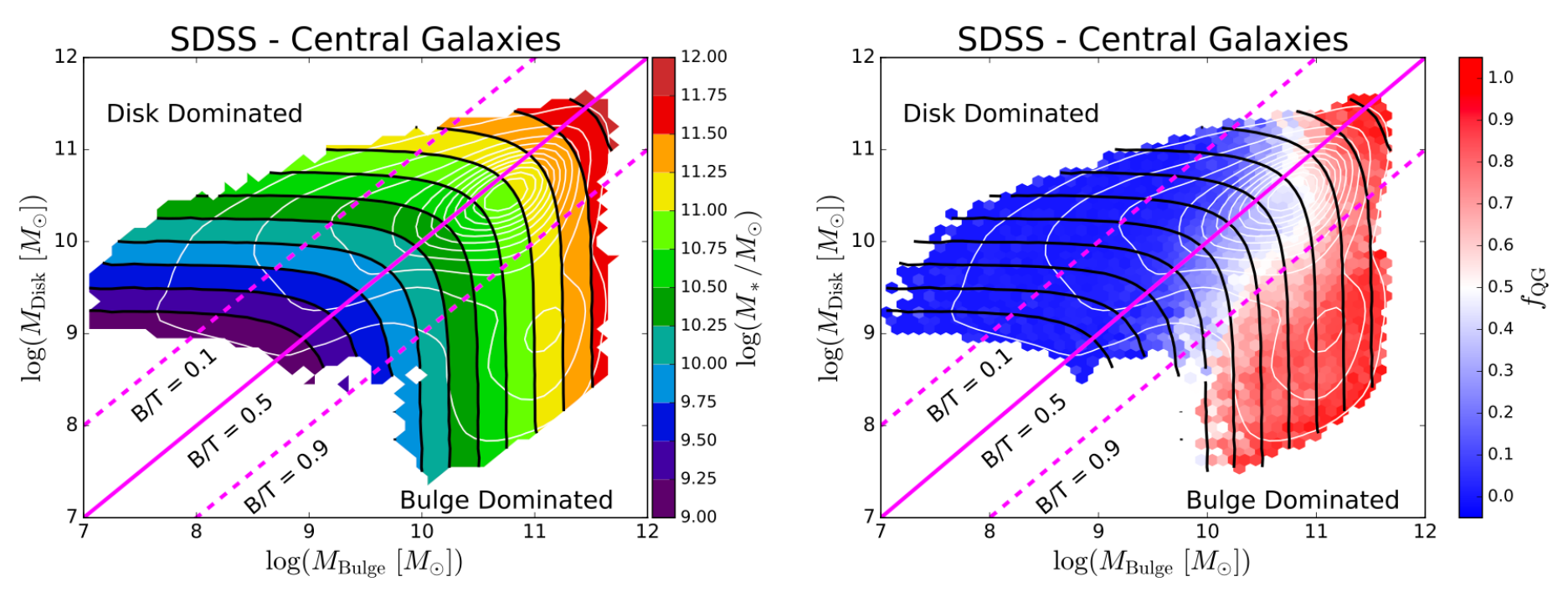}
\includegraphics[width=0.95\textwidth]{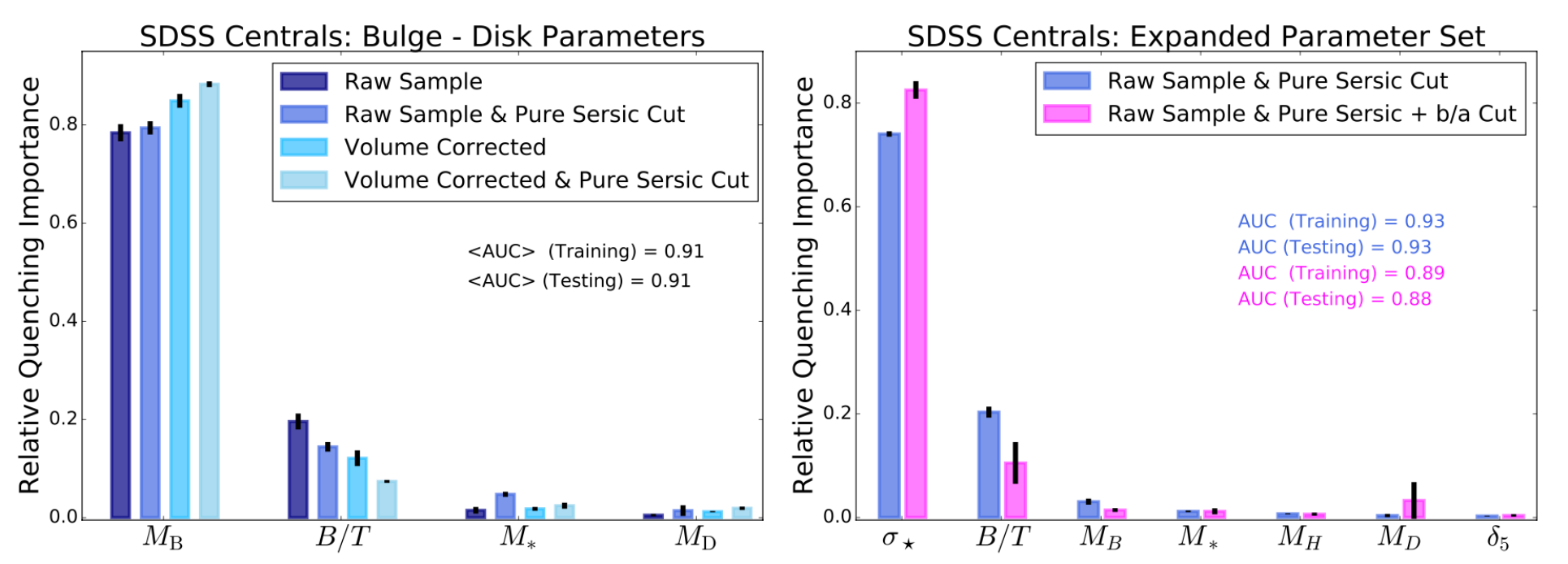}
\caption{\textit{Top left-hand panel:} The bulge\,--\,disc stellar mass plane for SDSS central galaxies, with stellar mass and bulge-to-total stellar mass ratio ($B/T$) indicated to guide the eye. \textit{Top right-hand panel:} The fraction of quenched galaxies across the bulge\,--\,disc plane. It is immediately clear that galaxies may be either star forming or quenched at fixed morphology or stellar mass, but that galaxies harboring high mass bulges are invariably quiescent. interestingly, this is independent of the mass of their disc structures. \textit{Bottom left-hand panel:} A Random Forest classification analysis to predict quenching in SDSS centrals. It is clear that bulge mass is the most predictive of the bulge\,--\,disc parameters for identifying quenched galaxies. Hence, knowing the properties at the center of galaxies is sufficient for ascertaining the level of star formation throughout the entire system. \textit{Bottom right-hand panel:} An expanded Random Forest classification quenching analysis, which demonstrates that central velocity dispersion is even more predictive than bulge mass for identifying quenched galaxies. Note the total lack of importance given to local galaxy density and halo mass, which is a strong indication that centrals quench independently of their environments. All panels are reproduced from \cite{Bluck2022}. }\label{f13}
\end{figure}

By combining morphological fitting with SED fitting, a number of works have gone beyond classical morphological measurements to reveal the stellar mass structure of galaxies, and their component bulges and discs (see, e.g., \citealt{Mendel2014, Lang2014, Dimauro2018}). Using these measurements, strong correlations between mass-based structure and star formation persist, which strongly imply that the reason for a close connection between galaxy structure and quenching (discussed in Sect.~\ref{s41}) is not a trivial consequence of star formation being primarily restricted to disc structures.

Moreover, the question of whether stellar mass or morphology are fundamentally linked to quenching appears to be resolved now. The answer is, no. In a series of works, the central mass density (e.g., \citealt{Cheung2012, Fang2013, Woo2015, Barro2017}) and bulge mass (e.g., \citealt{Bluck2014, Lang2014, Bluck2019, Bluck2022}) have been shown to be more strongly correlated to quenching than the total stellar mass, or the global morphology, of galaxies. Moreover, the strong correlations between stellar mass, morphology, and quenching have been shown to be spurious. This is established by revealing that the strong correlations between these parameters vanish when controlling for a more fundamental parameter -- that of the central mass density, or bulge mass.

By way of an example of this important result, in Fig.~\ref{f13} we reproduce work from \cite{Bluck2022}. In the top panels, the disc mass\,--\,bulge mass plane is displayed for central galaxies, color coded by total stellar mass (left panel) and the fraction of quenched galaxies (right panel). Central galaxies are defined here as the most massive galaxies within their dark matter haloes (see further discussion in Sect.~\ref{s4}; e.g., \citealt{Yang2007, Yang2009}). Central galaxies are now well established to have their quenching independent upon environment (e.g., \citealt{Peng2012}) and, hence, are an excellent population to utilize to investigate specifically intrinsic drivers to quenching.

It is clear that galaxies may be either star forming or quenched at a fixed stellar mass (as traced by solid black iso-mass contour lines). Similarly, galaxies may be either star forming or quenched at a fixed bulge-to-total stellar mass ratio ($B/T$) as well (as seen between the $B/T$ = 0.1 -- 0.9 dashed magenta lines). Additionally, varying disc mass at a fixed bulge mass has essentially no impact on the quenched fraction, except for a weak effect at the highest bulge masses. Conversely, varying bulge mass at a fixed disc mass leads to a very strong change in quenched fraction at all disc masses, whereby galaxies with the most massive bulges are most frequently quenched.

\begin{figure}[ht] 
\centering
\includegraphics[width=0.95\textwidth]{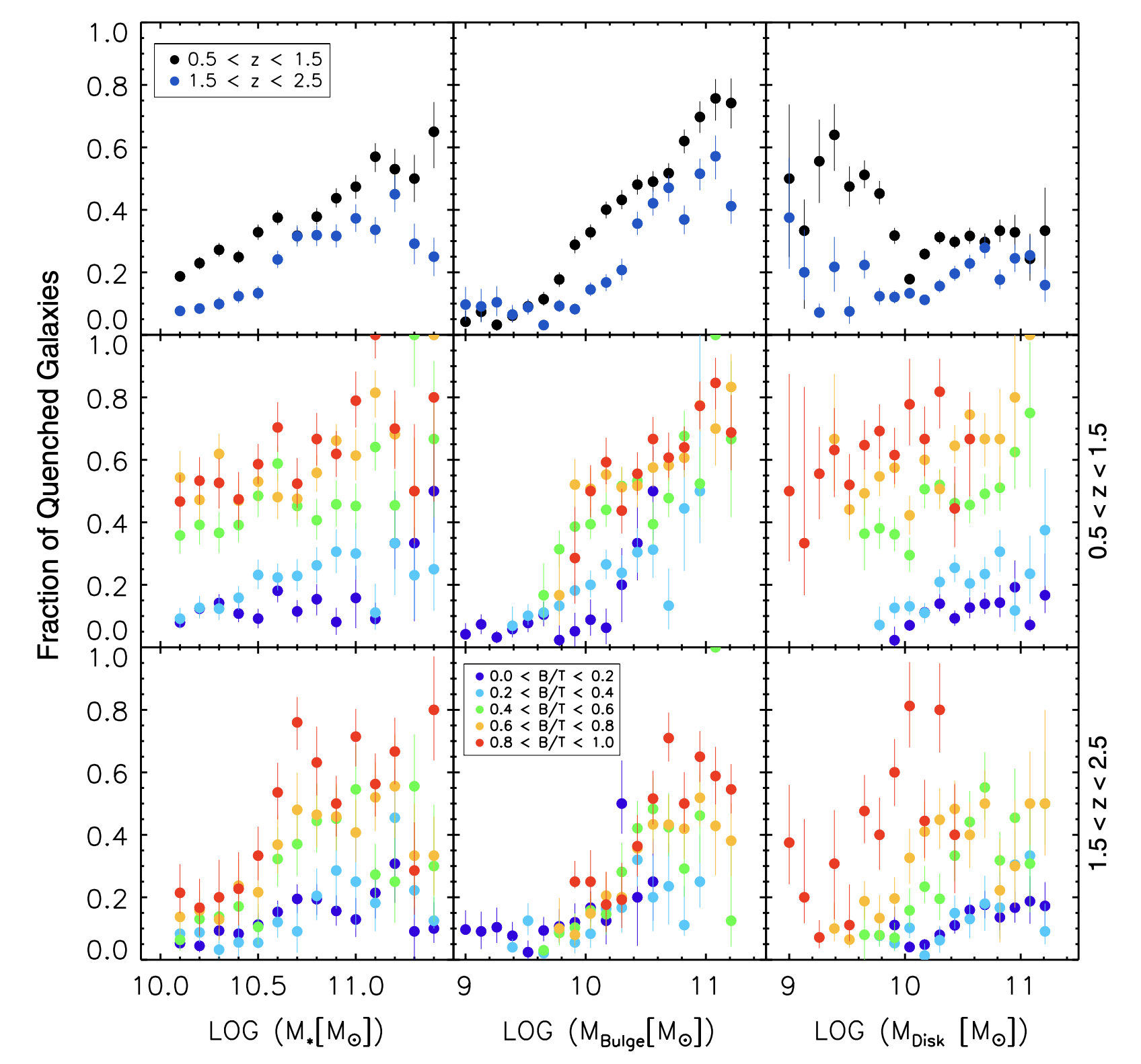}
\caption{The fraction of quenched galaxies from CANDELS (\citealt{Grogin2011, Koekemoer2011}) plot as a function of total stellar mass, bulge mass, and disc mass (from left to right). The top panel compares these relationships between redshift ranges. The lower panels compare these relationships across $B/T$ ratios (see legend). The center row shows results at intermediate redshifts ($0.5 < z < 1.5$) and the bottom row shows results at high redshifts ($1.5 < z < 2.5$). At all redshifts probed, the quenched fraction relationship with bulge mass is significantly tighter than with either total stellar mass or disc mass. This figure is reproduced from \cite{Lang2014}.}\label{f13b}
\end{figure}

In the bottom panels of Fig.~\ref{f13}, we show a series of random forest classification analyses to predict whether central galaxies will be star forming or quenched. See \cite{Bluck2022} for a detailed discussion on the method, including thorough testing with mock data and simulations. Briefly, the random forest framework pitches parameters against each other in a competition to reduce impurity in the sample. As such, this approach yields an effective route to control for nuisance parameters when establishing the casual links between any given parameter and the classification task at hand (here separating star forming from quenched galaxies). All random forest results should be interpreted as conditional on the included observables, sample selection, and central-galaxy focus here, rather than as a universal causal proof. 

On the left-hand panel, photometric parameters are used in central galaxies. It is clear that bulge mass is overwhelmingly the most predictive parameter of quenching at low-$z$, with only a very small importance given to $B/T$, $M_*$, and disc mass ($M_D$). On the right-hand panel, the number of parameters used to train the random forest is expanded by including halo mass (from abundance matching), local galaxy over-density, and central velocity dispersion. Here we see that central velocity dispersion is even more predictive of star formation quenching than bulge mass (consistent with \citealt{Bell2012, Wake2012, Cappellari2013, Bluck2016, Piotrowska2022}). Crucially, essentially no importance to quenching is found for stellar mass, morphological structure, environment, halo mass, or disc properties for central galaxies in the local Universe. This leads to a deep question: \textit{why should the star forming state of the whole central galaxy depend solely upon the kinematic conditions of its innermost region?}

One possible answer is a link to AGN feedback via the $M_\mathrm{BH}$\,--\,$\sigma_c$ relation (e.g., \citealt{Ferrarese2000, Saglia2016, Piotrowska2022}). We discuss this possibility in much detail in Part~II of this review series, from both a theoretical and observational perspective. Additionally, it is possible that the central region stabilizes the disk from gravitational collapse (e.g., \citealt{Martig2009, Gensior2020}). In either case, we now know that the strong empirical relationships between morphology and stellar mass and the quenching of galaxies arise from a deeper (more fundamental) connection between quenching and the central-most regions within galaxies, as probed by central density, central velocity dispersion, and bulge mass (see, e.g., \citealt{Bell2012, Wake2012, Cheung2012, Fang2013, Barro2013, Lang2014, Bluck2014, Bluck2016, Barro2017, Bluck2022}). Furthermore, these results are now known to hold up to at least cosmic noon ($z \sim 2$), and possibly up to the epoch of the very first quenched galaxies in the Universe at $z \sim 4 - 8$ (see, e.g., \citealt{Cheung2012, Fang2013, Barro2013, Lang2014, Barro2017, Bluck2022, Bluck2023, Bluck2024}).

As an example, in Fig.~\ref{f13b} we show an extension of the bulge mass results to higher redshifts, taken from \cite{Lang2014}. The fraction of quenched galaxies is plot as a function of stellar, bulge, and disc mass, separated out as a function of redshift (top panels) and $B/T$ structure (lower panels, split into intermediate [center panel] and high [bottom panel] redshifts). It is clear that the tightest relationship with quenched fraction is with bulge mass, with the least tight relationship being with disc mass. Hence, at all redshifts up to $z \sim 2$, bulge mass is a better predictor of the quenching of galaxies than disk mass or the total mass in stars. These conclusions are consistent with machine learning analyses of larger data sets (see, e.g., \citealt{Bluck2022, Bluck2023}) and are now known to hold up to $z \sim 8$ (see \citealt{Bluck2024}). Moreover, these results are also completely consistent with the extensive literature on the role of central stellar mass density in quenching central galaxies (see, e.g., \citealt{Cheung2012, Fang2013, Barro2013, Barro2014, Barro2017}).

\subsection{Galaxy size evolution}\label{s45}

The size of galaxies at a fixed stellar mass is known to decrease strongly as a function of redshift (e.g., \citealt{Franx2003, Toft2005, Trujillo2007, vanDokkum2008, Buitrago2008, Franx2008, Newman2012, vanderWel2014}). Moreover, at a fixed redshift and stellar mass, the sizes of quiescent galaxies are systematically smaller than that of their star forming progenitors (e.g., \citealt{Trujillo2006, vanderWel2014, Straatman2014, Straatman2015, Whitaker2017}). The latter result is consistent with the finding in the previous subsection that quenching proceeds as a stronger function of bulge mass (or central mass density) than with total stellar mass. Furthermore, these results on size evolution establish an evolutionary pathway in which all galaxies become less dense as cosmic time progresses.

\begin{figure}[ht] 
\centering
\includegraphics[width=1\textwidth]{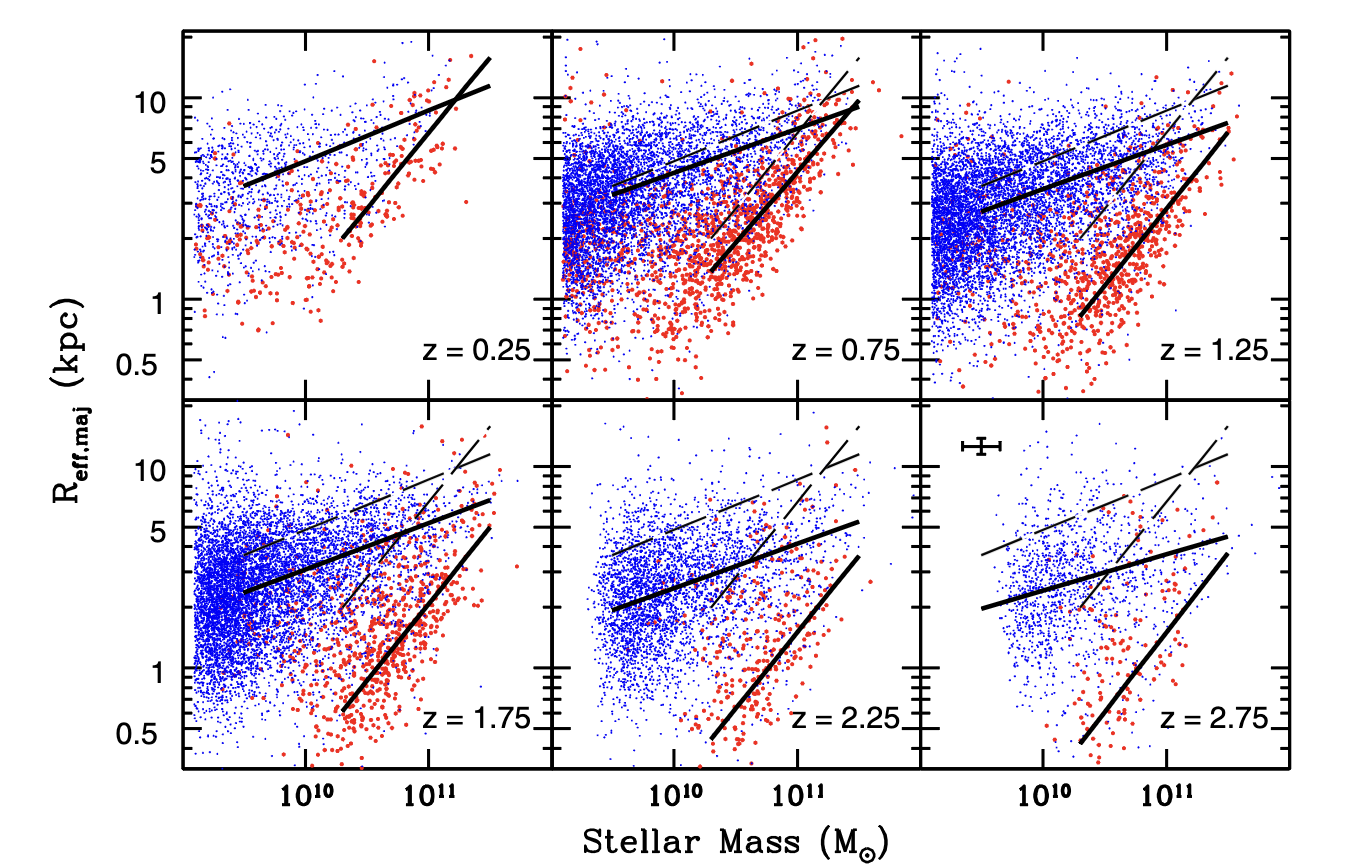}
\caption{Galaxy effective radius plot as a function of stellar mass for the HST-CANDELS survey, reproduced from \cite{vanderWel2014}. Star forming galaxies are shown in blue and quenched galaxies are shown in red. The evolution in this relationship is shown by varying bins of redshift, from low-$z$ in the top-left panel to high-$z$ in the bottom-right (as indicated within each panel). It is clear that quiescent systems are systematically smaller than their star forming counterparts at a fixed stellar mass and redshift. Moreover, both star forming and quiescent systems evolve to higher sizes as cosmic time progresses.}\label{f16b}
\end{figure}

In Fig.~\ref{f16b} we present the effective radius of galaxies as a function of stellar mass from HST-CANDELS (\citealt{Grogin2011, Koekemoer2011}), split between star forming and quiescent galaxies (shown in blue and red, respectively) and between redshift bins (shown in various panels from low-$z$ in the top-left panel to high-$z$ in the bottom-right panel). This figure is reproduced from \cite{vanderWel2014}. It is clear at all epochs up to $z = 2.75$ that quiescent galaxies are substantially smaller than their star forming counterparts, at a fixed stellar mass. Additionally, for both star forming and quiescent systems, there is marked evolution towards larger sizes as cosmic time progresses.

The strong size evolution observed in massive galaxies is generally interpreted as arising from a combination of dissipative early formation and subsequent dissipationless structural growth at late cosmic times. Within this framework, massive galaxies form compact stellar cores at high redshift through intense centrally concentrated star formation driven by gas-rich (`wet') mergers, violent disk instabilities, and/or compaction events (e.g., \citealt{Dekel2009, Dekel2014, Tacchella2015}). Following quenching, galaxies are thought to grow predominantly through gas-poor (`dry') minor mergers, which deposit stars at large radii and increase the effective radius more efficiently than the stellar mass. Ultimately, this scenario leads to inside-out growth and the gradual transformation of compact `red nuggets' into present-day massive ellipticals (e.g., \citealt{Naab2009, Oser2010, vanDokkum2010, Hilz2013}). 

Contemporary cosmological hydrodynamical simulations and semi-analytic models broadly reproduce this evolutionary picture, finding that ex-situ stellar accretion becomes increasingly important for the size growth of quenched massive galaxies at late times (e.g., \citealt{Wellons2016, Genel2018}).

\subsubsection{Can size evolution lead to confusion about quenching causes?}

Leveraging the insights on size evolution outlined above, \cite{Lilly2016} present an argument using an analytical model which seeks to explain the observed dependence of quenching on central density as a non-causal byproduct of size evolution. When galaxies stop forming stars (due to whatever quenching mechanism) their size and density is assumed to be set, modulo some minor post-quenching evolution via mergers. As such, one might expect quenched galaxies to be smaller, and hence denser, than star forming galaxies at a fixed stellar mass, irrespective of the actual physical mechanism of quenching.

This idea is important because it cautions against jumping prematurely to a physical interpretation of the results presented so far in this section. There are, however, numerous reasons to reject this hypothesis as the leading cause of the strong connection between star formation quenching and central mass density.

First, this type of argument applies equally to all classes of galaxies, yet (as we will see) satellites are not found to have their quenching strongly connected with central density, despite also quenching at earlier cosmic times, when they too are expected to be denser (see, e.g., \citealt{Bluck2016, Bluck2020b, Goubert2024, Goubert2025}). Additionally, a strong connection between contemporaneous quenching in the green valley and central density is also seen, whereby there is not a long enough time period to explain this phenomenology via size evolution (see \citealt{Barro2013, Zolotov2015, Bluck2016}). 

Perhaps most significantly, numerous works find that increased density in galaxies actually precedes quenching, with the progenitors of quenched systems being as compact (or even more so) than their quenched descendants. This is often described as the `compaction followed by quenching' sequence (see, e.g., \citealt{Barro2013, Lang2014, Zolotov2015, Barro2017, Fang2018, Tacchella2022}). This indicates that galaxies first evolve in morphology, structure, density, and kinematics (e.g., via a merger or violent disc instability) before quenching. Ultimately, this narrative reverses the direction of causality envisaged in \cite{Lilly2016}. That is, galaxies become more dense en route to quenching, rather than independent of it (via the global evolution in galaxy sizes). However, the best tests of the dependence of quenching on structure come from kinematics, which we discuss next.

\subsection{Kinematics}\label{s44}

The insights into quenching explored in this section have so far come from imaging, photometry, and SED fitting of galaxies (in addition to their component parts, i.e., bulges and discs). However, the most robust tracer of the mass distributions and structures within galaxies come from kinematics. For instance, \cite{Bell2012, Wake2012, Cappellari2013a, Cappellari2013b, Forster2006, Forster2009, Forster2020} find evidence for a connection between lack of ordered rotation (and/or presence of high stellar dispersion) and star formation quenching. This is broadly consistent with the insights gleaned from Fig.~\ref{f13} via photometry.

Integral-field spectroscopic surveys have revealed that the kinematic structure of galaxies is closely linked to quenching. In particular, the distinction between fast rotators (FRs) and slow rotators (SRs), established initially by the ATLAS3D survey (\citealt{Cappellari2011}), has transformed the classical picture of early- and late-type galaxies (e.g., \citealt{Emsellem2007, Cappellari2011, Cappellari2016}). Fast rotators are typically rotationally supported systems with disk-like morphologies, while slow rotators are dispersion-dominated system with spheroidal morphologies. The former are most frequently actively star forming, while the latter are most frequently (although not always) quenched.

\begin{figure}[ht]  
\centering
\includegraphics[width=1\textwidth]{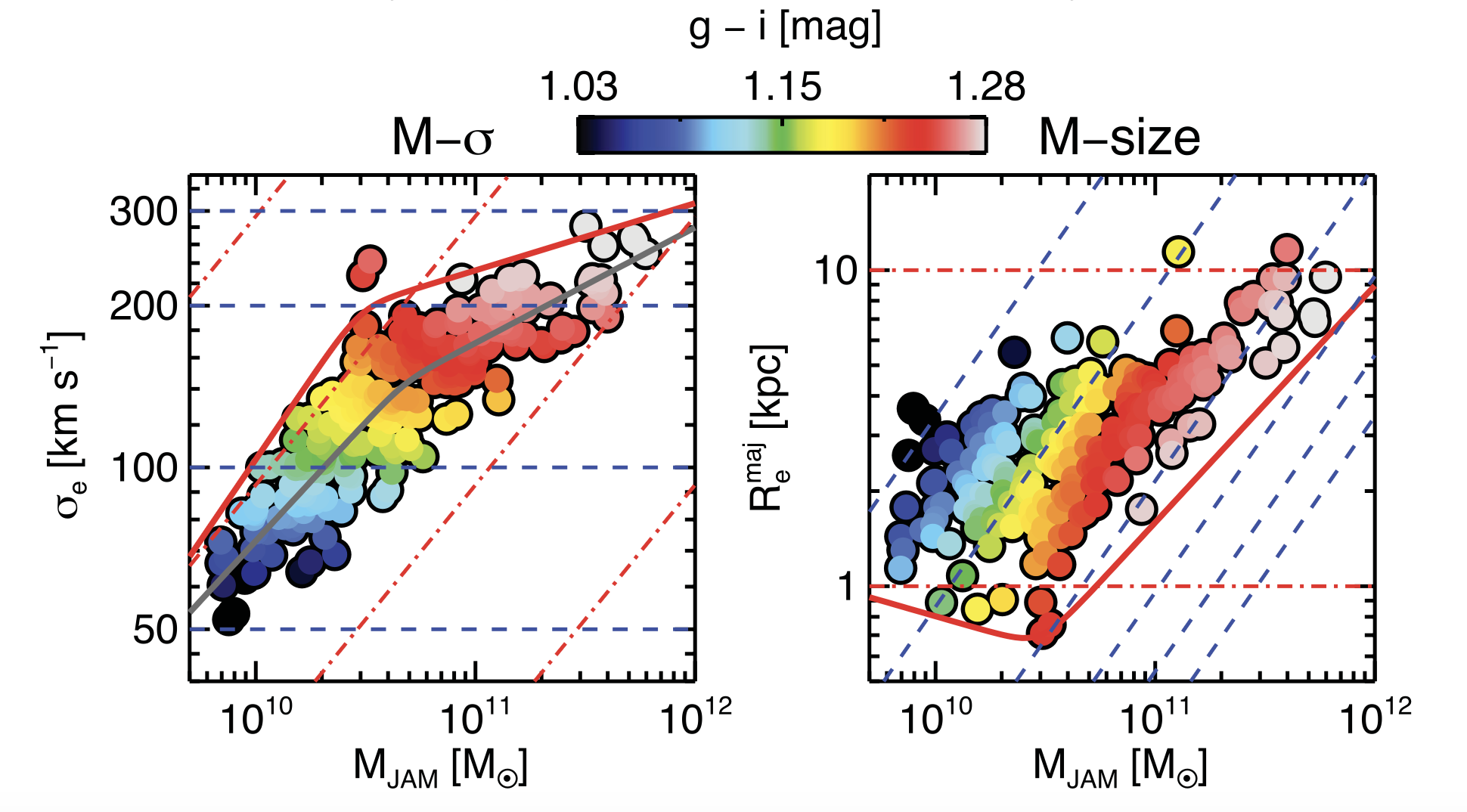}
\caption{Kinematic - quenching relations from ATLAS3D. {\it Left-hand panel:} Velocity dispersion within the effective radius plot as a function of dynamical mass (evaluated from kinematic fitting with the Jeans Anisotropic Modeling code, JAM; \citealt{Cappellari2008}). {\it Right-hand panel:} Effective radius plot as a function of dynamical mass. In both panels, data points are colored by the (g - i) restframe color of galaxies, which is a good proxy for sSFR. It is clear that quenching (i.e., transitioning from blue to red colors in this analysis) is more strongly connected to dynamical mass than size (see right panel). Moreover, quenching is also found to be much more connected to velocity dispersion than to total mass (see right panel). This figure is reproduced from \cite{Cappellari2013}.}\label{f17a}
\end{figure}

Observationally, quenched galaxies tend to exhibit elevated central stellar velocity dispersions, reduced rotational support, and higher bulge fractions relative to star-forming systems at a fixed stellar mass (e.g., \citealt{Franx2008, Wake2012, Cappellari2011, Cappellari2013, Brownson2020}). Surveys such as SAMI (\citealt{Bryant2015}, CALIFA (\citealt{Sanchez2012}), and MaNGA (\citealt{Bundy2015}) further demonstrate that quenching correlates strongly with increasing dynamical hotness, with slow rotators representing the extreme end of this trend (e.g., \citealt{vandeSande2017, vandeSande2018, Graham2018}). These results support a picture in which the buildup of dense, dispersion-dominated central regions (e.g., through mergers, compaction, or secular bulge growth) is closely connected to the suppression of star formation, consistent with both morphological quenching and SMBH-regulated feedback scenarios (e.g., \citealt{Martig2009, Gensior2020, Gensior2021, Cappellari2016, Bluck2022, Piotrowska2022, Brownson2022, Lim2025}).

The evolution of galaxy kinematics from cosmic noon to the present epoch provides additional insight into how quiescence emerges. IFU surveys such as SINS (\citealt{Forster2009}), KMOS3D (\citealt{Forster2019}), KROSS (\citealt{Stott2016}), and MOSDEF (\citealt{Reddy2015}) reveal that massive star-forming galaxies at high redshift are typically gas rich, turbulent, and dynamically thick systems, with elevated intrinsic velocity dispersions and lower $V/\sigma$ ratios than present-day star forming disks (e.g., \citealt{Forster2009, Wisnioski2015, Turner2017, Simons2017}). Over time, the star forming population evolves to greater rotational support (higher $V/\sigma$), while the quenched population evolves to greater pressure support (lower $V/\sigma)$.

Importantly, several studies now indicate that many high-redshift quiescent galaxies retain substantial rotational support and disc-like morphologies, implying that quenching rarely leads to the full destruction of disks (e.g., \citealt{vanDerWel2011, Newman2015, Belli2017, Newman2018}). Therefore, the emergence of quiescence at early cosmic times does not necessarily require an immediate transformation into classical slow-rotator ellipticals. Instead, observations favor an evolutionary pathway in which compact, turbulent star-forming discs first quench into compact rotating remnants, with subsequent dry merging and dynamical heating gradually producing the massive slow-rotator population observed in the local Universe (e.g., \citealt{Naab2014, Penoyre2017, Cappellari2016}). Ultimately, a high central velocity dispersion appears critical for quenching to occur, but the level of rotation is not directly constraining of quenching.

\begin{figure}[ht]  
\centering
\includegraphics[width=0.65\textwidth]{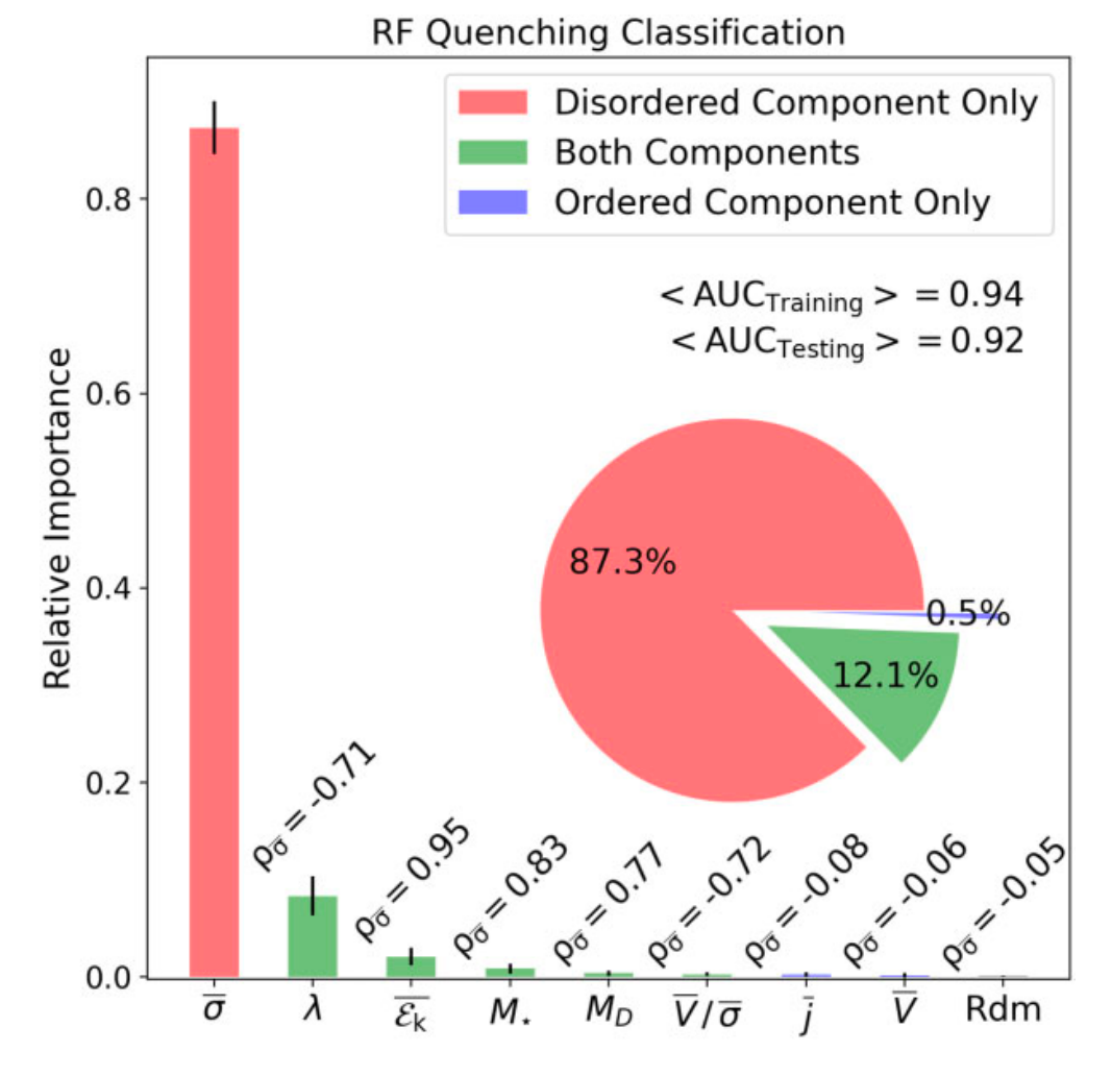}
\caption{A Random Forest classification analysis to predict whether central galaxies are star forming or quenched, based upon input kinematic parameters from the MaNGA survey. It is clear that central velocity dispersion is by far the most predictive parameter over quenching, indicating a close connection of galaxy quiescence to the disordered kinematic component. Other kinematical parameters related to the ordered rotation (e.g., the spin parameter, $\lambda$, or the angular momentum of the galaxy, $J$), or combinations of both (e.g., the total dynamical mass, $M_D$, or the specific kinetic energy of the system, $\mathscr{E}_k$), are far less constraining of central galaxy quenching. The overall dominance of the disordered component for predicting quenching is emphasized by the inset pie plot, which shows that the ordered component contributes negligibly to quenching prediction power. This figure is reproduced from \cite{Brownson2022}.}\label{f15}
\end{figure}

In Fig.~\ref{f17a}, we present the relationships between stellar velocity dispersion and dynamical mass (left-hand panel) and effective radius and dynamical mass (right-hand panel) from ATLAS3D, reproduced from \cite{Cappellari2013}. Both panels are color coded by (g-r) restframe color, which is a good proxy for sSFR. It is clear that the transition from blue to red galaxies is a stronger function of dynamical mass than galaxy size, and a stronger function of velocity dispersion than total dynamical mass. This important result clearly indicates the primacy of central velocity dispersion for constraining quenching, compared to both the total mass and effective size of galaxies.

Building upon this result, in Fig.~\ref{f15} we present results from a random forest classification analysis leveraging kinematic parameters derived from $\sim$2000 galaxies from the MaNGA survey (\citealt{Bundy2015}), reproduced from \cite{Brownson2022}. Briefly, this figure assesses the following kinematical parameters in their predictive power over quenching, all evaluated within 1\,$R_e$ from kinematic modeling: velocity dispersion ($\sigma$), the dimensionless spin parameter ($\lambda$), angular momentum ($J$), mean rotational velocity ($V$), the ratio of velocity dispersion to mean rotational velocity ($\sigma / V$), stellar mass of the galaxy ($M_*$), average kinetic energy ($\mathcal{E}_k$), and dynamical mass ($M_D$).

From the random forest analysis, it is clear that velocity dispersion is by far the most predictive kinematic parameter for quenching. This is consistent with the results from Fig.~\ref{f13}, which is focused primarily on photometric measurements. Furthermore, pressure supported kinematics dominate over rotational parameters (and parameters linked to both kinematic components within galaxies), as clearly seen by the pie plot inset within this figure. It is especially important to note the lack of importance given by the random forest classifier to the total dynamical mass (as seen previously found in Fig.~\ref{f17a}), stellar mass, and the $V_\mathrm{rot}/\sigma$ ratio, all of which have been previously considered as fundamental to quenching in the kinematics literature. As a reminder, results from random forest analyses are always limited by the chosen parameters to include and the sample selection. In this analysis, a wide variety of kinematical parameters are assessed for a heterogeneous sample of local galaxies taken from the MaNGA IFU survey.

As a result of the above discussion, it becomes apparent that central galaxy quenching is intimately linked to the disordered kinetic energy component of galaxies (rather than the ordered, rotational component). This fundamental observational property of quenching must be explained by any successful galaxy evolution model.

In summary of the entirety of this section, there are a host of galaxy parameters known to correlate strongly with central galaxy quenching, including morphology (e.g., \citealt{Driver2006, Cameron2009b, Wuyts2011, Omand2014}), stellar mass (e.g., \citealt{Baldry2006, Peng2010}), bulge mass (e.g., \citealt{Lang2014, Bluck2014, Bluck2022}), central mass density (e.g., \citealt{Cheung2012, Fang2013, Barro2017}), and central velocity dispersion (e.g., \citealt{Wake2012, Cappellari2013, Bluck2016}). However, the only parameter to survive rigorous control of nuisance variables is that of central velocity dispersion (see, \citealt{Bluck2022, Brownson2022, Piotrowska2022}). As such, it is critical to understand why the level of star formation within a galaxy is so tightly connected to it's central, disordered kinematics. We will revisit this question again periodically throughout this review series.

\section{Environmental correlators to quiescence}\label{s5}

An important observational result is that, once star forming and quiescent galaxies are considered separately, many of their internal scaling relations show surprisingly little dependence on environment. For star-forming galaxies, the mass - metallicity relation (e.g., \citealt{Cooper2008}), the star-forming main sequence (e.g., \citealt{Wijesinghe2012, Bluck2014}), and the stellar Tully - Fisher relation (e.g., \citealt{Mocz2012}) are all observed to vary only weakly with local galaxy environmental. Instead, the primary environmental trend appears in the relative abundance of star forming and quiescent systems, with dense environments hosting a substantially larger fraction of red, quenched galaxies \citep{Baldry2006, Peng2010, Peng2012, Wetzel2013}. 

This suggests that environmental processes act predominantly by causing galaxies to transition between the star forming and quenched populations, rather than by gradually modifying the internal properties of star forming (or quenched) systems. One notable exception is within the quenched population itself, where the relative fractions of fast and slow rotators vary significantly with environment, particularly in dense cluster cores (see, e.g., \citealt{Cappellari2011}). Taken together, these observations suggest that environmental quenching is often (though not necessarily always) a relatively rapid transformation process, compared to the secular evolution timescales of galaxy scaling relations.

\subsection{Quantifying the environments of galaxies}

In this subsection we review common methods used in the quenching literature to identify and quantify the environments in which galaxies reside. Since this is a vast topic, we focus here on only on the methods which have been most critical for the field of galaxy quenching (for broader environmental reviews, see \citealt{Blanton2009, Muldrew2012, Wechsler2018, Cortese2021}). 

Observed and simulated galaxy distributions are highly non-uniform, exhibiting voids, filaments, sheets, groups, and clusters -- collectively known as the cosmic web (e.g., \citealt{Bond1996, Hahn2007, Libeskind2018}). The local density of galaxies varies by over three orders of magnitude, and these environmental differences strongly impact galaxy formation, star formation, and, ultimately, quenching (e.g., \citealt{Baldry2006, Peng2012}).

\subsubsection{Local galaxy (over-)density}

Perhaps the simplest environmental metric is the local galaxy density (e.g., \citealt{Dressler1980, Baldry2006, Peng2010}). Because the mean density evolves with cosmic expansion, one generally measures an over-density relative to the average density at a given epoch (e.g., \citealt{Cooper2005, Scoville2013}). 

The most common approach is the `$n$th nearest-neighbor' method. In 3D simulations, the local over-density may be written as:

\begin{equation}
\delta_{n, \, \mathrm{3D}} = \log_{10}(\rho_n) - \langle \log_{10}(\rho_n) \rangle_z, \quad \rho_n = \frac{3n}{4\pi R_{n,\mathrm{3D}}^3}
\end{equation}

\noindent where $R_{n,\mathrm{3D}}$ is the distance to the $n$th nearest neighbor. Observationally, one is restricted to projected separations and redshift-space information. Hence, local densities are typically measured in 2D as follows:

\begin{equation}
\delta_{n, \, \mathrm{2D}} = \log_{10}(\Sigma_n) - \langle \log_{10}(\Sigma_n) \rangle_z, \quad \Sigma_n = \frac{n}{\pi R_{n,\mathrm{2D}}^2} \, ,
\end{equation}

\noindent with an additional line-of-sight velocity cut ($\Delta v \lesssim 1000$\,km/s) to reduce projection effects. Small values of $n$ probe close pairs and compact groups, while larger values trace cluster-scale environments (or the broader cosmic web). As such, one often needs to combine multiple scales for the nearest neighbor density metrics, to encapsulate the full range of environments. Alternative approaches include fixed apertures, Voronoi tessellation, and kernel density estimation (e.g., \citealt{Marinoni2002, Darvish2015, Cortese2021}).

\subsubsection{Identifying groups \& clusters}

Beyond local density measurements, one often seeks to identify physical galaxy associations such as groups and clusters. The most widely used method is the friends-of-friends (FoF) algorithm (e.g., \citealt{Davis1985, Berlind2006, Robotham2011}), which links galaxies separated by less than a chosen linking length.

In simulations, the 3D FoF criterion is generally written as:

\begin{equation}
\Delta X_\mathrm{3D} < l_\mathrm{3D}, \quad l_\mathrm{3D} = b \, \bar{n}_\mathrm{3D}(z)^{-1/3}
\end{equation}

\noindent where $\bar{n}_\mathrm{3D}$ is the mean galaxy number density and $b \sim 0.2$ is the dimensionless linking parameter (e.g., \citealt{Springel2005a}). 

For observational data, separate linking lengths are typically used in projected and redshift space to account for the `finger-of-God' effect. Explicitly, the condition is given by:

\begin{equation}
(\Delta X_\mathrm{2D} < l_\mathrm{sky}) \,\, \& \,\, (\Delta z < l_z) \,\,\,\,  {\rm where,} \,\,\,\, l_\mathrm{2D} = b_\perp \, \bar{n}_\mathrm{2D}(z)^{-1/2} \,\,\,\, \& \,\,\,\, l_z = b_{\parallel}(v) \, (1+z)/c
\end{equation}

\noindent For the perpendicular (sky) separation, the 2D average separation between galaxies at a given redshift is factored out (as in 3D). Then the fractional threshold, $b_\perp$, is set, with typical values of $\sim$0.07 - 0.14 (e.g., \citealt{Robotham2011}). Typical velocity cuts in the $z-$dimension are typically taken to be $b_{\parallel}(v) \sim 500$\,km/s (e.g., \citealt{Berlind2006, Robotham2011}). 

Alternative methods include virial group finding (e.g., \citealt{Yang2005, Tinker2011, Bluck2025}) and photometric cluster finders such as RedMapper, which leverage the red-sequence populations of clusters (e.g., \citealt{Gladders2000, Rykoff2014}).

Once groups are identified, galaxies are commonly classified as centrals or satellites. Centrals are usually defined as the most massive (or most luminous) galaxy in a dark matter halo, while all other galaxies are classified as satellites (e.g., \citealt{Yang2007, Yang2009, George2012}). This distinction is important because centrals and satellites experience different quenching pathways. We focused on centrals in the previous subsection at various points because these systems are observed to have little quenching dependence on environment. Conversely, satellites are observed to have a strong quenching dependence on environment, as we will explore in this section.

\subsubsection{Halo masses}

A major goal of observational studies of galaxy environmental is to estimate dark matter halo masses, which enables a connection between observations and the fundamental theory of $\Lambda$CDM. Direct methods include strong lensing, galaxy dynamics, satellite kinematics, X-ray measurements, and the Sunyaev-Zeldovich effect (e.g., \citealt{Rubin1978, Carlberg1996, Mandelbaum2006, Hoekstra2013}). However, these methods are observationally expensive and cannot generally be applied to all galaxies in large surveys.

Consequently, statistical approaches are widely employed and have been critical for the study of galaxy quenching. One of the most common is abundance matching (AM; e.g., \citealt{Moster2010, Behroozi2010}), which assumes a monotonic relationship between galaxy (or total group) stellar mass and halo mass. Schematically, the halo mass is estimated as follows:

\begin{equation}
M_{\mathrm{Halo-AM},i} = M_{\mathrm{Halo\,List}} [ \, n(M_{\mathrm{Halo\,List}}) = n(M_{\ast,i}) \, ]
\end{equation}

\noindent where halo masses are assigned by matching cumulative number densities ($n$) of galaxies and dark matter haloes. Other approaches include halo occupation distribution (HOD) modeling (e.g., \citealt{Berlind2002, Zheng2005}) and direct calibrations from simulations, often via machine learning methods (e.g., \citealt{Yang2009, Hahn2024, Bluck2025}).

Typical halo mass uncertainties for wide-field survey data are of order $\sim0.2$--0.5\,dex (depending on methodology). Halo masses additionally permit estimates of the virial radius and virial velocity, enabling the identification and analysis of satellite positions within the phase-space of the dark matter halo. This is a particularly important tool for understanding environmental quenching (e.g., \citealt{Woo2013, Goubert2024, Goubert2025}).

\subsection{The color--density relation}\label{s51}

The first works linking quenching to environment were \cite{Butcher1978, Dressler1980, Butcher1984}. In these works it is established that redder (and more spheroidal) galaxies lie preferentially closer to the center of high mass clusters than bluer (and more disc-dominated) galaxies. Moreover, these works also demonstrate that the optical color of satellite galaxies within clusters varies substantially with cosmic time. Ultimately, these pioneering works introduce the idea of star formation suppression in dense environments, emphasizing its link to cosmological evolution and the structural transition of satellite galaxies.

\begin{figure}[ht] 
\centering
\includegraphics[width=1\textwidth]{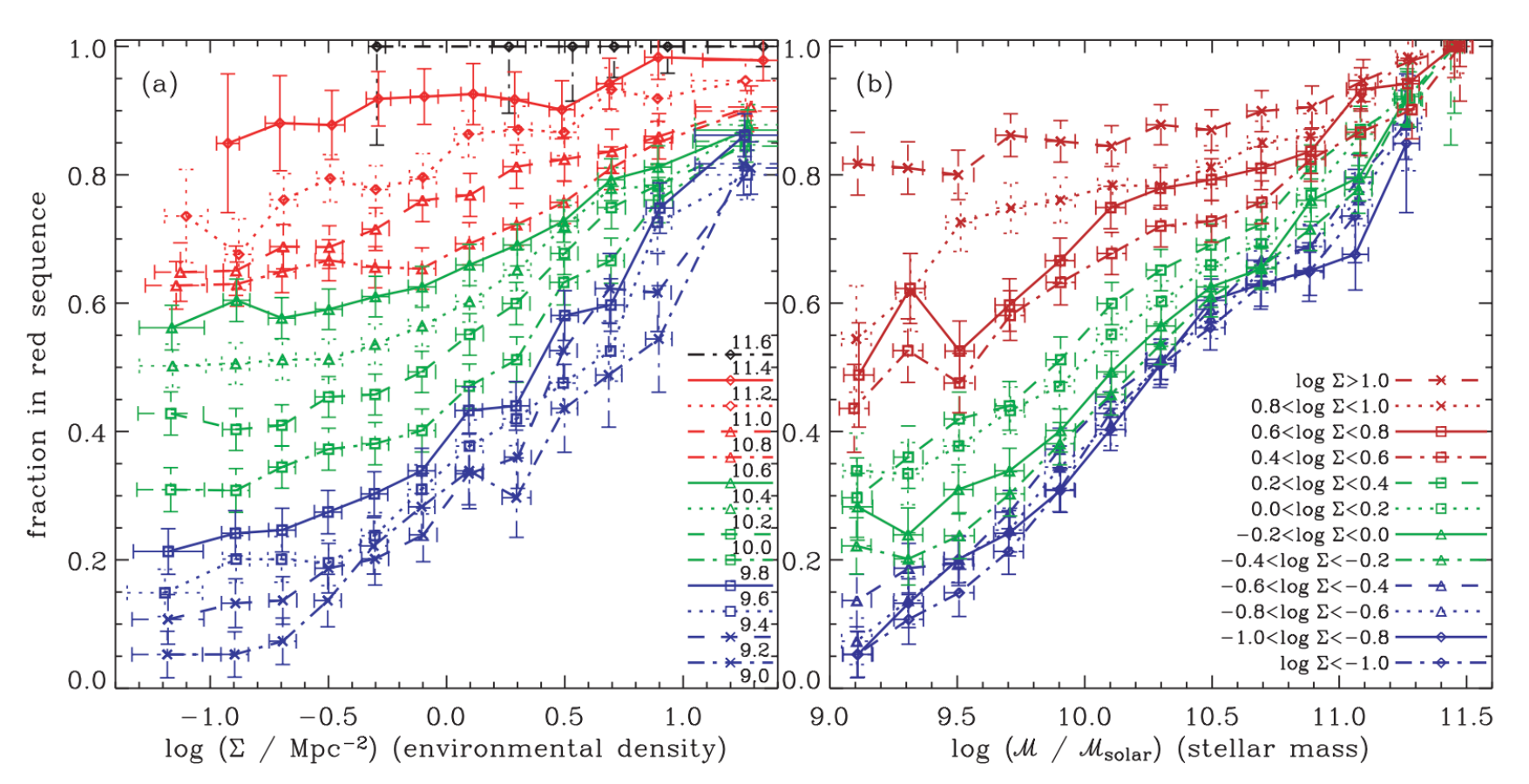}
\caption{The fraction of red sequence galaxies plot as a function of local galaxy density, split into ranges of stellar mass (left panel); and as a function of stellar mass, split into ranges of local galaxy density (right panel). It is clear that increasing density of cosmic environment is positively correlated with quiescence at a fixed stellar mass. Moreover, the impact of environment on quenching is noticeably more significant on low stellar mass systems. This figure is reproduced from \cite{Baldry2006}.}\label{f16}
\end{figure}

In Fig.~\ref{f16}, we reproduce a more modern analysis from \cite{Baldry2006}, which shows how the fraction of red sequence galaxies varies as a function of local galaxy density and stellar mass, across a wide range in both parameters. It is clear that at a fixed stellar mass, the fraction of red galaxies increases with galaxy density, but that this trend is steeper at lower stellar masses. Similarly, at a fixed local galaxy density, the fraction of quenched galaxies increases with stellar mass, but this trend is steeper at lower galaxy densities. These results strongly imply a deep connection between quenching and the environment in which galaxies reside. Additionally, see back to Fig.~\ref{f12} for a similar result from \cite{Peng2010}, who extend the analysis to demonstrate an important mathematical property of quenching. Explicitly, they demonstrate that mass and environmental quenching are separable and, to leading order, independent.

The environmental dependence of quenching has also been investigated at intermediate and high redshift using a variety of large spectroscopic and photometric surveys. Early spectroscopic work from DEEP2 demonstrated that the fraction of quenched galaxies increases with local galaxy density out to at least z $\sim$ 1, although stellar mass remains a dominant predictor of galaxy properties (e.g., \citealt{Cooper2010}). Subsequent studies using photometric-redshift-selected samples in UDS, ZFOURGE, and related surveys found that both stellar mass and environment contribute to quenching, with environmental effects becoming increasingly important for satellite galaxies at lower stellar masses (e.g., \citealt{Quadri2012, Tomczak2017}). 

Extending these analyses to ($z \gtrsim 1.5$), \citet{Kawinwanichakij2014, Kawinwanichakij2016, Kawinwanichakij2017} and \citet{Papovich2018} all report enhanced quenched fractions in over-dense environments. This implies that environmental quenching mechanisms must already operate during the epoch of peak cosmic star formation (i.e., at and around cosmic noon). More recently, the ORELSE galaxy survey has provided improved constraints in massive structures at intermediate redshifts, confirming elevated quenched fractions in dense environments (\citealt{Lemaux2022}). It is important to appreciate that many of these studies rely on photometric redshifts and, hence, the environmental measurements suffer from much greater uncertainties than low-redshift spectroscopic surveys. Nonetheless, collectively these works indicate that environmental quenching is already established by $z \sim 2$, operating alongside the strong dependence of quenching on stellar mass and internal structure/ kinematics.

Numerous mechanisms to account for enhanced quenching in high density environments have been proposed. These include gas removal through ram pressure stripping of the ISM (e.g., \citealt{Gunn1972, Balogh2000, vandenBosch2008, Kapferer2009, Cramer2019, Lotz2019}), gas stripping from tidal torques induced in galaxy--galaxy harassment (e.g., \citealt{Moore1996, Moore1998b, Moore1999, Peng2012}), and the starvation of galaxies through a loss of their CGM, due either to a low-density form of ram pressure stripping (e.g., \citealt{vandenBosch2008}) or else due more simply to the location within halo (e.g., \citealt{Cortese2009, Henriques2015, Fang2018}). The latter is a result of satellites orbiting the center of mass of groups and clusters, rather than residing near the gravitational minimum. Hence, satellites are no longer effectively fed by cold gas streams, or cooling flows (e.g., \citealt{Dekel2006, Dekel2009, Henriques2015}). We discuss all of these theoretical mechanisms for environment-triggered quenching in much more detail in Part~II of this review series. For now we concentrate on exploring further observational constraints on environmental quenching.

\subsection{The mass vs. environment dichotomy for centrals and satellites}\label{s52}

\begin{figure}[ht]  
\centering
\includegraphics[width=0.75\textwidth]{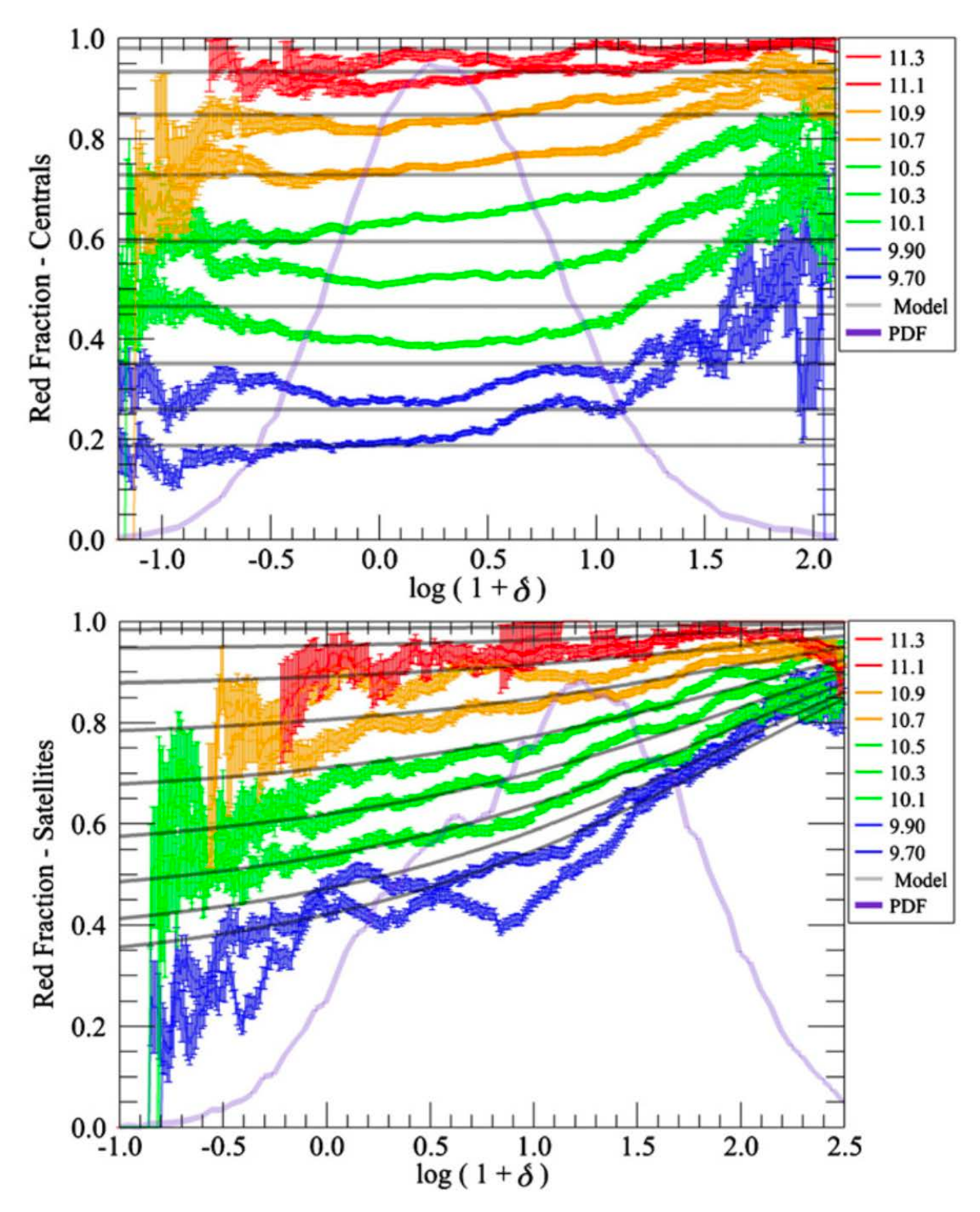}
\caption{The dependence of the red fraction of galaxies on local galaxy over-density in bins of stellar mass for centrals (top panel) and satellites (bottom panel). For central galaxies, there is a very weak dependence of quiescence on environment, except at the highest local densities and in lower stellar mass systems. Conversely, for satellites, environment is strongly correlated with quiescence across a much wider range in galaxy stellar masses and over the full range in over-densities. Both classes of galaxies exhibit a strong dependence on stellar mass, although for satellites this is notably weaker within higher density regions of the cosmic web. These results lead to the idea that central galaxies quench primarily through intrinsic (mass-correlating) means, yet satellites may quench either through environmental or intrinsic means, with environment being the dominant factor. This figure is reproduced from \cite{Peng2012}.}\label{f17}
\end{figure}

In \cite{Peng2012} an important insight into environmental quenching was revealed. Specifically, the dependence of quiescence on local galaxy over-density at a fixed stellar mass is much stronger for satellites than for centrals. In Fig.~\ref{f17}, we reproduce these key quenching results from \cite{Peng2012}. The top panel shows the red fraction\,--\,galaxy over-density relationship split into bins of stellar mass for centrals, with the bottom panel showing the same plot for satellites. For centrals, there is little dependence on environment at a fixed stellar mass at most values of over-density, with a ramp up only visible at the highest over-densities for low and intermediate stellar masses. Alternatively, for satellites, there is a strong dependence on over-density across the full range in over-density, although this dependence is steeper for lower mass satellites.

These insights lead to the central\,--\,satellite dichotomy of quenching. Canonically, it has been suggested that centrals `mass-quench' whereas satellites `environment-quench' (see \citealt{Peng2010, Peng2012}). This suggests the need for two distinct channels for quenching different classes of galaxies. Nonetheless, at a fixed over-density, satellites are progressively more quiescent at higher masses (although less so than for centrals). Hence, a slightly more accurate depiction would be that all galaxies may `mass-quench' but only satellites can `environment-quench'.

It should be noted, however, that in light of the results of Sect.~\ref{s4}, we now know that stellar mass is not fundamental to central galaxy quenching, since its predictive power over whether galaxies are star forming or quenched may be removed entirely by binning with central density, bulge mass, or central velocity dispersion (e.g., \citealt{Cheung2012, Fang2013, Bluck2014, Lang2014, Bluck2016, Bluck2022, Piotrowska2022, Brownson2022}). As such, it will be interesting to revisit the central\,--\,satellite dichotomy of quenching later in this review series, in order to assess whether this feature persists when a more fundamental view of `mass-quenching' is taken. Furthermore, there is nothing so far to suggest that local over-density of galaxies is the {\it fundamental} environmental factor at play here. We explore this question in relation to simulations in Part~II of this review.

\subsection{Halo mass and location within the halo as drivers of quenching}\label{s53}

Ideally, one wants to link the formation of galaxies to the formation of dark matter haloes, since the former reside and evolve within the latter. Moreover, this approach leads to important new insights on galaxy quenching. Using dark matter halo masses estimated from abundance matching (\citealt{Yang2007, Yang2009}), \cite{Woo2013} investigate the dependence of the quenching of central and satellite galaxies on dark matter halo mass, and the location within the halo for satellite galaxies.

\begin{figure}[ht]  
\centering
\includegraphics[width=0.49\textwidth]{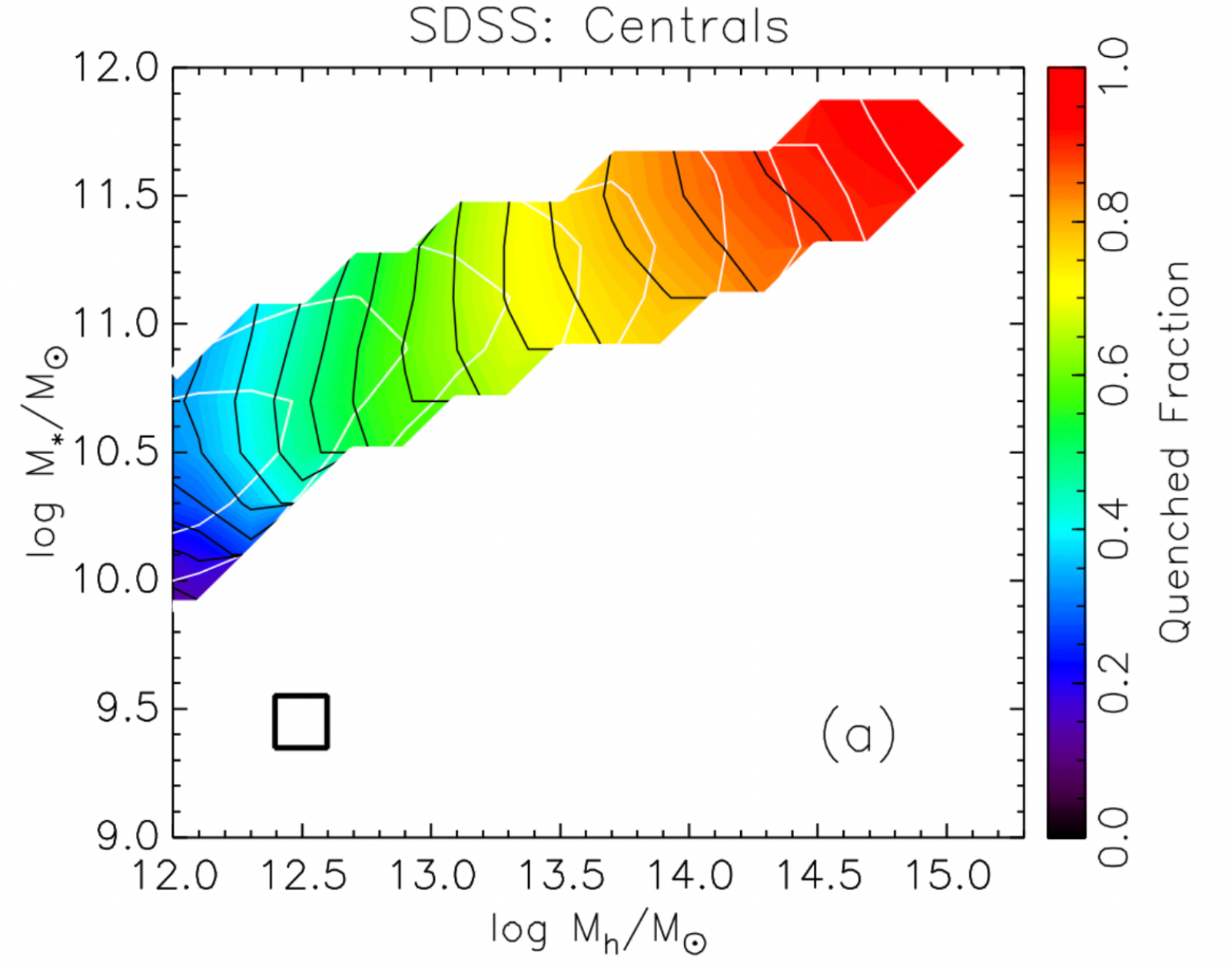}
\includegraphics[width=0.49\textwidth]{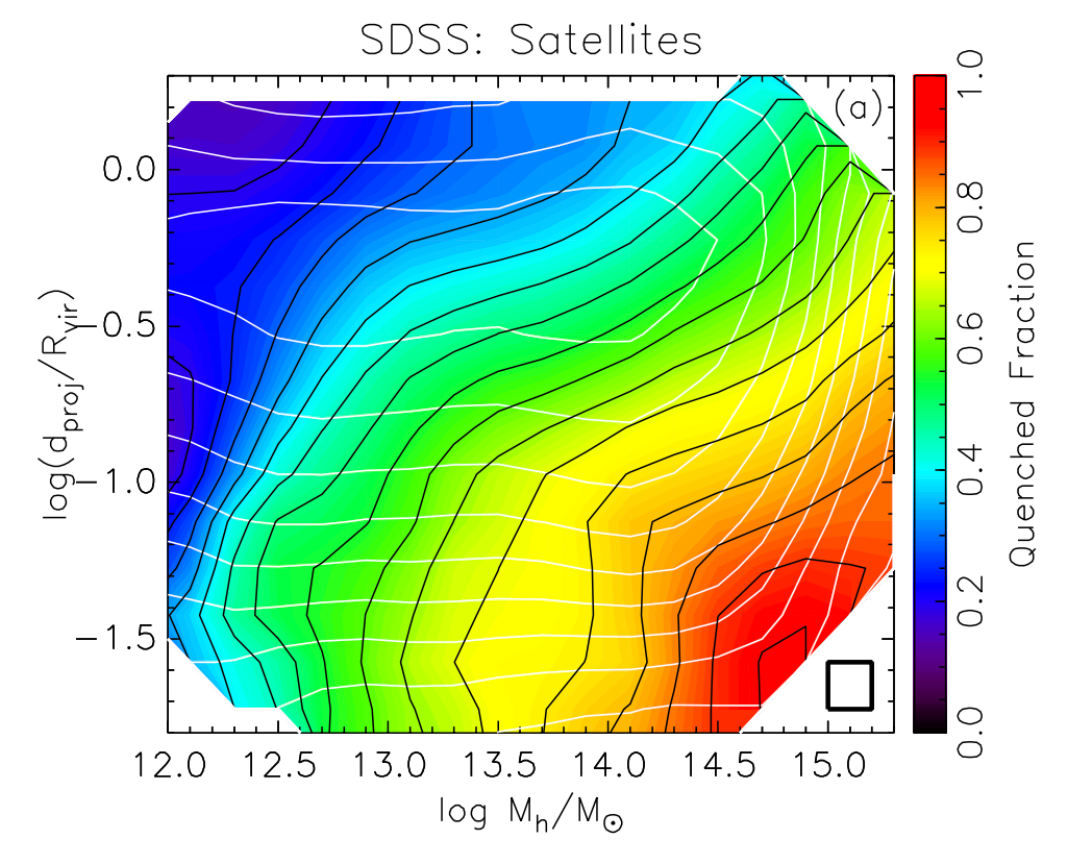}
\caption{\textit{Left-hand Panel: } The stellar mass--halo mass relationship for central galaxies, color coded by the quenched fraction within each region of the parameter space. It is clear from the progression in color throughout this figure that quenching proceeds more closely as a function of halo mass than stellar mass. \textit{Right-hand panel:} The distance of satellites to their nearest central galaxy plot as a function of group halo mass, color coded by the quenched fraction within each region of the parameter space. Here is is clear that satellite galaxies quench preferentially near the center of high mass haloes. These plots are reproduced from \cite{Woo2013}.}\label{f18}
\end{figure}

In the left-hand panel of Fig.~\ref{f18}, we reproduce the stellar mass--halo mass relationship for central galaxies from \cite{Woo2013}, which is color coded by the fraction of quenched galaxies. Contours of fixed quenched fraction are presented on this plot (displayed as black solid lines), which primarily present vertically. This implies that quenching progresses through the plane horizontally, which indicates a primary dependence of quenching on halo mass, not stellar mass. Ultimately, this analysis suggests that the mass of the halo is more important for quenching central galaxies than the total mass in stars. In \cite{Woo2013} this insight is attributed to halo mass quenching from virial shock heating (see, e.g., \citealt{Dekel2006, Dekel2009, Dekel2019}).

However, it is important to appreciate that halo mass is not fundamentally linked to quenching (e.g., \citealt{Bluck2014, Woo2015, Bluck2016, Piotrowska2022, Bluck2022}). Looking back to Fig.~\ref{f13} (bottom-right panel), we see that no quenching importance at all is given to halo mass for centrals, once a measurement of central velocity dispersion is available to the RF classifier (see \citealt{Bluck2022}). Hence, halo mass may be more strongly correlated to quenching than stellar mass, but both of these parameters are ultimately revealed to be spurious correlators to quenching in central galaxies (i.e., of no causal relationship despite the strong correlations). Note that the exact same halo masses and data sample (i.e., the Yang group catalogs and the SDSS; \citealt{Yang2007, Yang2009, York2000, Abazajian2009}) are used in both Figs.~\ref{f18} and \ref{f13}. The difference emerges because in the random forest method many other parameters of interest are controlled for simultaneously in the analysis.

In the right-hand panel of Fig.~\ref{f18} (also reproduced from \citealt{Woo2013}), the plane displays halo mass vs. projected distance to the nearest central galaxy for satellites (in units of the virial radius of the parent halo). Here quenching proceeds on the downward diagonal. This implies that satellite galaxies are more frequently quenched closer to the center of higher mass haloes. This is an important insight because it reveals that satellite galaxies have a strong quenching dependence upon both the halo mass in which they reside, and the specific location within that halo (see also \citealt{Wetzel2012, Wetzel2013, Hartley2015, Baxter2022, Goubert2024, Goubert2025} for similar conclusions).

Actually, the result for satellite galaxies in Fig.~\ref{f18} may have been anticipated from the results shown in Fig.~\ref{f17}, since local galaxy over-density rises as a strong function of \textit{both} increasing halo mass and proximity to the center of the group/ cluster. It remains to be seen whether it is ultimately local galaxy over-density, or a combination of halo mass and location within the halo, which fundamentally drives `environment-quenching'. We return to this important question later in Part~II of this review, in comparison to specific theoretical predictions.

By way of further examples, studies using spectroscopic surveys and group catalogs, including those from GAMA (\citealt{Driver2011,Robotham2011}), have shown that satellites residing in progressively more massive groups and clusters exhibit higher quenched fractions and lower specific star formation rates than comparable satellites in lower-mass haloes. For example, \citet{Brough2013} found significant environmental trends in galaxy morphology and star formation activity, while \citet{Haines2015} demonstrated that satellite star formation is progressively suppressed following infall into massive clusters, consistent with environmental processes such as starvation and ram-pressure stripping.

More recently, \citet{Davies2019} argued that halo mass is strongly correlated with satellite quenching, with the passive fraction rising systematically from low-mass groups to rich clusters, broadly consistent with earlier results from \citet{Woo2013, Wetzel2012, Wetzel2013}. Taken together, these studies support a picture in which environmental quenching becomes increasingly efficient within higher-mass haloes. This trend is commonly interpreted as reflecting the increasing effectiveness of environmental processes such as ram pressure stripping, tidal stripping, and galaxy harassment in progressively more massive environments.

In summary of this section, satellite galaxies experience strong environmental dependence on their probability of being quenched. In particular, local galaxy density, the halo mass of the group/ cluster, and the proximity to the center of the group/ cluster are all found to strongly correlate with quenching in satellites. Conversely, central galaxies have only a weak dependence on environment, except for halo mass (which is strongly correlated with internal properties in centrals). These results strongly suggest the existence of two fundamentally different modes of quenching -- (i) intrinsic, mass-correlating quenching (for all types of galaxies) and (ii) environmental, density-correlating quenching for satellites (especially at low stellar masses).

%%%%%%%%%%%%%%%%%%
%                %
%    Resolved    %
%                %
%%%%%%%%%%%%%%%%%%

\section{The spatially resolved view of galaxy quenching}\label{s6}

\begin{figure}[ht] 
\centering
\includegraphics[width=0.85\textwidth]{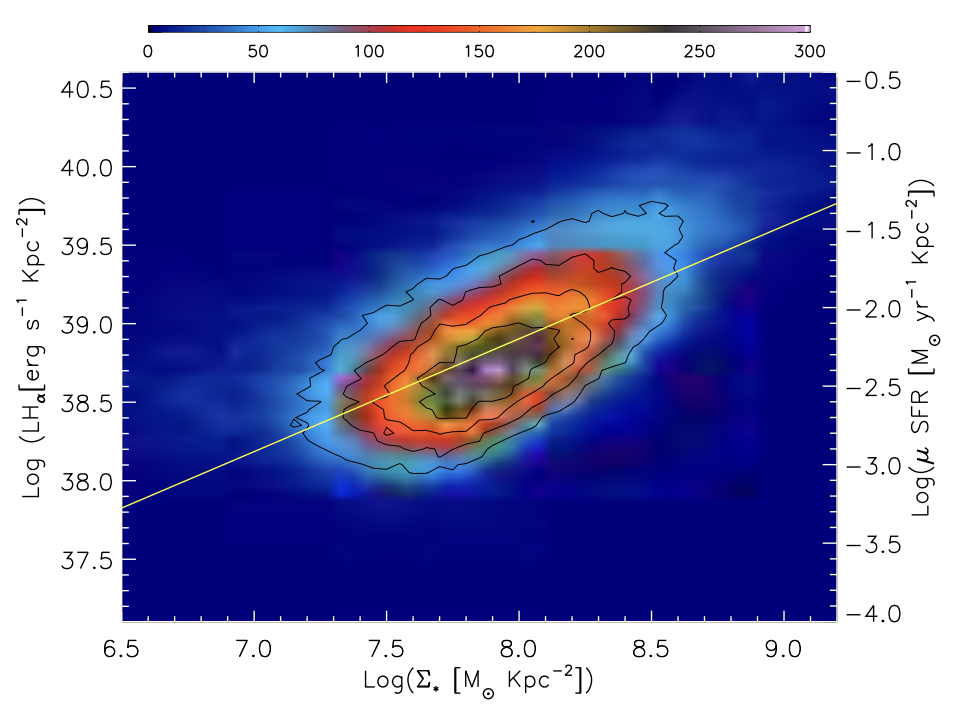}
\caption{The spatially resolved star forming main sequence relationship (rSFMS) from the CALIFA survey (\citealt{Sanchez2012}). The luminosity surface density of the dust-corrected H$\alpha$ line is plot against the stellar mass surface density ($\Sigma_*$) within each spectroscopic pixel (or spaxel). Additionally, a second Y-axis presents the surface density of star formation rate ($\Sigma_{\rm SFR}$), evaluated assuming the \cite{Kennicutt1998b} law. Contours and color shading indicate the density of points within the plane. A strong relationship between $\Sigma_{\rm SFR}$ and $\Sigma_*$ is evident for star forming regions within star forming galaxies. Overlaid is the best fit rSFMS relation (shown as a solid straight line). This figure is reproduced from \cite{Cano-Diaz2016}.}\label{f22a}
\end{figure}

So far in this review we have primarily focused on quenching on global (galaxy-wide) and environmental (super-galactic) scales. In this section we turn our attention to the important contributions to the field of galaxy quenching from spatially resolved (sub-galactic) measurements of star formation. This field was pioneered by the development of integral field unit (IFU) spectroscopy, and in particular the following crucial galaxy surveys: ATLAS$^{\rm 3D}$ (\citealt{Cappellari2011}), CALIFA (\citealt{Sanchez2012}), MaNGA (\citealt{Bundy2015}), and SAMI (\citealt{Bryant2015}).

\subsection{The resolved star forming main sequence}

\begin{figure}[t]  
\centering
\includegraphics[width=0.49\textwidth]{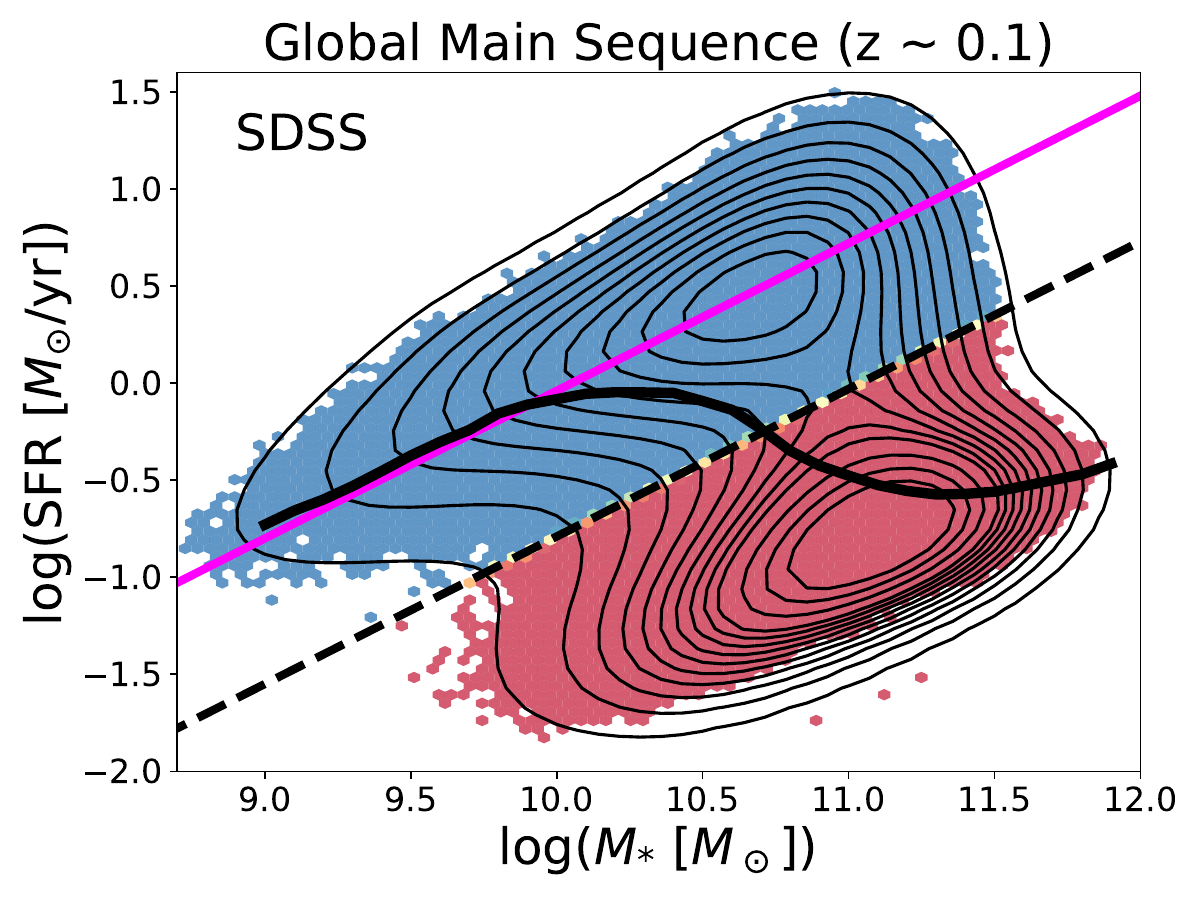}
\includegraphics[width=0.49\textwidth]{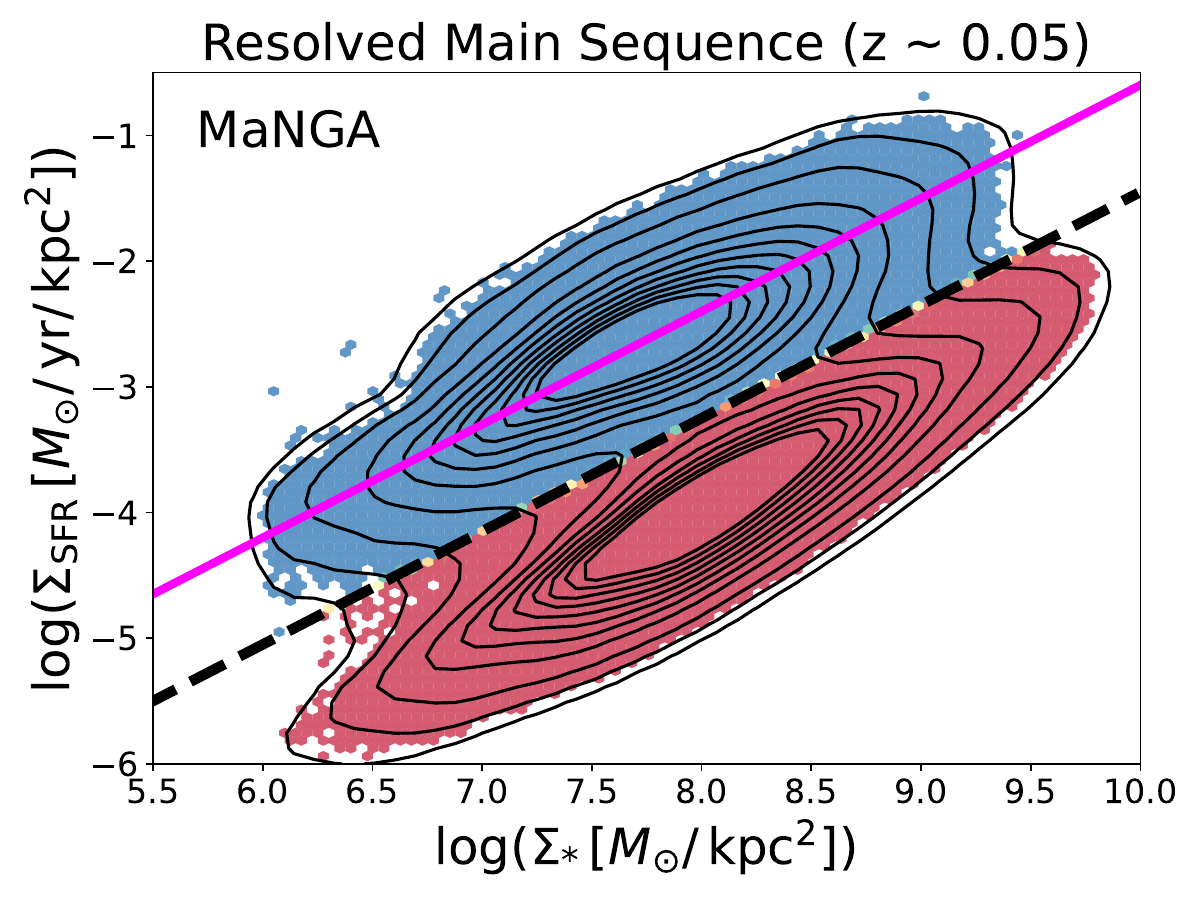}
\caption{\textit{Left-hand panel:} The global star forming main sequence (SFR - $M_*$ relation) from the SDSS (see \citealt{York2000, Abazajian2009}). Note the distinct bimodality visible in this plot. A ridge-line fit to the SFMS from \cite{Renzini2015} is shown as a magenta line. The median relation (shown as a solid black line) transitions from the star forming to the quenched population at $M_* \sim 10^{10.5} M_{\odot}$. \textit{Right-hand panel:} The kpc-resolved star forming main sequence ($\Sigma_{\rm SFR} - \Sigma_*$ relation) from the MaNGA survey (see \citealt{Bundy2015}). Note the similarity in structure to the global main sequence (discussed above). This figure is reproduced from \cite{Bluck2020b}.}\label{f23a}
\end{figure}

Among many other insights, IFU surveys have revealed that there exists a `resolved' ($\sim$kpc) analog of the global SFMS, i.e., a strong relationship between star formation rate surface density ($\Sigma_{\text{SFR}}$) and stellar mass surface density ($\Sigma_*$); e.g., \citealt{Sanchez2013, Cano-Diaz2016, GonzalezDelgado2016, Hsieh2017, Ellison2018, Medling2018}. In Fig.~\ref{f22a} we present the resolved SFMS (or, rSFMS) for star forming regions within star forming galaxies observed as part of the CALIFA survey, reproduced from \cite{Cano-Diaz2016}.

The existence of this relationship enables one to investigate what drives the local level of star formation and quenching within sub-galactic regions, in an analogous manner to studying the star formation and quenching of galaxies as a whole. Hence, one can also investigate whether galaxies vary in star forming and quenching properties within the same system. These techniques have opened up new avenues to explore how quenching takes place within galaxies, and refined our understanding of what mechanisms are ultimately responsible for the quenching of various types of systems.

As a further example, in Fig.~\ref{f23a} we present a comparison between the global SFMS for SDSS DR7 galaxies (\citealt{Abazajian2009}) and the rSFMS for MaNGA galaxies observed with FIU spectroscopy (\citealt{Bundy2015}). This figure is reproduced from \cite{Bluck2020b}. In this figure all galaxies, and regions within galaxies, are considered. Star formation rates are inferred via dust-corrected emission lines, where possible (i.e., for emission line systems which are not optically detected AGN). For all other regions or galaxies (depending on scale) an empirical constraint on sSFR is inferred from the D4000 index (see back to Sect.~\ref{s22}). This enables a fair comparison between the scales of SFMS measurements and, importantly, shows the location of quenched galaxies and regions within galaxies (unlike in Fig.~\ref{f22a}). 

By comparing the left and right panels of Fig.~\ref{f23a}, one immediately sees a remarkably similarity in structure between the SFMS on global (galaxy-wide) and resolved ($\sim$kpc) scales. This highlights a fundamental self-similarity at work within galaxies, whereby both galaxies as a whole and regions within galaxies separate out cleanly into star forming and quenched classes. Since these relationships are clearly interconnected, an important question becomes, {\it which (if either) is more fundamental?} We will consider this question further in Sect.~\ref{s61aa}, where we also bring measurements of molecular gas surface densities into the discussion.

The rSFMS is often parameterized (in direct analogy to the global SFMS) as:

\begin{equation}
\log_{10} \bigg( \Sigma_\mathrm{SFR} \, \big[M_{\odot}/\mathrm{yr}/\mathrm{kpc}^2\big] \bigg) = \alpha' \times \log_{10}\bigg( \Sigma_* \, \big[M_{\odot}/\mathrm{kpc}^2\big] \bigg) -\beta'
\end{equation}

\noindent where all terms are defined above. In Fig.~\ref{f23a} (right-panel) we show the best fit from \cite{Bluck2020a}, with: $\alpha' = (0.90\pm0.22)$ and $\beta' = (9.57\pm1.93)$. This is broadly consistent with previous measurements (including, \citealt{Cano-Diaz2016}). Resolved analogues of $\Delta$SFR and sSFR may be formed from the surface densities, leading to similar routes for segregating quenched from star forming regions within galaxies. For example, in Fig.~\ref{f23a} (right-panel), a cut at $\Delta\Sigma_\mathrm{SFR}$ $<$ -0.85\,dex is employed (which divides galactic regions at the minimum of the 2D distribution. Hence, one can classify regions within galaxies as being star forming or quenched, in direct analogy to galaxies as a whole.

\subsection{`Inside-out' vs. `outside-in' quenching }\label{s63}

\begin{figure}[ht] 
\centering
\includegraphics[width=1\textwidth]{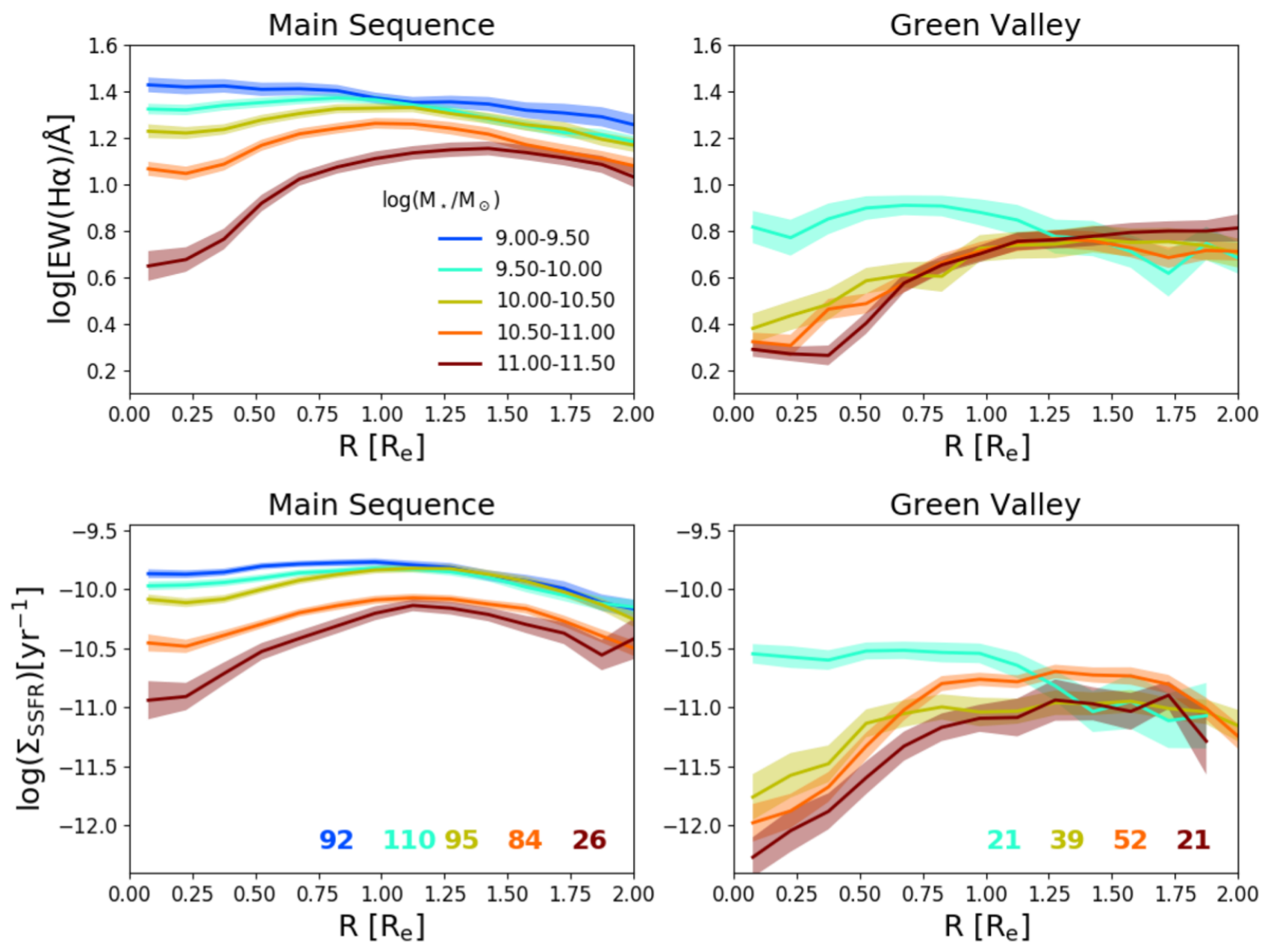}
\caption{The spatially resolved view of star formation and quenching within the MaNGA survey (\citealt{Bundy2015}), reproduced from \cite{Belfiore2018}. The left-hand panels show results for main sequence (globally star forming) systems, with the right-hand panels showing results for the global green valley (i.e., predominantly galaxies undergoing quenching contemporaneously). The top panels present radial profiles of the H$\alpha$ equivalent width (EW(H$\alpha$)), and the bottom panels present the radial profiles of resolved sSFR ($\Sigma_{\text{SSFR}}$). Within each panel, results are shown separately for different stellar mass bins (see legend). It is clear that for most stellar masses, galaxies quench `inside-out' (with a pronounced decline in star formation towards the center of the system). The only exception to this is in the lowest mass bin, where a hint of the reverse trend is seen, i.e. `outside-in' quenching. Interestingly, even among main sequence galaxies, at high stellar masses star formation is reduced towards the center of the system. This is most likely attributable to a quiescent bulge structure being present in many high-mass main sequence galaxies, even before the onset of quenching.  }\label{f22}
\end{figure}

Many studies have investigated how star formation is distributed throughout galaxies utilizing IFU surveys and, moreover, how quenching takes place on kpc-scales. There are a number of ways to quantify this, including via the resolved sSFR, offsets from the rSFMS, and the EW(H$\alpha$) statistic (e.g., \citealt{Tacchella2015, GonzalezDelgado2016, Belfiore2018, Ellison2018, Bluck2020b}). See back to Sect.~\ref{s2} for the definitions of these metrics and further discussion. 

To constrain star formation rates in non-emission line regions a variety of approaches have been applied. These include photometric SFRs from SED fitting, applying calibrations with the D4000 index, and assigning zero values to non-detection (see, e.g., \citealt{GonzalezDelgado2014, Belfiore2018, Wang2019, Bluck2020a, Bluck2020b}). Alternatively, many works have focused on the emission-line sub-sample (e.g., \citealt{Cano-Diaz2016, Schaefer2017, Ellison2018, Quai2019}). This has the advantage of acquiring the most accurate constraints on current star formation rates (when applying dust corrections via the Balmer decrement). However, this has the distinct disadvantage of missing the majority of regions within currently quenching (i.e., green valley) systems, and essentially all regions within fully quenched systems. Hence, given our focus on quenching, we will concentrate primarily on results incorporating star formation rate estimates for all regions within galaxies (as in Fig.~\ref{f23a}).

Utilizing a variety of the approaches discussed above, many works have found strong evidence for `inside-out' quenching (see, e.g., \citealt{Tacchella2015, Tacchella2016, GonzalezDelgado2016, Belfiore2017, Belfiore2018, Ellison2018, Sanchez2018, Medling2018, Bluck2020b}). Explicitly, this refers to rising sSFR (or similar parameters) with increasing radius in the global green valley population.

\begin{figure}[ht] 
\centering
\includegraphics[width=1\textwidth]{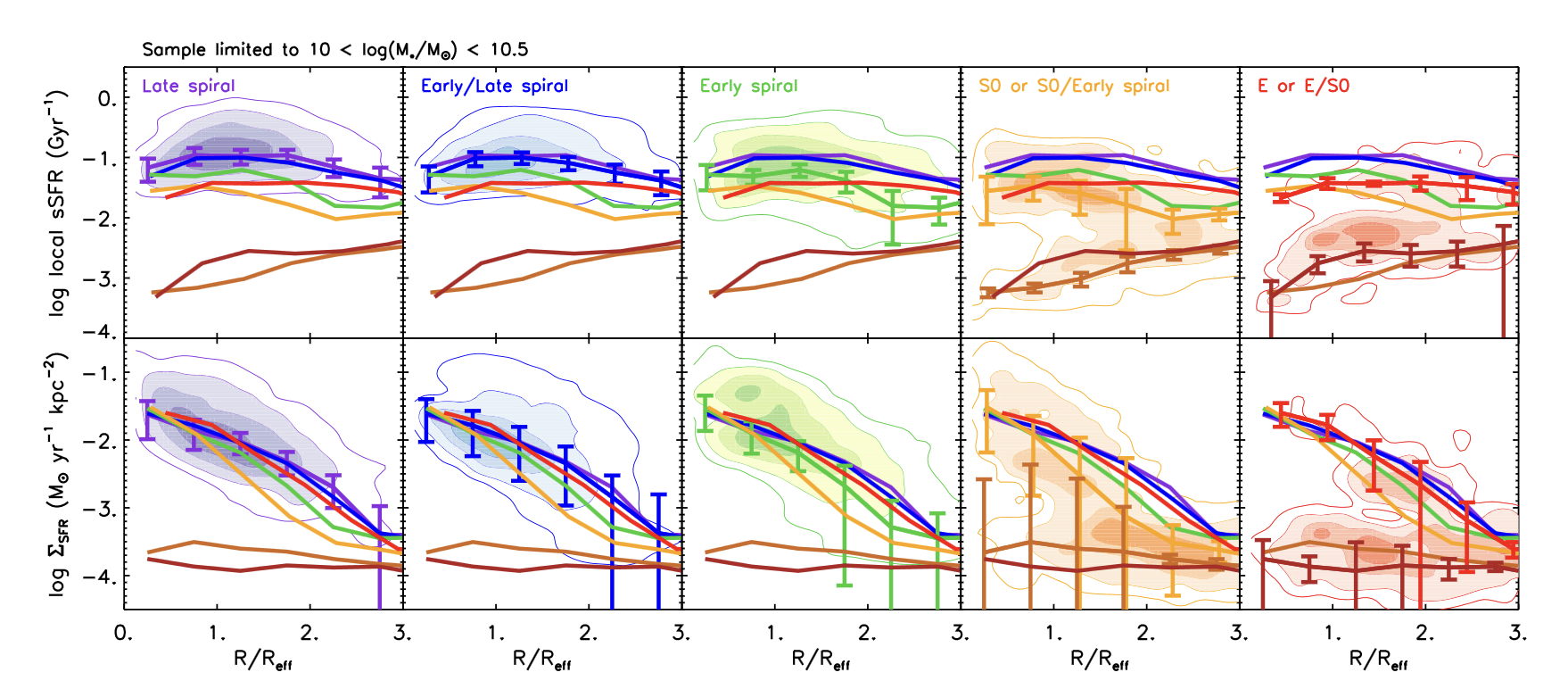}
\caption{Radial profiles in resolved sSFR (top panels) and $\Sigma_{\rm SFR}$ (bottom panels) for galaxies from the SAMI galaxy survey (\citealt{Bryant2015}). Radial profiles are split by morphological types, including: late spirals, early/late spirals, early spirals, S0, and elliptical systems. It is evident that most late-type systems are actively star forming, with a distinctive peak in $\Sigma_{\rm SFR}$ towards the center of the galaxy. In terms of sSFR, the star forming galaxies exhibit fairly flat radial profiles. Conversely, most early-type systems have flatter $\Sigma_{\rm SFR}$ profiles and weakly rising resolved sSFR profiles, indicative of inside-out quenching. This figure is reproduced from \cite{Medling2018}.}\label{f25a}
\end{figure}

In Fig.~\ref{f22}, reproduced from \cite{Belfiore2018}, we show profiles in resolved sSFR (top panels) and EW(H$\alpha$) (bottom panels) in various stellar mass bins (see legend). Additionally, this figure shows results separately for the global main sequence (left panels) and the global green valley (right panels). There is a very clear sign of inside-out quenching at most stellar mass ranges (as discussed above). However, at the lowest stellar masses there is a hint of the opposite trend, i.e. `outside-in' quenching, whereby the outskirts of low-mass green valley systems are more quiescent than their cores. Moreover, at high masses, even main sequence galaxies start to exhibit quiescent centers, likely a result of harboring classical bulge structures (e.g., \citealt{Belfiore2018, Wang2019}).

\begin{figure}[ht] 
\centering
\includegraphics[width=0.49\textwidth]{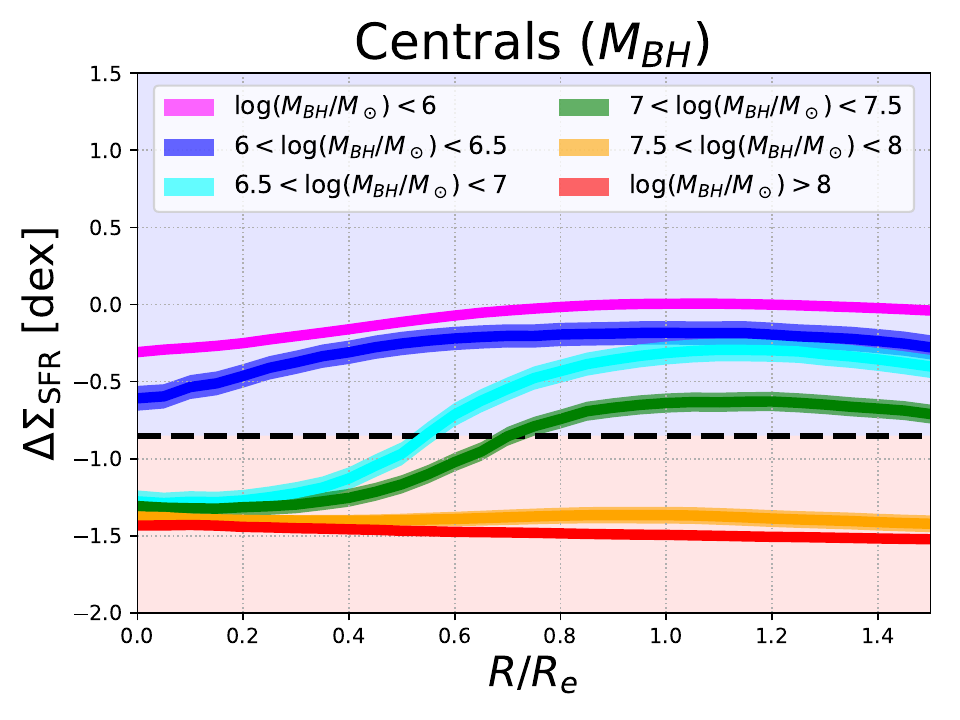}
\includegraphics[width=0.49\textwidth]{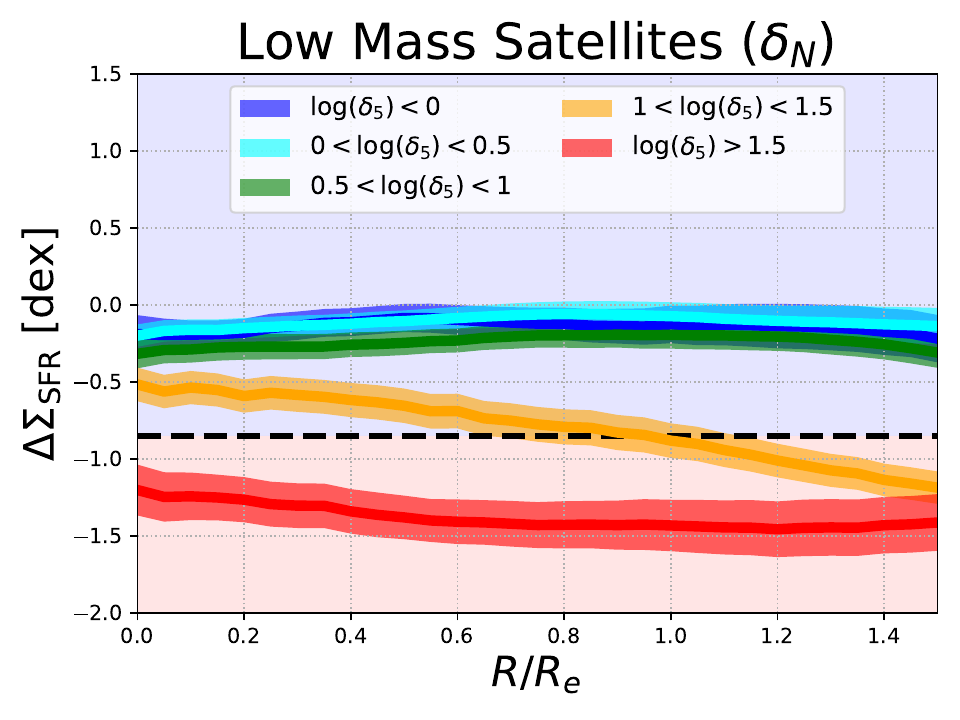}
\caption{Median averaged population offsets from the resolved SFMS ($\Delta \Sigma_{\text{SFR}}$) plot as a function of deprojected elliptical radius ($R/R_e$) for central galaxies (left panel; split into ranges of supermassive black hole mass, inferred via the $M_\mathrm{BH}$\,--\,$\sigma_c$ relation) and for low-mass satellite galaxies (right panel; split into ranges of local galaxy over-density, evaluated at the 5th nearest neighbor). This figure is reproduced from \cite{Bluck2020b}. For centrals, at low black hole masses galaxies are star forming throughout their full radial extent, whereas at high black hole masses they are quenched throughout their full radial extent. At intermediate black hole masses, galaxies exhibit quenched cores and star forming outskirts, indicating that AGN feedback results in `inside-out' quenching. For low-mass satellites, at low over-densities galaxies are star forming throughout their full radial extent, whereas at high over-densities galaxies are quenched throughout their full radial extent. At intermediate over-densities, low-mass satellites present with star forming cores and quiescent outskirts, indicating that environment quenches galaxies `outside-in'. Hence, intrinsic vs. environmental quenching maps onto inside-out vs. outside-in quenching on kpc-scales.}\label{f23}
\end{figure}

As another example, in Fig.~\ref{f25a} we present an analysis of galaxies observed by the SAMI survey (\citealt{Bryant2015}), showing their radial profiles in $\Sigma_{\rm SFR}$ (bottom panels) and resolved sSFR (top panels). This figure is reproduced from \cite{Medling2018}. Interestingly, in this figure profiles in star formation are split by the morphological types of galaxies (see caption for details). Broadly speaking, late-type systems are star forming and exhibit characteristically declining $\Sigma_{\rm SFR}$ profiles with flat local sSFR profiles. Alternatively, early-type systems are more frequently quenched, exhibiting with flat $\Sigma_{\rm SFR}$ profiles and weakly rising sSFR profiles, which is indicative of inside-out quenching.

Many studies find that central galaxies quench intrinsically, primarily as a function of SMBH mass (e.g., \citealt{Bluck2016, Terrazas2016, Terrazas2017, Terrazas2020, Piotrowska2022, Bluck2023, Bluck2024, Lim2025}), whereas low-mass satellites quench primarily as a function of environment, specifically local galaxy over-density in observations (e.g., \citealt{Wetzel2012, Peng2012, Bluck2020b, Goubert2024, Goubert2025}). The clearest evidence for this breakdown is presented in Sects.~4~\&~5 of Part~II of this review series, where we additionally compare these results to direct predictions from cosmological simulations to aid in establishing the cause of quenching. 

As a result of the above discussion, to explore how these different quenching causes exhibit on spatially resolved scales, in Fig.~\ref{f23} we present an analysis separating galaxies into centrals and low-mass satellites, reproduced from \cite{Bluck2020b}. On the left panel of Fig.~\ref{f23}, we show median averaged offsets from the resolved SFMS as a function of elliptical radius in bins of SMBH mass (estimated via the $M_\mathrm{BH}$\,--\,$\sigma_c$ relation) for central galaxies. On the right panel of Fig.~\ref{f23}, we show the same profiles for low-mass satellites, here split by local galaxy over-density. 

Centrals which harbor low mass SMBHs are typically star forming throughout their full radial extent, whereas centrals which harbor high mass SMBHs are typically quenched throughout their full radial extent. However, at intermediate SMBH masses, centrals exhibit a very strong sign of inside-out quenching, whereby galaxies exhibit with quiescent cores and star forming outskirts. Moreover, a clear progression is seen throughout the transitioning region as SMBH mass increases, whereby a greater fraction of the spatial extent of central galaxies is quenched with increasing SMBH mass.

Similarly, low-mass satellites in low density environments are found to be star forming at all radii, whereas low-mass satellites in high density environments are found to be quenched at all radii. Crucially, however, at intermediate over-densities, low mass satellites exhibit with quenched outskirts and star forming cores. That is, they experience `outside-in' quenching. This is the opposite result found for the intrinsic quenching of centrals. Hence, {\it different quenching causes give rise to different star formation profiles during transition.}

Ultimately, intrinsic quenching manifests via inside-out quenching, whereas environmental quenching manifests via outside-in quenching (see, e.g., \citealt{Schaefer2017, Belfiore2018, Bluck2020b}). This phenomenology makes sense given the probable origin of the quenching process in the two cases. For environmental quenching via stripping, the gas in the outskirts of a galaxy is more vulnerable to removal than in the core, due to it being much less tightly bound. Hence, it is reasonable to expect that as satellites move to progressively denser environments (e.g., on a radial cluster orbit), the outskirts will reduce in gas content (and hence star formation) first, followed eventually by the cores. 

Furthermore, since SMBHs reside at the very center of galaxies, it also reasonable to expect that AGN feedback would first impact the cores of galaxies (e.g., via driving multi-phase gas outflows and/ or injecting turbulence into the ISM) and then later impact the whole system, via stabilizing cooling flows from the halo into the ISM (see, e.g., \citealt{Weinberger2018, Zinger2020, Piotrowska2022}). These processes are discussed at length in Part II of this review, from both a theoretical and observational perspective.

\subsection{Global quenching vs. local star formation}\label{s64}

Given the results discussed in the previous sub-section, the title of this sub-section may appear a little counterintuitive. As we have seen, intrinsic quenching proceeds inside-out and environmental quenching proceeds outside-in (e.g., \citealt{Tacchella2015, GonzalezDelgado2016, Schaefer2017, Belfiore2018, Medling2018, Ellison2018, Bluck2020b}). Hence, both forms of quenching exhibit with strong spatially resolved features and, thus, may be considered to have a key local dependence. However, the vast majority of star forming galaxies are star forming throughout their entire structure. Similarly, the vast majority of quenched galaxies are quenched throughout their entire structure (e.g., \citealt{Belfiore2018, Bluck2020a, Bluck2020b}). Therefore, given that transitioning galaxies make up only a very small fraction of the total galaxy population at any epoch, whether regions within galaxies are star forming or quenched tends to follow the global state of the galaxy as a whole, as either an actively star forming or quiescent system. Put simply, it is rare to find quiescent regions in star forming galaxies, and even rarer to find star forming regions within quiescent galaxies (see further details in \citealt{Bluck2020a}).

\begin{figure}[htbp!] 
\centering
\includegraphics[width=0.85\textwidth]{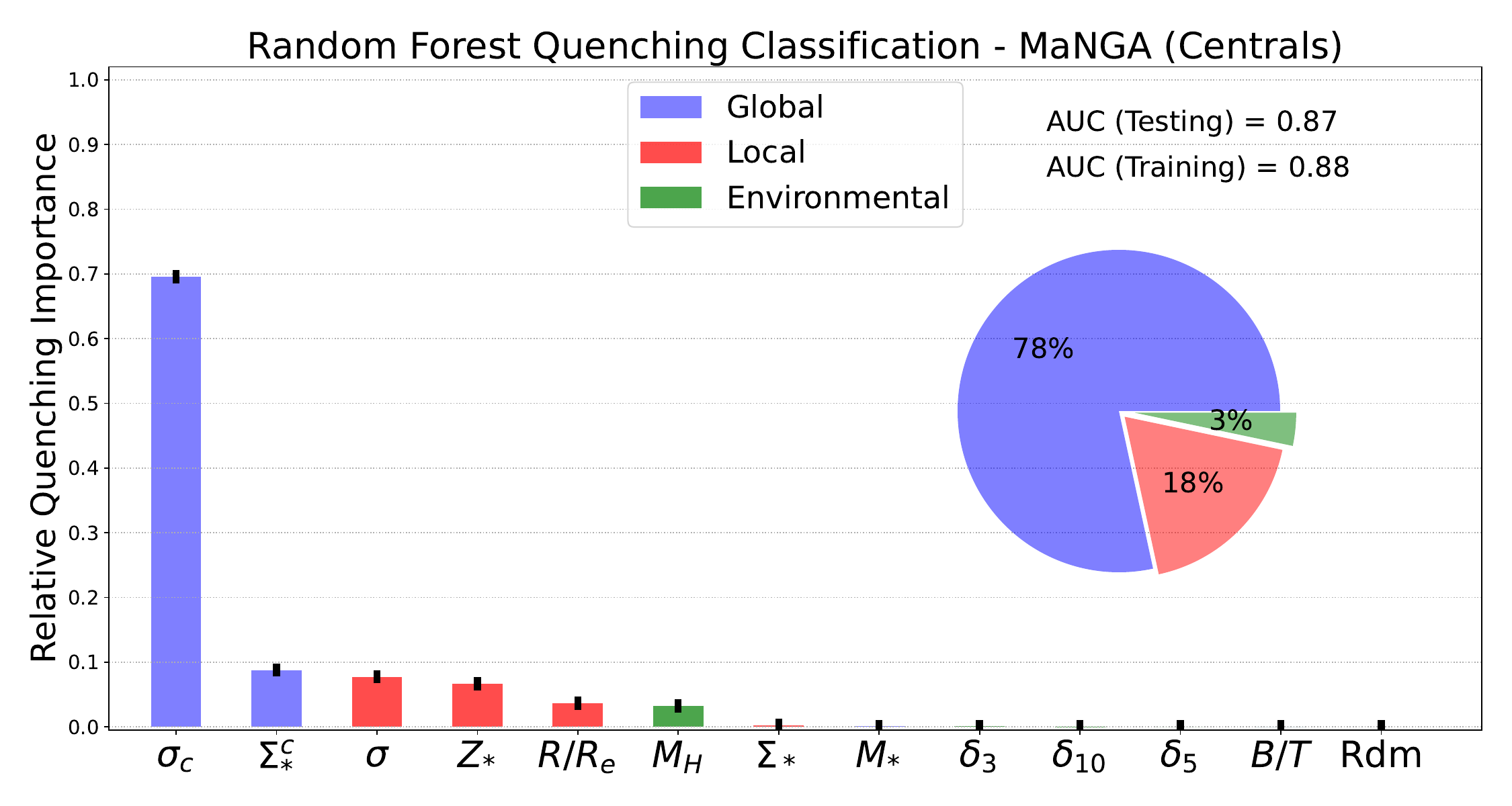}
\includegraphics[width=0.85\textwidth]{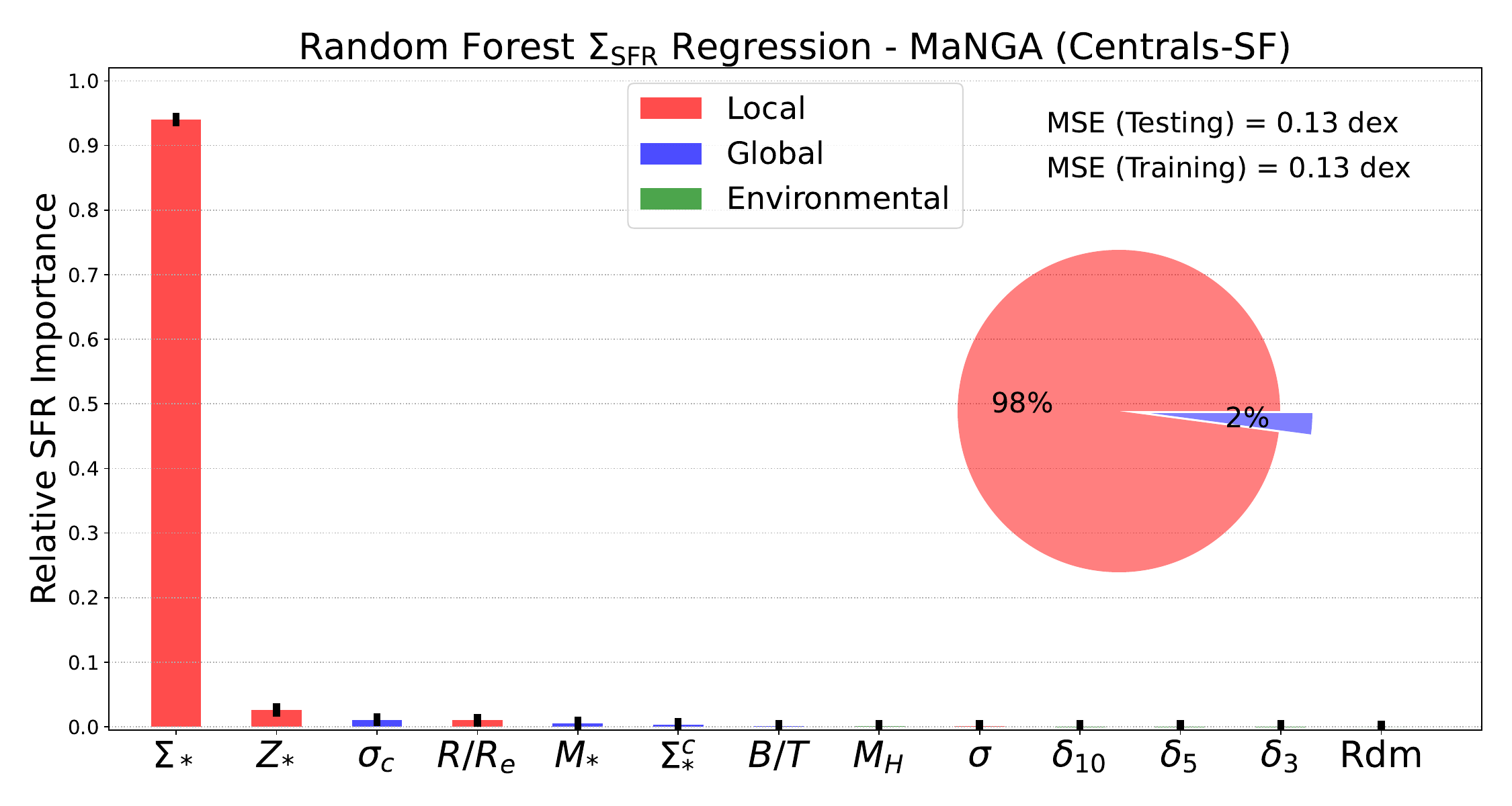}
\caption{Global vs. local quenching \& star formation. \textit{Top panel:} A random forest classification analysis to predict whether regions within central galaxies from the MaNGA survey (\citealt{Bundy2015}) are star forming or quenched, on the basis of various global, local, and environmental parameters (as listed along the X-axis). The Y-axis shows the relative importance of each parameter, with uncertainty given by the variance across multiple classification runs. For quenching, it is clear that global parameters dominate the predictive power of the classification (see inset pie plot), with central velocity dispersion ($\sigma_c$) being the most important single parameter. Moreover, these results establish that quenching operates globally within galaxies. This panel is reproduced from \cite{Bluck2022}. \textit{Bottom panel:} A random forest regression analysis to predict the star formation rate surface density ($\Sigma_\text{SFR}$) of regions within star forming galaxies from the MaNGA survey, on the basis of the same parameters outlined above. Unlike for quenching, star formation is governed locally by spatially resolved parameters (see pie plot inset), especially the stellar mass surface density ($\Sigma_*$). This panel is reproduced from \cite{Bluck2020a}. In concert, this figure reveals that quenching is fundamentally a global process, but the level of star formation within star forming systems is governed locally. }\label{f25}
\end{figure}

In Fig.~\ref{f25} we present the results from two complementary random forest analyses: (i)~a star forming vs. quenched classification for all spaxels within central galaxies from the MaNGA DR15 (top panel; reproduced from \citealt{Bluck2022}); and (ii)~a regression analysis to predict $\Sigma_\text{SFR}$ values within BPT identified star forming central galaxies (bottom panel; reproduced from \citealt{Bluck2020a}). A wide variety of parameters are used to train each random forest analysis, separated into: (i)~global parameters, defined as one parameter per galaxy which pertains to the galaxy itself; (ii)~local parameters, defined as one parameter per spaxel (or region); and (iii)~environmental parameters, defined as one parameter per galaxy which pertains to the environment within which the galaxy resides. Using this methodology one is able to assess which specific parameters (and, crucially, types of parameters) are most effective at predicting the level of star formation in star forming systems, and whether regions within galaxies will be star forming or quenched. 

These analyses follow on from similar machine learning approaches applied to global galaxy parameters (e.g., \citealt{Teimoorinia2016, Bluck2019}). Full details on the random forest method, including many detailed tests on performance, are provided in \cite{Bluck2022}. As a reminder, random forest analyses (like all statistical approaches) are limited by the choice of parameters included and the sample selection applied. Here we investigate a host of global, local, and environmental parameters to help ascertain the key scales at work in regulating star formation and quenching within local galaxies. Note that we do not include gas based properties here. These important measurements will be discussed in Sect.~\ref{s7a}.

In the star forming - quenched classification analysis (Fig.~\ref{f25}, top panel), we see that central velocity dispersion ($\sigma_c$, evaluated within 1\,kpc) is by far the most effective single parameter for predicting whether regions within central galaxies will be quenched or star forming. This is consistent with the results shown previously in Fig.~\ref{f13} on global scales, but expands significantly upon this prior result by considering quenching as a spatially resolved phenomenon and incorporating a host of spatially resolved parameters into the analysis.

Interestingly, {\it central velocity dispersion is far more predictive of quenching in a given region of a galaxy than the velocity dispersion computed within that region.} Hence, this result clearly implies that quenching on spatially resolved scales is governed globally, in a system-wide manner. The conditions at the very center of the galaxy are much more effective at predicting whether regions in the disc will be star forming or quenched than the conditions in the disc itself. We illustrate this point with a pie plot inset, which shows that, collectively, global parameters dominate the predictive power over quenching in the classification analysis.

Moving on to the resolved star formation rate regression analysis (Fig.~\ref{f25}; bottom panel), a completely different set of results are seen. Star formation on kpc-scales is found to be best predicted by the local stellar mass surface density ($\Sigma_*$), which is just the rSFMS (as discussed above). Hence, if one knows that a galaxy is globally star forming, the level of star formation within a given region is constrained most effectively by the mass surface density of that region. This leads to a pure local dependence of star formation in star forming galaxies (see inset pie plot). The total lack of importance given to the total galaxy stellar mass in this analysis essentially confirms that the rSFMS is more fundamental than the global SFMS. This result is supported by much prior work which also points in this direction (e.g., \citealt{Sanchez2013, Cano-Diaz2016, Hsieh2017, Cano-Diaz2019, Enia2020}).

Therefore, at a fundamental level, {\it star formation and quenching are distinct phenomena.} This statement is inherently confusing because in one sense quenching is just the absence of star formation. Nonetheless, it is crucial to understand the sense in which these processes are completely distinct, and even orthogonal. In actively star forming galaxies, with abundant fuel provided by cold gas streams or halo cooling flows (dependent upon halo mass), star formation is regulated by the local physics of the ISM. We will discuss the details of how gas content and star formation are related in the next section.

Conversely, in quenched galaxies, star formation is suppressed throughout the entire system and the local dependence is lost. Most probably, this global (galaxy-wide) quenching is a result of the surrounding halo no longer yielding efficient cooling flows into the ISM, resulting in starvation (which we discuss in detail in Part~II of this review). This may be achieved via preventative AGN feedback, which is a fundamentally global (halo-wide) process. Hence, quenching and star formation are dependent upon very different physics, which operate on vastly different astrophysical scales. To know if a region within a central galaxy is quenched, one must establish whether the halo is actively cooling or not; whereas, to know the level of star formation in a region within a star forming galaxy, one must know the conditions precisely within that region. It is this disconnect (in both scale and underlying physics) which is the explanation for why, in an important sense, {\it star formation and quenching are orthogonal processes in galaxy evolution.}

%%%%%%%%%%%%%%%%%%
%                %
%      GAS       %
%                %
%%%%%%%%%%%%%%%%%%

\section{Gas Physics: Fuel \& Efficiency}\label{s7a}

\subsection{Measuring gas properties \& efficiencies}

Ultimately, stars form out of the gravitational collapse of dense molecular gas clouds within galaxies, which themselves form out of the atomic gas reservoir via cooling and condensation. As such, it is critical to study the gas content within galaxies to reveal both star formation and quenching pathways.

\subsubsection{Inferring gas masses}

Atomic Hydrogen (often referred to as HI in astronomy) emits via the hyperfine transition at 21\,cm (which represents spontaneous or collisional transition from a state where the electron and proton have parallel spins to a state where they have anti-parallel spins). The flux across the surface area of a galaxy may be integrated to infer the total atomic Hydrogen mass bound by that surface as follows (see, e.g., \citealt{Roberts1975, Binney1998, Giovanelli1983, Haynes2011, Meyer2017}):

\begin{equation}
M_\mathrm{H\,I} \, [M_\odot] = 2.356 \times 10^5 \, \left( \frac{1}{1+z} \right) \left( \frac{D_L(z)}{\text{Mpc}} \right)^2 \, \left( \frac{\int S_{\rm 21cm}(v) \, dv}{\mathrm{Jy \, km \, s^{-1}}} \right)
\end{equation}

\noindent where the integral is performed in velocity space, $S_{\rm 21cm}(v)$ is the surface integrated specific flux per velocity within the 21\,cm emission line, $D_L(z)$ is the luminosity distance, and the normalization factor is determined via atomic theory (see \citealt{Binney1998} for a derivation).

The situation with molecular Hydrogen (H$_2$) is more subtle. The H$_2$ molecule has no permanent electric dipole moment and, hence, rotational transitions are forbidden. Vibrational transitions are permitted and do occur, leading to FIR emission, but this is typically only active at temperatures of T $\gtrsim 500$\,K. Molecular clouds in galaxies are much cooler ($\sim10 - 50$\,K), preventing this from being used as a useful probe of star forming gas. Carbon Monoxide (CO) is the second most common molecule in molecular clouds within the Milky Way and has strong dipole transitions. Hence, it is often used as a tracer of H$_2$ (see \citealt{Bolatto2013} for a review). 

The observed flux of the CO transition line is converted first to a `brightness-temperature' luminosity ($L'$) and then to a molecular gas mass as follows, which we show specifically for the rotational transition: J[1 $\rightarrow$ 0] (see, e.g., \citealt{Solomon1997, Bolatto2013, Tacconi2020}):

\begin{equation}
L'_{\mathrm{CO}} = 3.25 \times 10^7 \left( \frac{D_L(z)^2}{\nu_{\mathrm{obs}}^2 (1+z)^3} \right) \, \int S_\mathrm{CO \, J[1 \rightarrow 0]}(v) \, dv
\quad [\mathrm{K\,km\,s^{-1}\,pc^2}]
\end{equation}

\begin{equation}
M_{\mathrm{H}_2} = \alpha_{\mathrm{CO}}(Z, \Sigma_g,...) \times L'_{\mathrm{CO}} \quad [M_\odot]
\end{equation}

\noindent where the integral is performed in velocity space as before, and $\nu_{\rm obs}$ indicates the observed frequency of the CO transition line (here, CO J[1 $\rightarrow$ 0]). There are three factors of $(1+z)$ in the above expression, which account for bandwidth compression, photon energy redshift, and time dilation of detection. These effects do not occur in an energy-based total luminosity approach, but they are important here. 

In the final expression above, the mass is computed from the temperature-brightness luminosity via the $\alpha_{\rm CO}$ conversion factor. Excluding low-metallicity dwarf galaxies, this parameter is expected to be in the range: $\alpha_{\rm CO} = 0.8 - 5 \, [M_{\odot} / ({\rm K \, km/s \, pc^2})]$ (see \citealt{Bolatto2013}). Typical galaxies are thought to exhibit an $\alpha_{\rm CO} \sim 4.35 \, [M_{\odot} / ({\rm K \, km/s \, pc^2})]$ (\citealt{Bolatto2013}), which is often used as a simplifying assumption in the literature. However, variation is expected as a function of gas-phase metallicity and the surface density of the gas itself (see, e.g., \citealt{Narayanan2012, Genzel2012, Bolatto2013, Genzel2015}). Consequently, molecular Hydrogen masses are typically not inferred with as much certainty as HI masses. This is especially unfortunate for studies of star formation and quenching, since stars form out of the dense molecular gas (not the diffuse HI gas).

In addition to the two methods described above, the total neutral gas mass (and in some cases, the constituent atomic and molecular types) may be estimated via calibrations with dust mass, itself determined via FIR flux (e.g., \citealt{Leroy2011, Magdis2012, Scoville2016}), and/ or via optical extinction calibrations (e.g., \citealt{Gruver2009, Concas2019, Piotrowska2020}). These methods are useful when direct measurements of the 21cm radio emission and/or the sub-mm CO line emission cannot be obtained. Often this is the case for wide-field galaxy surveys.

\subsubsection{Scaling laws \& efficiencies}

Having discussed how to infer the mass of gas (in both atomic and molecular states) within the ISM, we are in a position to review how these parameters are connected to star formation. Since stars form from the dense gas within giant molecular clouds (GMCs), one expects a positive correlation between these components. The most important empirical gas scaling law is known as the Kennicutt-Schmidt, or KS, relation (see, \citealt{Schmidt1959, Kennicutt1998b}). This may be stated as: 

\begin{equation}
\Sigma_\mathrm{SFR} = {\epsilon_\mathrm{SF}} \,\, \cdot \,\, \big( \Sigma_{g}\big)^{\alpha_\mathrm{SF}}, \,\,\,\, {\rm where,} \,\,\,\, \alpha_\mathrm{SF} \approx 1.5 
\end{equation}

\noindent and where, $\Sigma_{\rm SFR}$ is the surface density of star formation rate (taken within a galaxy as a whole, or within a sub-region within a galaxy), and $\Sigma_g$ indicates the total (atomic + molecular) gas mass surface density (i.e., $\Sigma_g = \Sigma_{\text{HI}} + \Sigma_{\text{H}_2}$). The exponent ($\alpha_{\rm SF}$) is found to be greater than unity, indicating a super-linear relationship. A doubling in gas surface density yields a little under a tripling of star formation surface density. Finally, $\epsilon_{\rm SF}$ indicates a normalization constant, which is related to the efficiency of star formation. This is surprisingly low, with estimates of $\sim$10\% of gas converted into stars per orbital period of a galaxy disc (see, \citealt{Kennicutt1998b}).

The observed form of the KS relation may be understood physically as follows. Using simple dimensional analysis, one anticipates that the star formation rate density ($\rho_\mathrm{SFR}$) will be equal to the gas mass density ($\rho_g$) divided by the relevant dynamical time ($t_\mathrm{dyn}$) of the system. Explicitly, this leads to (see \citealt{Kennicutt1998b}):

\begin{equation}
\rho_\mathrm{SFR} \sim \frac{\rho_g}{t_\mathrm{dyn}} \sim \frac{\rho_g}{(G \rho_g)^{-1/2}} \propto \big( \rho_g \big)^{3/2}, \,\,\,\, {\rm where,} \,\, t_\mathrm{dyn} \sim  \frac{1}{\sqrt{G \rho_g}} \, .
\end{equation}

\noindent Hence, the expected exponent in 3D density is essentially identical to that observed in 2D. This makes perfect sense for a disc with fixed scale height, and most star formation is observed to occur within disc structures.

Later observational works find a much stronger (i.e., tighter) dependence of $\Sigma_\mathrm{SFR}$ on molecular gas mass surface density, rather than on HI surface density (or the sum of the two components), see \cite{Bigiel2008, Leroy2008}. Interestingly, the exponent on the KS relation ($\alpha_{\rm SF}$) for molecular gas lowers to approximately unity (see, e.g., \citealt{Bigiel2008, Leroy2008, Bigiel2011, Lin2019}). This is typically explained qualitatively by stars forming out of the dense molecular gas, in a fuel-limited manner. 

The physical processes which regulate the rate of star formation within GMCs are typically unresolved observationally. Theoretically, the GMC dynamical time is thought to be regulated via a complex interplay of gas turbulence, magnetic field pressure, and stellar feedback (see, e.g., \citealt{Bigiel2008, Bigiel2011, Leroy2008, Krumholz2009a, Krumholz2009b, Feldmann2011, Ostriker2011, Crutcher2012}). Whilst most observational studies find similar exponents for the KS relation, several studies find that its normalization (i.e., $\epsilon_{\rm SF})$ varies both within and between galaxies (see, e.g., \citealt{Feldmann2011, Leroy2013, Ellison2020c, Ellison2021a, Thorp2022}). 

A qualitatively different approach is to separate the specific star formation rate (${\rm sSFR \equiv SFR/}M_*$) of a galaxy into contributions from gas fraction ($f_g \equiv M_g/M_*$) and a new efficiency (${\rm SFE} \equiv {\rm SFR}/M_g$), see \cite{Saintonge2016, Saintonge2017, Lin2019, Ellison2020, Ellison2021, Piotrowska2020, Brownson2020}. Specifically,

\begin{equation}
\mathrm{sSFR} \big|_\mathrm{global} \equiv \frac{\mathrm{SFR}}{M_*} = \bigg( \frac{\mathrm{SFR}}{M_g} \bigg) \cdot \bigg( \frac{M_g}{M_*} \bigg) \equiv {\rm SFE}\big|_\mathrm{global} \cdot f_g \big|_\mathrm{global}
\end{equation}

\noindent where, $M_g$ usually refers specifically to the molecular gas mass in this case. The above expression is nothing more than a trivial mathematical identity. Nonetheless, it has value because it factorizes sSFR into a gas contribution and an efficiency contribution. 

Any reduction in sSFR can be divided into a reduction in gas fraction and a reduction in efficiency. Ultimately, this decomposition is useful for testing the mechanism(s) by which quenching occurs, but not necessarily for establishing the fundamental cause of quenching. This follows because the same cause (e.g., AGN feedback) can operate via multiple mechanisms (e.g., ejection of gas from the ISM leading to lower gas fractions {\it and} increasing turbulence in the remaining ISM gas leading to reduced efficiency). Even more confusingly, the same mechanism can be achieved by different causes (e.g., reduction in efficiency via dynamical stabilization or via turbulence injection from AGN feedback). Hence, to fully establish how quenching occurs one must constrain {\it both} the cause and the mechanism. This is an important theme, which we will revisit in Part II of this review.

In the previous expression, we show the decomposition of sSFR for a whole galaxy (the global version). However, this same approach may also be applied to a resolved sub-region within a galaxy as follows:

\begin{equation}
{\rm sSFR} \big|_\mathrm{local} \equiv \frac{\Sigma_{\rm SFR}}{\Sigma_*} = \bigg( \frac{\Sigma_\mathrm{SFR}}{\Sigma_g} \bigg) \cdot \bigg( \frac{\Sigma_g}{\Sigma_*} \bigg) \equiv \mathrm{SFE}\big|_\mathrm{local} \cdot f_g \big|_\mathrm{local}
\end{equation}

\noindent where $\Sigma_*$ is the stellar mass surface density, and $\Sigma_g$ is most often taken to be the molecular gas mass surface density. The value of the local approach is that the sSFR may vary across regions within a single galaxy and, furthermore, the relative contributions of $f_g$ and SFE to the local sSFR value may also vary as a function of location within a galaxy. Ultimately, reduction in sSFR during quenching must logically be separable into reduction in gas fraction and/or reduction in star forming efficiency. Disentangling these effects, as above, yields a powerful tool for extracting insight on the mechanisms which quench galaxies.

\subsection{The origin of the star forming main sequence}\label{s61aa}

As discussed in Sect.~\ref{s6}, there exist tight relationships between the global star formation rate and total stellar mass (i.e., the star forming main sequence; SFMS) and between the surface density of star formation rate and the surface density of stellar mass on sub-galactic scales (i.e., the resolved star forming main sequence; rSFMS). See back to Fig.~\ref{f23a} for a comparison between these relationships. Obviously, these two empirical relationships must be connected. In principle, a top-down process could set the local relationship from the global one, or else a bottom-up process could set the global relationship from the local one. Hence, an important question arises: which, if either, of these relationships is more fundamental? And, moreover, how do both of these relationships emerge physically?

\begin{figure}[ht] 
\centering
\includegraphics[width=0.55\textwidth]{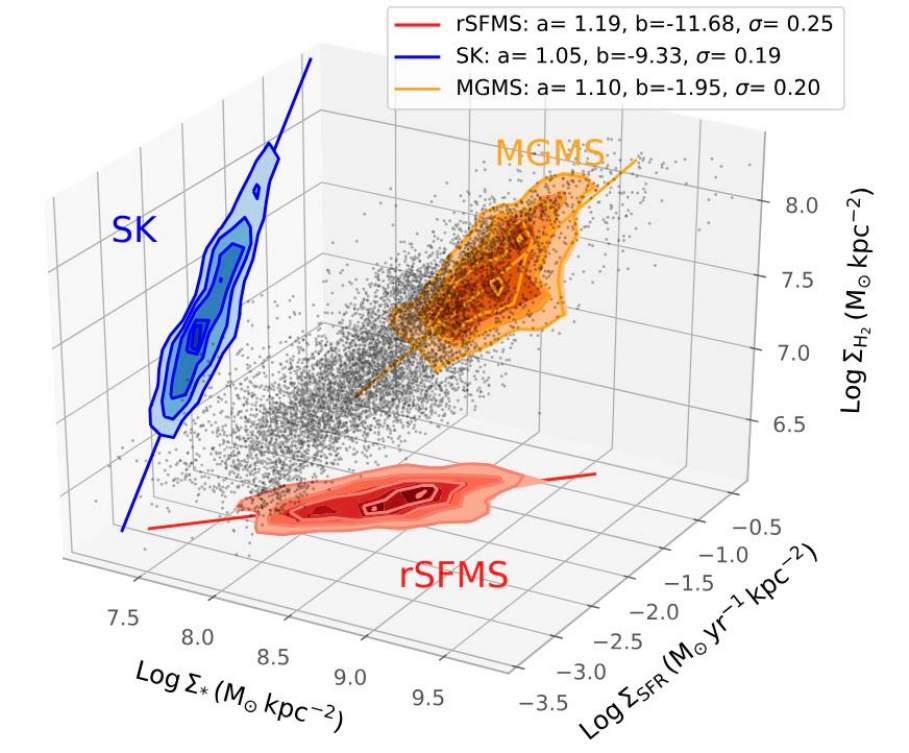}
\includegraphics[width=0.44\textwidth]{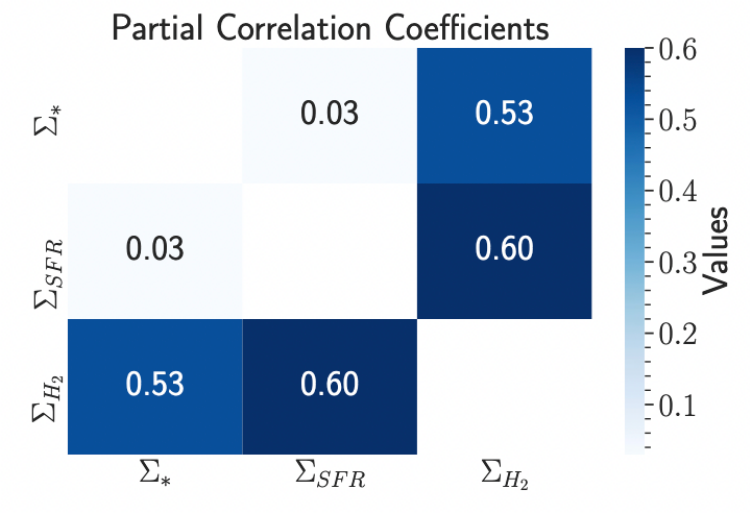}
\caption{\textit{Left-hand panel:} The tri-variate relationship between star formation rate surface density ($\Sigma_{\text{SFR}}$), molecular gas mass surface density ($\Sigma_{\text{H2}}$), and stellar mass surface density ($\Sigma_*$) presented in 3D, reproduced from \cite{Lin2019}. Additionally, the projections into the Schmidt-Kennicutt relation (SK; $\Sigma_{\text{SFR}}$\,--\,$\Sigma_{\text{H2}}$), the molecular gas main sequence (MGMS; $\Sigma_{\text{H2}}$\,--\,$\Sigma_*$), and the resolved star forming main sequence (rSFMS; $\Sigma_{\text{SFR}}$\,--\,$\Sigma_*$) are shown. Given that any two of these relationships completely specifies the final one, at least one of these relationships may be spurious (i.e., a direct product of the more fundamental relations). \textit{Right-hand panel:} The partial correlation coefficient matrix for the parameters studied on the left, reproduced from \cite{Baker2022}. At fixed values of the other parameters in the set, both the SK and MGMS retain strong correlation strengths. However, the rSFMS has a correlation strength tending to zero, when gas mass surface density is controlled for. This definitively demonstrates that it is the star forming main sequence which is spurious, i.e., arising solely out of the more fundamental relationships presented here. }\label{f19}
\end{figure}

The first clue in answering these questions comes from the fact that star formation varies substantially within individual star forming galaxies (e.g., \citealt{Sanchez2013, Wuyts2013, Tacchella2015, Belfiore2017, Ellison2018}). Therefore, the global SFMS cannot be the whole explanation for regulation of star formation within galaxies. Moreover, many studies find that the resolved SFMS is tighter, and varies less with external parameters, than the global SFMS (e.g., \citealt{Cano-Diaz2016, Hsieh2017, Bluck2020a, Bluck2020b}). In fact, once accounting for the rSFMS, \cite{Bluck2020a} demonstrate that global stellar mass has no predictive importance over local star formation at all in star forming systems (see back to Fig.~\ref{f25}). 

Therefore, it is reasonable to envisage a scenario in which the global SFMS emerges as the spatially integrated rSFMS, in a bottom-up manner (for further evidence, see \citealt{Wuyts2013, Cano-Diaz2016, Abdurrouf2018, Hsieh2017, Popesso2019, Bluck2020a}). This essentially answers our first question. The local/ spatially resolved SFMS is more fundamental than the global SFMS. However, this does not answer our second question, i.e., what causes the SFMS (on any scale) to exist in nature?

To answer this important question, \cite{Lin2019} investigate the tri-variate relationship between star formation rate surface density ($\Sigma_{\text{SFR}}$), molecular gas mass surface density ($\Sigma_{\text{H2}}$), and stellar mass surface density ($\Sigma_*$). This is displayed in Fig.~\ref{f19} (left-hand panel). It is clear that these data form a cone within the 3D parameter space. Moreover, the 3D relationship may be projected into three 2D relationships: (i)~the Kennicutt-Schmidt relation (SK; $\Sigma_{\text{SFR}}$\,--\,$\Sigma_{\text{H2}}$); (ii)~the resolved molecular gas main sequence (MGMS; $\Sigma_{\text{H2}}$\,--\,$\Sigma_*$), and (iii)~the resolved star forming main sequence (rSFMS; $\Sigma_{\text{SFR}}$\,--\,$\Sigma_*$). Any two of these relationships completely specifies the third. Hence, it becomes interesting to ask, which of these relationships are fundamental and which are a byproduct of the others? Of course, it also remains possible that none of these projections are truly fundamental, since one can make an infinite number of rotations within the 3D space.

To answer this question, \cite{Lin2019} explore the tightness of each relationship, concluding that, since the rSFMS is the least tight, it is most probably the least fundamental (see legend in Fig.~\ref{f19}, left-hand panel). However, the variation in scatter is mild and the possibility for differential measurement uncertainty to impact this conclusion is significant. To investigate this further, \cite{Baker2022} perform a partial correlation analysis of the same data, assessing the strength of correlation of each relationship, whilst holding the third variable fixed (see, e.g., \citealt{Bluck2020a} for full details on this statistical method). The results from this analysis are displayed on the right-hand panel of Fig.~\ref{f19}.

The SK and MGMS relationships have correlations between variables which remain strong at fixed values of the third variable. This clearly indicates that they cannot arise from the rSFMS. However, crucially, the rSFMS relationship has a correlation between variables which tends to zero, once $\Sigma_{\text{H2}}$ is controlled for. This clearly demonstrates that the rSFMS is not fundamental, but rather emerges as a byproduct of the more fundamental SK and MGMS relationships. Ultimately, the SK relation arises primarily as a result of fueling (the need for dense molecular gas to form stars) and potentially, secondarily, as a result of more dense gas having a shorter dynamical time scale (see \citealt{Kennicutt1998b}). 

However, the origin of the MGMS is less well agreed upon at a physical level within the literature. The most plausible explanation for its existence is that gas distributes within galaxies as a strong function of the local gravitational potential, which is itself set primarily by the distribution in stellar mass (see, e.g., \citealt{Scoville2016, Tacconi2018, Lin2019, Baker2022}). Hence, in actively star forming galaxies, which have unimpeded cold gas accretion (or cooling flows), gas distributes within the system according to the underlying stellar mass distribution. Subsequently, stars form out of the dense molecular gas by fuel-limited processes, i.e., the KS relationship. 

This narrative implies that the SFMS (on any scale) is a byproduct of more fundamental processes and relationships. Nonetheless, it remains highly useful as a tool for defining what is `typical' in terms of star formation within galaxies and, hence, for identifying atypical, quiescent galaxies (and regions within galaxies).

\subsubsection{Evolution in scaling laws}

If the above narrative is correct, it ought to additionally be able to account for the observed evolution in the SFMS (primarily seen on global scales; e.g. \citealt{Noeske2007a, Daddi2007, Elbaz2007, Whitaker2012, Schreiber2015}). To investigate this, a number of important works explore how the gas content of galaxies, and their star forming efficiency, vary with cosmic time (see, e.g., \citealt{Genzel2015, Scoville2017, Tacconi2018, Baker2023, Tacconi2020}). Generally speaking, these studies find that the ISM of galaxies become more gas rich for their stellar masses at higher redshifts (which is indicative of evolution to lower gas fractions over cosmic time), and have shorter depletion times at higher redshifts (which is indicative of evolution towards lower star formation efficiencies over cosmic time). In combination, this gives rise to the decline in sSFR within galaxies as a function of cosmic time (as discussed in Sect.~\ref{s2}).

\begin{figure}[ht] 
\centering
\includegraphics[width=1\textwidth]{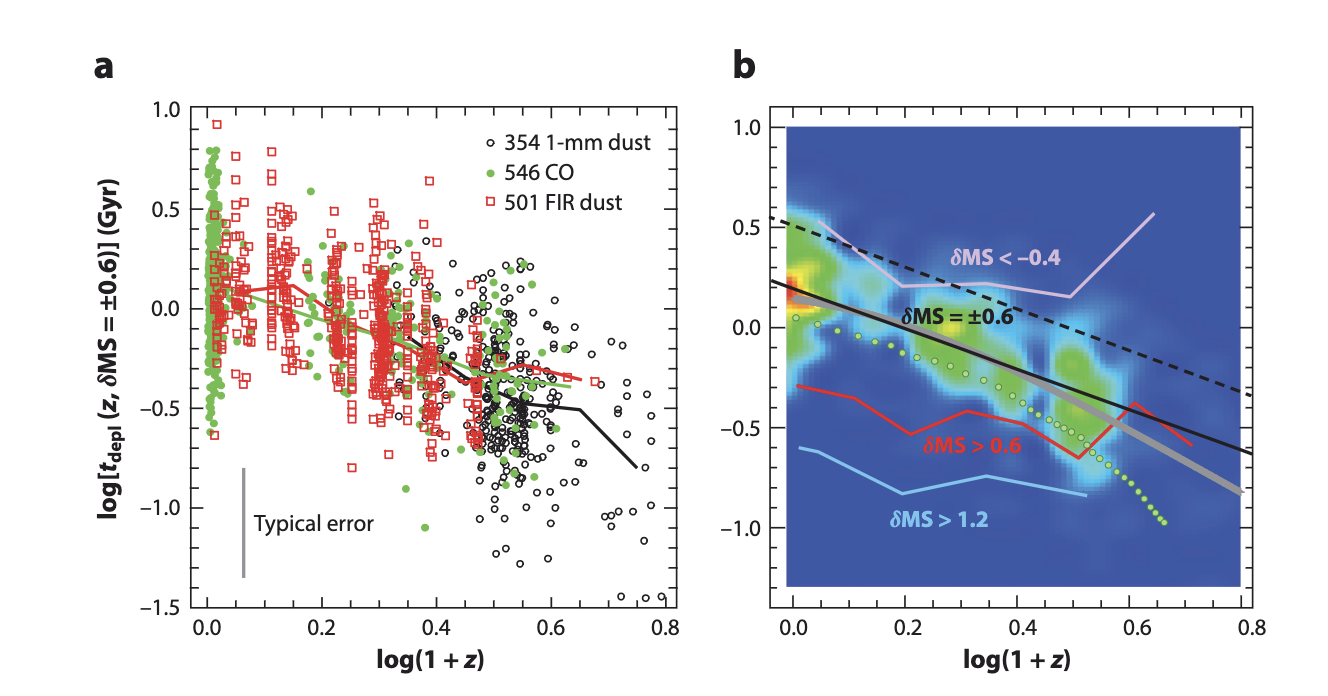}
\includegraphics[width=1\textwidth]{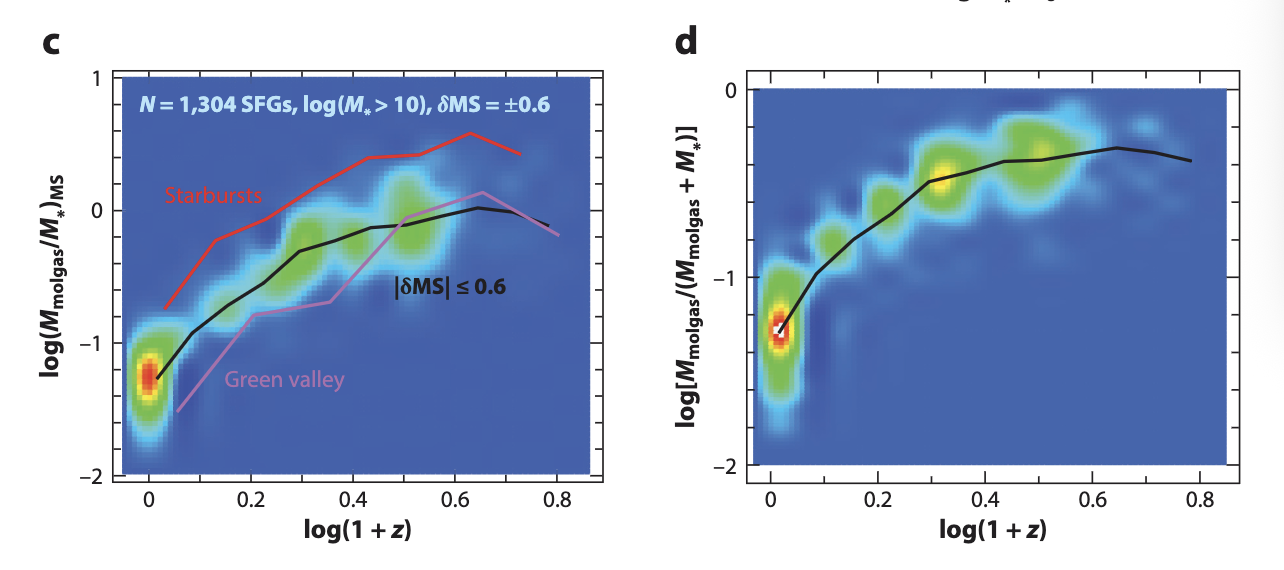}
\caption{The evolution in depletion time (top panels) and gas fraction (bottom panels) for over 1000 galaxies, reproduced from \cite{Tacconi2020}. {\it Top panels:} Panel (a) shows the results split by method to infer gas  (i.e., 1\,mm dust, CO, and FIR dust) and panel (b) combines the data into a density plot, with lines indicating the location of galaxies above, on, and blow the global SFMS. {\it Bottom panels:} Panel (a) shows the molecular gas fraction normalized by stellar mass and panel (b) shows the molecular gas fraction normalized by total baryonic mass. It is clear that the depletion time decreases with redshift (top panels) and the gas fraction rises with redshift (bottom panels). }\label{f20}
\end{figure}

As an example, in Fig.~\ref{f20} we present the evolution in the depletion time ($\tau_{\rm dep}~\equiv~1/{\rm SFE}$; top panels) and the evolution in the gas fraction (bottom panels). This figure is reproduced from several plots presented in \cite{Tacconi2020}. It is clear that the depletion time increases with cosmic time and the gas fraction decreases with cosmic time. Ultimately, this combination gives rise to the evolving SFMS. In \cite{Baker2023}, it is estimated that $\sim$70\% of the SFMS evolution is attributable to evolution in gas content, with the remaining $\sim$30\% attributable to evolution in SFE (or equivalently, $\tau_{\rm dep}$). For a dedicated review on the evolution of the ISM within galaxies, see \cite{Tacconi2020}.

The evolution in gas content within galaxies is an expected consequence of the declining accretion rate into haloes with cosmic time (see, e.g., \citealt{Neistein2008, McBride2009, Dekel2009, Fakhouri2010}). However, the evolution in the efficiency with which stars form out of their molecular gas reservoirs is more surprising. This implies that, at a fixed gas mass, the global gas content is more efficiently forming stars within galaxies at earlier cosmic times. This may potentially be explained by the size evolution of galaxies, whereby galaxies are smaller for their stellar mass, and hence denser, at earlier cosmic times (see, e.g., \citealt{Trujillo2007, Buitrago2008, Newman2012}). This further implies that for a constant \textit{local} KS relationship, the global relationship would still be expected to rise with redshift. Although a plausible explanation, at the time of writing no consensus exists in the literature to fully account for variation in the efficiency of star formation from molecular gas with cosmic time.

To summarize, the SFMS (which is critical for identifying quenched galaxies at all epochs) is not fundamental. It arises from local physics, with the resolved SFMS being a more accurate tracer of star formation within galaxies than the global version. Moreover, even the rSFMS is not a fundamental relationship. Instead, it emerges out of the (local) MGMS and KS relationships. While these relationships themselves may ultimately prove not to be fundamental, they have been convincingly shown to be more so than the SFMS, which emerges purely as a mathematical byproduct of the other two relationships.

The evolution in the SFMS is ultimately a result of evolution in molecular gas content (primarily) and evolution in the efficiency with which stars form out of the molecular gas reservoir (secondarily). Nonetheless, the SFMS is a fortunate byproduct of these more fundamental relationships because it establishes what is typical for galaxies in terms of star formation at a given epoch, stellar mass, and scale of measurement (i.e., global vs. local). Ultimately, this enables one to identify quenched galaxies, and regions within galaxies, without the need for expensive (and often uncertain; see \citealt{Bolatto2013}) molecular gas mass constraints.

\begin{figure}[ht] 
\centering
\includegraphics[width=1\textwidth]{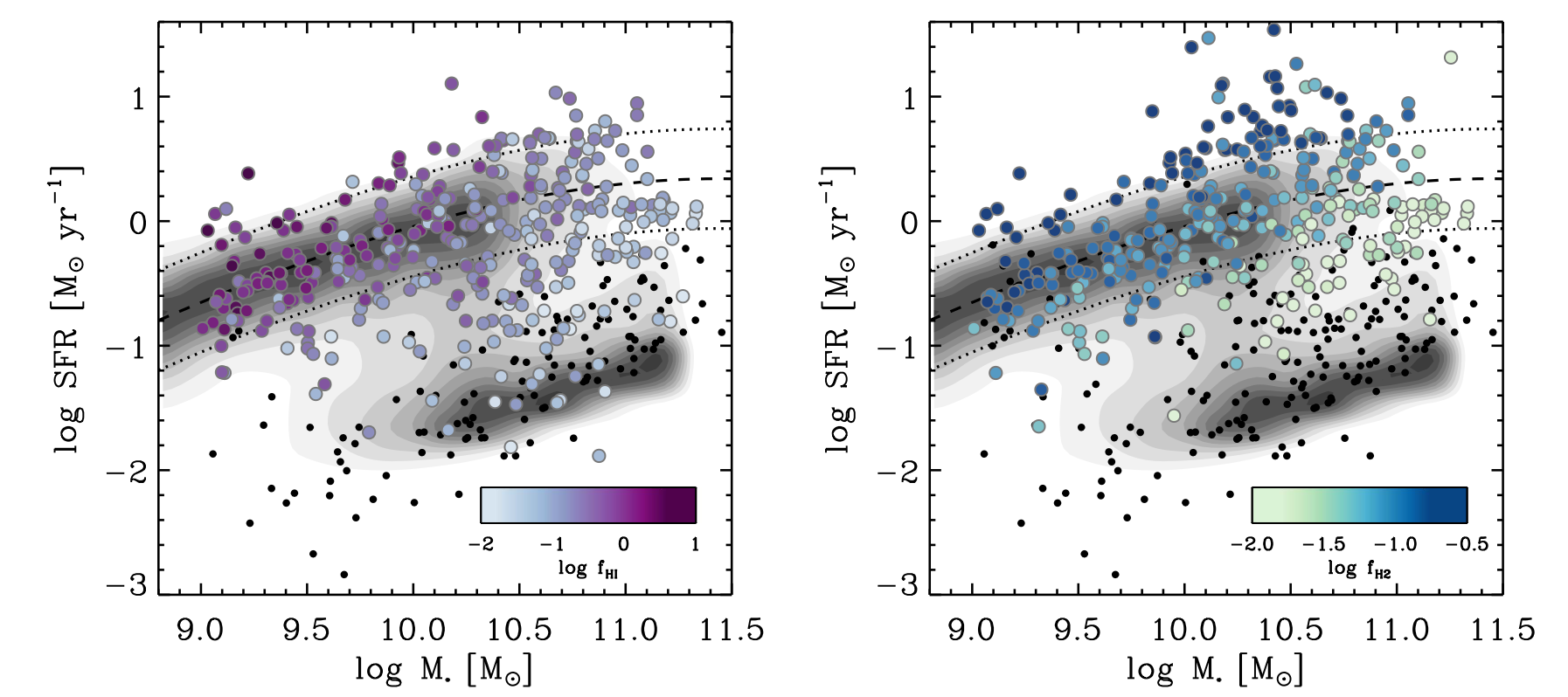}
\includegraphics[width=0.6\textwidth]{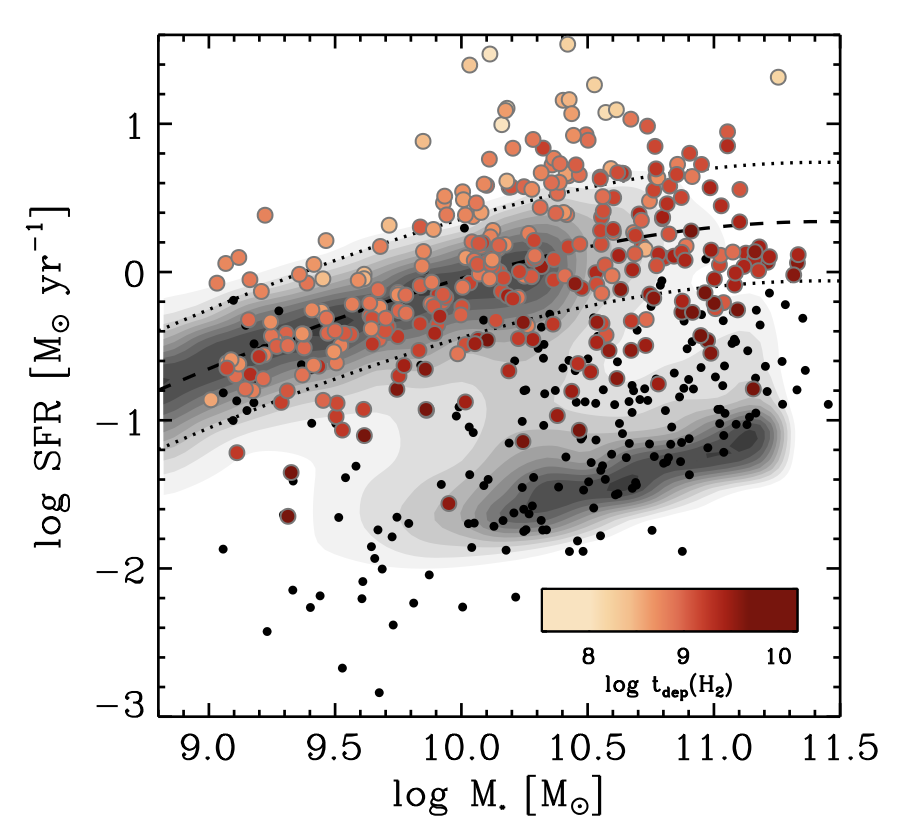}
\caption{The SFMS (grey-scale on all panels) with overlaid measurements of HI gas fraction (top-left panel), H$_2$ gas fraction (top-right panel), and depletion time (bottom panel), as indicated by the color bars. Non-detections are shown as black points. All panels are reproduced from \cite{Saintonge2017}. It is clear that as galaxies move towards quiescence (falling below the SFMS), the HI and H$_2$ gas fractions decline significantly (top panels). Furthermore, the depletion times increase as galaxies transition towards quiescence (bottom panel). Hence, quenching involves both the reduction in gas content and the reduction in star formation efficiency. The data in this figure come from the xCOLD GASS sample (\citealt{Saintonge2016, Saintonge2017}). }\label{f20aaa}
\end{figure}

\subsection{Declining fuel vs. efficiency as drivers of quenching}

\subsubsection{Global results}

Many works have investigated how the gas fraction and star formation efficiency within galaxies vary along the quenching sequence, i.e., as sSFR declines on global scales (see, e.g., \citealt{Saintonge2011, Saintonge2016, Saintonge2017, Catinella2018, Tacconi2018, Tacconi2020, Saintonge2022}). A broad consensus has emerged through these works that {\it both} $f_{\rm gas}$ and SFE decline as sSFR declines during quenching.

For example, in Fig.~\ref{f20aaa}, we present results from the xCOLD GASS survey, reproduced from \cite{Saintonge2017}. Each panel shows the local SFMS relationship as a grayscale, for comparison. The colored points indicate the HI gas fraction (top-left panel), the H$_2$ gas fraction (top-right panel), and the depletion time (i.e., the inverse of SFE), as labeled by the color bars within each panel. As galaxies move away from the SFMS towards the quenched population, a significant reduction in both HI and H$_2$ gas content is evident. Furthermore, a reduction in SFE (i.e., increase in $\tau_{\rm dep}$) is also evident.

Therefore, the quenching of galaxies in the local Universe must involve reduction in gas content across both the atomic and molecular phases, in addition to an increase in depletion time. These observational constraints pose stringent requirements for theoretical models of quenching. We discuss theoretical approaches to reproduce these results in Part~II of this review. 

Briefly, ejective feedback from starbursts and AGN can directly deplete the ISM of gas (e.g., \citealt{DiMatteo2005, Springel2005c, Hopkins2006, Hopkins2008}), radio-mode AGN feedback may prevent cooling flows into the ISM from the CGM leading to gas depletion via ongoing star formation (e.g., \citealt{Croton2006, Bower2006, Bower2008, Sijacki2007, Weinberger2017}), and local feedback from AGN (e.g., \citealt{Weinberger2018, Zinger2020, Piotrowska2022}) and/or morphological stabilization (e.g., \citealt{Martig2009, Gensior2020, Gensior2021}) may lead to a reduction in SFE. It is likely that some combination of these theoretical modes will be required to fully account for observations.

\subsubsection{Spatially resolved results}

In addition to the global constraints discussed above, several studies have investigated whether quenching on kpc scales (i.e., reduction in local sSFR) is driven by a reduction of gas content ($f_{\rm gas}$), a reduction in efficiency (SFE), or both processes acting in tandem (see, e.g., \citealt{Ellison2020, Ellison2021, Brownson2020, Piotrowska2020, Pan2024}). For instance, \cite{Ellison2020} find that while the level of star formation in star forming systems is regulated primarily by molecular gas content (in agreement with \citealt{Lin2019, Baker2022, Baker2023}), offsets from the rSFMS are primarily governed by varying SFE, with a secondary dependence on varying $f_{\rm gas}$. These spatially resolved results are in close alignment with the global trends (see, e.g., \citealt{Saintonge2016, Saintonge2017, Piotrowska2020}). 

Additionally, \cite{Pan2024} investigate whether the quenching mode varies with radius within transitioning galaxies. They find that SFE reduction dominates at the center of green valley galaxies, but that in the extended disc a mix of $f_{\rm gas}$ and SFE reduction is needed to explain the reduction in local sSFR. Interestingly, this conclusion is essentially the opposite of that reached previously in \cite{Lin2017}, who find that reduction in $f_{\rm gas}$ dominates the quenching of central regions within galaxies (albeit based on a much smaller sample). These discrepancies hint at a high diversity between galaxies, in addition to within galaxies. This is explored in detail in \cite{Ellison2021}, who find that the key scaling relations all vary substantially between galaxies, pointing towards non-universality. Hence, the scaling relationships (and offsets from them) emerge as statistical ensembles rather than as fundamental physical laws. Ultimately, any given galaxy may exhibit with unique trends that are discrepant from the population average.

Although highly interesting, all of the molecular gas based analyses suffer from an important limitation. To perform these investigations, the mass of H$_2$ must be estimated indirectly via calibrations based on CO transition lines. Yet, many fully quiescent galaxies (and regions within galaxies) present as non-detections in CO. Hence, it is most probably safest to interpret the importance of molecular gas depletion to quenching as a lower limit in most prior studies (see \citealt{Brownson2020} for further discussion). Nonetheless, that some regions within galaxies are suppressed in star formation without being devoid of molecular gas is clear and, hence, requires an explanation. This naturally implies that SFE must be lowered in these systems, resulting in a lack of star formation despite the presence of dense molecular gas reservoirs. 

The leading theories to account for this reduction in efficiency are: (i)~turbulence injection from AGN (e.g., \citealt{Weinberger2017, Zinger2020, Piotrowska2022}), and (ii)~morphological stabilization from galaxy dynamics (e.g., \citealt{Martig2009, Gensior2020, Gensior2021}). We will discuss these avenues for quenching in detail in Part~II of this review series. Both of these processes are expected to be most effective towards the center of galaxies and, hence, are consistent with the recent results pointing towards SFE-driven quenching dominating within central/ bulge regions within green valley galaxies (e.g., \citealt{Pan2024}).

%%%%%%%%%%%%%%%%%%
%                %
%    HIGH-Z      %
%                %
%%%%%%%%%%%%%%%%%%

\section{Quenching at the high-redshift frontier with JWST}\label{s7}

\noindent The James Webb Space Telescope (JWST; \citealt{Gardner2006, Pontoppidan2022, Gardner2023, Rigby2023, McElwain2023}) is NASA's premier infrared telescope, launched on 25 December, 2021, aboard an Ariane 5 rocket from Kourou, French Guiana. It currently resides at Sun\,--\,Earth Lagrange Point 2, approximately 1.5 million km from Earth. JWST is often seen as the successor to the Hubble Space Telescope (HST), achieving a six-fold increase in collecting area and vastly enhanced sensitivity in the infrared. 

JWST has transformed the study of quenching at high redshifts. Prior to JWST, identifying and characterizing quiescent galaxies beyond $z \sim 3$ was extremely challenging due to their intrinsically red spectral energy distributions, which are shifted further into the infrared with increasing redshift. The unprecedented near- and mid-infrared sensitivity of the JWST instrument suite enables the robust detection of massive quiescent systems during the first few Gyr of cosmic history. In particular, the spectroscopic capabilities onboard JWST enable absorption line measurements at redshifts previously inaccessible to any other facility (e.g., \citealt{Carnall2023, Valentino2023, Belli2024}). Furthermore, JWST provides spatial resolution comparable to HST at rest-frame optical wavelengths, enabling the structural and resolved spectroscopic analysis of compact high-redshift galaxies. Taken together, these advances have opened up a fundamentally new observational window on the formation, evolution, and quenching of galaxies in the early Universe.

Since its launch, a large number of extragalactic surveys have been conducted, including: JADES (\citealt{Eisenstein2023a, Eisenstein2023b, D'Eugenio2025c}), COSMOS-Web (\citealt{Casey2023}), NGDEEP (\citealt{Bagley2023}), JEMS (\citealt{Williams2023}), PRIMER (\citealt{Donnan2024}), PEARLS (\citealt{Windhorst2023a, Windhorst2023b}), and CEERS (\citealt{Finkelstein2025}). These surveys have discovered galaxies at earlier cosmic times than previously seen, as well as identifying extremely surprising characteristics of the first populations of galaxies. 

Indeed, the results from the first five years of JWST operations have been dramatic, even challenging the established $\Lambda$CDM cosmology (e.g., \citealt{Labbe2023, Boylan2023, Glazebrook2024, Carnall2024}). Moreover, the existence and abundance of high-mass, quenched galaxies in the very early Universe has challenged contemporary galaxy formation models, almost to breaking point (e.g., \citealt{Carnall2023, Carnall2023a, Carnall2024, Baker2025a, Zhang2025}). In addition to the results on galaxies, supermassive black holes (SMBHs) are identified in the early Universe at surprisingly high masses, and with clear evidence of AGN activity and feedback (e.g., \citealt{Matthee2024, Maiolino2024a, Maiolino2024b, Belli2024, D'Eugenio2024, Kocevski2025}). Taken together, these discoveries imply a Universe which is already surprisingly mature at extremely early cosmic times. 

In this section we provide a comprehensive overview of these discoveries, how they have impacted our understanding of galaxy formation and quenching, and how we might move forwards as a field. We start by giving an overview of the status of high-redshift quenching prior to JWST. We then discuss photometric and spectroscopic evidence for both high- and low-mass quenched systems in the very early Universe with JWST, in the era of HST. We go on to discuss in some detail the claims of `Universe breakers' and potential tensions with both cosmology and galaxy formation theory as a consequence of contemporary JWST observations. Finally, we discuss clear evidence for AGN-feedback operating within galaxies at very early cosmic times, potentially explaining the existence of the early quenched population.

\subsection{The status of high-{\it z} quenching in the HST era}\label{s71}

\noindent Prior to the advent of JWST, quiescent galaxies were already known to exist in the Universe out to at least redshifts of $z \sim 2 - 4$ (e.g., \citealt{Franx2003, Chen2004, McCarthy2004, Caputi2004, Kriek2008, Kriek2009, Carnall2018, Merlin2018, Merlin2019, Santini2019, Carnall2020, Santini2021}). Furthermore, there were spectroscopic confirmations of many individual quiescent galaxies at these relatively early cosmic times as well (see, e.g., \citealt{Cimatti1999, Cimatti2002, Cimatti2004, Cimatti2006, Kriek2006a, Kriek2006b, Bezanson2013, Glazebrook2017, Carnall2019b, Valentino2020}). 

A particularly noteworthy example is a massive ($M_* = 1.7 \times 10^{11}\,M_\odot$) fully quenched galaxy at $z = 3.72$, spectroscopically confirmed with Keck/MOSFIRE, reported by \cite{Glazebrook2017}. From full spectrum fitting, a formation and quenching time at $z > 5$ was inferred. This already implied that at least some galaxies formed very early and rapidly in the history of the Universe, and that quenching could occur before cosmic noon ($z \sim 2$). Nonetheless, the vast majority of quenched systems at high-$z$ were found to be at cosmic noon, or (substantially) later (e.g., \citealt{Brammer2011, Whitaker2012, Muzzin2013, Straatman2014}). For instance, \cite{Tomczak2014} find that the number density of quiescent galaxies increases by an order of magnitude from $z = 3 - 1$, and \cite{Davidzon2017} find that quiescent galaxies only dominate the high-mass end of the stellar mass function by $z \sim 1$.

In a large study of over 20,000 high-$z$ galaxies observed within HST-CANDELS, \cite{Merlin2019} provide strong evidence for a substantial population of massive, rapidly quenched galaxies already in place at $z \gtrsim 3$. This implies that at least some of the most massive galaxies in the Universe formed and ceased their star formation within the first $\sim$2\,Gyr after the Big Bang. Additionally, \cite{Merlin2019} provide a detailed comparison with cosmological simulations. Generally, they find that simulations struggle to reproduce the number density of quiescent systems at the earliest cosmic times. This important work foreshadows results to come from JWST, which we will discuss in detail in Sect.~\ref{s73}.

These early quiescent galaxies were found to be highly compact, often presenting with effective radii, $R_e~\lesssim~1\,$kpc (see \citealt{Trujillo2006, Trujillo2007, Buitrago2008, Franx2008, Cheung2012, Cimatti2012, Fang2013, Barro2013, Straatman2014, vanDokkum2015, Forrest2020}). Additionally, they are frequently found to host substantial bulge components, with bulge-to-total stellar mass ratios of $(B/T)_*~\gtrsim~0.5$ (see, e.g., \citealt{Lang2014, Bluck2022}). Leveraging the well established local relationships between SMBH mass and galaxy properties (e.g., \citealt{Maggorian1998, Ferrarese2000, Haring2004}), many studies point out that these systems likely contain substantial SMBHs (see, e.g., \citealt{Bell2012, Wake2012, Fang2013, Bluck2014, Lang2014, Bluck2016, Martin2018a, Martin2018b, Bluck2022, Bluck2023}). Hence, it is at least plausible that these early quiescent systems quench via AGN feedback in the early Universe. 

Additionally, the observations of quiescent galaxies at cosmic noon confirmed the downsizing scenario (e.g., \citealt{Thomas2005, Thomas2010}), whereby the most massive galaxies tend to form their stars earlier and faster than less massive systems (see, e.g., \citealt{Whitaker2013, Barro2013, Barro2014, Fumagalli2016, Belli2019, Kriek2019}). The downsizing scenario, which will be discussed more in Part II in relation to simulations, was derived from the local galaxy population and suggests that massive quiescent galaxies should be found at high redshifts. These results match with the presence of quiescent galaxies at cosmic noon, and also the discovery of quiescent galaxies at even higher redshifts. 

Given that early quenched galaxies are highly compact, an important question arises: \textit{how do these systems grow into present-day giant elliptical galaxies, without forming a substantial amount of new stars?} The leading explanation appears to be via dry (gas-poor) minor mergers, which are effective at growing the sizes of galaxies without forming (or adding) a large amount of new stars, whilst simultaneously leaving the high central density and velocity dispersion largely unchanged (see, e.g., \citealt{Naab2009, Newman2012, Cimatti2012, Trujillo2011, Hilz2013}). Although, alternative scenarios such as `puffing up' via ejective AGN feedback have also been proposed (e.g., \citealt{Fan2008, Ragone-Figueroa2011, Trujillo2011}).

Prior to JWST, a broad consensus emerged whereby galaxy evolution is roughly described by three fundamental stages (see, e.g., \citealt{Naab2009, Oser2010, Barro2013, Naab2014, Barro2014, Zolotov2015, Tacchella2015, Tacchella2016, Barro2017}): (i)~in situ star formation and disc growth in the (very) early Universe; (ii)~wet-compaction events via gas-rich mergers and/or violent disc instabilities at around cosmic noon and below; and (iii)~quenching via AGN feedback at late cosmic times, as a consequence of SMBHs reaching a critical mass for sustained preventative feedback (and potentially aided by ejective feedback in fast-quenching events; e.g., \citealt{Feruglio2010, Maiolino2012, Cicone2012, Cicone2014, Baron2018}). 

Most galaxy formation and evolution models prior to 2020 were broadly in line with this view of galaxy formation and evolution (see, e.g., \citealt{Croton2006, Bower2008, Henriques2015, Vogelsberger2014a, Vogelsberger2014b, Schaye2015, Somerville2015, Nelson2018, Pillepich2018, Dave2019}). However, it is fair to acknowledge that many of these models still struggle to match several aspects of the local galaxy population, in particular the age and metal-content of the most massive galaxies.

Therefore, prior to the advent of JWST, many extragalactic astrophysicists expected the very early Universe to be mostly composed of low-mass, star forming galaxies. Certainly by cosmic noon, massive quiescent galaxies were expected, but not at the very earliest cosmic times and certainly not in great abundance (although there were already some claims of tension here; e.g., \citealt{Merlin2019}). Therefore, at $z > 4$, in the simplest possible terms, one anticipated the early Universe to be a galactic nursery --- filled with proto-galaxies and dwarf systems, slowly bursting into their star forming phase. The reality (which we turn to next) has been quite surprising for many in the community.

\subsection{Early quenched galaxies observed with JWST}\label{s72}

\noindent There are two primary routes to identifying quiescent galaxies at any epoch: photometry and spectroscopy. Generally, photometry is the faster route, as it may be applied to large samples of galaxies from imaging surveys without the need for detailed spectroscopic follow-up. Yet, spectroscopy is much more accurate. In the first year of JWST observations, a host of photometrically selected quiescent galaxy candidates were found at $2 < z < 5$ (see \citealt{Carnall2023, Valentino2023, Ferreira2023, Nanayakkara2024, delaVega2025}). Within the photometric approach two main variants exist: (i)~the location of galaxies on rest-frame color diagnostic plots (e.g., UVJ and near-UV\,--\,optical\,--\,near-IR diagrams; see \citealt{Valentino2023}); and (ii)~the extraction of star formation rates and star formation histories (SFHs) via CSP SED fitting (see \citealt{Carnall2023}). Both of these approaches may identify systems which are likely forming stars at rates much lower than their contemporaneous counterparts, at similar stellar masses.

\begin{figure}[htbp]  
\centering
\includegraphics[width=0.85\textwidth]{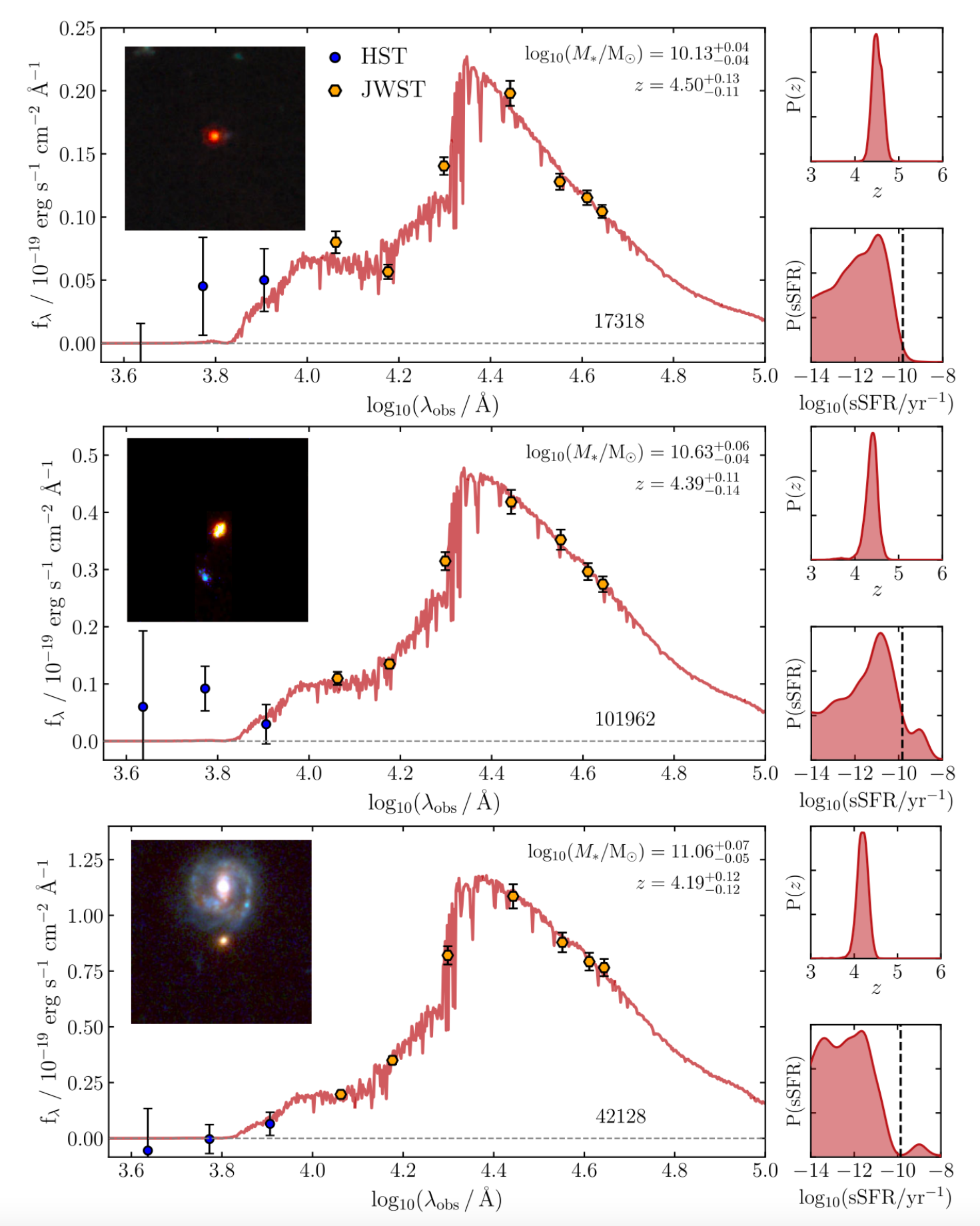}
\caption{Examples of three high-redshift quiescent galaxies determined from SED fitting of photometry from JWST and HST multi-band observations. This figure is reproduced from \cite{Carnall2023}. Photometry from HST is shown as blue circles, with photometry from JWST shown as gold circles. The posterior median model from {\small BAGPIPES} is shown in red on the main panels, with the posterior distributions for redshift ($z$) and specific star formation rate (sSFR) shown as histograms to the right of the main panels. Inset within the main figure panels are RGB cutouts, composed of the F444W, F200W and F150W JWST images (respectively). The posterior median stellar mass and redshifts are displayed as text on the top right of the main panels. These systems represent the highest redshift photometric quiescent galaxy candidates from a sample of ten at $z > 3$ discovered by \cite{Carnall2023}.}\label{f26}
\end{figure}

In Fig.~\ref{f26}, we show photometry (data points with error bars, shown in blue from HST and gold from JWST) for the three highest redshift potential quenched galaxies identified in \cite{Carnall2023}. The posterior median CSP model is shown as the red-colored spectra. Additionally, to the right of each panel the posterior distributions for redshift and sSFR are shown (with the dashed vertical line indicating the quenched threshold at each redshift). Within the main panel, an RGB image constructed from JWST bands is displayed. These systems show a very clear Balmer break, indicative of an old stellar population with no young (O- and B-type) stars.

The above photometric results are highly valuable because they select possibly quenched galaxies at early cosmic times. However, there are a number of potential issues with pure photometric selection (see, e.g., \citealt{Kauffmann2020, Haro2023, Atek2023a, Atek2023b, Matthee2024, Turner2025, Adams2025, Harvey2025}). First, there are age\,--\,dust\,--\,metallicity degeneracies inherent in the SED fitting of photometry, which may be mitigated by probing larger wavelength ranges and increasing the resolution of the observed SED (e.g., \citealt{Worthey1994, Conroy2013, Robotham2020}). Second, for deep observations probing the high-$z$ Universe, there is a potential risk of misidentifying Balmer and Lyman breaks, leading to catastrophic errors in redshift determination (e.g., \citealt{Ilbert2006, Kauffmann2020, Adams2023, Adams2025}). Third, unresolved emission lines may be interpreted as a Balmer (or Lyman) break in the absence of spectroscopy, again leading to catastrophic errors in the SED fitting (e.g., \citealt{Haro2023, Atek2023a, Atek2023b, Zavala2023, Clausen2025}). Fourth, the possibility of AGN contamination is high in photometric fitting, often leading to over-estimates in stellar mass and erroneous star formation histories (e.g., \citealt{Ciesla2015, Thorne2022, Durodola2025}). Fifth, there remain issues with `outshining' whereby recent star formation obscures the bulk of the historic star formation history (e.g., \citealt{Maraston2010, Leja2019, Tacchella2022}). 

In addition to the above, systematics arising from the assumed SFH and the fitting code itself are prevalent and lead to severe discrepancies on occasion (e.g., \citealt{Turner2025, Harvey2025}). Consequently, spectroscopic confirmation is the `gold standard' for the identification of both high-$z$ galaxies in general, and high-$z$ quenched systems in particular.

\begin{figure}[htbp]
\centering
\includegraphics[width=0.9\textwidth]{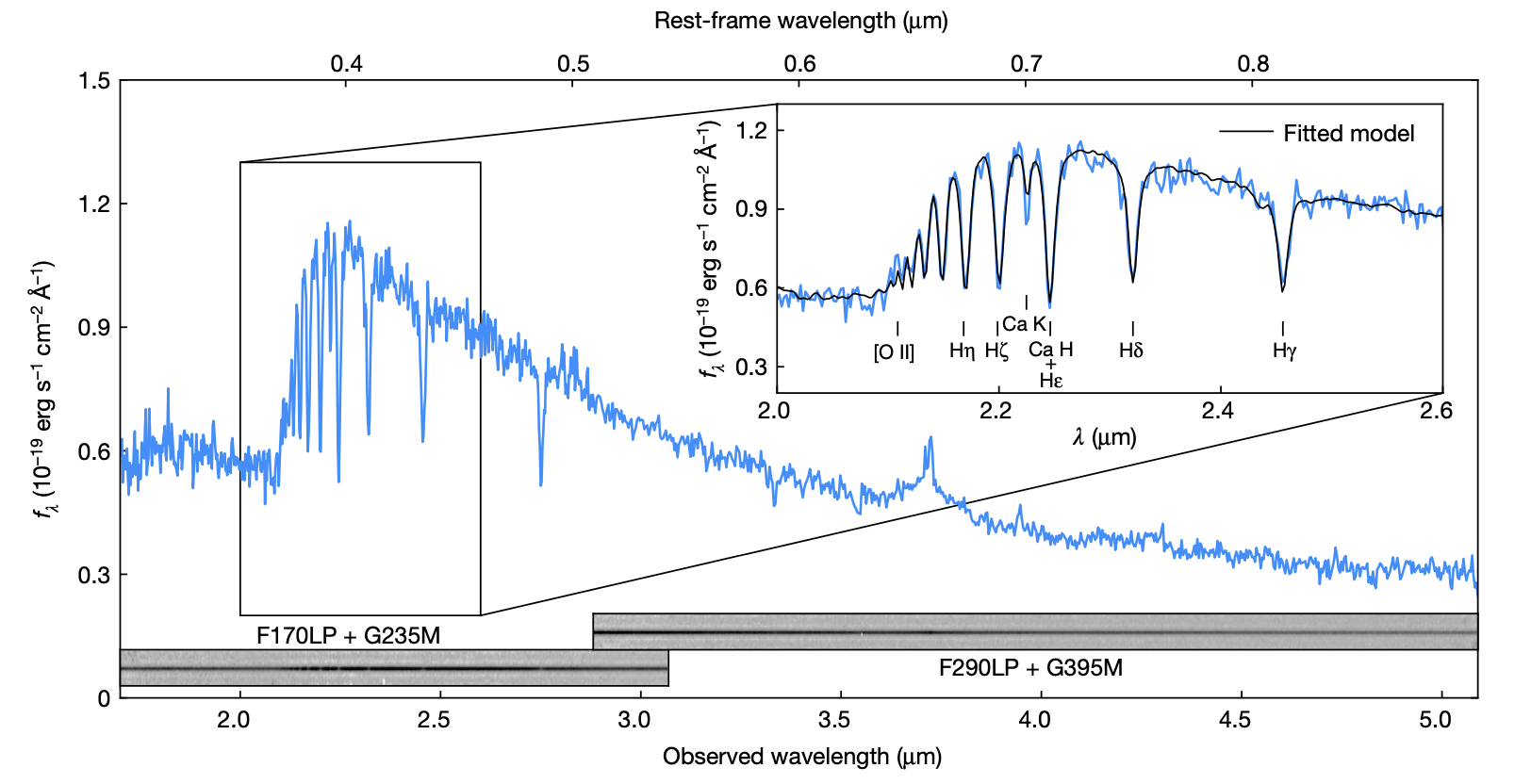}
\includegraphics[width=0.8\textwidth]{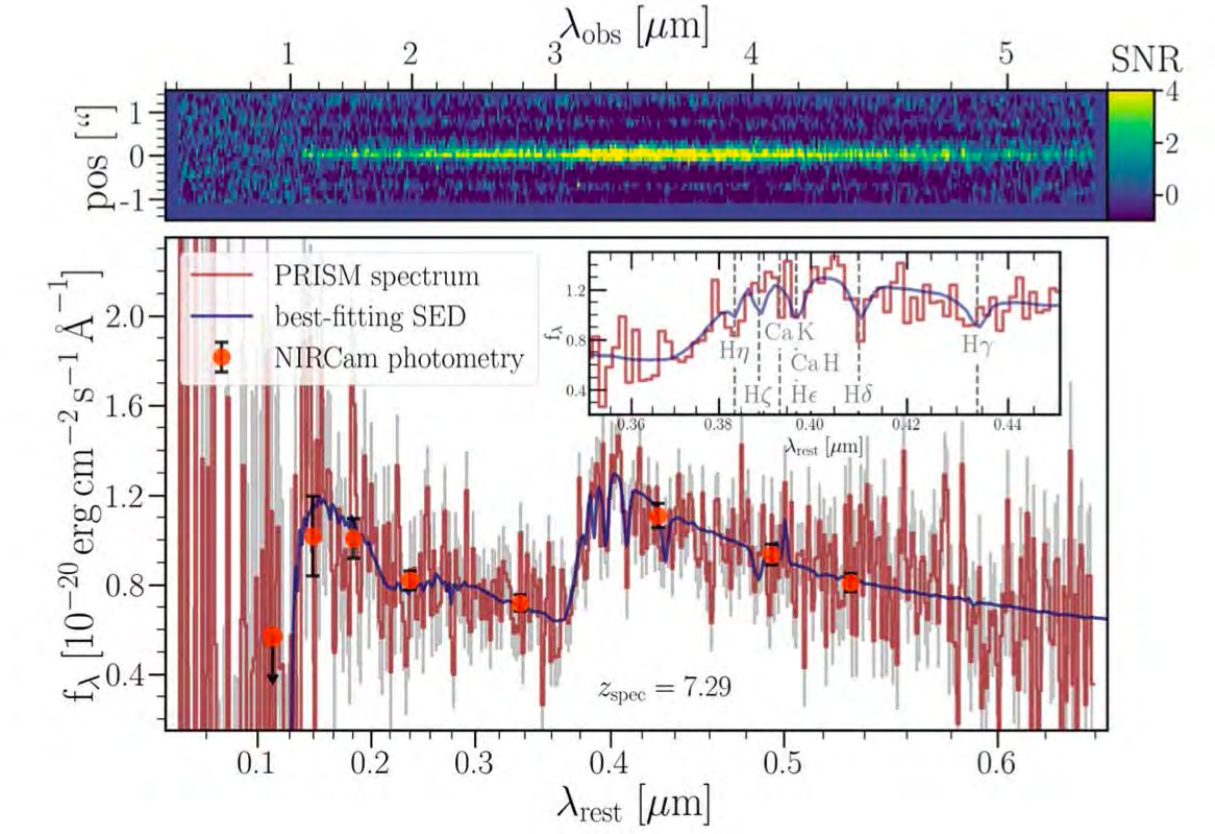}
\caption{Spectroscopic confirmation of high-$z$ massive quenched galaxies. \textit{Top Panel:} JWST/NIRSpec spectrum of GS-9209, reproduced from \cite{Carnall2023a}. This galaxy is observed at $z_\mathrm{spec} = 4.6582 \pm 0.0002$, with a stellar mass of $\log(M_*/M_\odot) = 10.58 \pm 0.02$. The spectrum clearly shows a Balmer break, a lack of most nebular emission lines, and evidence of deep absorption features, consistent with a recently quenched (post starburst-like) system. The inset panel shows a zoom-in region around the Balmer break, clearly showing deep absorption features (consistent with A-type stellar spectra), with the posterior median {\small BAGPIPES} model overlaid in black. The SFH is recovered via full spectrum fitting, indicating that this galaxy underwent quenching at $z_Q \sim 6.5$. Although weak, a clear broad H$\alpha$ emission line is evident, signaling that this system is most probably a Type~I AGN. \textit{Bottom Panel: } NIRSpec/PRISM spectrum of RUBIES-UDS-QG-z7, reproduced from \cite{Weibel2025}. NIRCam photometry is overlaid as large orange dots, with the best-fitting SED from {\small PROSPECTOR} shown in blue. This galaxy is observed at $z_\mathrm{spec} = 7.29 \pm 0.01$, with a stellar mass of $\log(M_*/M_\odot) = 10.23 \pm 0.04$. A clear and strong Balmer (and Lyman) break is evident, with no evidence of nebular emission lines. The best fit SFH reveals quenching occurred 50 -- 100 Myr prior to observation. At the time of writing, this is the highest redshift massive quenched system known.}\label{f27}
\end{figure}

Many high redshift quiescent galaxies have now been spectroscopically confirmed (e.g., \citealt{Carnall2023a, Glazebrook2024, Nanayakkara2024, deGraaff2024, Weibel2025}). In Fig.~\ref{f27} (top panel), we show the spectrum of a quenched galaxy at $z \sim 4.7$, reproduced from \cite{Carnall2023a}. A clear and strong Balmer break is evident in the spectra, which is unambiguously identified via the presence of multiple absorption lines. This clearly indicates an older stellar population, with no evidence for massive O- and B-type stars. Moreover, this spectrum presents with little-to-no emission lines in the rest-frame optical, confirming this narrative. The only exception to this is a relatively weak (compared to many star forming galaxies at similar redshifts), but very broad, H$\alpha$ line, which is indicative of this system hosting a Type I AGN. A zoom-in region of the spectrum around the Balmer break is shown as an inset on the main panel, which clearly shows deep Balmer absorption lines consistent with the presence of a substantial A-type stellar population. From full spectrum SED fitting with {\small BAGPIPES}, \cite{Carnall2023a} infer that this system has a stellar mass of $M_* \sim 10^{10.6}\,M_\odot$ and underwent quenching at the staggeringly early time of $z_Q \sim 6.5$. However, it is important to appreciate that SED fitting methods all have inherent degeneracies and biases and different methods typically lead to different results.

The current record holder (at the time of writing) for the earliest quiescent massive galaxy is at $z = 7.3$ (just 700\,Myr after the Big Bang), reported in \cite{Weibel2025}. We reproduce the spectrum of this object in the lower-panel of Fig.~\ref{f27}. Both a Balmer and Lyman break are clear in the spectrum, unambiguously constraining its redshift. Moreover, no strong nebular emission lines are evident. Taken together, the lack of emission lines and clear Balmer break imply a quiescent system with an old stellar population. From full spectrum SED fitting, \cite{Weibel2025} infer rapid quenching at 50 -- 100\,Myr prior to observation. Additionally, the authors report a central density for this system of $\Sigma_* \sim 10^{10.85}\,M_\odot$/kpc$^2$, comparable to the cores of modern-day elliptical galaxies. However, the effective radius of this galaxy is just, $R_e \sim 200\,$pc (well over an order of magnitude smaller than typical modern-day quenched ellipticals).

At the time of writing, dozens of high-$z$ ($z > 3$), massive ($M_* > 10^{10}\,M_\odot$) quiescent galaxies have been confirmed spectroscopically by JWST (e.g., \citealt{Carnall2023a, Glazebrook2024, Nanayakkara2024, deGraaff2024, Weibel2025, Baker2025, Zhang2025}). We will explore the statistical properties of these systems, and consider whether they pose a challenge to contemporary cosmology, in the next sub-section. 

It is important to recall that at high redshifts ($z \gtrsim 3$), the principal spectroscopic age diagnostic for quiescent and recently quenched galaxies is generally the Balmer break, rather than the narrower $D_n(4000)$ index. Although both features arise from stellar absorption, the Balmer break is strongest in stellar populations dominated by A-type stars and is therefore particularly sensitive to post-starburst (PSB) populations with ages of $\sim$0.1 - 0.7\,Gyr, whereas a strong 4000\AA \, break typically requires older stellar populations and a substantial contribution from evolved stars (e.g., \citealt{Bruzual1983, Kauffmann2003b, Wild2009}). 

Consequently, most spectroscopically confirmed quiescent galaxies at $z>3$ are identified through prominent Balmer absorption features and Balmer breaks rather than large $D_n(4000)$ values. An important exception is the remarkably mature quiescent galaxy reported by \citealt{Glazebrook2024}, whose spectrum exhibits a clear 4000\AA \, break, implying an unusually old stellar population for its epoch. This discovery provides an instructive illustration of this distinction, especially through comparison with a younger post-starburst spectrum and the galaxy's disk-like morphology.

\begin{figure}[htbp]
\centering
\includegraphics[width=0.8\textwidth]{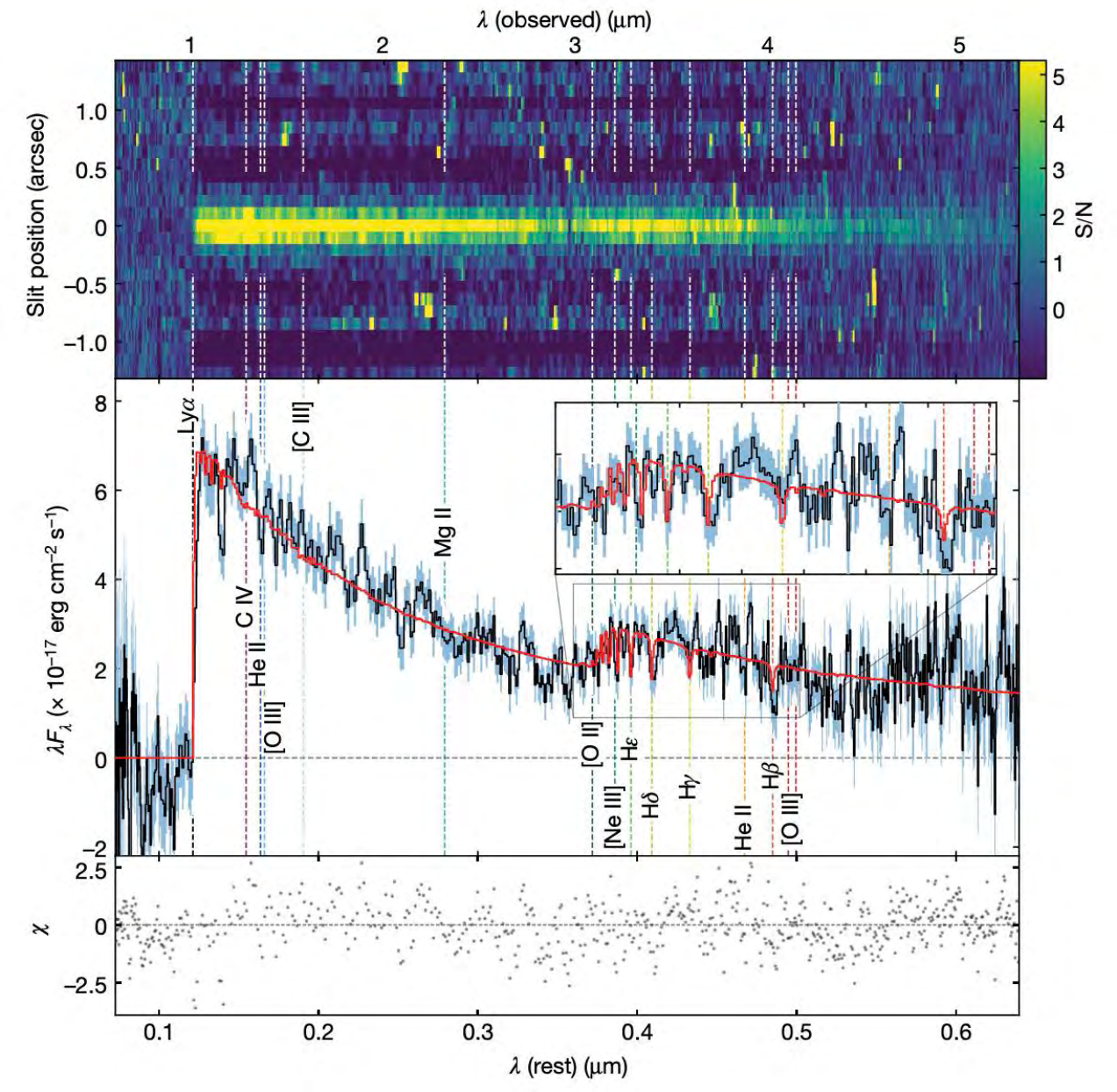}
\caption{An example of a high-$z$ `mini-quenched' galaxy. NIRSpec R100/prism spectrum of JADES-GS-z7-01-QU, reproduced from \cite{Looser2024}. This galaxy is observed at $z_\mathrm{spec} = 7.29 \pm 0.01$, with a stellar mass of just $\log(M_*/M_\odot) = 8.5 - 8.7$ (dependent upon fitting code). A clear Balmer break (in addition to a strong Lyman break) is evident, with no indication of strong emission lines. Hence, this galaxy appears to be quenched (at least temporarily) at the time of observation. The black line indicates the observed spectrum (with the blue region indicating uncertainty) and the red line indicates the {\small pPXF} full-spectrum fit. The upper panel shows the signal-to-noise ratio (S/N) in the 2D prism spectrum, with the bottom panel showing the ratio between the residuals. Note that the rising spectrum towards the UV is indicative of the presence of massive B-type stars, whereas the absence of emission lines (and presence of a Balmer break) suggests a lack of O-type stars. Hence, this system must have quenched very recently in its history.}\label{f28}
\end{figure}

In addition to massive quenched galaxies in the early Universe, an interesting population of `mini-quenched' systems have been identified (e.g., \citealt{Strait2023, Looser2024, Dome2024, Looser2025, Baker2025b, Witten2025}). These are high-$z$ galaxies that present with quiescent spectra, but have very low stellar masses and are highly likely to rejuvenate post observation. 

In Fig.~\ref{f28}, we show the spectrum from a striking example of a mini-quenched galaxy at $z = 7.3$, reported in \cite{Looser2024}. This system is determined via full-spectrum SED fitting to have a stellar mass of $M_*~<~10^{9}\,M_\odot$, marking it out as a distinct type of galaxy from the high-mass quenched galaxies discussed above. Both a (very strong) Lyman break and (mild, but clear) Balmer break are evident in the spectrum. Moreover, no strong rest-frame optical emission lines are evident. This strongly suggests that this galaxy is a quiescent system. However, the rising spectrum to the UV also indicates the presence of massive (most probably B-type) stars. Hence, this galaxy must have `quenched' very recently, $\sim 10 - 50$\,Myr prior to observation.

The example mini-quenched galaxy spectra in Fig.~\ref{f28} is typical of the population, which present with very low current star formation rates but young stellar population ages in low-mass systems, which are highly likely to rejuvenate over time (e.g., \citealt{Strait2023, Looser2025, Baker2025b, Witten2025}). This population contrasts with the much more massive quiescent galaxy population that present with old stellar populations and are likely to remain quenched for substantial periods post observation. 

In the case of mini-quenched galaxies, it is often proposed that these systems may be experiencing the extremes of main sequence oscillations (e.g., \citealt{Looser2024, Looser2025}), likely as a result of stellar and supernova feedback, but also potentially impacted by AGN (e.g., \citealt{Strait2023}), and/or environmental effects (e.g., \citealt{Sandles2023}). Given their low stellar mass, it is extremely unlikely that these systems have high enough halo masses to sustain a hot static atmosphere (e.g., \citealt{Dekel2006, Dekel2009}) and hence are unlikely to be decoupled from gas accretion into the system from the IGM, which is expected to ramp up towards cosmic noon (e.g., \citealt{Dekel2009, Lilly2013, Behroozi2013}). Therefore, mini-quenched galaxies are highly likely to be only temporarily quiescent and are expected to undergo rejuvenation within timescales short compared to the Hubble time (see, e.g., \citealt{Witten2025}).

Finally, using JWST-NIRSpec spectroscopy of 14 quiescent galaxies at $3 < z < 5$, \cite{Leung2026} demonstrate that the stellar ages of quiescent galaxies increase systematically with stellar mass, confirming that downsizing was already in place in the very early Universe. Hence, the most massive galaxies form their stellar populations earliest, while lower-mass quiescent systems are typically much younger, and may frequently rejuvenate. Taken in concert, this implies that long-term quenching is first established in the highest-mass galaxies.

In summary of this sub-section, observed quenched galaxies at high-$z$ come in two broad classes: (i)~high-mass quenched systems with old stellar populations (e.g., \citealt{Carnall2023a, Glazebrook2024, Nanayakkara2024, deGraaff2024, Weibel2025, Baker2025, Zhang2025}); and (ii)~low-mass `mini-quenched' galaxies which present with young stellar populations (e.g., \citealt{Strait2023, Looser2025, Baker2025b, Witten2025}). The former exhibit several characteristics in line with local quenched galaxies, but are much more compact (e.g., \citealt{Carnall2023a, deGraaff2025, Carnall2024, Wright2024}) and are more frequently identified as hosting an AGN (e.g., \citealt{Carnall2024, deGraaff2025, Baker2025a}). Additionally, many of these systems are found to have quenched very rapidly in the early Universe, resulting in a large fraction of their stars formed in short bursts in the primordial Universe (e.g., \citealt{Carnall2023a, Carnall2024, deGraaff2025}). Conversely, mini-quenched systems appear to have quenched extremely recently in their histories, and consequently retain very blue spectra, despite a lack of emission lines (e.g., \citealt{Strait2023, Looser2024, Looser2025}). There are no clear analogs to these systems in the local Universe, although they do bear some resemblance to fast-quenched, post starburst galaxies and to quenched dwarf galaxies.

\subsection{The challenge of massive quiescent galaxies in the very early Universe}\label{s73}

\noindent So far we have established that massive, quiescent galaxies exist in the early Universe, raising a number of interesting questions about their origin and potential fate. In this sub-section we explore the ways in which these novel observations with JWST have challenged previous ideas of galaxy formation theory, and even cosmology. In \cite{Labbe2023}, a potential population of six massive ($M* > 10^{10}\,M_\odot$) galaxies at extremely early cosmic times ($z \sim 7.5 - 9$) were presented, utilizing JWST photometry. One of these systems was even claimed to have a stellar mass of $M_* \sim 10^{11}\,M_\odot$ at $z \approx 7.5$. Immediately these galaxies caused a lot of speculation in the field: \textit{how could these galaxies get so massive so early in the history of the Universe?} This question was brought into sharp focus by \cite{Boylan2023}, who present a rigorous method to test the $\Lambda$CDM cosmology utilizing high-$z$ observations of the most massive galaxies at each early epoch from JWST. This approach is based on foundational theoretical and observational work pre-JWST by \cite{Steinhardt2016, Glazebrook2017, Behroozi2018}.

\begin{figure}[htbp]
\centering
\includegraphics[width=0.85\textwidth]{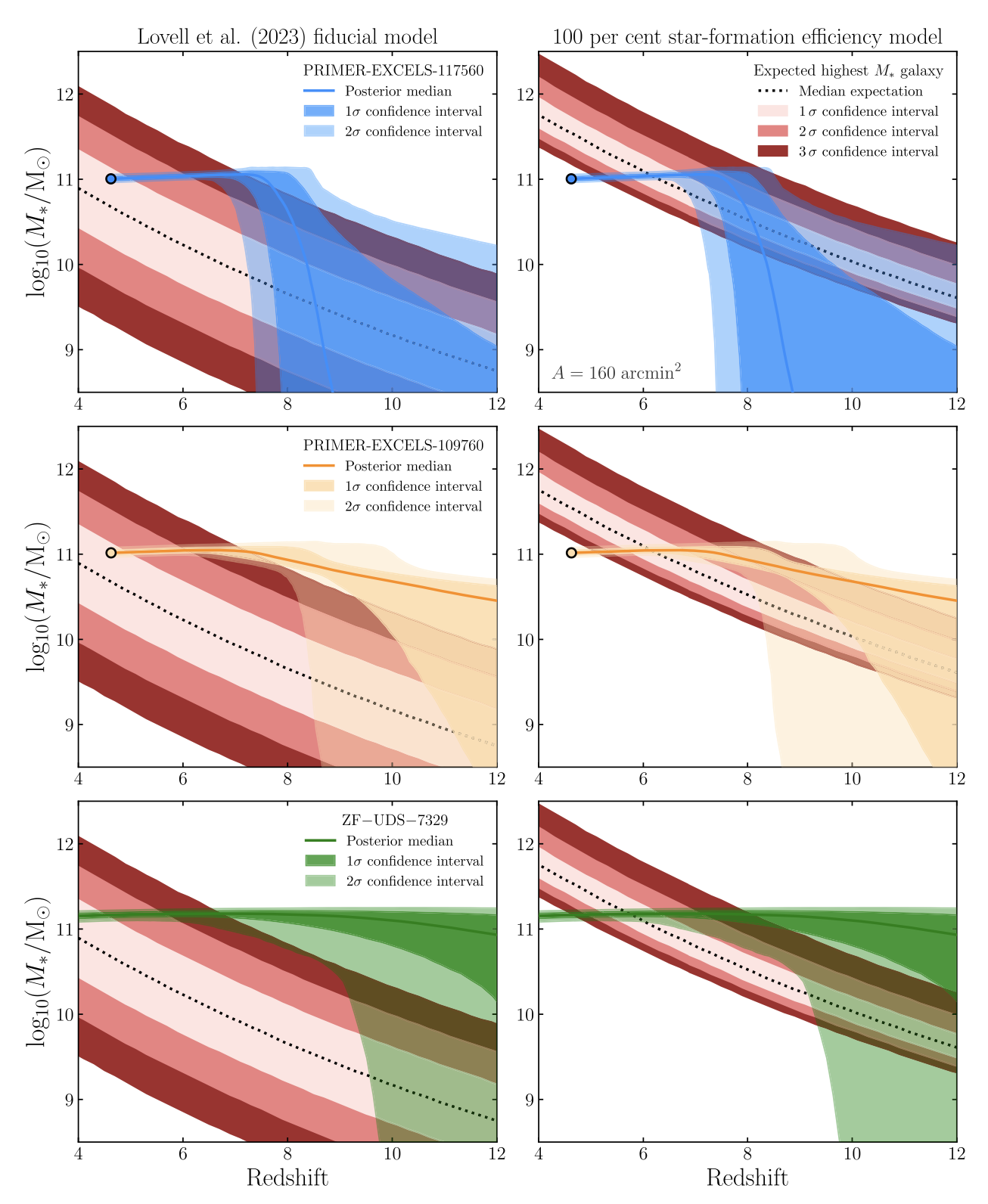}
\caption{Star formation histories (SFHs) for three high-$z$, ultra-massive quenched galaxies observed spectroscopically with JWST in the PRIMER UDS field, reproduced from \cite{Carnall2024}. The SFHs are determined via spectro-photometric SED fitting with {\small BAGPIPES} (with probability distributions displayed as blue, orange, and green colored regions; see legends). The SFHs are compared to the maximum allowable stellar masses of galaxies within the PRIMER UDS area at each epoch, as inferred from the extreme value statistics approach of \cite{Lovell2023}. On the left-hand panels, the fiducial model is shown, which assumes a log-normal distribution in star formation efficiencies in the range, $0 < \epsilon_* \, (\equiv M_*/f_b M_H) \, < 1$. On the right-hand panels, a maximal star formation efficiency of $\epsilon_* = 1$ is applied. The inferred star formation histories present a significant tension with the fiducial model, but may be accommodated by the extreme star formation efficiency model (at least within $2\sigma$). Hence, the latest results on the star formation build-up of massive quiescent galaxies do not challenge the $\Lambda$CDM background cosmology, but do require more efficient star formation in the early Universe than realized in most standard galaxy formation models. }\label{f29}
\end{figure}

The essential strategy of \cite{Boylan2023} is to compute the expected halo mass function as a function of cosmic time (direct from theory), apply the universal baryon fraction ($f_b \equiv \Omega_b / \Omega_M$) to estimate the total expected baryon content of any given halo, and then apply a 100\% baryon-to-star conversion factor to infer the most massive galaxies allowable in $\Lambda$CDM at any given epoch (and for any given observational volume). Utilizing this approach, \cite{Boylan2023} conclude that some of the galaxies reported in \cite{Labbe2023} are right at the limit (or possibly even beyond the limit) of what can be accommodated by $\Lambda$CDM. These galaxies, and similar very high-mass early galaxies, have been dubbed as `Universe breakers' by the media, due to the potentially serious tension they pose with the core cosmological theory. Several other publications present similar tensions in photometric JWST data, as well as pointing out the potential need to modify $\Lambda$CDM as a result of these observations (see, e.g., \citealt{Melia2023, Carnall2023, Ziegler2025}).

However, extraordinary claims (like the need to modify $\Lambda$CDM) require extraordinary evidence, and without spectroscopic confirmation the sample of `Universe breakers' were not decisive. Indeed, the majority of these systems have been downgraded in both mass and redshift (e.g., \citealt{Baggen2023, Steinhardt2023, Desprez2024}). Crucially, the majority of the \cite{Labbe2023} sources have now been observed with JWST/NIRSpec and all of these spectra show clear signs of AGN activity (e.g., \citealt{D'Eugenio2025a, Fujimoto2024, deGraaff2025b, Hviding2025, Heintz2024}). Hence, the consensus in the field is that these objects had erroneously high stellar masses before, and therefore do not challenge the $\Lambda$CDM paradigm.

Consequently, the field has turned en masse towards spectroscopic samples of high-mass early galaxies to stress test $\Lambda$CDM, as well as to test the theory of galaxy formation within the $\Lambda$CDM context. Moreover, it is particularly massive \textit{quiescent} galaxies which are most conducive to these types of analyses because one can more accurately reconstruct their star formation histories (SFHs), testing whether their inferred stellar masses violate constraints from either cosmology or galaxy formation theory (see, e.g., \citealt{Carnall2024, Weibel2025, deGraaff2024}). 

As an example of this approach, in Fig.~\ref{f29} we present the probability distributions for the posterior SFHs for three high-mass, spectroscopically confirmed quiescent galaxies, reproduced from \cite{Carnall2024}. It is crucial to stress that this population contains exclusively spectroscopically confirmed passive systems, without AGN contamination. Overlaid in Fig.~\ref{f29} are the probability distributions for the most massive galaxy expected in the volume from $\Lambda$CDM using two approaches: (i)~a log-normal distribution in star formation efficiency from \cite{Lovell2023}; and (ii)~a maximal star formation efficiency of $\epsilon_* \equiv M_*/f_bM_H = 1$. The inferred SFHs pose a significant tension with the default \cite{Lovell2023} model. However, the tension is reduced to the $\sim 2\,\sigma$ confidence level with the maximal efficiency model. This highlights that the latest constraints on the build-up of stellar mass in galaxies are not in serious tension with $\Lambda$CDM itself, but do require very high baryon-to-stellar conversion in the very early Universe, which is itself in tension with most contemporary galaxy formation models (see, e.g., \citealt{Furlong2015, Wilkins2017, Lagos2018, Pillepich2018, Dave2019, Yung2019, Yung2022}). Many other observational works based on JWST spectroscopy are in general agreement with the need for (very) high star formation efficiencies in the early Universe, but present no serious tension with the $\Lambda$CDM cosmology, broadly in agreement with the results discussed above (e.g., \citealt{Urbano2024, Weibel2025, deGraaff2025}).

\begin{figure}[htbp]
\centering
\includegraphics[width=1\textwidth]{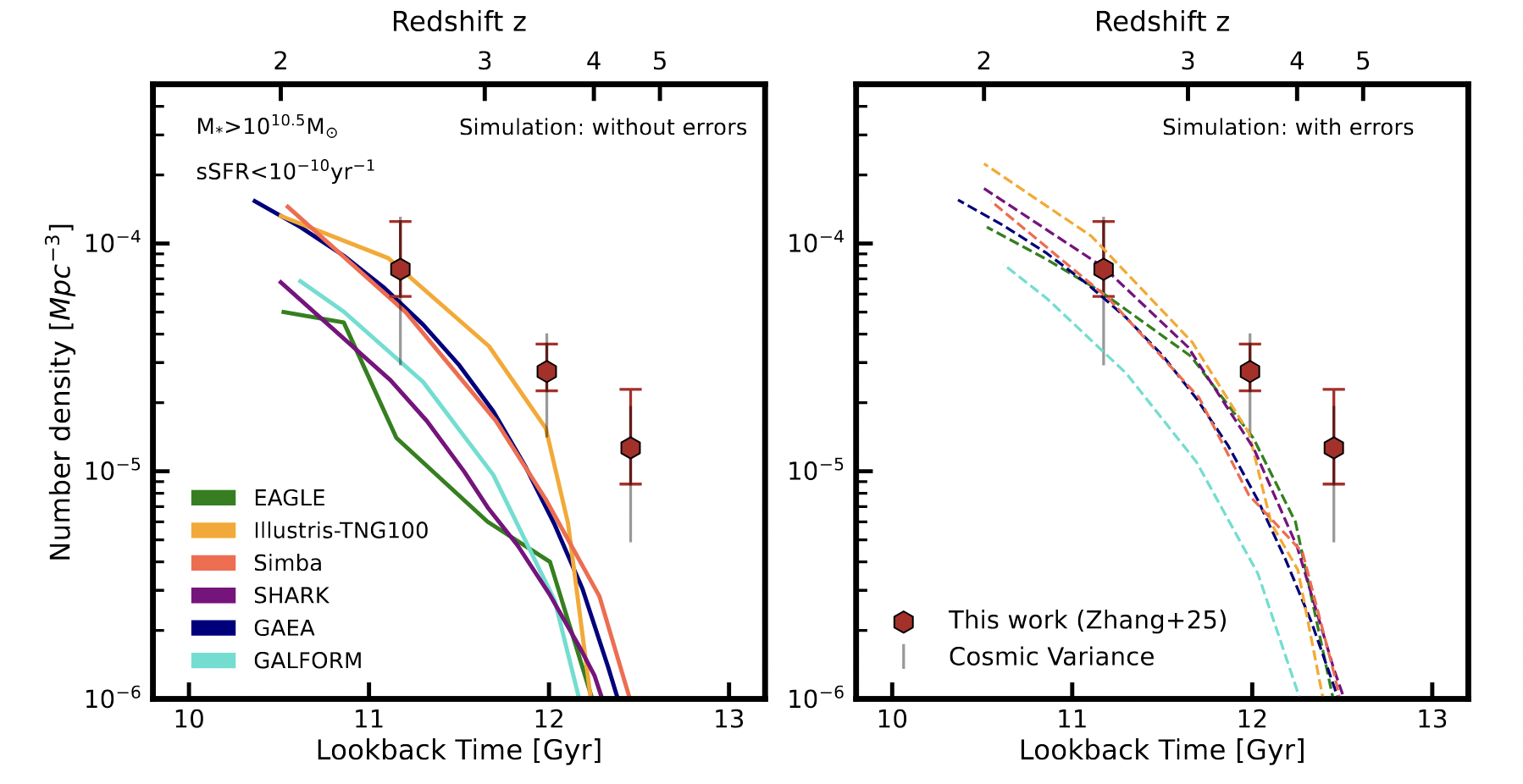}
\includegraphics[width=0.65\textwidth]{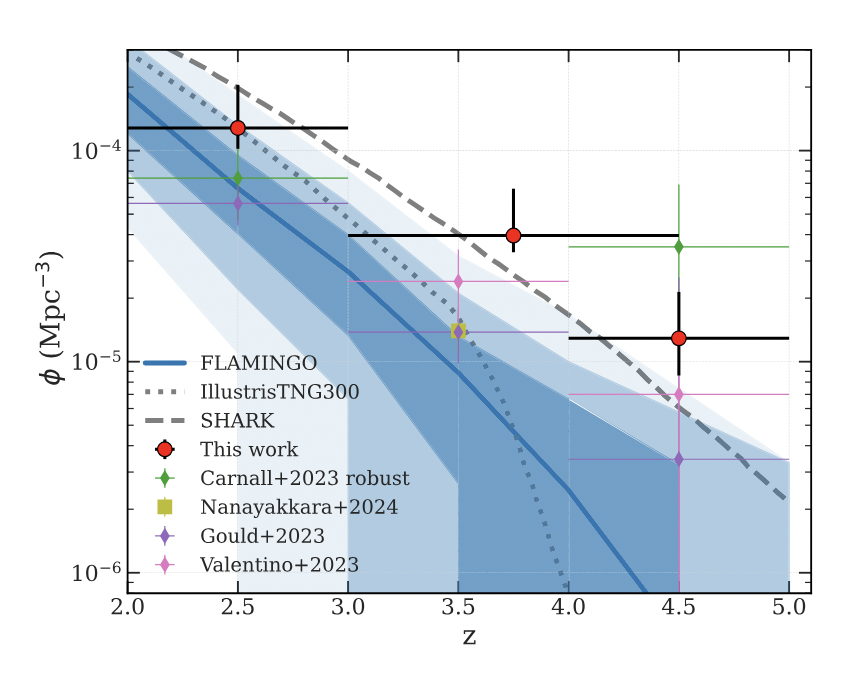}
\caption{An over-abundance of observed quenched galaxies in the very early Universe, compared to contemporary models of galaxy formation. \textit{Top panels}: The number density of massive ($M_*~>~10^{10.5} M_\odot$) quenched galaxies from the RUBIES survey (Area $\approx$ 150 arcmin$^2$) plot as a function of lookback time, reproduced from \cite{Zhang2025}. Observational results are shown as large red circles with error bars. These are compared to a host of contemporary simulations and models (see legends), with the left panel showing raw results from the models and the right panel showing the simulation results with realistic error modeling. The observed number densities of quenched, high mass galaxies at high redshifts are in significant tension with the predictions from essentially all galaxy formation models considered. \textit{Bottom panel:} The number density of quenched $M_* > 10^{10} M_\odot$ galaxies from JWST observations in the GOODS fields (Area $\approx$ 800 arcmin$^2$) plot as a function of redshift, reproduced from \cite{Baker2025a}. Observational data are shown as circles with error bars. The shaded blue regions show the probability distribution in number densities of massive quenched galaxies from the FLAMINGO simulation (the largest volume cosmological hydrodynamical simulation to date). The observed number densities of quenched galaxies are in significant tension with the FLAMINGO simulation at high redshifts. However, the SHARK semi-analytic model (which incorporates constraints from JWST observations) is reasonably effective at reproducing the observed trend, indicating that these observations can be explained within the $\Lambda$CDM cosmological framework.}\label{f30}
\end{figure}

Even if the observations of high-mass galaxies in general, and the inferred SFHs of massive quiescent galaxies in particular, do not pose significant problems to the $\Lambda$CDM cosmology, they are nonetheless highly challenging to account for in standard galaxy formation theory. This is best illustrated by summative works, which combine reasonably large samples of spectroscopically confirmed high-mass quenched galaxies in the early Universe (see especially, \citealt{Zhang2025, Baker2025a}). 

In Fig.~\ref{f30}, we present the number density of observed massive quenched galaxies compared to those predicted in galaxy formation models from \cite{Zhang2025} (top panels) and \cite{Baker2025b} (bottom panel). In the top panels, the number density of quiescent galaxies with $M_* > 10^{10.5}\,M_\odot$ from the RUBIES survey is plot as a function of lookback time and compared to simulations, both raw results (left panel) and after careful error modeling (right panel). Explicitly, observational-like uncertainty is added to the simulated SFR and $M_*$ values in the right-hand panel to assess the impact on the statistical results via Gaussian random sampling. In both cases the abundance of high-mass quenched galaxies far exceeds the simulation predictions at the earliest cosmic times probed. This is a clear indication that both the formation and quenching of galaxies occurs earlier in nature than predicted in the vast majority of contemporary cosmological simulations. 

In the bottom panel of Fig.~\ref{f30}, the number density of quiescent galaxies with $M_*~>~10^{10}\,M_\odot$ from the JWST-GOODS fields is plot as a function of increasing redshift (reproduced from \citealt{Baker2025b}), with several other observational constraints also displayed (see legends). This is compared against the probability distributions for the appropriate survey area, drawn from the FLAMINGO simulation (which is the largest volume cosmological hydrodynamical simulation to date). The advantage of this approach is that the full impact of cosmic variance can be assessed directly within the simulation, and any discrepancy can be fully quantified at a probabilistic level. 

At the highest-redshifts probed, there is a significant tension with the FLAMINGO simulation at the $\sim 3\sigma$ level. An even more severe tension is seen with the IllustrisTNG-300 simulation (shown as a dotted line). Additionally, \cite{Stevenson2026} find similar results whereby the number density of high-mass quenched galaxies at $z > 3$ is higher than previously expected, and hard to accommodate within most contemporary models. The latter work utilizes PRIMER and JADES imaging over $\sim$ 300 arcmin$^2$. 

However, the SHARK semi-analytic model (see \citealt{Lagos2018, Lagos2024}), with free parameters tuned to contemporary JWST data, is in much better agreement with the observational constraints (see dashed line in the bottom panel of Fig.~\ref{f30}). Crucially, these models also retain a good fit to the low-$z$ stellar mass functions as well. This is important because it clearly demonstrates that a galaxy formation model built within the $\Lambda$CDM cosmological framework can account for the latest JWST high-$z$ observations and the established low-$z$ observations, simultaneously.

Nevertheless, SFH modeling of these early quenched galaxies frequently predict bursts of $\sim 100+ M_\odot/\mathrm{yr}$ at $z \sim 8$ -- 11 (e.g., \citealt{Glazebrook2024, Carnall2023a, Carnall2024, deGraaff2025, Weibel2025}). Given the relatively high density of these passive galaxies, as evidenced from the fact they are seen at all in relatively small volumes (e.g., \citealt{Carnall2023, Baker2025a, Zhang2025}), there ought be a similarly high density of their `super-starburst' progenitors at $z \sim 8$ -- 11. Yet, this is not observed, as clearly seen by: (i)~the UV luminosity functions at very early cosmic times (see, e.g., \citealt{Donnan2023, Bouwens2023, Harikane2023}), (ii)~the relative fraction of dust obscured ultra-high-$z$ starbursts (see, e.g., \citealt{Fudamoto2022, Fujimoto2023}), and (iii)~via direct and stacked spectroscopic measurements of SFR at very early cosmic times (e.g., \citealt{Tang2023, Curtis-Lake2023, Tacchella2022a}). Indeed, even the most luminous galaxies observed at z $\sim$ 10, such as GN-z11 (\citealt{Oesch2016, Jiang2021, Tacchella2023a}), would have an SFR $\sim$ 10 -- 20 $M_\odot/\mathrm{yr}$ (and accounting for AGN contamination, even less; see \citealt{Maiolino2024b}). This is nowhere close to the levels needed to satisfy the effective predictions from the reconstructed SFHs of several high-mass quenched galaxies at $z > 3$.

This leads to a serious unsolved problem: \textit{where are the putative progenitors of the high-mass quenched galaxy population at $z > 3$?} Irrespective of whether galaxy formation models can create the high-mass quiescent galaxy population at early cosmic times, one must have consistency between the inferred SFHs from SED fitting and the direct constraints on the evolution of the star formation rate density.

Various possible solutions have been suggested, including the use of carefully constructed SFH priors in SED fitting, motivated by the global star formation rate density evolution at very early cosmic times (e.g., \citealt{Turner2025}). That is, unlike in many SED fitting codes, a flat SFH prior is discarded in favor of an evolution motivated by direct observational constraints at high-$z$ (see further details in \citealt{Turner2025}). This tests whether consistency is {\it possible} between the observational constraints across various methodologies and redshifts. Additionally, it is possible that a high number of minor mergers take place in these early galaxies, obscuring the true formation history in star formation histories from SED fitting (e.g., \citealt{Cochrane2025}).

Alternatively, accounting for variable metallicity throughout galaxy formation, and the impact of $\alpha$-enhancements on SED fitting, may further improve the agreement with the $z > 8$ observations (e.g., \citealt{Park2024b}). Allowing for variable metallicity and $\alpha$-enhancement introduces an important modification to SED fitting at high redshift. In the early Universe, rapid star formation leads to enrichment dominated by $\alpha$-elements from core-collapse supernovae, resulting in stellar populations with super-solar [$\alpha$/Fe] ratios and evolving metallicities. These abundance patterns alter stellar spectra, reducing metal-line opacity and yielding bluer continua and lower mass-to-light ratios at fixed age. As a result, fits based on solar-scaled templates can bias stellar masses and ages to higher values. Hence, incorporating more realistic chemical enrichment histories can therefore reduce inferred stellar masses and formation redshifts, improving agreement between observations and theoretical expectations at z $>$ 8. However, at the time of writing this issue remains an open question in the field, with active ongoing research.

Finally, one must seek to understand why these quiescent galaxies quench so early in the history of the Universe. There are a number of clues in their observed properties. For instance, a high fraction of these early quenched galaxies are identified as hosting an AGN (e.g., \citealt{Carnall2024, deGraaff2025, Baker2025a}), clearly suggesting a plausible link between AGN-feedback and quenching. Additionally, the vast majority of these early quenched galaxies are ultra-compact (e.g., \citealt{Carnall2023a, deGraaff2025, Carnall2024, Wright2024}), often presenting with central densities as high as modern-day quenched ellipticals. This implies that they are likely to harbor high-mass SMBHs, following the logic of local scaling laws (e.g., \citealt{Ferrarese2000, Maggorian1998, Haring2004}). 

Additionally, the SFHs of these early quenched systems are often consistent with rapid cessation of star formation (e.g., \citealt{Carnall2023a, Carnall2024, deGraaff2025}), which in the local Universe is often associated with quasar-mode, ejective AGN feedback (e.g., \citealt{DiMatteo2005, Hopkins2006, Maiolino2012, Cicone2014}). Consequently, there is reasonably strong circumstantial evidence for a plausible link between AGN feedback and quenching in the very early Universe. As an interesting example from the theoretical literature, \cite{Chittenden2026} demonstrate that one can enhance the production of massive quenched galaxies in the very early Universe by lowering the strength of supermassive black hole feedback, enabling rapid growth of supermassive black holes. In the remaining part of this section we review direct evidence for AGN-feedback impacting the first populations of galaxies.

\subsection{Evidence for AGN feedback in the early Universe}\label{s74}

There exists a host of evidence for AGN-driven outflows in both neutral and ionized gas phases from JWST/NIRSpec observations at very early cosmic times (e.g., \citealt{Marshall2023, D'Eugenio2024, Davies2024, Belli2024, Wu2025, Wang2025, Valentino2025}). This type of AGN feedback is often described in the theoretical literature as `quasar-mode' because it originates primarily in high-Eddington ratio accretion discs around supermassive black holes (see, e.g., \citealt{DiMatteo2005, Springel2005c, Hopkins2006, Hopkins2008}). Ionized outflows are traced most frequently with kinematically offset emission lines, most notably [O~III], H$\alpha$, H$\beta$, [S II], and [N II]. Neutral outflows are detected primarily via blueshifted (from the rest-frame of the galaxy) metal absorption lines, most frequently Na ID, Mg II, and Fe II. The interpretation is that the neutral gas outflows are metal enriched (perhaps by the first populations of supernovae at very high-$z$), with the metals then acting as tracer particles for the neutral gas outflow, which is hard to observe directly with JWST. 

Whilst ionizing outflows may reach up to very high velocities of $v_\mathrm{out} \sim $1500\,km/s or so, they typically contain relatively low masses of $M_\mathrm{out} \sim 10^{7-8}\,M_\odot$ (e.g., \citealt{Ubler2023, Marshall2023, Belli2024}). On the other hand, neutral outflows tend to contain much higher masses of $M_\mathrm{out} \sim 10^{9-10}\,M_\odot$, but typically have more modest velocities of $v_\mathrm{out} \sim 500 - 1000$\,km/s (e.g., \citealt{Belli2024, Davies2024, Wu2025}). Consequently, the contemporary literature favors neutral (and potentially molecular) outflows as the most probable quenching trigger at very high redshifts, in line with lower-$z$ observations. Additionally, there is evidence of luminous AGN photo-ionizing the ISM within the central regions of a high-$z$ galaxies, which is also likely to impact star formation and induce local quiescence (e.g., \citealt{Ubler2024}).

\begin{figure}[htbp]
\centering
\includegraphics[width=1\textwidth]{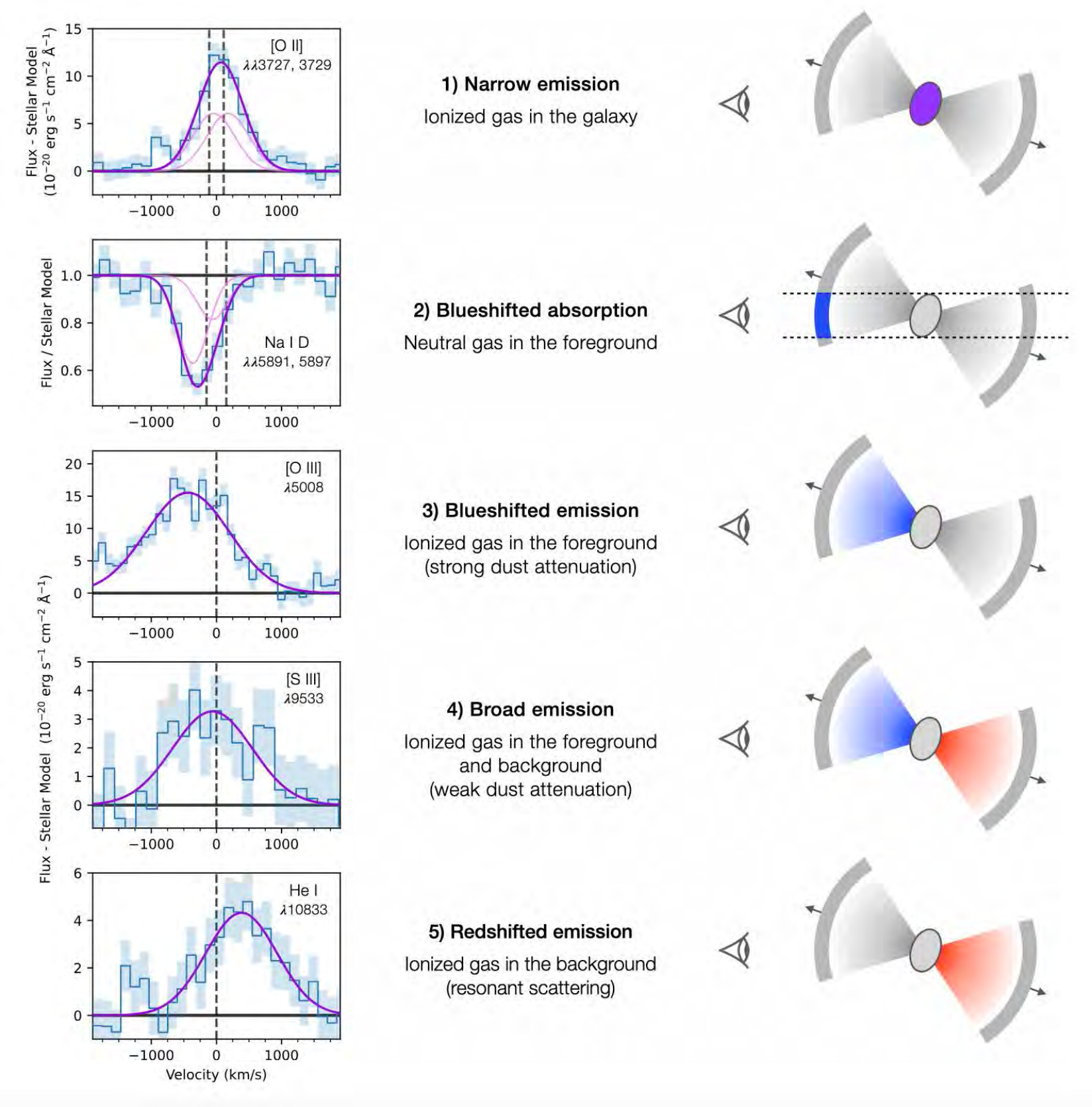}
\caption{Kinematic evidence for both neutral and ionized outflows powered by an AGN in a massive ($\log(M_*/M_\odot) \approx 10.9$) galaxy at $z_\mathrm{spec} = 2.445$, reproduced from \cite{Belli2024}. Each row shows (from left to right): a zoom-in of an emission or absorption line, including its best fit; a description of the line type and origin; and a cartoon image of where in the galaxy / outflow the feature originates. The top row shows the [O II] narrow line doublet, which originates from the galaxy itself. The second row shows a blueshifted Na ID doublet, indicative of metals in a neutral outflow absorbing photons along the line-of-sight. The third row shows a blueshifted [O III] emission line, interpreted as ionized gas in the outflow along the leading edge. The fourth row shows a symmetric [S III] line, likely originating from both the foreground and background of the ionized outflow component. The bottom row shows a redshifted He I line, which is interpreted as resonant scattering off the receding part of the outflow. From modeling, ionized outflow velocities are found up to $\sim1700$ km/s, indicating that this outflow is likely powered by an AGN rather than SN driven winds. Moreover, the presence of an AGN is confirmed by high [N II]/H$\alpha$ and [O III]/H$\beta$ line ratios (see \citealt{Baldwin1981}). Mass outflow rates are estimated to be $\dot{M}_\mathrm{out} \sim 35 \, M_{\odot}/\text{yr}$ for the neutral outflow and $\dot{M}_\mathrm{out} \sim 1 \, M_{\odot}/\text{yr}$ for the ionized outflow. Taken together, this figure provides clear evidence for ejective AGN feedback operating in the early Universe.}\label{f31}
\end{figure}

Ejective feedback is often thought about theoretically from the merger\,--\,quasar\,--\,quench paradigm (e.g., \citealt{Springel2005c, DiMatteo2005, Hopkins2006, Hopkins2008}). Briefly, this is a scenario in which galaxy--galaxy mergers lead to funneling of gas to the center of the system via tidal torques, which in turn drive a central starburst and enhanced AGN activity, sometimes igniting full blown quasars. The ignited quasar may then drive powerful ejective feedback from the ISM, leading to triggering quenching via gas depletion. However, galaxies are expected to rejuvenate star formation thereafter on long time scales, unless there is additional preventative feedback to stabilize the hot gaseous halo (see, e.g., \citealt{Croton2006, Fabian2006, Bower2006, Bower2008, Sijacki2007, Fabian2012, Henriques2015, Weinberger2017}). This picture suggests two important things. First, in order for the observed quenched massive systems to maintain quiescence, one may expect there to be accompanying `radio-mode' feedback observed at early cosmic times (which turns out to be the case, see later in this sub-section and \citealt{Cresci2023, Roy2025}). This type of feedback originates primarily from low-Eddington ratio accretion onto supermassive black holes, and frequently exhibits with relativistic jets (see, e.g., \citealt{Croton2006, Fabian2006, Fabian2012}). Second, one might anticipate recently quenched systems (and/or those experiencing strong ejective feedback) to be ongoing mergers or merger remnants. 

Interestingly, \cite{D'Eugenio2024} present observations of a dynamically cool rotating disc galaxy at $z = 3.1$, which is clearly undergoing both ionized and neutral outflows, likely driven by the detected AGN within the system. Yet, this galaxy shows no sign at all of being a merger or a merger remnant. Additionally, \cite{Pascalau2025} find similar characteristics for a $z = 4.66$ quenched galaxy. These observations bring into question the necessity of merging to launch quasar-mode feedback in the early Universe, which may instead be triggered my more subtle secular evolution (e.g., via disc instabilities).

As an example of ejective feedback in the early Universe, in Fig.~\ref{f31} we reproduce the observational evidence for a multi-phase AGN-driven outflow from a massive galaxy at $z = 2.45$ (i.e., significantly before cosmic noon), reproduced from \cite{Belli2024}. The left-hand column shows JWST/NIRSpec observations of various emission and absorption features in the galaxy spectrum, the central column provides a description and interpretation of these, and the right-hand column presents a cartoon illustration of where in the galaxy or outflow the feature is believed to originate. Of particular interest is a clear blueshifted Na ID absorption line, indicating the presence of a metal enriched neutral gas outflow viewed along the leading edge towards the observer (second row). Additionally, there is a clear blueshifted [O III] line (third row), indicating the presence of an ionized outflow viewed on the side of the outflow oriented towards the observer. There is also evidence of a redshifted resonant He I line (bottom row), most probably originating from scattering off the receding part of the outflow (relative to the observer). From dynamical modeling, \cite{Belli2024} infer a weak mass outflow rate of $\dot{M}_\mathrm{out} \sim 1 \, M_{\odot}/\text{yr}$ for the ionized component, and a strong mass outflow rate of $\dot{M}_\mathrm{out} \sim 35 \, M_{\odot}/\text{yr}$ for the neutral component, with the latter likely able to quench the system. Interestingly, this galaxy likely hosts an AGN (as determined via emission line diagnostics), but is not detected in either radio or X-ray.

\begin{figure}[htbp] 
\centering
\includegraphics[width=1\textwidth]{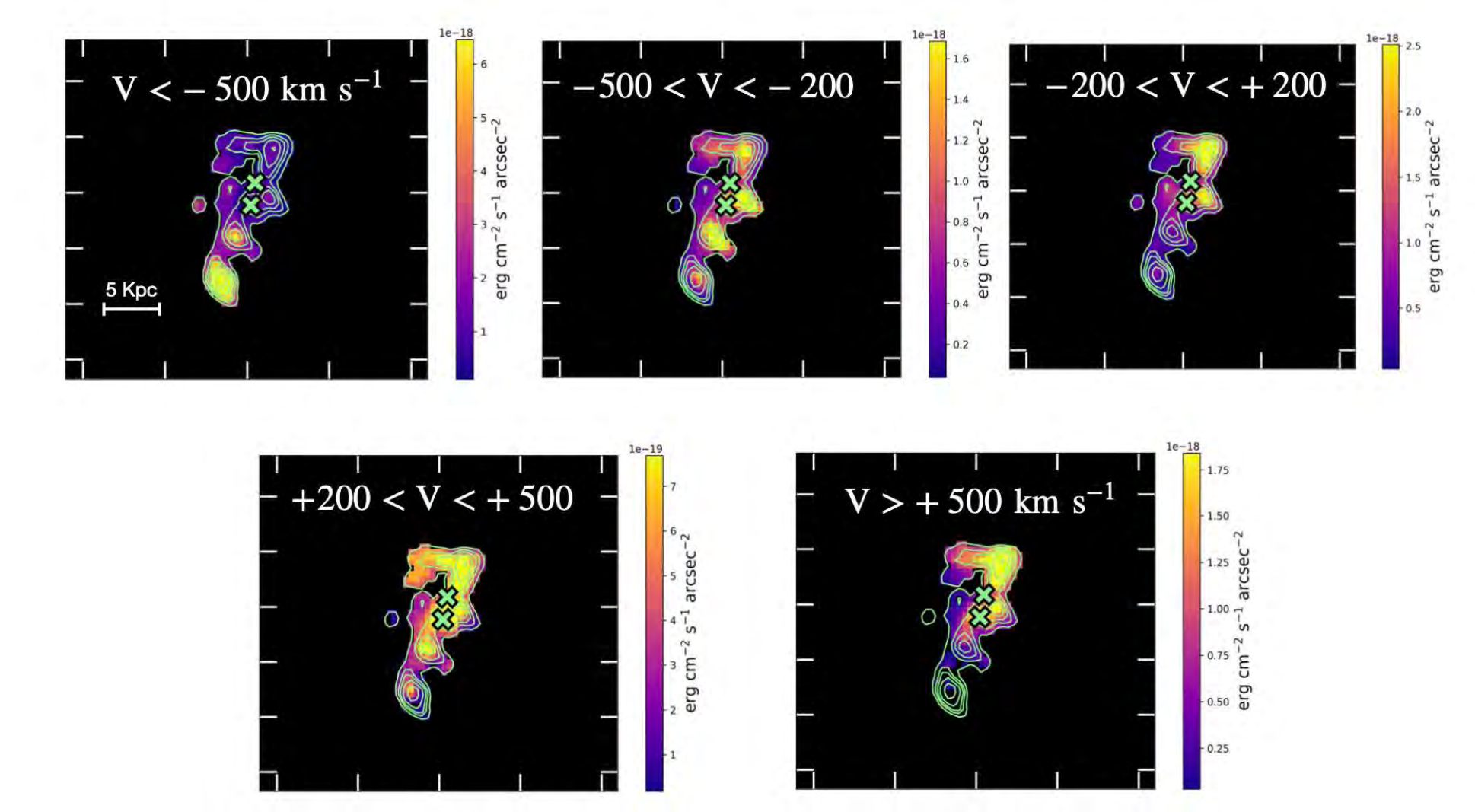}
\caption{Velocity channel maps of the spatially resolved H$\alpha$ emission line from JWST/NIRSpec IFU observations of TGSS1530, reproduced from \cite{Roy2025}. TGSS1530 was first detected as a luminous radio galaxy, and has a redshift of $z = 4.0$. From top-left to bottom-right: slices in velocity space from maximally blueshifted to maximally redshifted, respectively. Surface brightness (as indicated by the color bar) shifts from the lower-left to the upper-right systematically in line with the velocity cut. This clearly indicates the presence of a bi-conical outflow. Moreover, the outflow is very tight and jet-like in structure, hinting at a potential relativistic jet origin to this ionized outflow. Supporting this interpretation, the two green crosses displayed on each panel indicate the compact double radio source detected in this system, derived from VLBI radio observations. The ionized outflow is oriented closely in alignment with the inferred radio lobe structure.}\label{f32}
\end{figure}

In addition to ejective feedback, there is also some important observational evidence for relativistic jet (or `radio-mode') feedback in the very early Universe (see especially, \citealt{Cresci2023, Roy2025}). This suggests that the preventative mode may also be active at early cosmic times, potentially enabling the maintenance of long-term quiescence for the high-mass quenched galaxies observed by JWST. For instance, \cite{Roy2025} present a sample of six powerful radio galaxies at $3.5 < z < 4$ (observed with VLBI), which show strong feedback via turbulence injection and tight bi-conical outflows associated with the radio jets, as seen in JWST/NIRSpec IFU follow-up observations. 

In Fig.~\ref{f32}, we show one particularly interesting example, that of TGSS1530 observed at $z = 4.0$, reproduced from \cite{Roy2025}. From top-left to bottom-right, H$\alpha$ surface brightness is displayed in velocity bins, from most blueshifted to most redshifted (relative to the galaxy). The clear shift in flux as a function of velocity cut demonstrates the presence of a tight bi-conical outflow. Additionally, the location of dual radio source detections from VLBI observations are presented as green crosses. The ionized outflow aligns closely with the inferred orientation of the radio lobes, highly favoring a jet origin to this outflow. The sample as a whole yields kinetic powers between, $L_k = 10^{43.2} - 10^{45.0}\,\mathrm{erg\, s}^{-1}$, demonstrating that radio-mode feedback can input significant energy into the ISM and CGM of massive galaxies at very early times.

In addition to JWST observations, there is also evidence from ALMA observations for at least some high-$z$ quenched galaxies to be essentially void of molecular gas (see \citealt{Whitaker2021}). This may be explained in the AGN-feedback paradigm via a mix of removing gas from the ISM via quasar-mode feedback (triggering fast, early quenching) and prevention of gas replenishment from the CGM, likely achieved via CGM heating through late-time, low luminosity radio-mode feedback. 

Additionally, prior and contemporaneously to JWST observations, powerful molecular outflows associated with quasars and very luminous AGN have been discovered at very high redshifts (see, e.g., \citealt{Maiolino2012, Bischetti2019, Salak2024, Spilker2025}), complementing the discoveries from JWST (which are mostly limited to neutral and ionized phases). Hence, there is now a consensus in the literature that multi-phase outflows are launched via quasar-mode AGN feedback in the very early Universe, and that radio jets exist in the very early universe, significantly impacting their host galaxies' ISM and CGM. 

In summary of this sub-section, AGN feedback is directly observed to impact galaxies and their surrounding haloes in the very early Universe via both ejective multi-phase outflows (i.e., the `quasar mode') and via relativistic jets (i.e., the `radio mode'). This is much like the feedback seen at lower redshifts as well, which we discuss in depth in relation to simulations in Part II (see, e.g., \citealt{Fabian2006, Schawinski2007, Fabian2012, Maiolino2012, Cicone2012, Cicone2014, Fluetsch2019, Fluetsch2021}). Hence, it is highly likely that AGN feedback is at least partially responsible for the quenching of the first population of massive, quiescent galaxies in the Universe. To move beyond this conclusion, we need to look at population level analyses of star forming and quenched galaxies in the very early Universe, which we will explore in Part~II of this review in relation to direct predictions from theoretical models.

%%%%%%%%%%%%%%%%%%
%                %
%    SUMMARY     %
%                %
%%%%%%%%%%%%%%%%%%

\section{Summary of Part I}\label{s8}

In Sect.~\ref{s1}, we introduce three deep problems with $\Lambda$CDM as a theory of galaxy formation and evolution -- the cosmological problem (why is star formation so inefficient?),  the clusters problem (why is the hot gas halo stable to cooling and collapse?), and the bimodality problem (why are there two fundamental classes of galaxies?). We posit at the outset that a single solution, that of galaxy quenching, may offer an explanation to these pressing concerns. Furthermore, we briefly assess a host of potential candidates for quenching routes, separated by their fundamental cause (i.e., the underlying physical process) and the mechanism (i.e., how quenching is achieved in practice). These possibilities are summarized in Fig.~\ref{f4}, which provides an idea map of quenching. 

In Sect.~\ref{s2}, we begin by describing the key methods to infer the star formation rates and stellar masses of galaxies. We then introduce the star forming main sequence (SFMS: ${\rm SFR} - M_*$ relation). We also discuss the evolution in the star forming main sequence. We on on to review methods to identify quenched galaxies, including utilizing the SFMS, galaxy colors, and spectral indices. Additionally, we discuss methods to identify transitioning galaxies, via the `green valley' and post-starburst systems. 

In Sect.~\ref{s3}, we review two key observational results: (i) the evolution in the star formation rate density, and (ii) the halo mass\,--\,stellar mass relationship. We discuss how the features of both of these fundamental relations are intimately connected to quenching, requiring a mass-dependence on galactic star formation efficiency.

In Sect.~\ref{s4}, we discuss intrinsic observational correlators to quenching, including: morphology, stellar mass, central density, and central velocity dispersion. We pause briefly to consider whether size evolution could lead to confusion in the interpretation of these intrinsic quenching correlators. We conclude that central velocity dispersion is the best intrinsic predictor of central galaxy quenching, which is potentially explained via historic AGN feedback (given the $M_\mathrm{BH}$\,--\,$\sigma_c$ relation, and the expectation for the integrated feedback energy from AGN to scale with supermassive black hole mass). 

In Sect.~\ref{s5}, we discuss environmental correlators to quenching, including: local galaxy (over-)density, dark matter halo mass, and the location of satellite galaxies within their parent group haloes. We note an important disconnect between central galaxies (which are quenched intrinsically, independent of environment) and low-mass satellites (which quench environmentally, independent of intrinsic properties).

In Sect.~\ref{s6}, we move on to discuss the observational constraints on galaxy quenching from spatially resolved spectroscopy (primarily in the local Universe). We start by presenting evidence for a resolved analog of the SFMS relation, which exists on $\sim$kpc scales within galaxies. We go on to discuss the observational evidence for `inside-out' quenching of centrals (and high-mass galaxies more generally) and compare this to evidence for `outside-in' quenching of low-mass satellites. Finally in this section, we present evidence for quenching being a global (galaxy/ halo wide) process, with star formation being governed locally. This suggests a fundamental disconnect (or orthogonality) between star formation and quenching, which is perhaps surprising. We explain this as a result of the physics (and physical scales) of quenching and star formation being distinct. 

In Sect.~\ref{s7a}, we discuss the gas physics of quenching. We begin with a description of methods used to constrain the atomic and molecular abundance of Hydrogen gas within galaxies. We then discuss various definitions of star formation efficiency out of the gas reservoir. We go on to consider the causal origin of the star forming main sequence, which we provide evidence for originating from the local/ spatially resolved Kennicutt--Schmidt (KS) and molecular gas main sequence (MGMS) relations. We also discuss how the gas fraction and star formation efficiency reduce during quenching, on both global and local scales.

Finally, in Sect.~\ref{s7}, we review the latest discoveries on galaxy quenching from JWST observations at the high-$z$ frontier. We discuss the surprising observational evidence for massive, quenched galaxies at extremely early cosmic times, from both photometry and spectroscopy. We go on to consider the challenges these observations pose for theoretical models of galaxy evolution within the $\Lambda$CDM cosmological framework. Additionally, we discuss evidence for both quasar-mode/ ejective and radio-mode/ relativistic jet AGN feedback in the very early Universe, which may provide a plausible explanation for the early quenching of massive galaxies.

In Part~II of this review series we move on to exploring galaxy quenching in theoretical models and simulations. Additionally, we present several observational tests of the specific causes and mechanisms of galaxy quenching introduced by these theoretical models. At the end of this review series, we present the most probable explanations for galaxy quenching, as a function of galaxy type.

\backmatter

\section*{Declarations}
\bmhead{Competing Interests} The author declares no competing interests.


%%\newpage

\phantomsection
\addcontentsline{toc}{section}{References}
\bibliography{Quench_Review_submitted}% common bib file

@article{Baldry2006,
  author = {I. K. Baldry and M. L. Balogh and R. G. Bower and others},
  title = {{Galaxy bimodality versus stellar mass and environment}},
  journal = {\mnras},
  volume = {373},
  number = {2},
  pages = {469},
  year = {2006},
  doi = {10.1111/j.1365-2966.2006.11081.x}
}

@article{Behroozi2010,
  author = {P. S. Behroozi and C. Conroy and R. H. Wechsler},
  title = {{A comprehensive analysis of uncertainties affecting the stellar mass–halo mass relation for $0 < z < 4$}},
  journal = {\apj},
  volume = {717},
  number = {1},
  pages = {379},
  year = {2010},
  doi = {10.1088/0004-637X/717/1/379}
}

@article{Belfiore2017,
  author = {F. Belfiore and R. Maiolino and C. Maraston and others},
  title = {{SDSS-IV MaNGA–the spatially resolved transition from star formation to quiescence}},
  journal = {\mnras},
  volume = {466},
  number = {3},
  pages = {2570},
  year = {2017},
  doi = {10.1093/mnras/stw3211}
}

@article{Belfiore2018,
  author = {F. Belfiore and R. Maiolino and K. Bundy and others},
  title = {{SDSS IV MaNGA–sSFR profiles and the slow quenching of discs in green valley galaxies}},
  journal = {\mnras},
  volume = {477},
  number = {3},
  pages = {3014},
  year = {2018},
  doi = {10.1093/mnras/sty768}
}

@article{Bell2012,
  author = {E. F. Bell and A. van der Wel and C. Papovich and others},
  title = {{What turns galaxies off? The different morphologies of star-forming and quiescent galaxies since $z \sim 2$ from CANDELS}},
  journal = {\apj},
  volume = {753},
  number = {2},
  pages = {167},
  year = {2012},
  doi = {10.1088/0004-637X/753/2/167}
}

@article{Bluck2014,
  author = {A. F. L. Bluck and J. T. Mendel and S. L. Ellison and others},
  title = {{Bulge mass is king: the dominant role of the bulge in determining the fraction of passive galaxies in the Sloan Digital Sky Survey}},
  journal = {\mnras},
  volume = {441},
  number = {1},
  pages = {599},
  year = {2014},
  doi = {10.1093/mnras/stu594}
}

@article{Bluck2016,
  author = {A. F. L. Bluck and J. T. Mendel and S. L. Ellison and others},
  title = {{The impact of galactic properties and environment on the quenching of central and satellite galaxies: a comparison between SDSS, Illustris and L-Galaxies}},
  journal = {\mnras},
  volume = {462},
  number = {3},
  pages = {2559},
  year = {2016},
  doi = {10.1093/mnras/stw1665}
}

@article{Bluck2020a,
  author = {A. F. L. Bluck and R. Maiolino and S. F. S{\'a}nchez and others},
  title = {{Are galactic star formation and quenching governed by local, global, or environmental phenomena?}},
  journal = {\mnras},
  volume = {492},
  number = {1},
  pages = {96},
  year = {2020},
  doi = {10.1093/mnras/stz3264}
}

@article{Bluck2020b,
  author = {A. F. L. Bluck and R. Maiolino and J. M. Piotrowska and others},
  title = {{How do central and satellite galaxies quench?–Insights from spatially resolved spectroscopy in the MaNGA survey}},
  journal = {\mnras},
  volume = {499},
  number = {1},
  pages = {230},
  year = {2020},
  doi = {10.1093/mnras/staa2806}
}

@article{Bluck2022,
  author = {A. F. L. Bluck and R. Maiolino and S. Brownson and others},
  title = {{The quenching of galaxies, bulges, and disks since cosmic noon–A machine learning approach for identifying causality in astronomical data}},
  journal = {\aap},
  volume = {659},
  pages = {160},
  year = {2022},
  doi = {10.1051/0004-6361/202142643}
}

@article{Bluck2023,
  author = {A. F. L. Bluck and J. M. Piotrowska and R. Maiolino},
  title = {{The Fundamental Signature of Star Formation Quenching from AGN Feedback: A Critical Dependence of Quiescence on Supermassive Black Hole Mass, Not Halo Mass or Environment}},
  journal = {\apj},
  volume = {944},
  number = {2},
  pages = {108},
  year = {2023},
  doi = {10.3847/1538-4357/acac7c}
}

@article{Bluck2024,
  author = {A. F. L. Bluck and C. J. Conselice and K. Ormerod and others},
  title = {{The impact of supermassive black holes on galaxy evolution from cosmic noon to the present day}},
  journal = {\apj},
  volume = {961},
  number = {1},
  pages = {163},
  year = {2024},
  doi = {10.1088/0004-637X/961/1/163}
}

@article{Bower2006,
  author = {R. G. Bower and A. J. Benson and R. Malbon and others},
  title = {{Breaking the hierarchy of galaxy formation}},
  journal = {\mnras},
  volume = {370},
  number = {2},
  pages = {645},
  year = {2006},
  doi = {10.1111/j.1365-2966.2006.10519.x}
}

@article{Bower2008,
  author = {R. G. Bower and I. G. McCarthy and A. J. Benson},
  title = {{The flip side of galaxy formation: a combined model of galaxy formation and cluster heating}},
  journal = {\mnras},
  volume = {390},
  number = {4},
  pages = {1399},
  year = {2008},
  doi = {10.1111/j.1365-2966.2008.13838.x}
}

@article{Brinchmann2004,
  author = {J. Brinchmann and S. Charlot and S. D. M. White and others},
  title = {{The physical properties of star-forming galaxies in the low-redshift Universe}},
  journal = {\mnras},
  volume = {351},
  number = {4},
  pages = {1151},
  year = {2004},
  doi = {10.1111/j.1365-2966.2004.07881.x}
}

@article{Brownson2020,
  author = {S. Brownson and F. Belfiore and R. Maiolino and L. Lin and S. Carniani},
  title = {{Molecular gas and star formation in green valley galaxies: A case study}},
  journal = {MNRAS Lett.},
  volume = {498},
  number = {1},
  pages = {L66},
  year = {2020},
  doi = {10.1093/mnrasl/slaa107}
}

@article{Brownson2022,
  author = {S. Brownson and A. F. L. Bluck and R. Maiolino and G. C. Jones},
  title = {{What drives galaxy quenching? A deep connection between galaxy kinematics and quenching in the local Universe}},
  journal = {\mnras},
  volume = {511},
  number = {2},
  pages = {1913},
  year = {2022},
  doi = {10.1093/mnras/stab3749}
}

@article{Cameron2009a,
  author = {E. Cameron and S. P. Driver},
  title = {{Galaxy evolution by color-log(n) type since redshift unity in the Hubble Ultra Deep Field}},
  journal = {\aap},
  volume = {493},
  number = {2},
  pages = {489},
  year = {2009},
  doi = {10.1051/0004-6361:20078558}
}

@article{Cameron2009b,
  author = {E. Cameron and S. P. Driver and A. W. Graham and J. Liske},
  title = {{The Millennium Galaxy Catalogue: Exploring the color-concentration bimodality via bulge-disk decomposition}},
  journal = {\apj},
  volume = {699},
  number = {1},
  pages = {105},
  year = {2009},
  doi = {10.1088/0004-637X/699/1/105}
}

@article{Carnall2023,
  author = {A. C. Carnall and D. J. McLeod and R. J. McLure and others},
  title = {{A surprising abundance of massive quiescent galaxies at $3 < z < 5$ in the first data from JWST CEERS}},
  journal = {\mnras},
  volume = {520},
  number = {3},
  pages = {3974},
  year = {2023},
  doi = {10.1093/mnras/stad369}
}

@article{Cheung2012,
  author = {E. Cheung and S. M. Faber and D. C. Koo and A. A. Dutton and others},
  title = {{The dependence of quenching upon the inner structure of galaxies at $0.5 \leq z < 0.8$ in the DEEP2/AEGIS survey}},
  journal = {\apj},
  volume = {760},
  number = {2},
  pages = {131},
  year = {2012},
  doi = {10.1088/0004-637X/760/2/131}
}

@article{Cicone2012,
  author = {C. Cicone and C. Feruglio and R. Maiolino and F. Fiore and others},
  title = {{The physics and the structure of the quasar-driven outflow in Mrk 231}},
  journal = {\aap},
  volume = {543},
  pages = {A99},
  year = {2012},
  doi = {10.1051/0004-6361/201218793}
}

@article{Cicone2014,
  author = {C. Cicone and R. Maiolino and E. Sturm and J. Graci{\'a}-Carpio and others},
  title = {{Massive molecular outflows and evidence for AGN feedback from CO observations}},
  journal = {\aap},
  volume = {562},
  pages = {A21},
  year = {2014},
  doi = {10.1051/0004-6361/201322464}
}

@article{Cicone2015,
  author = {C. Cicone and R. Maiolino and S. Gallerani and R. Neri and others},
  title = {{Very extended cold gas, star formation and outflows in the halo of a bright quasar at $z > 6$}},
  journal = {\aap},
  volume = {574},
  pages = {A14},
  year = {2015},
  doi = {10.1051/0004-6361/201424980}
}

@article{Crain2015,
  author = {R. A. Crain and J. Schaye and R. G. Bower and others},
  title = {{The EAGLE simulations of galaxy formation: calibration of subgrid physics and model variations}},
  journal = {\mnras},
  volume = {450},
  number = {2},
  pages = {1937},
  year = {2015},
  doi = {10.1093/mnras/stv725}
}

@article{Croton2006,
  author = {D. J. Croton and V. Springel and S. D. M. White and G. De Lucia and others},
  title = {{The many lives of active galactic nuclei: cooling flows, black holes and the luminosities and colours of galaxies}},
  journal = {\mnras},
  volume = {365},
  number = {1},
  pages = {11},
  year = {2006},
  doi = {10.1111/j.1365-2966.2005.09675.x}
}

@article{Dave2019,
  author = {R. Dav{\'e} and D. Angl{\'e}s-Alc{\'a}zar and D. Narayanan and others},
  title = {{Simba: Cosmological simulations with black hole growth and feedback}},
  journal = {\mnras},
  volume = {486},
  number = {2},
  pages = {2827},
  year = {2019},
  doi = {10.1093/mnras/stz937}
}

@article{Dekel2006,
  author = {A. Dekel and Y. Birnboim},
  title = {{Galaxy bimodality due to cold flows and shock heating}},
  journal = {\mnras},
  volume = {368},
  number = {1},
  pages = {2},
  year = {2006},
  doi = {10.1111/j.1365-2966.2006.10145.x}
}

@article{Dekel2014,
  author = {A. Dekel and A. Burkert},
  title = {{Wet disc contraction to galactic blue nuggets and quenching to red nuggets}},
  journal = {\mnras},
  volume = {438},
  number = {2},
  pages = {1870},
  year = {2014},
  doi = {10.1093/mnras/stt2331}
}

@article{Dekel2009,
  author = {A. Dekel and Y. Birnboim and G. Engel and J. Freundlich and T. Goerdt and others},
  title = {{Cold streams in early massive hot haloes as the main mode of galaxy formation}},
  journal = {Nature},
  volume = {457},
  pages = {451},
  year = {2009},
  doi = {10.1038/nature07648}
}

@article{Dressler1980,
  author = {A. Dressler},
  title = {{Galaxy morphology in rich clusters: Implications for the formation and evolution of galaxies}},
  journal = {\apj},
  volume = {236},
  pages = {351},
  year = {1980},
  doi = {10.1086/157753}
}

@article{Driver2006,
  author = {S. P. Driver and P. D. Allen and A. W. Graham and others},
  title = {{The Millennium Galaxy Catalogue: Morphological classification and bimodality in the colour–concentration plane}},
  journal = {\mnras},
  volume = {368},
  number = {1},
  pages = {414},
  year = {2006},
  doi = {10.1111/j.1365-2966.2006.10114.x}
}

@article{Driver2011,
  author = {S. P. Driver and D. T. Hill and L. S. Kelvin and others},
  title = {{Galaxy and Mass Assembly (GAMA): Survey diagnostics and core data release}},
  journal = {\mnras},
  volume = {413},
  number = {2},
  pages = {971},
  year = {2011},
  doi = {10.1111/j.1365-2966.2010.18188.x}
}

@article{Ebeling2014,
  author = {H. Ebeling and L. N. Stephenson and A. C. Edge},
  title = {{Jellyfish: evidence of extreme ram-pressure stripping in massive galaxy clusters}},
  journal = {\apjl},
  volume = {781},
  number = {2},
  pages = {L40},
  year = {2014},
  doi = {10.1088/2041-8205/781/2/L40}
}

@article{Ellison2018,
  author = {S. L. Ellison and S. F. S{\'a}nchez and H. Ibarra-Medel and others},
  title = {{Star formation is boosted (and quenched) from the inside-out: radial star formation profiles from MaNGA}},
  journal = {\mnras},
  volume = {474},
  number = {2},
  pages = {2039},
  year = {2018},
  doi = {10.1093/mnras/stx2882}
}

@article{Ellison2021a,
  author = {S. L. Ellison and L. Lin and M. D. Thorp and others},
  title = {{The ALMaQUEST Survey–V. The non-universality of kpc-scale star formation relations and the factors that drive them}},
  journal = {\mnras},
  volume = {501},
  number = {4},
  pages = {4777},
  year = {2021},
  doi = {10.1093/mnras/staa3822}
}

@article{Fabian1994,
  author = {A. C. Fabian},
  title = {{Cooling flows in clusters of galaxies}},
  journal = {\araa},
  volume = {32},
  pages = {277},
  year = {1994},
  doi = {10.1146/annurev.aa.32.090194.001425}
}

@article{Fabian1999,
  author = {A. C. Fabian},
  title = {{The obscured growth of massive black holes}},
  journal = {\mnras},
  volume = {308},
  number = {4},
  pages = {L39},
  year = {1999},
  doi = {10.1046/j.1365-8711.1999.02941.x}
}

@article{Fabian2012,
  author = {A. C. Fabian},
  title = {{Observational evidence of active galactic nuclei feedback}},
  journal = {\araa},
  volume = {50},
  pages = {455},
  year = {2012},
  doi = {10.1146/annurev-astro-081811-125521}
}

@article{Fabian2006,
  author = {A. C. Fabian and J. S. Sanders and G. B. Taylor and others},
  title = {{A very deep Chandra observation of the Perseus cluster: shocks, ripples and conduction}},
  journal = {\mnras},
  volume = {366},
  number = {2},
  pages = {417},
  year = {2006},
  doi = {10.1111/j.1365-2966.2005.09896.x}
}

@article{Fang2013,
  author = {J. J. Fang and S. M. Faber and D. C. Koo and A. Dekel},
  title = {{A link between star formation quenching and inner stellar mass density in Sloan Digital Sky Survey central galaxies}},
  journal = {\apj},
  volume = {776},
  number = {1},
  pages = {63},
  year = {2013},
  doi = {10.1088/0004-637X/776/1/63}
}

@article{Feruglio2010,
  author = {C. Feruglio and R. Maiolino and E. Piconcelli and N. Menci and others},
  title = {{Quasar feedback revealed by giant molecular outflows}},
  journal = {\aap},
  volume = {518},
  pages = {L155},
  year = {2010},
  doi = {10.1051/0004-6361/201015164}
}

@article{Fluetsch2019,
  author = {A. Fluetsch and R. Maiolino and S. Carniani and others},
  title = {{Cold molecular outflows in the local Universe and their feedback effect on galaxies}},
  journal = {\mnras},
  volume = {483},
  number = {4},
  pages = {4586},
  year = {2019},
  doi = {10.1093/mnras/sty3449}
}

@article{Fukugita2004,
  author = {M. Fukugita and P. J. E. Peebles},
  title = {{The cosmic energy inventory}},
  journal = {\apj},
  volume = {616},
  number = {2},
  pages = {643},
  year = {2004},
  doi = {10.1086/425155}
}

@article{GonzalezDelgado2014,
  author = {R. M. Gonzalez Delgado and E. Perez and R. C. Fernandes and others},
  title = {{The star formation history of CALIFA galaxies: Radial structures}},
  journal = {\aap},
  volume = {562},
  pages = {A47},
  year = {2014},
  doi = {10.1051/0004-6361/201322011}
}

@article{GonzalezDelgado2016,
  author = {R. M. Gonzalez Delgado and R. C. Fernandes and E. Perez and others},
  title = {{Star formation along the Hubble sequence: Radial structure of the star formation of CALIFA galaxies}},
  journal = {\aap},
  volume = {590},
  pages = {A44},
  year = {2016},
  doi = {10.1051/0004-6361/201628174}
}

@article{Henriques2013,
  author = {B. M. B. Henriques and S. D. M. White and P. A. Thomas and others},
  title = {{Simulations of the galaxy population constrained by observations from z = 3 to the present day: implications for galactic winds and the fate of their ejecta}},
  journal = {\mnras},
  volume = {431},
  number = {4},
  pages = {3373},
  year = {2013},
  doi = {10.1093/mnras/stt415}
}

@article{Henriques2015,
  author = {B. M. B. Henriques and S. D. M. White and P. A. Thomas and others},
  title = {{Galaxy formation in the Planck cosmology – I. Matching the observed evolution of star formation rates, colours and stellar masses}},
  journal = {\mnras},
  volume = {451},
  number = {3},
  pages = {2663},
  year = {2015},
  doi = {10.1093/mnras/stv705}
}

@article{Henriques2019,
  author = {B. M. B. Henriques and S. D. M. White and S. J. Lilly and others},
  title = {{The origin of the mass scales for maximal star formation efficiency and quenching: the critical role of supernovae}},
  journal = {\mnras},
  volume = {485},
  number = {3},
  pages = {3446},
  year = {2019},
  doi = {10.1093/mnras/stz668}
}

@article{HlavacekLarrondo2012,
  author = {J. Hlavacek-Larrondo and A. C. Fabian and A. C. Edge},
  title = {{Extreme AGN feedback in the MAssive Cluster Survey: a detailed study of X-ray cavities at $z > 0.3$}},
  journal = {\mnras},
  volume = {421},
  number = {2},
  pages = {1360},
  year = {2012},
  doi = {10.1111/j.1365-2966.2012.20403.x}
}

@article{HlavacekLarrondo2015,
  author = {J. Hlavacek-Larrondo and M. McDonald and B. A. Benson and others},
  title = {{X-Ray Cavities in a Sample of 83 SPT-selected Clusters of Galaxies: Tracing the Evolution of AGN Feedback in Clusters of Galaxies out to z=1.2}},
  journal = {\apj},
  volume = {805},
  number = {1},
  pages = {35},
  year = {2015},
  doi = {10.1088/0004-637X/805/1/35}
}

@article{HlavacekLarrondo2018,
  author = {J. Hlavacek-Larrondo and others},
  title = {{Mystery solved: discovery of extended radio emission in the merging galaxy cluster Abell 2146}},
  journal = {\mnras},
  volume = {475},
  number = {2},
  pages = {2743},
  year = {2018},
  doi = {10.1093/mnras/stx3160}
}

@article{Hopkins2006,
  author = {P. F. Hopkins and L. Hernquist and T. J. Cox and others},
  title = {{A unified, merger-driven model of the origin of starbursts, quasars, the cosmic X-ray background, supermassive black holes, and galaxy spheroids}},
  journal = {\apjs},
  volume = {163},
  number = {1},
  pages = {1},
  year = {2006},
  doi = {10.1086/499298}
}

@article{Hopkins2008,
  author = {P. F. Hopkins and L. Hernquist and T. J. Cox and others},
  title = {{A cosmological framework for the co-evolution of quasars, supermassive black holes, and elliptical galaxies. I. Galaxy mergers and quasar activity}},
  journal = {\apjs},
  volume = {175},
  number = {2},
  pages = {356},
  year = {2008},
  doi = {10.1086/524362}
}

@article{Hopkins2010,
  author = {P. F. Hopkins and K. Bundy and D. Croton and others},
  title = {{Mergers and bulge formation in $\Lambda$CDM: which mergers matter?}},
  journal = {\apj},
  volume = {715},
  number = {1},
  pages = {202},
  year = {2010},
  doi = {10.1088/0004-637X/715/1/202}
}

@article{Kennicutt1998,
  author = {R. C. Kennicutt},
  title = {{Star formation in galaxies along the Hubble sequence}},
  journal = {\araa},
  volume = {36},
  pages = {189},
  year = {1998},
  doi = {10.1146/annurev.astro.36.1.189}
}

@article{Lang2014,
  author = {P. Lang and S. Wuyts and R. S. Somerville and others},
  title = {{Bulge Growth and Quenching since z = 2.5 in CANDELS/3D-HST}},
  journal = {\apj},
  volume = {788},
  number = {1},
  pages = {11},
  year = {2014},
  doi = {10.1088/0004-637X/788/1/11}
}

@article{Lilly1996,
  author = {S. J. Lilly and O. Le Fevre and F. Hammer and D. Crampton},
  title = {{The Canada-France Redshift Survey: the luminosity density and star formation history of the Universe to $z \sim 1$}},
  journal = {\apj},
  volume = {460},
  pages = {L1},
  year = {1996},
  doi = {10.1086/309975}
}

@article{Madau2014,
  author = {P. Madau and M. Dickinson},
  title = {{Cosmic star-formation history}},
  journal = {\araa},
  volume = {52},
  pages = {415},
  year = {2014},
  doi = {10.1146/annurev-astro-081811-125615}
}

@article{Madau1996,
  author = {P. Madau and H. C. Ferguson and M. E. Dickinson and others},
  title = {{High-redshift galaxies in the Hubble Deep Field: colour selection and star formation history to z ~ 4}},
  journal = {\mnras},
  volume = {283},
  number = {4},
  pages = {1388},
  year = {1996},
  doi = {10.1093/mnras/283.4.1388}
}

@article{Maiolino2012,
  author = {R. Maiolino and S. Gallerani and R. Neri and others},
  title = {{Evidence of strong quasar feedback in the early Universe}},
  journal = {MNRAS Lett.},
  volume = {425},
  number = {1},
  pages = {L66},
  year = {2012},
  doi = {10.1111/j.1745-3933.2012.01306.x}
}

@article{Martig2009,
  author = {M. Martig and F. Bournaud and R. Teyssier and A. Dekel},
  title = {{Morphological quenching of star formation: making early-type galaxies red}},
  journal = {\apj},
  volume = {707},
  number = {1},
  pages = {250},
  year = {2009},
  doi = {10.1088/0004-637X/707/1/250}
}

@article{Moster2010,
  author = {B. P. Moster and R. S. Somerville and C. Maulbetsch and others},
  title = {{Constraints on the relationship between stellar mass and halo mass at low and high redshift}},
  journal = {\apj},
  volume = {710},
  number = {2},
  pages = {903},
  year = {2010},
  doi = {10.1088/0004-637X/710/2/903}
}

@article{Moster2013,
  author = {B. P. Moster and T. Naab and S. D. M. White},
  title = {{Galactic star formation and accretion histories from matching galaxies to dark matter haloes}},
  journal = {\mnras},
  volume = {428},
  number = {4},
  pages = {3121},
  year = {2013},
  doi = {10.1093/mnras/sts261}
}

@article{Nelson2018,
  author = {D. Nelson and V. Springel and R. Pakmor and others},
  title = {{The IllustrisTNG Simulations: Public Data Release}},
  journal = {\mnras},
  volume = {475},
  number = {1},
  pages = {624},
  year = {2018},
  doi = {10.1093/mnras/stx3040}
}

@article{Peng2010,
  author = {Y.-J. Peng and S. J. Lilly and K. Kovac and others},
  title = {{Mass and Environment as Drivers of Galaxy Evolution in SDSS and zCOSMOS and the Origin of the Schechter Function}},
  journal = {\apj},
  volume = {721},
  number = {1},
  pages = {193},
  year = {2010},
  doi = {10.1088/0004-637X/721/1/193}
}

@article{Peng2012,
  author = {Y.-J. Peng and S. J. Lilly and A. Renzini and M. Carollo},
  title = {{A Mass-dependent Quenching Model for Star Formation Histories of Galaxies}},
  journal = {\apj},
  volume = {757},
  number = {1},
  pages = {4},
  year = {2012},
  doi = {10.1088/0004-637X/757/1/4}
}

@article{Pillepich2018,
  author = {A. Pillepich and D. Nelson and L. Hernquist and others},
  title = {{First results from the IllustrisTNG simulations: the stellar mass content of groups and clusters of galaxies}},
  journal = {\mnras},
  volume = {475},
  number = {1},
  pages = {648},
  year = {2018},
  doi = {10.1093/mnras/stx3112}
}

@article{Piotrowska2020,
  author = {J. M. Piotrowska and A. F. L. Bluck and R. Maiolino and A. Concas and Y. Peng},
  title = {{Towards a deeper understanding of the physics driving galaxy quenching–inferring trends in the gas content via extinction}},
  journal = {MNRAS Lett.},
  volume = {492},
  number = {1},
  pages = {L6},
  year = {2020},
  doi = {10.1093/mnrasl/slz167}
}

@article{Piotrowska2022,
  author = {J. M. Piotrowska and A. F. L. Bluck and R. Maiolino and Y. Peng},
  title = {{On the quenching of star formation in observed and simulated central galaxies: evidence for the role of integrated AGN feedback}},
  journal = {\mnras},
  volume = {512},
  number = {1},
  pages = {1052},
  year = {2022},
  doi = {10.1093/mnras/stab3673}
}

@article{Poggianti2017,
  author = {B. M. Poggianti and A. Moretti and M. Gullieuszik and others},
  title = {{GASP. I. Gas stripping phenomena in galaxies with MUSE}},
  journal = {\apj},
  volume = {844},
  number = {1},
  pages = {48},
  year = {2017},
  doi = {10.3847/1538-4357/aa78ed}
}

@article{Saintonge2016,
  author = {A. Saintonge and B. Catinella and L. Cortese and others},
  title = {{Molecular and atomic gas along and across the main sequence of star-forming galaxies}},
  journal = {\mnras},
  volume = {462},
  number = {2},
  pages = {1749},
  year = {2016},
  doi = {10.1093/mnras/stw1715}
}

@article{Saintonge2017,
  author = {A. Saintonge and B. Catinella and L. J. Tacconi and others},
  title = {{xCOLD GASS: the complete IRAM 30 m legacy survey of molecular gas for galaxy evolution studies}},
  journal = {\apjs},
  volume = {233},
  number = {2},
  pages = {22},
  year = {2017},
  doi = {10.3847/1538-4365/aa97e0}
}

@article{Sanchez2018,
  author = {S. F. Sanchez and others},
  title = {{SDSS IV-MaNGA: Properties of AGN Host Galaxies}},
  journal = {Revista Mexicana de Astronomía y Astrofísica},
  volume = {54},
  pages = {217},
  year = {2018}
}

@article{Sanchez2019b,
  author = {S. F. Sanchez and V. Avila-Reese and others},
  title = {{SDSS-IV MaNGA: An Archaeological View of the Cosmic Star Formation History}},
  journal = {\mnras},
  volume = {482},
  number = {2},
  pages = {1557},
  year = {2019},
  doi = {10.1093/mnras/sty2730}
}

@article{Schaye2015,
  author = {J. Schaye and R. A. Crain and R. G. Bower and others},
  title = {{The EAGLE Project: Simulating the Evolution and Assembly of Galaxies and Their Environments}},
  journal = {\mnras},
  volume = {446},
  number = {1},
  pages = {521},
  year = {2015},
  doi = {10.1093/mnras/stu2058}
}

@article{Shull2012,
  author = {J. M. Shull and B. D. Smith and C. W. Danforth},
  title = {{The Baryon Census in a Multiphase Intergalactic Medium: 30\% of the Baryons May Still Be Missing}},
  journal = {\apj},
  volume = {759},
  number = {1},
  pages = {23},
  year = {2012},
  doi = {10.1088/0004-637X/759/1/23}
}

@article{Sijacki2007,
  author = {D. Sijacki and V. Springel and T. Di Matteo and L. Hernquist},
  title = {{A unified model for AGN feedback in cosmological simulations of structure formation}},
  journal = {\mnras},
  volume = {380},
  number = {3},
  pages = {877},
  year = {2007},
  doi = {10.1111/j.1365-2966.2007.12153.x}
}

@article{Somerville2015,
  author = {R. S. Somerville and R. Dav{\'e}},
  title = {{Physical Models of Galaxy Formation in a Cosmological Framework}},
  journal = {\araa},
  volume = {53},
  pages = {51},
  year = {2015},
  doi = {10.1146/annurev-astro-082812-140951}
}

@article{Strateva2001,
  author = {I. Strateva and {\v{Z}}. {Ivezi{\'c}} and G. R. Knapp and others},
  title = {{Color separation of galaxy types in the Sloan Digital Sky Survey imaging data}},
  journal = {\aj},
  volume = {122},
  number = {4},
  pages = {1861},
  year = {2001},
  doi = {10.1086/323301}
}

@article{Tacchella2015,
  author = {S. Tacchella and C. M. Carollo and A. Renzini and others},
  title = {{Evidence for mature bulges and an inside-out quenching phase 3 billion years after the Big Bang}},
  journal = {Science},
  volume = {348},
  number = {6232},
  pages = {314},
  year = {2015},
  doi = {10.1126/science.1261094}
}

@article{Terrazas2016,
  author = {B. A. Terrazas and E. F. Bell and B. M. B. Henriques and others},
  title = {{Quiescence correlates strongly with directly measured black hole mass in central galaxies}},
  journal = {\apjl},
  volume = {830},
  number = {1},
  pages = {L12},
  year = {2016},
  doi = {10.3847/2041-8205/830/1/L12}
}

@article{Terrazas2017,
  author = {B. A. Terrazas and E. F. Bell and J. Woo and B. M. B. Henriques},
  title = {{Supermassive black holes as the regulators of star formation in central galaxies}},
  journal = {\apj},
  volume = {844},
  number = {2},
  pages = {170},
  year = {2017},
  doi = {10.3847/1538-4357/aa7d07}
}

@article{Terrazas2020,
  author = {B. A. Terrazas and E. F. Bell and A. Pillepich and others},
  title = {{The relationship between black hole mass and galaxy properties: examining the black hole feedback model in IllustrisTNG}},
  journal = {\mnras},
  volume = {493},
  number = {2},
  pages = {1888},
  year = {2020},
  doi = {10.1093/mnras/staa374}
}

@article{vandenBosch2008,
  author = {F. C. van den Bosch and D. Aquino and X. Yang and others},
  title = {{The importance of satellite quenching for the build-up of the red sequence of present-day galaxies}},
  journal = {\mnras},
  volume = {387},
  number = {1},
  pages = {79},
  year = {2008},
  doi = {10.1111/j.1365-2966.2008.13230.x}
}

@article{Vogelsberger2014a,
  author = {M. Vogelsberger and S. Genel and V. Springel and P. Torrey and others},
  title = {{Properties of galaxies reproduced by a hydrodynamic simulation}},
  journal = {Nature},
  volume = {509},
  pages = {177},
  year = {2014},
  doi = {10.1038/nature13316}
}

@article{Vogelsberger2014b,
  author = {M. Vogelsberger and S. Genel and V. Springel and others},
  title = {{Introducing the Illustris Project: simulating the coevolution of dark and visible matter in the Universe}},
  journal = {\mnras},
  volume = {444},
  number = {2},
  pages = {1518},
  year = {2014},
  doi = {10.1093/mnras/stu1536}
}

@article{Wake2012,
  author = {D. A. Wake and P. G. van Dokkum and M. Franx},
  title = {{Revealing Velocity Dispersion as the Best Indicator of a Galaxy's Color, Compared to Stellar Mass, Surface Mass Density, or Morphology}},
  journal = {\apjl},
  volume = {751},
  number = {2},
  pages = {L44},
  year = {2012},
  doi = {10.1088/2041-8205/751/2/L44}
}

@article{Weinberger2017,
  author = {R. Weinberger and V. Springel and L. Hernquist and others},
  title = {{The IllustrisTNG Simulations: AGN Feedback as a Driver of Galaxy Evolution}},
  journal = {\mnras},
  volume = {465},
  number = {1},
  pages = {329},
  year = {2017},
  doi = {10.1093/mnras/stx1187}
}

@article{Weinberger2018,
  author = {R. Weinberger and V. Springel and R. Pakmor and others},
  title = {{Supermassive black holes and their feedback effects in the IllustrisTNG simulation}},
  journal = {\mnras},
  volume = {479},
  number = {3},
  pages = {4056},
  year = {2018},
  doi = {10.1093/mnras/sty1733}
}

@article{Wild2009,
  author = {V. Wild and S. J. Charlot and R. Kauffmann},
  title = {{Understanding the physics of star formation quenching in the green valley}},
  journal = {\mnras},
  volume = {395},
  pages = {144},
  year = {2009},
  doi = {10.1111/j.1365-2966.2009.14668.x}
}

@article{Wild2010,
  author = {V. Wild and T. Heckman and S. Charlot},
  title = {{Timing the starburst–AGN connection}},
  journal = {\mnras},
  volume = {405},
  number = {2},
  pages = {933},
  year = {2010},
  doi = {10.1111/j.1365-2966.2010.16536.x}
}

@article{Wild2016,
  author = {V. Wild and O. Almaini and J. Dunlop and C. Simpson},
  title = {{The evolution of post-starburst galaxies from z = 2 to 0.5}},
  journal = {\mnras},
  volume = {463},
  number = {1},
  pages = {832},
  year = {2016},
  doi = {10.1093/mnras/stw1996}
}

@article{Woo2013,
  author = {J. Woo and A. Dekel and S. M. Faber and K. Noeske},
  title = {{Dependence of galaxy quenching on halo mass and distance from its centre}},
  journal = {\mnras},
  volume = {428},
  number = {4},
  pages = {3306},
  year = {2013},
  doi = {10.1093/mnras/sts274}
}

@article{Woo2015,
  author = {J. Woo and A. Dekel and S. M. Faber and D. C. Koo},
  title = {{Two conditions for galaxy quenching: compact centres and massive haloes}},
  journal = {\mnras},
  volume = {448},
  number = {1},
  pages = {237},
  year = {2015},
  doi = {10.1093/mnras/stu2755}
}

@article{Wuyts2013,
  author = {S. Wuyts and N. M. Förster Schreiber and E. J. Nelson and others},
  title = {{Resolved star formation patterns in star-forming galaxies at $z \sim 1$}},
  journal = {\apj},
  volume = {779},
  number = {2},
  pages = {135},
  year = {2013},
  doi = {10.1088/0004-637X/779/2/135}
}

@article{Yang2009,
  author = {X. Yang and H. J. Mo and F. C. van den Bosch},
  title = {{Galaxy Groups in the SDSS DR4: III. The Luminosity and Stellar Mass Functions}},
  journal = {\apj},
  volume = {695},
  number = {2},
  pages = {900},
  year = {2009},
  doi = {10.1088/0004-637X/695/2/900}
}

@article{Zinger2020,
  author = {E. Zinger and A. Pillepich and D. Nelson and others},
  title = {{Ejective and preventative: the IllustrisTNG black hole feedback and its effects on the thermodynamics of the gas within and around galaxies}},
  journal = {\mnras},
  volume = {499},
  number = {1},
  pages = {768},
  year = {2020},
  doi = {10.1093/mnras/staa2607}
}

@BOOK{Misner1973,
       author = {{Misner}, Charles W. and {Thorne}, Kip S. and {Wheeler}, John Archibald},
        title = "{Gravitation}",
      address = {San Francisco},
     publisher = {W.H. Freeman},
         year = 1973,
       adsurl = {https://ui.adsabs.harvard.edu/abs/1973grav.book.....M}
}

@ARTICLE{Springel2005c,
       author = {{Springel}, Volker and {Di Matteo}, Tiziana and {Hernquist}, Lars},
        title = "{Black Holes in Galaxy Mergers: The Formation of Red Elliptical Galaxies}",
      journal = {\apjl},
         year = 2005,
        month = feb,
       volume = {620},
       number = {2},
        pages = {L79-L82},
          doi = {10.1086/428772},
archivePrefix = {arXiv},
       eprint = {astro-ph/0409436},
 primaryClass = {astro-ph},
       adsurl = {https://ui.adsabs.harvard.edu/abs/2005ApJ...620L..79S}
}

@ARTICLE{Springel2006,
       author = {{Springel}, Volker and {Frenk}, Carlos S. and {White}, Simon D.~M.},
        title = "{The large-scale structure of the Universe}",
      journal = {\nat},
         year = 2006,
        month = apr,
       volume = {440},
       number = {7088},
        pages = {1137-1144},
          doi = {10.1038/nature04805},
archivePrefix = {arXiv},
       eprint = {astro-ph/0604561},
 primaryClass = {astro-ph},
       adsurl = {https://ui.adsabs.harvard.edu/abs/2006Natur.440.1137S}
}

@ARTICLE{Salpeter1955,
       author = {{Salpeter}, Edwin E.},
        title = "{The Luminosity Function and Stellar Evolution}",
      journal = {\apj},
         year = 1955,
        month = jan,
       volume = {121},
        pages = {161},
          doi = {10.1086/145971},
       adsurl = {https://ui.adsabs.harvard.edu/abs/1955ApJ...121..161S}
}

@ARTICLE{Chabrier2003,
       author = {{Chabrier}, Gilles},
        title = "{Galactic Stellar and Substellar Initial Mass Function}",
      journal = {\pasp},
         year = 2003,
        month = jul,
       volume = {115},
       number = {809},
        pages = {763-795},
          doi = {10.1086/376392},
archivePrefix = {arXiv},
       eprint = {astro-ph/0304382},
 primaryClass = {astro-ph},
       adsurl = {https://ui.adsabs.harvard.edu/abs/2003PASP..115..763C}
}

@ARTICLE{Kroupa2001,
       author = {{Kroupa}, Pavel},
        title = "{On the variation of the initial mass function}",
      journal = {\mnras},
         year = 2001,
        month = apr,
       volume = {322},
       number = {2},
        pages = {231-246},
          doi = {10.1046/j.1365-8711.2001.04022.x},
archivePrefix = {arXiv},
       eprint = {astro-ph/0009005},
 primaryClass = {astro-ph},
       adsurl = {https://ui.adsabs.harvard.edu/abs/2001MNRAS.322..231K}
}

@ARTICLE{Bell2003,
       author = {{Bell}, Eric F. and {McIntosh}, Daniel H. and {Katz}, Neal and {Weinberg}, Martin D.},
        title = "{The Optical and Near-Infrared Properties of Galaxies. I. Luminosity and Stellar Mass Functions}",
      journal = {\apjs},
         year = 2003,
        month = dec,
       volume = {149},
       number = {2},
        pages = {289-312},
          doi = {10.1086/378847},
archivePrefix = {arXiv},
       eprint = {astro-ph/0302543},
 primaryClass = {astro-ph},
       adsurl = {https://ui.adsabs.harvard.edu/abs/2003ApJS..149..289B}
}

@ARTICLE{Muzzin2013,
       author = {{Muzzin}, Adam and {Marchesini}, Danilo and {Stefanon}, Mauro and {Franx}, Marijn and {McCracken}, Henry J. and {Milvang-Jensen}, Bo and {Dunlop}, James S. and {Fynbo}, J.~P.~U. and {Brammer}, Gabriel and {Labb{\'e}}, Ivo and {van Dokkum}, Pieter G.},
        title = "{The Evolution of the Stellar Mass Functions of Star-forming and Quiescent Galaxies to z = 4 from the COSMOS/UltraVISTA Survey}",
      journal = {\apj},
         year = 2013,
        month = nov,
       volume = {777},
       number = {1},
          eid = {18},
        pages = {18},
          doi = {10.1088/0004-637X/777/1/18},
archivePrefix = {arXiv},
       eprint = {1303.4409},
 primaryClass = {astro-ph.CO},
       adsurl = {https://ui.adsabs.harvard.edu/abs/2013ApJ...777...18M}
}

@ARTICLE{Baldry2012,
       author = {{Baldry}, I.~K. and {Driver}, S.~P. and {Loveday} and J. and {Taylor}, E.~N. and {Kelvin}, L.~S. and {Liske} and J. and {Norberg} and P. and {Robotham}, A.~S.~G. and {Brough} and S. and {Hopkins}, A.~M. and {Bamford}, S.~P. and {Peacock}, J.~A. and {Bland-Hawthorn} and J. and {Conselice}, C.~J. and {Croom}, S.~M. and {Jones}, D.~H. and {Parkinson}, H.~R. and {Popescu}, C.~C. and {Prescott} and M. and {Sharp}, R.~G. and {Tuffs}, R.~J.},
        title = "{Galaxy And Mass Assembly (GAMA): the galaxy stellar mass function at $z < 0.06$}",
      journal = {\mnras},
         year = 2012,
        month = mar,
       volume = {421},
       number = {1},
        pages = {621-634},
          doi = {10.1111/j.1365-2966.2012.20340.x},
archivePrefix = {arXiv},
       eprint = {1111.5707},
 primaryClass = {astro-ph.CO},
       adsurl = {https://ui.adsabs.harvard.edu/abs/2012MNRAS.421..621B}
}

@ARTICLE{Marchesini2009,
       author = {{Marchesini}, Danilo and {van Dokkum}, Pieter G. and {F{\"o}rster Schreiber}, Natascha M. and {Franx}, Marijn and {Labb{\'e}}, Ivo and {Wuyts}, Stijn},
        title = "{The Evolution of the Stellar Mass Function of Galaxies from z = 4.0 and the First Comprehensive Analysis of its Uncertainties: Evidence for Mass-Dependent Evolution}",
      journal = {\apj},
         year = 2009,
        month = aug,
       volume = {701},
       number = {2},
        pages = {1765-1796},
          doi = {10.1088/0004-637X/701/2/1765},
archivePrefix = {arXiv},
       eprint = {0811.1773},
 primaryClass = {astro-ph},
       adsurl = {https://ui.adsabs.harvard.edu/abs/2009ApJ...701.1765M}
}

@article{Cole2001,
  author = {Cole and S. and Norberg and P. and Baugh and C. M. and Frenk and C. S. and Bland-Hawthorn and J. and Bridges and T. and Cannon and R. and Colless and M. and Collins and C. and Couch and W. and others},
  title = {The 2dF galaxy redshift survey: near-infrared galaxy luminosity functions},
  journal = {\mnras},
  volume = {326},
  pages = {255--273},
  year = {2001},
  doi = {10.1046/j.1365-8711.2001.04623.x}
}

@article{Somerville1999,
  author = {Somerville and R. S. and Primack and J. R.},
  title = {Semi-analytic modelling of galaxy formation: the local Universe},
  journal = {\mnras},
  volume = {310},
  pages = {1087--1110},
  year = {1999},
  doi = {10.1046/j.1365-8711.1999.03012.x}
}

@article{Naab2017,
  author = {Naab and T. and Ostriker and J. P.},
  title = {Theoretical Challenges in Galaxy Formation},
  journal = {\araa},
  volume = {55},
  pages = {59--109},
  year = {2017},
  doi = {10.1146/annurev-astro-081913-040019}
}

@ARTICLE{Cole2000,
       author = {{Cole}, Shaun and {Lacey}, Cedric G. and {Baugh}, Carlton M. and {Frenk}, Carlos S.},
        title = "{Hierarchical galaxy formation}",
      journal = {\mnras},
         year = 2000,
        month = nov,
       volume = {319},
       number = {1},
        pages = {168-204},
          doi = {10.1046/j.1365-8711.2000.03879.x},
archivePrefix = {arXiv},
       eprint = {astro-ph/0007281},
 primaryClass = {astro-ph},
       adsurl = {https://ui.adsabs.harvard.edu/abs/2000MNRAS.319..168C}
}

@ARTICLE{Schawinski2007,
       author = {{Schawinski}, Kevin and {Thomas}, Daniel and {Sarzi}, Marc and {Maraston}, Claudia and {Kaviraj}, Sugata and {Joo}, Seok-Joo and {Yi}, Sukyoung K. and {Silk}, Joseph},
        title = "{Observational evidence for AGN feedback in early-type galaxies}",
      journal = {\mnras},
         year = 2007,
        month = dec,
       volume = {382},
       number = {4},
        pages = {1415-1431},
          doi = {10.1111/j.1365-2966.2007.12487.x},
archivePrefix = {arXiv},
       eprint = {0709.3015},
 primaryClass = {astro-ph},
       adsurl = {https://ui.adsabs.harvard.edu/abs/2007MNRAS.382.1415S}
}

@ARTICLE{Schawinski2014,
       author = {{Schawinski}, Kevin and {Urry} and C. Megan and {Simmons}, Brooke D. and {Fortson}, Lucy and {Kaviraj}, Sugata and {Keel}, William C. and {Lintott}, Chris J. and {Masters}, Karen L. and {Nichol}, Robert C. and {Sarzi}, Marc and {Skibba}, Ramin and {Treister}, Ezequiel and {Willett}, Kyle W. and {Wong} and O. Ivy and {Yi}, Sukyoung K.},
        title = "{The green valley is a red herring: Galaxy Zoo reveals two evolutionary pathways towards quenching of star formation in early- and late-type galaxies}",
      journal = {\mnras},
         year = 2014,
        month = may,
       volume = {440},
       number = {1},
        pages = {889-907},
          doi = {10.1093/mnras/stu327},
archivePrefix = {arXiv},
       eprint = {1402.4814},
 primaryClass = {astro-ph.GA},
       adsurl = {https://ui.adsabs.harvard.edu/abs/2014MNRAS.440..889S}
}

@ARTICLE{Bruzual2003,
       author = {{Bruzual} and G. and {Charlot}, S.},
        title = "{Stellar population synthesis at the resolution of 2003}",
      journal = {\mnras},
         year = 2003,
        month = oct,
       volume = {344},
       number = {4},
        pages = {1000-1028},
          doi = {10.1046/j.1365-8711.2003.06897.x},
archivePrefix = {arXiv},
       eprint = {astro-ph/0309134},
 primaryClass = {astro-ph},
       adsurl = {https://ui.adsabs.harvard.edu/abs/2003MNRAS.344.1000B}
}

@ARTICLE{Toomre1972,
       author = {{Toomre}, Alar and {Toomre}, Juri},
        title = "{Galactic Bridges and Tails}",
      journal = {\apj},
         year = 1972,
        month = dec,
       volume = {178},
        pages = {623-666},
          doi = {10.1086/151823},
       adsurl = {https://ui.adsabs.harvard.edu/abs/1972ApJ...178..623T}
}

@ARTICLE{Gensior2020,
       author = {{Gensior}, Jindra and {Kruijssen}, J.~M. Diederik and {Keller}, Benjamin W.},
        title = "{Heart of darkness: the influence of galactic dynamics on quenching star formation in galaxy spheroids}",
      journal = {\mnras},
         year = 2020,
        month = jun,
       volume = {495},
       number = {1},
        pages = {199-223},
          doi = {10.1093/mnras/staa1184},
archivePrefix = {arXiv},
       eprint = {2002.01484},
 primaryClass = {astro-ph.GA},
       adsurl = {https://ui.adsabs.harvard.edu/abs/2020MNRAS.495..199G}
}

@ARTICLE{York2000,
       author = {{York}, Donald G. and {Adelman} and J. and {Anderson}, Jr., John E. and {Anderson}, Scott F. and {Annis}, James and {Bahcall}, Neta A. and {Bakken}, J.~A. and {Barkhouser}, Robert and {Bastian}, Steven and {Berman}, Eileen and {Boroski}, William N. and {Bracker}, Steve and {Briegel}, Charlie and {Briggs}, John W. and {Brinkmann} and J. and {Brunner}, Robert and {Burles}, Scott and {Carey}, Larry and {Carr}, Michael A. and {Castander}, Francisco J. and {Chen}, Bing and {Colestock}, Patrick L. and {Connolly}, A.~J. and {Crocker}, J.~H. and {Csabai}, Istv{\'a}n and {Czarapata}, Paul C. and {Davis}, John Eric and {Doi}, Mamoru and {Dombeck}, Tom and {Eisenstein}, Daniel and {Ellman}, Nancy and {Elms}, Brian R. and {Evans}, Michael L. and {Fan}, Xiaohui and {Federwitz}, Glenn R. and {Fiscelli}, Larry and {Friedman}, Scott and {Frieman}, Joshua A. and {Fukugita}, Masataka and {Gillespie}, Bruce and {Gunn}, James E. and {Gurbani}, Vijay K. and {de Haas}, Ernst and {Haldeman}, Merle and {Harris}, Frederick H. and {Hayes} and J. and {Heckman}, Timothy M. and {Hennessy}, G.~S. and {Hindsley}, Robert B. and {Holm}, Scott and {Holmgren}, Donald J. and {Huang}, Chi-hao and {Hull}, Charles and {Husby}, Don and {Ichikawa}, Shin-Ichi and {Ichikawa}, Takashi and {Ivezi{\'c}}, {\v{Z}}eljko and {Kent}, Stephen and {Kim}, Rita S.~J. and {Kinney} and E. and {Klaene}, Mark and {Kleinman}, A.~N. and {Kleinman} and S. and {Knapp}, G.~R. and {Korienek}, John and {Kron}, Richard G. and {Kunszt}, Peter Z. and {Lamb}, D.~Q. and {Lee} and B. and {Leger} and R. French and {Limmongkol}, Siriluk and {Lindenmeyer}, Carl and {Long}, Daniel C. and {Loomis}, Craig and {Loveday}, Jon and {Lucinio}, Rich and {Lupton}, Robert H. and {MacKinnon}, Bryan and {Mannery}, Edward J. and {Mantsch}, P.~M. and {Margon}, Bruce and {McGehee}, Peregrine and {McKay}, Timothy A. and {Meiksin}, Avery and {Merelli}, Aronne and {Monet}, David G. and {Munn}, Jeffrey A. and {Narayanan}, Vijay K. and {Nash}, Thomas and {Neilsen}, Eric and {Neswold}, Rich and {Newberg}, Heidi Jo and {Nichol}, R.~C. and {Nicinski}, Tom and {Nonino}, Mario and {Okada}, Norio and {Okamura}, Sadanori and {Ostriker}, Jeremiah P. and {Owen}, Russell and {Pauls} and A. George and {Peoples}, John and {Peterson}, R.~L. and {Petravick}, Donald and {Pier}, Jeffrey R. and {Pope}, Adrian and {Pordes}, Ruth and {Prosapio}, Angela and {Rechenmacher}, Ron and {Quinn}, Thomas R. and {Richards}, Gordon T. and {Richmond}, Michael W. and {Rivetta}, Claudio H. and {Rockosi}, Constance M. and {Ruthmansdorfer}, Kurt and {Sandford}, Dale and {Schlegel}, David J. and {Schneider}, Donald P. and {Sekiguchi}, Maki and {Sergey}, Gary and {Shimasaku}, Kazuhiro and {Siegmund}, Walter A. and {Smee}, Stephen and {Smith} and J. Allyn and {Snedden} and S. and {Stone} and R. and {Stoughton}, Chris and {Strauss}, Michael A. and {Stubbs}, Christopher and {SubbaRao}, Mark and {Szalay}, Alexander S. and {Szapudi}, Istvan and {Szokoly}, Gyula P. and {Thakar}, Anirudda R. and {Tremonti}, Christy and {Tucker}, Douglas L. and {Uomoto}, Alan and {Vanden Berk}, Dan and {Vogeley}, Michael S. and {Waddell}, Patrick and {Wang}, Shu-i. and {Watanabe}, Masaru and {Weinberg}, David H. and {Yanny}, Brian and {Yasuda}, Naoki and {SDSS Collaboration}},
        title = "{The Sloan Digital Sky Survey: Technical Summary}",
      journal = {\aj},
         year = 2000,
        month = sep,
       volume = {120},
       number = {3},
        pages = {1579-1587},
          doi = {10.1086/301513},
archivePrefix = {arXiv},
       eprint = {astro-ph/0006396},
 primaryClass = {astro-ph},
       adsurl = {https://ui.adsabs.harvard.edu/abs/2000AJ....120.1579Y}
}

@ARTICLE{Abazajian2009,
       author = {{Abazajian}, Kevork N. and {Adelman-McCarthy}, Jennifer K. and {Ag{\"u}eros}, Marcel A. and {Allam}, Sahar S. and {Allende Prieto}, Carlos and {An}, Deokkeun and {Anderson}, Kurt S.~J. and {Anderson}, Scott F. and {Annis}, James and {Bahcall}, Neta A. and {Bailer-Jones}, C.~A.~L. and {Barentine}, J.~C. and {Bassett}, Bruce A. and {Becker}, Andrew C. and {Beers}, Timothy C. and {Bell}, Eric F. and {Belokurov}, Vasily and {Berlind}, Andreas A. and {Berman}, Eileen F. and {Bernardi}, Mariangela and {Bickerton}, Steven J. and {Bizyaev}, Dmitry and {Blakeslee}, John P. and {Blanton}, Michael R. and {Bochanski}, John J. and {Boroski}, William N. and {Brewington}, Howard J. and {Brinchmann}, Jarle and {Brinkmann} and J. and {Brunner}, Robert J. and {Budav{\'a}ri}, Tam{\'a}s and {Carey}, Larry N. and {Carliles}, Samuel and {Carr}, Michael A. and {Castander}, Francisco J. and {Cinabro}, David and {Connolly}, A.~J. and {Csabai}, Istv{\'a}n and {Cunha}, Carlos E. and {Czarapata}, Paul C. and {Davenport}, James R.~A. and {de Haas}, Ernst and {Dilday}, Ben and {Doi}, Mamoru and {Eisenstein}, Daniel J. and {Evans}, Michael L. and {Evans}, N.~W. and {Fan}, Xiaohui and {Friedman}, Scott D. and {Frieman}, Joshua A. and {Fukugita}, Masataka and {G{\"a}nsicke}, Boris T. and {Gates}, Evalyn and {Gillespie}, Bruce and {Gilmore} and G. and {Gonzalez}, Belinda and {Gonzalez}, Carlos F. and {Grebel}, Eva K. and {Gunn}, James E. and {Gy{\"o}ry}, Zsuzsanna and {Hall}, Patrick B. and {Harding}, Paul and {Harris}, Frederick H. and {Harvanek}, Michael and {Hawley}, Suzanne L. and {Hayes}, Jeffrey J.~E. and {Heckman}, Timothy M. and {Hendry}, John S. and {Hennessy}, Gregory S. and {Hindsley}, Robert B. and {Hoblitt} and J. and {Hogan}, Craig J. and {Hogg}, David W. and {Holtzman}, Jon A. and {Hyde}, Joseph B. and {Ichikawa}, Shin-ichi and {Ichikawa}, Takashi and {Im}, Myungshin and {Ivezi{\'c}}, {\v{Z}}eljko and {Jester}, Sebastian and {Jiang}, Linhua and {Johnson}, Jennifer A. and {Jorgensen}, Anders M. and {Juri{\'c}}, Mario and {Kent}, Stephen M. and {Kessler} and R. and {Kleinman}, S.~J. and {Knapp}, G.~R. and {Konishi}, Kohki and {Kron}, Richard G. and {Krzesinski}, Jurek and {Kuropatkin}, Nikolay and {Lampeitl}, Hubert and {Lebedeva}, Svetlana and {Lee}, Myung Gyoon and {Lee}, Young Sun and {French Leger} and R. and {L{\'e}pine}, S{\'e}bastien and {Li}, Nolan and {Lima}, Marcos and {Lin}, Huan and {Long}, Daniel C. and {Loomis}, Craig P. and {Loveday}, Jon and {Lupton}, Robert H. and {Magnier}, Eugene and {Malanushenko}, Olena and {Malanushenko}, Viktor and {Mandelbaum}, Rachel and {Margon}, Bruce and {Marriner}, John P. and {Mart{\'\i}nez-Delgado}, David and {Matsubara}, Takahiko and {McGehee}, Peregrine M. and {McKay}, Timothy A. and {Meiksin}, Avery and {Morrison}, Heather L. and {Mullally}, Fergal and {Munn}, Jeffrey A. and {Murphy}, Tara and {Nash}, Thomas and {Nebot}, Ada and {Neilsen}, Jr., Eric H. and {Newberg}, Heidi Jo and {Newman}, Peter R. and {Nichol}, Robert C. and {Nicinski}, Tom and {Nieto-Santisteban}, Maria and {Nitta}, Atsuko and {Okamura}, Sadanori and {Oravetz}, Daniel J. and {Ostriker}, Jeremiah P. and {Owen}, Russell and {Padmanabhan}, Nikhil and {Pan}, Kaike and {Park}, Changbom and {Pauls}, George and {Peoples}, Jr., John and {Percival}, Will J. and {Pier}, Jeffrey R. and {Pope}, Adrian C. and {Pourbaix}, Dimitri and {Price}, Paul A. and {Purger}, Norbert and {Quinn}, Thomas and {Raddick} and M. Jordan and {Re Fiorentin}, Paola and {Richards}, Gordon T. and {Richmond}, Michael W. and {Riess}, Adam G. and {Rix}, Hans-Walter and {Rockosi}, Constance M. and {Sako}, Masao and {Schlegel}, David J. and {Schneider}, Donald P. and {Scholz}, Ralf-Dieter and {Schreiber}, Matthias R. and {Schwope}, Axel D. and {Seljak}, Uro{\v{s}} and {Sesar}, Branimir and {Sheldon}, Erin and {Shimasaku}, Kazu and {Sibley}, Valena C. and {Simmons}, A.~E. and {Sivarani}, Thirupathi and {Allyn Smith} and J. and {Smith}, Martin C. and {Smol{\v{c}}i{\'c}}, Vernesa and {Snedden}, Stephanie A. and {Stebbins}, Albert and {Steinmetz}, Matthias and {Stoughton}, Chris and {Strauss}, Michael A. and {SubbaRao}, Mark and {Suto}, Yasushi and {Szalay}, Alexander S. and {Szapudi}, Istv{\'a}n and {Szkody}, Paula and {Tanaka}, Masayuki and {Tegmark}, Max and {Teodoro}, Luis F.~A. and {Thakar}, Aniruddha R. and {Tremonti}, Christy A. and {Tucker}, Douglas L. and {Uomoto}, Alan and {Vanden Berk}, Daniel E. and {Vandenberg}, Jan and {Vidrih} and S. and {Vogeley}, Michael S. and {Voges}, Wolfgang and {Vogt}, Nicole P. and {Wadadekar}, Yogesh and {Watters}, Shannon and {Weinberg}, David H. and {West}, Andrew A. and {White}, Simon D.~M. and {Wilhite}, Brian C. and {Wonders}, Alainna C. and {Yanny}, Brian and {Yocum}, D.~R.},
        title = "{The Seventh Data Release of the Sloan Digital Sky Survey}",
      journal = {\apjs},
         year = 2009,
        month = jun,
       volume = {182},
       number = {2},
        pages = {543-558},
          doi = {10.1088/0067-0049/182/2/543},
archivePrefix = {arXiv},
       eprint = {0812.0649},
 primaryClass = {astro-ph},
       adsurl = {https://ui.adsabs.harvard.edu/abs/2009ApJS..182..543A}
}

@ARTICLE{Giavalisco2004,
       author = {{Giavalisco} and M. and {Ferguson}, H.~C. and {Koekemoer}, A.~M. and {Dickinson} and M. and {Alexander}, D.~M. and {Bauer}, F.~E. and {Bergeron} and J. and {Biagetti} and C. and {Brandt}, W.~N. and {Casertano} and S. and {Cesarsky} and C. and {Chatzichristou} and E. and {Conselice} and C. and {Cristiani} and S. and {Da Costa} and L. and {Dahlen} and T. and {de Mello} and D. and {Eisenhardt} and P. and {Erben} and T. and {Fall}, S.~M. and {Fassnacht} and C. and {Fosbury} and R. and {Fruchter} and A. and {Gardner}, J.~P. and {Grogin} and N. and {Hook}, R.~N. and {Hornschemeier}, A.~E. and {Idzi} and R. and {Jogee} and S. and {Kretchmer} and C. and {Laidler} and V. and {Lee}, K.~S. and {Livio} and M. and {Lucas} and R. and {Madau} and P. and {Mobasher} and B. and {Moustakas}, L.~A. and {Nonino} and M. and {Padovani} and P. and {Papovich} and C. and {Park} and Y. and {Ravindranath} and S. and {Renzini} and A. and {Richardson} and M. and {Riess} and A. and {Rosati} and P. and {Schirmer} and M. and {Schreier} and E. and {Somerville}, R.~S. and {Spinrad} and H. and {Stern} and D. and {Stiavelli} and M. and {Strolger} and L. and {Urry}, C.~M. and {Vandame} and B. and {Williams} and R. and {Wolf}, C.},
        title = "{The Great Observatories Origins Deep Survey: Initial Results from Optical and Near-Infrared Imaging}",
      journal = {\apjl},
         year = 2004,
        month = jan,
       volume = {600},
       number = {2},
        pages = {L93-L98},
          doi = {10.1086/379232},
archivePrefix = {arXiv},
       eprint = {astro-ph/0309105},
 primaryClass = {astro-ph},
       adsurl = {https://ui.adsabs.harvard.edu/abs/2004ApJ...600L..93G}
}

@ARTICLE{Grogin2011,
       author = {{Grogin}, Norman A. and {Kocevski}, Dale D. and {Faber}, S.~M. and {Ferguson}, Henry C. and {Koekemoer}, Anton M. and {Riess}, Adam G. and {Acquaviva}, Viviana and {Alexander}, David M. and {Almaini}, Omar and {Ashby}, Matthew L.~N. and {Barden}, Marco and {Bell}, Eric F. and {Bournaud}, Fr{\'e}d{\'e}ric and {Brown}, Thomas M. and {Caputi}, Karina I. and {Casertano}, Stefano and {Cassata}, Paolo and {Castellano}, Marco and {Challis}, Peter and {Chary}, Ranga-Ram and {Cheung}, Edmond and {Cirasuolo}, Michele and {Conselice}, Christopher J. and {Roshan Cooray}, Asantha and {Croton}, Darren J. and {Daddi}, Emanuele and {Dahlen}, Tomas and {Dav{\'e}}, Romeel and {de Mello}, Du{\'\i}lia F. and {Dekel}, Avishai and {Dickinson}, Mark and {Dolch}, Timothy and {Donley}, Jennifer L. and {Dunlop}, James S. and {Dutton}, Aaron A. and {Elbaz}, David and {Fazio}, Giovanni G. and {Filippenko}, Alexei V. and {Finkelstein}, Steven L. and {Fontana}, Adriano and {Gardner}, Jonathan P. and {Garnavich}, Peter M. and {Gawiser}, Eric and {Giavalisco}, Mauro and {Grazian}, Andrea and {Guo}, Yicheng and {Hathi}, Nimish P. and {H{\"a}ussler}, Boris and {Hopkins}, Philip F. and {Huang}, Jia-Sheng and {Huang}, Kuang-Han and {Jha}, Saurabh W. and {Kartaltepe}, Jeyhan S. and {Kirshner}, Robert P. and {Koo}, David C. and {Lai}, Kamson and {Lee}, Kyoung-Soo and {Li}, Weidong and {Lotz}, Jennifer M. and {Lucas}, Ray A. and {Madau}, Piero and {McCarthy}, Patrick J. and {McGrath}, Elizabeth J. and {McIntosh}, Daniel H. and {McLure}, Ross J. and {Mobasher}, Bahram and {Moustakas}, Leonidas A. and {Mozena}, Mark and {Nandra}, Kirpal and {Newman}, Jeffrey A. and {Niemi}, Sami-Matias and {Noeske}, Kai G. and {Papovich}, Casey J. and {Pentericci}, Laura and {Pope}, Alexandra and {Primack}, Joel R. and {Rajan}, Abhijith and {Ravindranath}, Swara and {Reddy}, Naveen A. and {Renzini}, Alvio and {Rix}, Hans-Walter and {Robaina}, Aday R. and {Rodney}, Steven A. and {Rosario}, David J. and {Rosati}, Piero and {Salimbeni}, Sara and {Scarlata}, Claudia and {Siana}, Brian and {Simard}, Luc and {Smidt}, Joseph and {Somerville}, Rachel S. and {Spinrad}, Hyron and {Straughn}, Amber N. and {Strolger}, Louis-Gregory and {Telford}, Olivia and {Teplitz}, Harry I. and {Trump}, Jonathan R. and {van der Wel}, Arjen and {Villforth}, Carolin and {Wechsler}, Risa H. and {Weiner}, Benjamin J. and {Wiklind}, Tommy and {Wild}, Vivienne and {Wilson}, Grant and {Wuyts}, Stijn and {Yan}, Hao-Jing and {Yun}, Min S.},
        title = "{CANDELS: The Cosmic Assembly Near-infrared Deep Extragalactic Legacy Survey}",
      journal = {\apjs},
         year = 2011,
        month = dec,
       volume = {197},
       number = {2},
          eid = {35},
        pages = {35},
          doi = {10.1088/0067-0049/197/2/35},
archivePrefix = {arXiv},
       eprint = {1105.3753},
 primaryClass = {astro-ph.CO},
       adsurl = {https://ui.adsabs.harvard.edu/abs/2011ApJS..197...35G}
}

@ARTICLE{Koekemoer2011,
       author = {{Koekemoer}, Anton M. and {Faber}, S.~M. and {Ferguson}, Henry C. and {Grogin}, Norman A. and {Kocevski}, Dale D. and {Koo}, David C. and {Lai}, Kamson and {Lotz}, Jennifer M. and {Lucas}, Ray A. and {McGrath}, Elizabeth J. and {Ogaz}, Sara and {Rajan}, Abhijith and {Riess}, Adam G. and {Rodney}, Steve A. and {Strolger}, Louis and {Casertano}, Stefano and {Castellano}, Marco and {Dahlen}, Tomas and {Dickinson}, Mark and {Dolch}, Timothy and {Fontana}, Adriano and {Giavalisco}, Mauro and {Grazian}, Andrea and {Guo}, Yicheng and {Hathi}, Nimish P. and {Huang}, Kuang-Han and {van der Wel}, Arjen and {Yan}, Hao-Jing and {Acquaviva}, Viviana and {Alexander}, David M. and {Almaini}, Omar and {Ashby}, Matthew L.~N. and {Barden}, Marco and {Bell}, Eric F. and {Bournaud}, Fr{\'e}d{\'e}ric and {Brown}, Thomas M. and {Caputi}, Karina I. and {Cassata}, Paolo and {Challis}, Peter J. and {Chary}, Ranga-Ram and {Cheung}, Edmond and {Cirasuolo}, Michele and {Conselice}, Christopher J. and {Roshan Cooray}, Asantha and {Croton}, Darren J. and {Daddi}, Emanuele and {Dav{\'e}}, Romeel and {de Mello}, Duilia F. and {de Ravel}, Loic and {Dekel}, Avishai and {Donley}, Jennifer L. and {Dunlop}, James S. and {Dutton}, Aaron A. and {Elbaz}, David and {Fazio}, Giovanni G. and {Filippenko}, Alexei V. and {Finkelstein}, Steven L. and {Frazer}, Chris and {Gardner}, Jonathan P. and {Garnavich}, Peter M. and {Gawiser}, Eric and {Gruetzbauch}, Ruth and {Hartley}, Will G. and {H{\"a}ussler}, Boris and {Herrington}, Jessica and {Hopkins}, Philip F. and {Huang}, Jia-Sheng and {Jha}, Saurabh W. and {Johnson}, Andrew and {Kartaltepe}, Jeyhan S. and {Khostovan}, Ali A. and {Kirshner}, Robert P. and {Lani}, Caterina and {Lee}, Kyoung-Soo and {Li}, Weidong and {Madau}, Piero and {McCarthy}, Patrick J. and {McIntosh}, Daniel H. and {McLure}, Ross J. and {McPartland}, Conor and {Mobasher}, Bahram and {Moreira}, Heidi and {Mortlock}, Alice and {Moustakas}, Leonidas A. and {Mozena}, Mark and {Nandra}, Kirpal and {Newman}, Jeffrey A. and {Nielsen}, Jennifer L. and {Niemi}, Sami and {Noeske}, Kai G. and {Papovich}, Casey J. and {Pentericci}, Laura and {Pope}, Alexandra and {Primack}, Joel R. and {Ravindranath}, Swara and {Reddy}, Naveen A. and {Renzini}, Alvio and {Rix}, Hans-Walter and {Robaina}, Aday R. and {Rosario}, David J. and {Rosati}, Piero and {Salimbeni}, Sara and {Scarlata}, Claudia and {Siana}, Brian and {Simard}, Luc and {Smidt}, Joseph and {Snyder}, Diana and {Somerville}, Rachel S. and {Spinrad}, Hyron and {Straughn}, Amber N. and {Telford}, Olivia and {Teplitz}, Harry I. and {Trump}, Jonathan R. and {Vargas}, Carlos and {Villforth}, Carolin and {Wagner}, Cory R. and {Wandro}, Pat and {Wechsler}, Risa H. and {Weiner}, Benjamin J. and {Wiklind}, Tommy and {Wild}, Vivienne and {Wilson}, Grant and {Wuyts}, Stijn and {Yun}, Min S.},
        title = "{CANDELS: The Cosmic Assembly Near-infrared Deep Extragalactic Legacy Survey{\textemdash}The Hubble Space Telescope Observations, Imaging Data Products, and Mosaics}",
      journal = {\apjs},
         year = 2011,
        month = dec,
       volume = {197},
       number = {2},
          eid = {36},
        pages = {36},
          doi = {10.1088/0067-0049/197/2/36},
archivePrefix = {arXiv},
       eprint = {1105.3754},
 primaryClass = {astro-ph.CO},
       adsurl = {https://ui.adsabs.harvard.edu/abs/2011ApJS..197...36K}
}

@ARTICLE{Gardner2006,
       author = {{Gardner}, Jonathan P. and {Mather}, John C. and {Clampin}, Mark and {Doyon}, Rene and {Greenhouse}, Matthew A. and {Hammel}, Heidi B. and {Hutchings}, John B. and {Jakobsen}, Peter and {Lilly}, Simon J. and {Long}, Knox S. and {Lunine}, Jonathan I. and {McCaughrean}, Mark J. and {Mountain}, Matt and {Nella}, John and {Rieke}, George H. and {Rieke}, Marcia J. and {Rix}, Hans-Walter and {Smith}, Eric P. and {Sonneborn}, George and {Stiavelli}, Massimo and {Stockman}, H.~S. and {Windhorst}, Rogier A. and {Wright}, Gillian S.},
        title = "{The James Webb Space Telescope}",
      journal = {\ssr},
         year = 2006,
        month = apr,
       volume = {123},
       number = {4},
        pages = {485-606},
          doi = {10.1007/s11214-006-8315-7},
archivePrefix = {arXiv},
       eprint = {astro-ph/0606175},
 primaryClass = {astro-ph},
       adsurl = {https://ui.adsabs.harvard.edu/abs/2006SSRv..123..485G}
}

@ARTICLE{Bundy2015,
       author = {{Bundy}, Kevin and {Bershady}, Matthew A. and {Law}, David R. and {Yan}, Renbin and {Drory}, Niv and {MacDonald}, Nicholas and {Wake}, David A. and {Cherinka}, Brian and {S{\'a}nchez-Gallego}, Jos{\'e} R. and {Weijmans}, Anne-Marie and {Thomas}, Daniel and {Tremonti}, Christy and {Masters}, Karen and {Coccato}, Lodovico and {Diamond-Stanic}, Aleksandar M. and {Arag{\'o}n-Salamanca}, Alfonso and {Avila-Reese}, Vladimir and {Badenes}, Carles and {Falc{\'o}n-Barroso}, J{\'e}sus and {Belfiore}, Francesco and {Bizyaev}, Dmitry and {Blanc}, Guillermo A. and {Bland-Hawthorn}, Joss and {Blanton}, Michael R. and {Brownstein}, Joel R. and {Byler}, Nell and {Cappellari}, Michele and {Conroy}, Charlie and {Dutton}, Aaron A. and {Emsellem}, Eric and {Etherington}, James and {Frinchaboy}, Peter M. and {Fu}, Hai and {Gunn}, James E. and {Harding}, Paul and {Johnston}, Evelyn J. and {Kauffmann}, Guinevere and {Kinemuchi}, Karen and {Klaene}, Mark A. and {Knapen}, Johan H. and {Leauthaud}, Alexie and {Li}, Cheng and {Lin}, Lihwai and {Maiolino}, Roberto and {Malanushenko}, Viktor and {Malanushenko}, Elena and {Mao}, Shude and {Maraston}, Claudia and {McDermid}, Richard M. and {Merrifield}, Michael R. and {Nichol}, Robert C. and {Oravetz}, Daniel and {Pan}, Kaike and {Parejko}, John K. and {Sanchez}, Sebastian F. and {Schlegel}, David and {Simmons}, Audrey and {Steele}, Oliver and {Steinmetz}, Matthias and {Thanjavur}, Karun and {Thompson}, Benjamin A. and {Tinker}, Jeremy L. and {van den Bosch}, Remco C.~E. and {Westfall}, Kyle B. and {Wilkinson}, David and {Wright}, Shelley and {Xiao}, Ting and {Zhang}, Kai},
        title = "{Overview of the SDSS-IV MaNGA Survey: Mapping nearby Galaxies at Apache Point Observatory}",
      journal = {\apj},
         year = 2015,
        month = jan,
       volume = {798},
       number = {1},
          eid = {7},
        pages = {7},
          doi = {10.1088/0004-637X/798/1/7},
archivePrefix = {arXiv},
       eprint = {1412.1482},
 primaryClass = {astro-ph.GA},
       adsurl = {https://ui.adsabs.harvard.edu/abs/2015ApJ...798....7B}
}

@ARTICLE{Sanchez2012,
       author = {{S{\'a}nchez}, S.~F. and {Kennicutt}, R.~C. and {Gil de Paz}, A. and {van de Ven}, G. and {V{\'\i}lchez}, J.~M. and {Wisotzki}, L. and {Walcher}, C.~J. and {Mast}, D. and {Aguerri}, J.~A.~L. and {Albiol-P{\'e}rez}, S. and {Alonso-Herrero}, A. and {Alves}, J. and {Bakos}, J. and {Bart{\'a}kov{\'a}}, T. and {Bland-Hawthorn}, J. and {Boselli}, A. and {Bomans}, D.~J. and {Castillo-Morales}, A. and {Cortijo-Ferrero}, C. and {de Lorenzo-C{\'a}ceres}, A. and {Del Olmo}, A. and {Dettmar}, R.-J. and {D{\'\i}az}, A. and {Ellis}, S. and {Falc{\'o}n-Barroso}, J. and {Flores}, H. and {Gallazzi}, A. and {Garc{\'\i}a-Lorenzo}, B. and {Gonz{\'a}lez Delgado}, R. and {Gruel}, N. and {Haines}, T. and {Hao}, C. and {Husemann}, B. and {Igl{\'e}sias-P{\'a}ramo}, J. and {Jahnke}, K. and {Johnson}, B. and {Jungwiert}, B. and {Kalinova}, V. and {Kehrig}, C. and {Kupko}, D. and {L{\'o}pez-S{\'a}nchez}, {\'A}. R. and {Lyubenova}, M. and {Marino}, R.~A. and {M{\'a}rmol-Queralt{\'o}}, E. and {M{\'a}rquez}, I. and {Masegosa}, J. and {Meidt}, S. and {Mendez-Abreu}, J. and {Monreal-Ibero}, A. and {Montijo}, C. and {Mour{\~a}o}, A.~M. and {Palacios-Navarro}, G. and {Papaderos}, P. and {Pasquali}, A. and {Peletier}, R. and {P{\'e}rez}, E. and {P{\'e}rez}, I. and {Quirrenbach}, A. and {Rela{\~n}o}, M. and {Rosales-Ortega}, F.~F. and {Roth}, M.~M. and {Ruiz-Lara}, T. and {S{\'a}nchez-Bl{\'a}zquez}, P. and {Sengupta}, C. and {Singh}, R. and {Stanishev}, V. and {Trager}, S.~C. and {Vazdekis}, A. and {Viironen}, K. and {Wild}, V. and {Zibetti}, S. and {Ziegler}, B.},
        title = "{CALIFA, the Calar Alto Legacy Integral Field Area survey. I. Survey presentation}",
      journal = {\aap},
         year = 2012,
        month = feb,
       volume = {538},
          eid = {A8},
        pages = {A8},
          doi = {10.1051/0004-6361/201117353},
archivePrefix = {arXiv},
       eprint = {1111.0962},
 primaryClass = {astro-ph.CO},
       adsurl = {https://ui.adsabs.harvard.edu/abs/2012A&A...538A...8S}
}

@ARTICLE{Bryant2015,
       author = {{Bryant}, J.~J. and {Owers}, M.~S. and {Robotham}, A.~S.~G. and {Croom}, S.~M. and {Driver}, S.~P. and {Drinkwater}, M.~J. and {Lorente}, N.~P.~F. and {Cortese} and L. and {Scott} and N. and {Colless} and M. and {Schaefer} and A. and {Taylor}, E.~N. and {Konstantopoulos}, I.~S. and {Allen}, J.~T. and {Baldry} and I. and {Barnes} and L. and {Bauer}, A.~E. and {Bland-Hawthorn} and J. and {Bloom}, J.~V. and {Brooks}, A.~M. and {Brough} and S. and {Cecil} and G. and {Couch} and W. and {Croton} and D. and {Davies} and R. and {Ellis} and S. and {Fogarty}, L.~M.~R. and {Foster} and C. and {Glazebrook} and K. and {Goodwin} and M. and {Green} and A. and {Gunawardhana}, M.~L. and {Hampton} and E. and {Ho} and I. -T. and {Hopkins}, A.~M. and {Kewley} and L. and {Lawrence}, J.~S. and {Leon-Saval}, S.~G. and {Leslie} and S. and {McElroy} and R. and {Lewis} and G. and {Liske} and J. and {L{\'o}pez-S{\'a}nchez}, {\'A}. R. and {Mahajan} and S. and {Medling}, A.~M. and {Metcalfe} and N. and {Meyer} and M. and {Mould} and J. and {Obreschkow} and D. and {O'Toole} and S. and {Pracy} and M. and {Richards}, S.~N. and {Shanks} and T. and {Sharp} and R. and {Sweet}, S.~M. and {Thomas}, A.~D. and {Tonini} and C. and {Walcher}, C.~J.},
        title = "{The SAMI Galaxy Survey: instrument specification and target selection}",
      journal = {\mnras},
         year = 2015,
        month = mar,
       volume = {447},
       number = {3},
        pages = {2857-2879},
          doi = {10.1093/mnras/stu2635},
archivePrefix = {arXiv},
       eprint = {1407.7335},
 primaryClass = {astro-ph.GA},
       adsurl = {https://ui.adsabs.harvard.edu/abs/2015MNRAS.447.2857B}
}

@ARTICLE{Maiolino2020,
       author = {{Maiolino} and R. and {Cirasuolo} and M. and {Afonso} and J. and {Bauer}, F.~E. and {Bowler} and R. and {Cucciati} and O. and {Daddi} and E. and {De Lucia} and G. and {Evans} and C. and {Flores} and H. and {Gargiulo} and A. and {Garilli} and B. and {Jablonka} and P. and {Jarvis} and M. and {Kneib} and J. -P. and {Lilly} and S. and {Looser} and T. and {Magliocchetti} and M. and {Man} and Z. and {Mannucci} and F. and {Maurogordato} and S. and {McLure}, R.~J. and {Norberg} and P. and {Oesch} and P. and {Oliva} and E. and {Paltani} and S. and {Pappalardo} and C. and {Peng} and Y. and {Pentericci} and L. and {Pozzetti} and L. and {Renzini} and A. and {Rodrigues} and M. and {Royer} and F. and {Serjeant} and S. and {Vanzi} and L. and {Wild} and V. and {Zamorani}, G.},
        title = "{MOONRISE: The Main MOONS GTO Extragalactic Survey}",
      journal = {The Messenger},
         year = 2020,
        month = jun,
       volume = {180},
        pages = {24-29},
          doi = {10.18727/0722-6691/5197},
archivePrefix = {arXiv},
       eprint = {2009.00644},
 primaryClass = {astro-ph.GA},
       adsurl = {https://ui.adsabs.harvard.edu/abs/2020Msngr.180...24M}
}

@ARTICLE{Cirasuolo2020,
       author = {{Cirasuolo} and M. and {Fairley} and A. and {Rees} and P. and {Gonzalez}, O.~A. and {Taylor} and W. and {Maiolino} and R. and {Afonso} and J. and {Evans} and C. and {Flores} and H. and {Lilly} and S. and {Oliva} and E. and {Paltani} and S. and {Vanzi} and L. and {Abreu} and M. and {Accardo} and M. and {Adams} and N. and {{\'A}lvarez M{\'e}ndez} and D. and {Amans} and J. -P. and {Amarantidis} and S. and {Atek} and H. and {Atkinson} and D. and {Banerji} and M. and {Barrett} and J. and {Barrientos} and F. and {Bauer} and F. and {Beard} and S. and {B{\'e}chet} and C. and {Belfiore} and A. and {Bellazzini} and M. and {Benoist} and C. and {Best} and P. and {Biazzo} and K. and {Black} and M. and {Boettger} and D. and {Bonifacio} and P. and {Bowler} and R. and {Bragaglia} and A. and {Brierley} and S. and {Brinchmann} and J. and {Brinkmann} and M. and {Buat} and V. and {Buitrago} and F. and {Burgarella} and D. and {Burningham} and B. and {Buscher} and D. and {Cabral} and A. and {Caffau} and E. and {Cardoso} and L. and {Carnall} and A. and {Carollo} and M. and {Castillo} and R. and {Castignani} and G. and {Catelan} and M. and {Cicone} and C. and {Cimatti} and A. and {Cioni} and M. -R.~L. and {Clementini} and G. and {Cochrane} and W. and {Coelho} and J. and {Colling} and M. and {Contini} and T. and {Contreras} and R. and {Conzelmann} and R. and {Cresci} and G. and {Cropper} and M. and {Cucciati} and O. and {Cullen} and F. and {Cumani} and C. and {Curti} and M. and {Da Silva} and A. and {Daddi} and E. and {Dalessandro} and E. and {Dalessio} and F. and {Dauvin} and L. and {Davidson} and G. and {de Laverny} and P. and {Delplancke-Str{\"o}bele} and F. and {De Lucia} and G. and {Del Vecchio} and C. and {Dessauges-Zavadsky} and M. and {Di Matteo} and P. and {Dole} and H. and {Drass} and H. and {Dunlop} and J. and {D{\"u}nner} and R. and {Eales} and S. and {Ellis} and R. and {Enriques} and B. and {Fasola} and G. and {Ferguson} and A. and {Ferruzzi} and D. and {Fisher} and M. and {Flores} and M. and {Fontana} and A. and {Forchi} and V. and {Francois} and P. and {Franzetti} and P. and {Gargiulo} and A. and {Garilli} and B. and {Gaudemard} and J. and {Gieles} and M. and {Gilmore} and G. and {Ginolfi} and M. and {Gomes}, J.~M. and {Guinouard} and I. and {Gutierrez} and P. and {Haigron} and R. and {Hammer} and F. and {Hammersley} and P. and {Haniff} and C. and {Harrison} and C. and {Haywood} and M. and {Hill} and V. and {Hubin} and N. and {Humphrey} and A. and {Ibata} and R. and {Infante} and L. and {Ives} and D. and {Ivison} and R. and {Iwert} and O. and {Jablonka} and P. and {Jakob} and G. and {Jarvis} and M. and {King} and D. and {Kneib} and J. -P. and {Laporte} and P. and {Lawrence} and A. and {Lee} and D. and {Li Causi} and G. and {Lorenzoni} and S. and {Lucatello} and S. and {Luco} and Y. and {Macleod} and A. and {Magliocchetti} and M. and {Magrini} and L. and {Mainieri} and V. and {Maire} and C. and {Mannucci} and F. and {Martin} and N. and {Matute} and I. and {Maurogordato} and S. and {McGee} and S. and {Mcleod} and D. and {McLure} and R. and {McMahon} and R. and {Melse} and B. -T. and {Messias} and H. and {Mucciarelli} and A. and {Nisini} and B. and {Nix} and J. and {Norberg} and P. and {Oesch} and P. and {Oliveira} and A. and {Origlia} and L. and {Padilla} and N. and {Palsa} and R. and {Pancino} and E. and {Papaderos} and P. and {Pappalardo} and C. and {Parry} and I. and {Pasquini} and L. and {Peacock} and J. and {Pedichini} and F. and {Pello} and R. and {Peng} and Y. and {Pentericci} and L. and {Pfuhl} and O. and {Piazzesi} and R. and {Popovic} and D. and {Pozzetti} and L. and {Puech} and M. and {Puzia} and T. and {Raichoor} and A. and {Randich} and S. and {Recio-Blanco} and A. and {Reis} and S. and {Reix} and F. and {Renzini} and A. and {Rodrigues} and M. and {Rojas} and F. and {Rojas-Arriagada}, {\'A}. and {Rota} and S. and {Royer} and F. and {Sacco} and G. and {Sanchez-Janssen} and R. and {Sanna} and N. and {Santos} and P. and {Sarzi} and M. and {Schaerer} and D. and {Schiavon} and R. and {Schnell} and R. and {Schultheis} and M. and {Scodeggio} and M. and {Serjeant} and S. and {Shen} and T. -C. and {Simmonds} and C. and {Smoker} and J. and {Sobral} and D. and {Sordet} and M. and {Sp{\'e}rone}, D.},
        title = "{MOONS: The New Multi-Object Spectrograph for the VLT}",
      journal = {The Messenger},
         year = 2020,
        month = jun,
       volume = {180},
        pages = {10-17},
          doi = {10.18727/0722-6691/5195},
archivePrefix = {arXiv},
       eprint = {2009.00628},
 primaryClass = {astro-ph.IM},
       adsurl = {https://ui.adsabs.harvard.edu/abs/2020Msngr.180...10C}
}

@ARTICLE{Thanjavur2016,
       author = {{Thanjavur}, Karun and {Simard}, Luc and {Bluck}, Asa F.~L. and {Mendel}, Trevor},
        title = "{Stellar mass functions of galaxies, discs and spheroids at $z \sim 0.1$}",
      journal = {\mnras},
         year = 2016,
        month = jun,
       volume = {459},
       number = {1},
        pages = {44-69},
          doi = {10.1093/mnras/stw495},
archivePrefix = {arXiv},
       eprint = {1602.08674},
 primaryClass = {astro-ph.GA},
       adsurl = {https://ui.adsabs.harvard.edu/abs/2016MNRAS.459...44T}
}

@ARTICLE{Wuyts2011,
       author = {{Wuyts}, Stijn and {F{\"o}rster Schreiber}, Natascha M. and {van der Wel}, Arjen and {Magnelli}, Benjamin and {Guo}, Yicheng and {Genzel}, Reinhard and {Lutz}, Dieter and {Aussel}, Herv{\'e} and {Barro}, Guillermo and {Berta}, Stefano and {Cava}, Antonio and {Graci{\'a}-Carpio}, Javier and {Hathi}, Nimish P. and {Huang}, Kuang-Han and {Kocevski}, Dale D. and {Koekemoer}, Anton M. and {Lee}, Kyoung-Soo and {Le Floc'h}, Emeric and {McGrath}, Elizabeth J. and {Nordon}, Raanan and {Popesso}, Paola and {Pozzi}, Francesca and {Riguccini}, Laurie and {Rodighiero}, Giulia and {Saintonge}, Amelie and {Tacconi}, Linda},
        title = "{Galaxy Structure and Mode of Star Formation in the SFR-Mass Plane from $z \sim 2.5$ to $z \sim 0.1$}",
      journal = {\apj},
         year = 2011,
        month = dec,
       volume = {742},
       number = {2},
          eid = {96},
        pages = {96},
          doi = {10.1088/0004-637X/742/2/96},
archivePrefix = {arXiv},
       eprint = {1107.0317},
 primaryClass = {astro-ph.CO},
       adsurl = {https://ui.adsabs.harvard.edu/abs/2011ApJ...742...96W}
}

@ARTICLE{Mandelbaum2006,
       author = {{Mandelbaum}, Rachel and {Seljak}, Uro{\v{s}} and {Kauffmann}, Guinevere and {Hirata}, Christopher M. and {Brinkmann}, Jonathan},
        title = "{Galaxy halo masses and satellite fractions from galaxy-galaxy lensing in the Sloan Digital Sky Survey: stellar mass, luminosity, morphology and environment dependencies}",
      journal = {\mnras},
         year = 2006,
        month = may,
       volume = {368},
       number = {2},
        pages = {715-731},
          doi = {10.1111/j.1365-2966.2006.10156.x},
archivePrefix = {arXiv},
       eprint = {astro-ph/0511164},
 primaryClass = {astro-ph},
       adsurl = {https://ui.adsabs.harvard.edu/abs/2006MNRAS.368..715M}
}

@ARTICLE{Behroozi2013,
       author = {{Behroozi}, Peter S. and {Wechsler}, Risa H. and {Conroy}, Charlie},
        title = "{The Average Star Formation Histories of Galaxies in Dark Matter Halos from z = 0-8}",
      journal = {\apj},
         year = 2013,
        month = jun,
       volume = {770},
       number = {1},
          eid = {57},
        pages = {57},
          doi = {10.1088/0004-637X/770/1/57},
archivePrefix = {arXiv},
       eprint = {1207.6105},
 primaryClass = {astro-ph.CO},
       adsurl = {https://ui.adsabs.harvard.edu/abs/2013ApJ...770...57B}
}

@ARTICLE{Tinker2008,
       author = {{Tinker}, Jeremy and {Kravtsov}, Andrey V. and {Klypin}, Anatoly and {Abazajian}, Kevork and {Warren}, Michael and {Yepes}, Gustavo and {Gottl{\"o}ber}, Stefan and {Holz}, Daniel E.},
        title = "{Toward a Halo Mass Function for Precision Cosmology: The Limits of Universality}",
      journal = {\apj},
         year = 2008,
        month = dec,
       volume = {688},
       number = {2},
        pages = {709-728},
          doi = {10.1086/591439},
archivePrefix = {arXiv},
       eprint = {0803.2706},
 primaryClass = {astro-ph},
       adsurl = {https://ui.adsabs.harvard.edu/abs/2008ApJ...688..709T}
}

@ARTICLE{Moster2018,
       author = {{Moster}, Benjamin P. and {Naab}, Thorsten and {White}, Simon D.~M.},
        title = "{EMERGE - an empirical model for the formation of galaxies since $z \sim 10$}",
      journal = {\mnras},
         year = 2018,
        month = jun,
       volume = {477},
       number = {2},
        pages = {1822-1852},
          doi = {10.1093/mnras/sty655},
archivePrefix = {arXiv},
       eprint = {1705.05373},
 primaryClass = {astro-ph.GA},
       adsurl = {https://ui.adsabs.harvard.edu/abs/2018MNRAS.477.1822M}
}

@ARTICLE{Sheth1999,
       author = {{Sheth}, Ravi K. and {Tormen}, Giuseppe},
        title = "{Large-scale bias and the peak background split}",
      journal = {\mnras},
         year = 1999,
        month = sep,
       volume = {308},
       number = {1},
        pages = {119-126},
          doi = {10.1046/j.1365-8711.1999.02692.x},
archivePrefix = {arXiv},
       eprint = {astro-ph/9901122},
 primaryClass = {astro-ph},
       adsurl = {https://ui.adsabs.harvard.edu/abs/1999MNRAS.308..119S}
}

@ARTICLE{Fakhouri2010,
       author = {{Fakhouri}, Onsi and {Ma}, Chung-Pei and {Boylan-Kolchin}, Michael},
        title = "{The merger rates and mass assembly histories of dark matter haloes in the two Millennium simulations}",
      journal = {\mnras},
         year = 2010,
        month = aug,
       volume = {406},
       number = {4},
        pages = {2267-2278},
          doi = {10.1111/j.1365-2966.2010.16859.x},
archivePrefix = {arXiv},
       eprint = {1001.2304},
 primaryClass = {astro-ph.CO},
       adsurl = {https://ui.adsabs.harvard.edu/abs/2010MNRAS.406.2267F}
}

@ARTICLE{Press1974,
       author = {{Press}, William H. and {Schechter}, Paul},
        title = "{Formation of Galaxies and Clusters of Galaxies by Self-Similar Gravitational Condensation}",
      journal = {\apj},
         year = 1974,
        month = feb,
       volume = {187},
        pages = {425-438},
          doi = {10.1086/152650},
       adsurl = {https://ui.adsabs.harvard.edu/abs/1974ApJ...187..425P}
}

@ARTICLE{Bond1991,
       author = {{Bond}, J.~R. and {Cole} and S. and {Efstathiou} and G. and {Kaiser}, N.},
        title = "{Excursion Set Mass Functions for Hierarchical Gaussian Fluctuations}",
      journal = {\apj},
         year = 1991,
        month = oct,
       volume = {379},
        pages = {440},
          doi = {10.1086/170520},
       adsurl = {https://ui.adsabs.harvard.edu/abs/1991ApJ...379..440B}
}

@ARTICLE{Butcher1978,
       author = {{Butcher} and H. and {Oemler}, Jr., A.},
        title = "{The evolution of galaxies in clusters. I. ISIT photometry of Cl 0024+1654 and 3C 295.}",
      journal = {\apj},
         year = 1978,
        month = jan,
       volume = {219},
        pages = {18-30},
          doi = {10.1086/155751},
       adsurl = {https://ui.adsabs.harvard.edu/abs/1978ApJ...219...18B}
}

@ARTICLE{Visvanathan1977,
       author = {{Visvanathan} and N. and {Sandage}, A.},
        title = "{The color - absolute magnitude relation for E and S0 galaxies. I. Calibration and tests for universality using Virgo and eight other nearby clusters.}",
      journal = {\apj},
         year = 1977,
        month = aug,
       volume = {216},
        pages = {214-226},
          doi = {10.1086/155464},
       adsurl = {https://ui.adsabs.harvard.edu/abs/1977ApJ...216..214V}
}

@ARTICLE{deVaucouleurs1948,
       author = {{de Vaucouleurs}, Gerard},
        title = "{Recherches sur les Nebuleuses Extragalactiques}",
      journal = {Annales d'Astrophysique},
         year = 1948,
        month = jan,
       volume = {11},
        pages = {247},
       adsurl = {https://ui.adsabs.harvard.edu/abs/1948AnAp...11..247D}
}

@ARTICLE{Peng2002,
       author = {{Peng}, Chien Y. and {Ho}, Luis C. and {Impey}, Chris D. and {Rix}, Hans-Walter},
        title = "{Detailed Structural Decomposition of Galaxy Images}",
      journal = {\aj},
         year = 2002,
        month = jul,
       volume = {124},
       number = {1},
        pages = {266-293},
          doi = {10.1086/340952},
archivePrefix = {arXiv},
       eprint = {astro-ph/0204182},
 primaryClass = {astro-ph},
       adsurl = {https://ui.adsabs.harvard.edu/abs/2002AJ....124..266P}
}

@ARTICLE{Sersic1963,
       author = {{S{\'e}rsic}, J.~L.},
        title = "{Influence of the atmospheric and instrumental dispersion on the brightness distribution in a galaxy}",
      journal = {Boletin Asoc. Argentina Astron.},
         year = 1963,
        month = feb,
       volume = {6},
        pages = {41-43},
       adsurl = {https://ui.adsabs.harvard.edu/abs/1963BAAA....6...41S}
}

@ARTICLE{Simard2011,
       author = {{Simard}, Luc and {Mendel} and J. Trevor and {Patton}, David R. and {Ellison}, Sara L. and {McConnachie}, Alan W.},
        title = "{A Catalog of Bulge+disk Decompositions and Updated Photometry for 1.12 Million Galaxies in the Sloan Digital Sky Survey}",
      journal = {\apjs},
         year = 2011,
        month = sep,
       volume = {196},
       number = {1},
          eid = {11},
        pages = {11},
          doi = {10.1088/0067-0049/196/1/11},
archivePrefix = {arXiv},
       eprint = {1107.1518},
 primaryClass = {astro-ph.CO},
       adsurl = {https://ui.adsabs.harvard.edu/abs/2011ApJS..196...11S}
}

@ARTICLE{Lintott2008,
       author = {{Lintott}, Chris J. and {Schawinski}, Kevin and {Slosar}, An{\v{z}}e and {Land}, Kate and {Bamford}, Steven and {Thomas}, Daniel and {Raddick} and M. Jordan and {Nichol}, Robert C. and {Szalay}, Alex and {Andreescu}, Dan and {Murray}, Phil and {Vandenberg}, Jan},
        title = "{Galaxy Zoo: morphologies derived from visual inspection of galaxies from the Sloan Digital Sky Survey}",
      journal = {\mnras},
         year = 2008,
        month = sep,
       volume = {389},
       number = {3},
        pages = {1179-1189},
          doi = {10.1111/j.1365-2966.2008.13689.x},
archivePrefix = {arXiv},
       eprint = {0804.4483},
 primaryClass = {astro-ph},
       adsurl = {https://ui.adsabs.harvard.edu/abs/2008MNRAS.389.1179L}
}

@ARTICLE{Lintott2011,
       author = {{Lintott}, Chris and {Schawinski}, Kevin and {Bamford}, Steven and {Slosar}, An{\r{a}}{\textthreequarters}e and {Land}, Kate and {Thomas}, Daniel and {Edmondson}, Edd and {Masters}, Karen and {Nichol}, Robert C. and {Raddick} and M. Jordan and {Szalay}, Alex and {Andreescu}, Dan and {Murray}, Phil and {Vandenberg}, Jan},
        title = "{Galaxy Zoo 1: data release of morphological classifications for nearly 900 000 galaxies}",
      journal = {\mnras},
         year = 2011,
        month = jan,
       volume = {410},
       number = {1},
        pages = {166-178},
          doi = {10.1111/j.1365-2966.2010.17432.x},
archivePrefix = {arXiv},
       eprint = {1007.3265},
 primaryClass = {astro-ph.GA},
       adsurl = {https://ui.adsabs.harvard.edu/abs/2011MNRAS.410..166L}
}

@ARTICLE{Papovich2012,
       author = {{Papovich} and C. and {Bassett} and R. and {Lotz}, J.~M. and {van der Wel} and A. and {Tran} and K. -V. and {Finkelstein}, S.~L. and {Bell}, E.~F. and {Conselice}, C.~J. and {Dekel} and A. and {Dunlop}, J.~S. and {Guo}, Yicheng and {Faber}, S.~M. and {Farrah} and D. and {Ferguson}, H.~C. and {Finkelstein}, K.~D. and {H{\"a}ussler} and B. and {Kocevski}, D.~D. and {Koekemoer}, A.~M. and {Koo}, D.~C. and {McGrath}, E.~J. and {McLure}, R.~J. and {McIntosh}, D.~H. and {Momcheva} and I. and {Newman}, J.~A. and {Rudnick} and G. and {Weiner} and B. and {Willmer}, C.~N.~A. and {Wuyts}, S.},
        title = "{CANDELS Observations of the Structural Properties of Cluster Galaxies at z = 1.62}",
      journal = {\apj},
         year = 2012,
        month = may,
       volume = {750},
       number = {2},
          eid = {93},
        pages = {93},
          doi = {10.1088/0004-637X/750/2/93},
archivePrefix = {arXiv},
       eprint = {1110.3794},
 primaryClass = {astro-ph.CO},
       adsurl = {https://ui.adsabs.harvard.edu/abs/2012ApJ...750...93P}
}

@ARTICLE{Szomoru2011,
       author = {{Szomoru}, Daniel and {Franx}, Marijn and {Bouwens}, Rychard J. and {van Dokkum}, Pieter G. and {Labb{\'e}}, Ivo and {Illingworth}, Garth D. and {Trenti}, Michele},
        title = "{Morphological Evolution of Galaxies from Ultra-deep Hubble Space Telescope Wide Field Camera 3 Imaging: The Hubble Sequence at $z \sim 2$}",
      journal = {\apjl},
         year = 2011,
        month = jul,
       volume = {735},
       number = {1},
          eid = {L22},
        pages = {L22},
          doi = {10.1088/2041-8205/735/1/L22},
archivePrefix = {arXiv},
       eprint = {1106.1641},
 primaryClass = {astro-ph.CO},
       adsurl = {https://ui.adsabs.harvard.edu/abs/2011ApJ...735L..22S}
}

@ARTICLE{Mortlock2013,
       author = {{Mortlock}, Alice and {Conselice}, Christopher J. and {Hartley}, William G. and {Ownsworth}, Jamie R. and {Lani}, Caterina and {Bluck}, Asa F.~L. and {Almaini}, Omar and {Duncan}, Kenneth and {van der Wel}, Arjen and {Koekemoer}, Anton M. and {Dekel}, Avishai and {Dav{\'e}}, Romeel and {Ferguson}, Harry C. and {de Mello}, Duilia F. and {Newman}, Jeffrey A. and {Faber}, Sandra M. and {Grogin}, Norman A. and {Kocevski}, Dale D. and {Lai}, Kamson},
        title = "{The redshift and mass dependence on the formation of the Hubble sequence at $z > 1$ from CANDELS/UDS}",
      journal = {\mnras},
         year = 2013,
        month = aug,
       volume = {433},
       number = {2},
        pages = {1185-1201},
          doi = {10.1093/mnras/stt793},
archivePrefix = {arXiv},
       eprint = {1305.2204},
 primaryClass = {astro-ph.CO},
       adsurl = {https://ui.adsabs.harvard.edu/abs/2013MNRAS.433.1185M}
}

@ARTICLE{Swinbank2017,
       author = {{Swinbank}, A.~M. and {Harrison}, C.~M. and {Trayford} and J. and {Schaller} and M. and {Smail}, Ian and {Schaye} and J. and {Theuns} and T. and {Smit} and R. and {Alexander}, D.~M. and {Bacon} and R. and {Bower}, R.~G. and {Contini} and T. and {Crain}, R.~A. and {de Breuck} and C. and {Decarli} and R. and {Epinat} and B. and {Fumagalli} and M. and {Furlong} and M. and {Galametz} and A. and {Johnson}, H.~L. and {Lagos} and C. and {Richard} and J. and {Vernet} and J. and {Sharples}, R.~M. and {Sobral} and D. and {Stott}, J.~P.},
        title = "{Angular momentum evolution of galaxies over the past 10 Gyr: a MUSE and KMOS dynamical survey of 400 star-forming galaxies from z = 0.3 to 1.7}",
      journal = {\mnras},
         year = 2017,
        month = may,
       volume = {467},
       number = {3},
        pages = {3140-3159},
          doi = {10.1093/mnras/stx201},
archivePrefix = {arXiv},
       eprint = {1701.07448},
 primaryClass = {astro-ph.GA},
       adsurl = {https://ui.adsabs.harvard.edu/abs/2017MNRAS.467.3140S}
}

@ARTICLE{Peng2020,
       author = {{Peng}, Ying-jie and {Renzini}, Alvio},
        title = "{Disc growth and quenching}",
      journal = {\mnras},
         year = 2020,
        month = jan,
       volume = {491},
       number = {1},
        pages = {L51-L55},
          doi = {10.1093/mnrasl/slz163},
archivePrefix = {arXiv},
       eprint = {1910.10446},
 primaryClass = {astro-ph.GA},
       adsurl = {https://ui.adsabs.harvard.edu/abs/2020MNRAS.491L..51P}
}

@ARTICLE{Barro2017,
       author = {{Barro}, Guillermo and {Faber}, S.~M. and {Koo}, David C. and {Dekel}, Avishai and {Fang}, Jerome J. and {Trump}, Jonathan R. and {P{\'e}rez-Gonz{\'a}lez}, Pablo G. and {Pacifici}, Camilla and {Primack}, Joel R. and {Somerville}, Rachel S. and {Yan}, Haojing and {Guo}, Yicheng and {Liu}, Fengshan and {Ceverino}, Daniel and {Kocevski}, Dale D. and {McGrath}, Elizabeth},
        title = "{Structural and Star-forming Relations since $z \sim 3$: Connecting Compact Star-forming and Quiescent Galaxies}",
      journal = {\apj},
         year = 2017,
        month = may,
       volume = {840},
       number = {1},
          eid = {47},
        pages = {47},
          doi = {10.3847/1538-4357/aa6b05},
archivePrefix = {arXiv},
       eprint = {1509.00469},
 primaryClass = {astro-ph.GA},
       adsurl = {https://ui.adsabs.harvard.edu/abs/2017ApJ...840...47B}
}

@ARTICLE{Bluck2019,
       author = {{Bluck}, Asa F.~L. and {Bottrell}, Connor and {Teimoorinia}, Hossen and {Henriques}, Bruno M.~B. and {Mendel} and J. Trevor and {Ellison}, Sara L. and {Thanjavur}, Karun and {Simard}, Luc and {Patton}, David R. and {Conselice}, Christopher J. and {Moreno}, Jorge and {Woo}, Joanna},
        title = "{What shapes a galaxy? - unraveling the role of mass, environment, and star formation in forming galactic structure}",
      journal = {\mnras},
         year = 2019,
        month = may,
       volume = {485},
       number = {1},
        pages = {666-696},
          doi = {10.1093/mnras/stz363},
archivePrefix = {arXiv},
       eprint = {1902.01665},
 primaryClass = {astro-ph.GA},
       adsurl = {https://ui.adsabs.harvard.edu/abs/2019MNRAS.485..666B}
}

@ARTICLE{Dimauro2018,
       author = {{Dimauro}, Paola and {Huertas-Company}, Marc and {Daddi}, Emanuele and {P{\'e}rez-Gonz{\'a}lez}, Pablo G. and {Bernardi}, Mariangela and {Barro}, Guillermo and {Buitrago}, Fernando and {Caro}, Fernando and {Cattaneo}, Andrea and {Dominguez-S{\'a}nchez}, Helena and {Faber}, Sandra M. and {H{\"a}u{\ss}ler}, Boris and {Kocevski}, Dale D. and {Koekemoer}, Anton M. and {Koo}, David C. and {Lee}, Christoph T. and {Mei}, Simona and {Margalef-Bentabol}, Berta and {Primack}, Joel and {Rodriguez-Puebla}, Aldo and {Salvato}, Mara and {Shankar}, Francesco and {Tuccillo}, Diego},
        title = "{A catalog of polychromatic bulge-disc decompositions of {\ensuremath{\sim}}17.600 galaxies in CANDELS}",
      journal = {\mnras},
         year = 2018,
        month = aug,
       volume = {478},
       number = {4},
        pages = {5410-5426},
          doi = {10.1093/mnras/sty1379},
archivePrefix = {arXiv},
       eprint = {1803.10234},
 primaryClass = {astro-ph.GA},
       adsurl = {https://ui.adsabs.harvard.edu/abs/2018MNRAS.478.5410D}
}

@ARTICLE{Mendel2014,
       author = {{Mendel} and J. Trevor and {Simard}, Luc and {Palmer}, Michael and {Ellison}, Sara L. and {Patton}, David R.},
        title = "{A Catalog of Bulge, Disk, and Total Stellar Mass Estimates for the Sloan Digital Sky Survey}",
      journal = {\apjs},
         year = 2014,
        month = jan,
       volume = {210},
       number = {1},
          eid = {3},
        pages = {3},
          doi = {10.1088/0067-0049/210/1/3},
archivePrefix = {arXiv},
       eprint = {1310.8304},
 primaryClass = {astro-ph.CO},
       adsurl = {https://ui.adsabs.harvard.edu/abs/2014ApJS..210....3M}
}

@ARTICLE{Saglia2016,
       author = {{Saglia}, R.~P. and {Opitsch} and M. and {Erwin} and P. and {Thomas} and J. and {Beifiori} and A. and {Fabricius} and M. and {Mazzalay} and X. and {Nowak} and N. and {Rusli}, S.~P. and {Bender}, R.},
        title = "{The SINFONI Black Hole Survey: The Black Hole Fundamental Plane Revisited and the Paths of (Co)evolution of Supermassive Black Holes and Bulges}",
      journal = {\apj},
         year = 2016,
        month = feb,
       volume = {818},
       number = {1},
          eid = {47},
        pages = {47},
          doi = {10.3847/0004-637X/818/1/47},
archivePrefix = {arXiv},
       eprint = {1601.00974},
 primaryClass = {astro-ph.GA},
       adsurl = {https://ui.adsabs.harvard.edu/abs/2016ApJ...818...47S}
}

@ARTICLE{Ferrarese2000,
       author = {{Ferrarese}, Laura and {Merritt}, David},
        title = "{A Fundamental Relation between Supermassive Black Holes and Their Host Galaxies}",
      journal = {\apjl},
         year = 2000,
        month = aug,
       volume = {539},
       number = {1},
        pages = {L9-L12},
          doi = {10.1086/312838},
archivePrefix = {arXiv},
       eprint = {astro-ph/0006053},
 primaryClass = {astro-ph},
       adsurl = {https://ui.adsabs.harvard.edu/abs/2000ApJ...539L...9F}
}

@ARTICLE{Cappellari2013a,
       author = {{Cappellari}, Michele and {Scott}, Nicholas and {Alatalo}, Katherine and {Blitz}, Leo and {Bois}, Maxime and {Bournaud}, Fr{\'e}d{\'e}ric and {Bureau} and M. and {Crocker}, Alison F. and {Davies}, Roger L. and {Davis}, Timothy A. and {de Zeeuw}, P.~T. and {Duc}, Pierre-Alain and {Emsellem}, Eric and {Khochfar}, Sadegh and {Krajnovi{\'c}}, Davor and {Kuntschner}, Harald and {McDermid}, Richard M. and {Morganti}, Raffaella and {Naab}, Thorsten and {Oosterloo}, Tom and {Sarzi}, Marc and {Serra}, Paolo and {Weijmans}, Anne-Marie and {Young}, Lisa M.},
        title = "{The ATLAS$^{3D}$ project - XV. Benchmark for early-type galaxies scaling relations from 260 dynamical models: mass-to-light ratio, dark matter, Fundamental Plane and Mass Plane}",
      journal = {\mnras},
         year = 2013,
        month = jul,
       volume = {432},
       number = {3},
        pages = {1709-1741},
          doi = {10.1093/mnras/stt562},
archivePrefix = {arXiv},
       eprint = {1208.3522},
 primaryClass = {astro-ph.CO},
       adsurl = {https://ui.adsabs.harvard.edu/abs/2013MNRAS.432.1709C}
}

@ARTICLE{Cappellari2013b,
       author = {{Cappellari}, Michele and {McDermid}, Richard M. and {Alatalo}, Katherine and {Blitz}, Leo and {Bois}, Maxime and {Bournaud}, Fr{\'e}d{\'e}ric and {Bureau} and M. and {Crocker}, Alison F. and {Davies}, Roger L. and {Davis}, Timothy A. and {de Zeeuw}, P.~T. and {Duc}, Pierre-Alain and {Emsellem}, Eric and {Khochfar}, Sadegh and {Krajnovi{\'c}}, Davor and {Kuntschner}, Harald and {Morganti}, Raffaella and {Naab}, Thorsten and {Oosterloo}, Tom and {Sarzi}, Marc and {Scott}, Nicholas and {Serra}, Paolo and {Weijmans}, Anne-Marie and {Young}, Lisa M.},
        title = "{The ATLAS$^{3D}$ project - XX. Mass-size and mass-{\ensuremath{\sigma}} distributions of early-type galaxies: bulge fraction drives kinematics, mass-to-light ratio, molecular gas fraction and stellar initial mass function}",
      journal = {\mnras},
         year = 2013,
        month = jul,
       volume = {432},
       number = {3},
        pages = {1862-1893},
          doi = {10.1093/mnras/stt644},
archivePrefix = {arXiv},
       eprint = {1208.3523},
 primaryClass = {astro-ph.CO},
       adsurl = {https://ui.adsabs.harvard.edu/abs/2013MNRAS.432.1862C}
}

@ARTICLE{Forster2009,
       author = {{F{\"o}rster Schreiber}, N.~M. and {Genzel} and R. and {Bouch{\'e}} and N. and {Cresci} and G. and {Davies} and R. and {Buschkamp} and P. and {Shapiro} and K. and {Tacconi}, L.~J. and {Hicks}, E.~K.~S. and {Genel} and S. and {Shapley}, A.~E. and {Erb}, D.~K. and {Steidel}, C.~C. and {Lutz} and D. and {Eisenhauer} and F. and {Gillessen} and S. and {Sternberg} and A. and {Renzini} and A. and {Cimatti} and A. and {Daddi} and E. and {Kurk} and J. and {Lilly} and S. and {Kong} and X. and {Lehnert}, M.~D. and {Nesvadba} and N. and {Verma} and A. and {McCracken} and H. and {Arimoto} and N. and {Mignoli} and M. and {Onodera}, M.},
        title = "{The SINS Survey: SINFONI Integral Field Spectroscopy of $z \sim 2$ Star-forming Galaxies}",
      journal = {\apj},
         year = 2009,
        month = dec,
       volume = {706},
       number = {2},
        pages = {1364-1428},
          doi = {10.1088/0004-637X/706/2/1364},
archivePrefix = {arXiv},
       eprint = {0903.1872},
 primaryClass = {astro-ph.CO},
       adsurl = {https://ui.adsabs.harvard.edu/abs/2009ApJ...706.1364F}
}

@ARTICLE{Forster2006,
       author = {{F{\"o}rster Schreiber}, N.~M. and {Genzel} and R. and {Lehnert}, M.~D. and {Bouch{\'e}} and N. and {Verma} and A. and {Erb}, D.~K. and {Shapley}, A.~E. and {Steidel}, C.~C. and {Davies} and R. and {Lutz} and D. and {Nesvadba} and N. and {Tacconi}, L.~J. and {Eisenhauer} and F. and {Abuter} and R. and {Gilbert} and A. and {Gillessen} and S. and {Sternberg}, A.},
        title = "{SINFONI Integral Field Spectroscopy of $z \sim 2$ UV-selected Galaxies: Rotation Curves and Dynamical Evolution}",
      journal = {\apj},
         year = 2006,
        month = jul,
       volume = {645},
       number = {2},
        pages = {1062-1075},
          doi = {10.1086/504403},
archivePrefix = {arXiv},
       eprint = {astro-ph/0603559},
 primaryClass = {astro-ph},
       adsurl = {https://ui.adsabs.harvard.edu/abs/2006ApJ...645.1062F}
}

@ARTICLE{Forster2020,
       author = {{F{\"o}rster Schreiber}, Natascha M. and {Wuyts}, Stijn},
        title = "{Star-Forming Galaxies at Cosmic Noon}",
      journal = {\araa},
         year = 2020,
        month = aug,
       volume = {58},
        pages = {661-725},
          doi = {10.1146/annurev-astro-032620-021910},
archivePrefix = {arXiv},
       eprint = {2010.10171},
 primaryClass = {astro-ph.GA},
       adsurl = {https://ui.adsabs.harvard.edu/abs/2020ARA&A..58..661F}
}

@ARTICLE{Lilly2016,
       author = {{Lilly}, Simon J. and {Carollo} and C. Marcella},
        title = "{Surface Density Effects in Quenching: Cause or Effect?}",
      journal = {\apj},
         year = 2016,
        month = dec,
       volume = {833},
       number = {1},
          eid = {1},
        pages = {1},
          doi = {10.3847/0004-637X/833/1/1},
archivePrefix = {arXiv},
       eprint = {1604.06459},
 primaryClass = {astro-ph.GA},
       adsurl = {https://ui.adsabs.harvard.edu/abs/2016ApJ...833....1L}
}

@ARTICLE{Trujillo2007,
       author = {{Trujillo}, Ignacio and {Conselice}, C.~J. and {Bundy}, Kevin and {Cooper}, M.~C. and {Eisenhardt} and P. and {Ellis}, Richard S.},
        title = "{Strong size evolution of the most massive galaxies since $z \sim 2$}",
      journal = {\mnras},
         year = 2007,
        month = nov,
       volume = {382},
       number = {1},
        pages = {109-120},
          doi = {10.1111/j.1365-2966.2007.12388.x},
archivePrefix = {arXiv},
       eprint = {0709.0621},
 primaryClass = {astro-ph},
       adsurl = {https://ui.adsabs.harvard.edu/abs/2007MNRAS.382..109T}
}

@ARTICLE{Buitrago2008,
       author = {{Buitrago}, Fernando and {Trujillo}, Ignacio and {Conselice}, Christopher J. and {Bouwens}, Rychard J. and {Dickinson}, Mark and {Yan}, Haojing},
        title = "{Size Evolution of the Most Massive Galaxies at $1.7 < z < 3$ from GOODS NICMOS Survey Imaging}",
      journal = {\apjl},
         year = 2008,
        month = nov,
       volume = {687},
       number = {2},
        pages = {L61},
          doi = {10.1086/592836},
archivePrefix = {arXiv},
       eprint = {0807.4141},
 primaryClass = {astro-ph},
       adsurl = {https://ui.adsabs.harvard.edu/abs/2008ApJ...687L..61B}
}

@ARTICLE{vanderWel2014,
       author = {{van der Wel} and A. and {Franx} and M. and {van Dokkum}, P.~G. and {Skelton}, R.~E. and {Momcheva}, I.~G. and {Whitaker}, K.~E. and {Brammer}, G.~B. and {Bell}, E.~F. and {Rix} and H. -W. and {Wuyts} and S. and {Ferguson}, H.~C. and {Holden}, B.~P. and {Barro} and G. and {Koekemoer}, A.~M. and {Chang}, Yu-Yen and {McGrath}, E.~J. and {H{\"a}ussler} and B. and {Dekel} and A. and {Behroozi} and P. and {Fumagalli} and M. and {Leja} and J. and {Lundgren}, B.~F. and {Maseda}, M.~V. and {Nelson}, E.~J. and {Wake}, D.~A. and {Patel}, S.~G. and {Labb{\'e}} and I. and {Faber}, S.~M. and {Grogin}, N.~A. and {Kocevski}, D.~D.},
        title = "{3D-HST+CANDELS: The Evolution of the Galaxy Size-Mass Distribution since z = 3}",
      journal = {\apj},
         year = 2014,
        month = jun,
       volume = {788},
       number = {1},
          eid = {28},
        pages = {28},
          doi = {10.1088/0004-637X/788/1/28},
archivePrefix = {arXiv},
       eprint = {1404.2844},
 primaryClass = {astro-ph.GA},
       adsurl = {https://ui.adsabs.harvard.edu/abs/2014ApJ...788...28V}
}

@ARTICLE{Goubert2024,
       author = {{Goubert}, Paul H. and {Bluck}, Asa F.~L. and {Piotrowska}, Joanna M. and {Maiolino}, Roberto},
        title = "{The role of environment and AGN feedback in quenching local galaxies: comparing cosmological hydrodynamical simulations to the SDSS}",
      journal = {\mnras},
         year = 2024,
        month = mar,
       volume = {528},
       number = {3},
        pages = {4891-4921},
          doi = {10.1093/mnras/stae269},
archivePrefix = {arXiv},
       eprint = {2401.12953},
 primaryClass = {astro-ph.GA},
       adsurl = {https://ui.adsabs.harvard.edu/abs/2024MNRAS.528.4891G}
}

@ARTICLE{Newman2012,
       author = {{Newman}, Andrew B. and {Ellis}, Richard S. and {Bundy}, Kevin and {Treu}, Tommaso},
        title = "{Can Minor Merging Account for the Size Growth of Quiescent Galaxies? New Results from the CANDELS Survey}",
      journal = {\apj},
         year = 2012,
        month = feb,
       volume = {746},
       number = {2},
          eid = {162},
        pages = {162},
          doi = {10.1088/0004-637X/746/2/162},
archivePrefix = {arXiv},
       eprint = {1110.1637},
 primaryClass = {astro-ph.CO},
       adsurl = {https://ui.adsabs.harvard.edu/abs/2012ApJ...746..162N}
}

@ARTICLE{Naab2009,
       author = {{Naab}, Thorsten and {Johansson}, Peter H. and {Ostriker}, Jeremiah P.},
        title = "{Minor Mergers and the Size Evolution of Elliptical Galaxies}",
      journal = {\apjl},
         year = 2009,
        month = jul,
       volume = {699},
       number = {2},
        pages = {L178-L182},
          doi = {10.1088/0004-637X/699/2/L178},
archivePrefix = {arXiv},
       eprint = {0903.1636},
 primaryClass = {astro-ph.CO},
       adsurl = {https://ui.adsabs.harvard.edu/abs/2009ApJ...699L.178N}
}

@ARTICLE{Gunn1972,
       author = {{Gunn}, James E. and {Gott}, III and J. Richard},
        title = "{On the Infall of Matter Into Clusters of Galaxies and Some Effects on Their Evolution}",
      journal = {\apj},
         year = 1972,
        month = aug,
       volume = {176},
        pages = {1},
          doi = {10.1086/151605},
       adsurl = {https://ui.adsabs.harvard.edu/abs/1972ApJ...176....1G}
}

@ARTICLE{Balogh2000,
       author = {{Balogh}, Michael L. and {Navarro}, Julio F. and {Morris}, Simon L.},
        title = "{The Origin of Star Formation Gradients in Rich Galaxy Clusters}",
      journal = {\apj},
         year = 2000,
        month = sep,
       volume = {540},
       number = {1},
        pages = {113-121},
          doi = {10.1086/309323},
archivePrefix = {arXiv},
       eprint = {astro-ph/0004078},
 primaryClass = {astro-ph},
       adsurl = {https://ui.adsabs.harvard.edu/abs/2000ApJ...540..113B}
}

@ARTICLE{Kapferer2009,
       author = {{Kapferer} and W. and {Sluka} and C. and {Schindler} and S. and {Ferrari} and C. and {Ziegler}, B.},
        title = "{The effect of ram pressure on the star formation, mass distribution and morphology of galaxies}",
      journal = {\aap},
         year = 2009,
        month = may,
       volume = {499},
       number = {1},
        pages = {87-102},
          doi = {10.1051/0004-6361/200811551},
archivePrefix = {arXiv},
       eprint = {0903.3818},
 primaryClass = {astro-ph.CO},
       adsurl = {https://ui.adsabs.harvard.edu/abs/2009A&A...499...87K}
}

@ARTICLE{Cramer2019,
       author = {{Cramer}, W.~J. and {Kenney}, J.~D.~P. and {Sun} and M. and {Crowl} and H. and {Yagi} and M. and {J{\'a}chym} and P. and {Roediger} and E. and {Waldron}, W.},
        title = "{Spectacular Hubble Space Telescope Observations of the Coma Galaxy D100 and Star Formation in Its Ram Pressure-stripped Tail}",
      journal = {\apj},
         year = 2019,
        month = jan,
       volume = {870},
       number = {2},
          eid = {63},
        pages = {63},
          doi = {10.3847/1538-4357/aaefff},
archivePrefix = {arXiv},
       eprint = {1811.04916},
 primaryClass = {astro-ph.GA},
       adsurl = {https://ui.adsabs.harvard.edu/abs/2019ApJ...870...63C}
}

@ARTICLE{Lotz2019,
       author = {{Lotz}, Marcel and {Remus}, Rhea-Silvia and {Dolag}, Klaus and {Biviano}, Andrea and {Burkert}, Andreas},
        title = "{Gone after one orbit: How cluster environments quench galaxies}",
      journal = {\mnras},
         year = 2019,
        month = oct,
       volume = {488},
       number = {4},
        pages = {5370-5389},
          doi = {10.1093/mnras/stz2070},
archivePrefix = {arXiv},
       eprint = {1810.02382},
 primaryClass = {astro-ph.GA},
       adsurl = {https://ui.adsabs.harvard.edu/abs/2019MNRAS.488.5370L}
}

@ARTICLE{Moore1996,
       author = {{Moore}, Ben and {Katz}, Neal and {Lake}, George and {Dressler}, Alan and {Oemler}, Augustus},
        title = "{Galaxy harassment and the evolution of clusters of galaxies}",
      journal = {\nat},
         year = 1996,
        month = feb,
       volume = {379},
       number = {6566},
        pages = {613-616},
          doi = {10.1038/379613a0},
archivePrefix = {arXiv},
       eprint = {astro-ph/9510034},
 primaryClass = {astro-ph},
       adsurl = {https://ui.adsabs.harvard.edu/abs/1996Natur.379..613M}
}

@ARTICLE{Moore1998b,
       author = {{Moore}, Ben and {Lake}, George and {Katz}, Neal},
        title = "{Morphological Transformation from Galaxy Harassment}",
      journal = {\apj},
         year = 1998,
        month = mar,
       volume = {495},
       number = {1},
        pages = {139-151},
          doi = {10.1086/305264},
archivePrefix = {arXiv},
       eprint = {astro-ph/9701211},
 primaryClass = {astro-ph},
       adsurl = {https://ui.adsabs.harvard.edu/abs/1998ApJ...495..139M}
}

@ARTICLE{Moore1999,
       author = {{Moore}, Ben and {Lake}, George and {Quinn}, Thomas and {Stadel}, Joachim},
        title = "{On the survival and destruction of spiral galaxies in clusters}",
      journal = {\mnras},
         year = 1999,
        month = apr,
       volume = {304},
       number = {3},
        pages = {465-474},
          doi = {10.1046/j.1365-8711.1999.02345.x},
archivePrefix = {arXiv},
       eprint = {astro-ph/9811127},
 primaryClass = {astro-ph},
       adsurl = {https://ui.adsabs.harvard.edu/abs/1999MNRAS.304..465M}
}

@ARTICLE{Fang2018,
       author = {{Fang}, Jerome J. and {Faber}, S.~M. and {Koo}, David C. and {Rodr{\'\i}guez-Puebla}, Aldo and {Guo}, Yicheng and {Barro}, Guillermo and {Behroozi}, Peter and {Brammer}, Gabriel and {Chen}, Zhu and {Dekel}, Avishai and {Ferguson}, Henry C. and {Gawiser}, Eric and {Giavalisco}, Mauro and {Kartaltepe}, Jeyhan and {Kocevski}, Dale D. and {Koekemoer}, Anton M. and {McGrath}, Elizabeth J. and {McIntosh}, Daniel and {Newman}, Jeffrey A. and {Pacifici}, Camilla and {Pandya}, Viraj and {P{\'e}rez-Gonz{\'a}lez}, Pablo G. and {Primack}, Joel R. and {Salmon}, Brett and {Trump}, Jonathan R. and {Weiner}, Benjamin and {Willner}, S.~P. and {Acquaviva}, Viviana and {Dahlen}, Tomas and {Finkelstein}, Steven L. and {Finlator}, Kristian and {Fontana}, Adriano and {Galametz}, Audrey and {Grogin}, Norman A. and {Gruetzbauch}, Ruth and {Johnson}, Seth and {Mobasher}, Bahram and {Papovich}, Casey J. and {Pforr}, Janine and {Salvato}, Mara and {Santini} and P. and {van der Wel}, Arjen and {Wiklind}, Tommy and {Wuyts}, Stijn},
        title = "{Demographics of Star-forming Galaxies since $z \sim 2.5$. I. The UVJ Diagram in CANDELS}",
      journal = {\apj},
         year = 2018,
        month = may,
       volume = {858},
       number = {2},
          eid = {100},
        pages = {100},
          doi = {10.3847/1538-4357/aabcba},
archivePrefix = {arXiv},
       eprint = {1710.05489},
 primaryClass = {astro-ph.GA},
       adsurl = {https://ui.adsabs.harvard.edu/abs/2018ApJ...858..100F}
}

@ARTICLE{Cortese2009,
       author = {{Cortese} and L. and {Hughes}, T.~M.},
        title = "{Evolutionary paths to and from the red sequence: star formation and HI properties of transition galaxies at z \raisebox{-0.5ex}\textasciitilde 0}",
      journal = {\mnras},
         year = 2009,
        month = dec,
       volume = {400},
       number = {3},
        pages = {1225-1240},
          doi = {10.1111/j.1365-2966.2009.15548.x},
archivePrefix = {arXiv},
       eprint = {0908.3564},
 primaryClass = {astro-ph.CO},
       adsurl = {https://ui.adsabs.harvard.edu/abs/2009MNRAS.400.1225C}
}

@ARTICLE{Dekel2019,
       author = {{Dekel}, Avishai and {Lapiner}, Sharon and {Dubois}, Yohan},
        title = "{Origin of the Golden Mass of Galaxies and Black Holes}",
      journal = {arXiv e-prints},
         year = 2019,
        month = apr,
          eid = {arXiv:1904.08431},
        pages = {arXiv:1904.08431},
          doi = {10.48550/arXiv.1904.08431},
archivePrefix = {arXiv},
       eprint = {1904.08431},
 primaryClass = {astro-ph.GA},
       adsurl = {https://ui.adsabs.harvard.edu/abs/2019arXiv190408431D}
}

@ARTICLE{DiMatteo2005,
       author = {{Di Matteo}, Tiziana and {Springel}, Volker and {Hernquist}, Lars},
        title = "{Energy input from quasars regulates the growth and activity of black holes and their host galaxies}",
      journal = {\nat},
         year = 2005,
        month = feb,
       volume = {433},
       number = {7026},
        pages = {604-607},
          doi = {10.1038/nature03335},
archivePrefix = {arXiv},
       eprint = {astro-ph/0502199},
 primaryClass = {astro-ph},
       adsurl = {https://ui.adsabs.harvard.edu/abs/2005Natur.433..604D}
}

@ARTICLE{Guo2011,
       author = {{Guo}, Qi and {White}, Simon and {Boylan-Kolchin}, Michael and {De Lucia}, Gabriella and {Kauffmann}, Guinevere and {Lemson}, Gerard and {Li}, Cheng and {Springel}, Volker and {Weinmann}, Simone},
        title = "{From dwarf spheroidals to cD galaxies: simulating the galaxy population in a {\ensuremath{\Lambda}}CDM cosmology}",
      journal = {\mnras},
         year = 2011,
        month = may,
       volume = {413},
       number = {1},
        pages = {101-131},
          doi = {10.1111/j.1365-2966.2010.18114.x},
archivePrefix = {arXiv},
       eprint = {1006.0106},
 primaryClass = {astro-ph.CO},
       adsurl = {https://ui.adsabs.harvard.edu/abs/2011MNRAS.413..101G}
}

@ARTICLE{Springel2005a,
       author = {{Springel}, Volker and {White}, Simon D.~M. and {Jenkins}, Adrian and {Frenk}, Carlos S. and {Yoshida}, Naoki and {Gao}, Liang and {Navarro}, Julio and {Thacker}, Robert and {Croton}, Darren and {Helly}, John and {Peacock}, John A. and {Cole}, Shaun and {Thomas}, Peter and {Couchman}, Hugh and {Evrard}, August and {Colberg}, J{\"o}rg and {Pearce}, Frazer},
        title = "{Simulations of the formation, evolution and clustering of galaxies and quasars}",
      journal = {\nat},
         year = 2005,
        month = jun,
       volume = {435},
       number = {7042},
        pages = {629-636},
          doi = {10.1038/nature03597},
archivePrefix = {arXiv},
       eprint = {astro-ph/0504097},
 primaryClass = {astro-ph},
       adsurl = {https://ui.adsabs.harvard.edu/abs/2005Natur.435..629S}
}

@ARTICLE{Kauffmann1993,
       author = {{Kauffmann} and G. and {White}, S.~D.~M. and {Guiderdoni}, B.},
        title = "{The formation and evolution of galaxies within merging dark matter haloes.}",
      journal = {\mnras},
         year = 1993,
        month = sep,
       volume = {264},
        pages = {201-218},
          doi = {10.1093/mnras/264.1.201},
       adsurl = {https://ui.adsabs.harvard.edu/abs/1993MNRAS.264..201K}
}

@ARTICLE{Moreno2015,
       author = {{Moreno}, Jorge and {Torrey}, Paul and {Ellison}, Sara L. and {Patton}, David R. and {Bluck}, Asa F.~L. and {Bansal}, Gunjan and {Hernquist}, Lars},
        title = "{Mapping galaxy encounters in numerical simulations: the spatial extent of induced star formation}",
      journal = {\mnras},
         year = 2015,
        month = apr,
       volume = {448},
       number = {2},
        pages = {1107-1117},
          doi = {10.1093/mnras/stv094},
archivePrefix = {arXiv},
       eprint = {1501.03573},
 primaryClass = {astro-ph.GA},
       adsurl = {https://ui.adsabs.harvard.edu/abs/2015MNRAS.448.1107M}
}

@ARTICLE{Torrey2014,
       author = {{Torrey}, Paul and {Vogelsberger}, Mark and {Genel}, Shy and {Sijacki}, Debora and {Springel}, Volker and {Hernquist}, Lars},
        title = "{A model for cosmological simulations of galaxy formation physics: multi-epoch validation}",
      journal = {\mnras},
         year = 2014,
        month = mar,
       volume = {438},
       number = {3},
        pages = {1985-2004},
          doi = {10.1093/mnras/stt2295},
archivePrefix = {arXiv},
       eprint = {1305.4931},
 primaryClass = {astro-ph.CO},
       adsurl = {https://ui.adsabs.harvard.edu/abs/2014MNRAS.438.1985T}
}

@ARTICLE{Thorne1974,
       author = {{Thorne}, Kip S.},
        title = "{Disk-Accretion onto a Black Hole. II. Evolution of the Hole}",
      journal = {\apj},
         year = 1974,
        month = jul,
       volume = {191},
        pages = {507-520},
          doi = {10.1086/152991},
       adsurl = {https://ui.adsabs.harvard.edu/abs/1974ApJ...191..507T}
}

@book{Eddington1926,
  author = {Eddington, A. S.},
  title = {The Internal Constitution of the Stars},
  publisher = {Cambridge University Press},
  year = {1926}
}

@ARTICLE{Haring2004,
       author = {{H{\"a}ring}, Nadine and {Rix}, Hans-Walter},
        title = "{On the Black Hole Mass-Bulge Mass Relation}",
      journal = {\apjl},
         year = 2004,
        month = apr,
       volume = {604},
       number = {2},
        pages = {L89-L92},
          doi = {10.1086/383567},
archivePrefix = {arXiv},
       eprint = {astro-ph/0402376},
 primaryClass = {astro-ph},
       adsurl = {https://ui.adsabs.harvard.edu/abs/2004ApJ...604L..89H}
}

@ARTICLE{Maggorian1998,
       author = {{Magorrian}, John and {Tremaine}, Scott and {Richstone}, Douglas and {Bender}, Ralf and {Bower}, Gary and {Dressler}, Alan and {Faber}, S.~M. and {Gebhardt}, Karl and {Green}, Richard and {Grillmair}, Carl and {Kormendy}, John and {Lauer}, Tod},
        title = "{The Demography of Massive Dark Objects in Galaxy Centers}",
      journal = {\aj},
         year = 1998,
        month = jun,
       volume = {115},
       number = {6},
        pages = {2285-2305},
          doi = {10.1086/300353},
archivePrefix = {arXiv},
       eprint = {astro-ph/9708072},
 primaryClass = {astro-ph},
       adsurl = {https://ui.adsabs.harvard.edu/abs/1998AJ....115.2285M}
}

@ARTICLE{Fluetsch2021,
       author = {{Fluetsch} and A. and {Maiolino} and R. and {Carniani} and S. and {Arribas} and S. and {Belfiore} and F. and {Bellocchi} and E. and {Cazzoli} and S. and {Cicone} and C. and {Cresci} and G. and {Fabian}, A.~C. and {Gallagher} and R. and {Ishibashi} and W. and {Mannucci} and F. and {Marconi} and A. and {Perna} and M. and {Sturm} and E. and {Venturi}, G.},
        title = "{Properties of the multiphase outflows in local (ultra)luminous infrared galaxies}",
      journal = {\mnras},
         year = 2021,
        month = aug,
       volume = {505},
       number = {4},
        pages = {5753-5783},
          doi = {10.1093/mnras/stab1666},
archivePrefix = {arXiv},
       eprint = {2006.13232},
 primaryClass = {astro-ph.GA},
       adsurl = {https://ui.adsabs.harvard.edu/abs/2021MNRAS.505.5753F}
}

@ARTICLE{Tacchella2016,
       author = {{Tacchella}, Sandro and {Dekel}, Avishai and {Carollo} and C. Marcella and {Ceverino}, Daniel and {DeGraf}, Colin and {Lapiner}, Sharon and {Mandelker}, Nir and {Primack Joel}, R.},
        title = "{The confinement of star-forming galaxies into a main sequence through episodes of gas compaction, depletion and replenishment}",
      journal = {\mnras},
         year = 2016,
        month = apr,
       volume = {457},
       number = {3},
        pages = {2790-2813},
          doi = {10.1093/mnras/stw131},
archivePrefix = {arXiv},
       eprint = {1509.02529},
 primaryClass = {astro-ph.GA},
       adsurl = {https://ui.adsabs.harvard.edu/abs/2016MNRAS.457.2790T}
}

@ARTICLE{Lilly2013,
       author = {{Lilly}, Simon J. and {Carollo} and C. Marcella and {Pipino}, Antonio and {Renzini}, Alvio and {Peng}, Yingjie},
        title = "{Gas Regulation of Galaxies: The Evolution of the Cosmic Specific Star Formation Rate, the Metallicity-Mass-Star-formation Rate Relation, and the Stellar Content of Halos}",
      journal = {\apj},
         year = 2013,
        month = aug,
       volume = {772},
       number = {2},
          eid = {119},
        pages = {119},
          doi = {10.1088/0004-637X/772/2/119},
archivePrefix = {arXiv},
       eprint = {1303.5059},
 primaryClass = {astro-ph.CO},
       adsurl = {https://ui.adsabs.harvard.edu/abs/2013ApJ...772..119L}
}

@ARTICLE{Omand2014,
       author = {{Omand}, Conor M.~B. and {Balogh}, Michael L. and {Poggianti}, Bianca M.},
        title = "{The connection between galaxy structure and quenching efficiency}",
      journal = {\mnras},
         year = 2014,
        month = may,
       volume = {440},
       number = {1},
        pages = {843-858},
          doi = {10.1093/mnras/stu331},
archivePrefix = {arXiv},
       eprint = {1402.3394},
 primaryClass = {astro-ph.GA},
       adsurl = {https://ui.adsabs.harvard.edu/abs/2014MNRAS.440..843O}
}

@ARTICLE{Vogelsberger2013,
       author = {{Vogelsberger}, Mark and {Genel}, Shy and {Sijacki}, Debora and {Torrey}, Paul and {Springel}, Volker and {Hernquist}, Lars},
        title = "{A model for cosmological simulations of galaxy formation physics}",
      journal = {\mnras},
         year = 2013,
        month = dec,
       volume = {436},
       number = {4},
        pages = {3031-3067},
          doi = {10.1093/mnras/stt1789},
archivePrefix = {arXiv},
       eprint = {1305.2913},
 primaryClass = {astro-ph.CO},
       adsurl = {https://ui.adsabs.harvard.edu/abs/2013MNRAS.436.3031V}
}

@ARTICLE{Gensior2021,
       author = {{Gensior}, Jindra and {Kruijssen}, J.~M. Diederik},
        title = "{The elephant in the bathtub: when the physics of star formation regulate the baryon cycle of galaxies}",
      journal = {\mnras},
         year = 2021,
        month = jan,
       volume = {500},
       number = {2},
        pages = {2000-2011},
          doi = {10.1093/mnras/staa3453},
archivePrefix = {arXiv},
       eprint = {2011.01235},
 primaryClass = {astro-ph.GA},
       adsurl = {https://ui.adsabs.harvard.edu/abs/2021MNRAS.500.2000G}
}

@ARTICLE{Lin2019,
       author = {{Lin}, Lihwai and {Pan}, Hsi-An and {Ellison}, Sara L. and {Belfiore}, Francesco and {Shi}, Yong and {S{\'a}nchez}, Sebasti{\'a}n F. and {Hsieh}, Bau-Ching and {Rowlands}, Kate and {Ramya} and S. and {Thorp}, Mallory D. and {Li}, Cheng and {Maiolino}, Roberto},
        title = "{The ALMaQUEST Survey: The Molecular Gas Main Sequence and the Origin of the Star-forming Main Sequence}",
      journal = {\apjl},
         year = 2019,
        month = oct,
       volume = {884},
       number = {2},
          eid = {L33},
        pages = {L33},
          doi = {10.3847/2041-8213/ab4815},
archivePrefix = {arXiv},
       eprint = {1909.11243},
 primaryClass = {astro-ph.GA},
       adsurl = {https://ui.adsabs.harvard.edu/abs/2019ApJ...884L..33L}
}

@ARTICLE{Cooper2010,
       author = {{Cooper}, A.~P. and {Cole} and S. and {Frenk}, C.~S. and {White}, S.~D.~M. and {Helly} and J. and {Benson}, A.~J. and {De Lucia} and G. and {Helmi} and A. and {Jenkins} and A. and {Navarro}, J.~F. and {Springel} and V. and {Wang}, J.},
        title = "{Galactic stellar haloes in the CDM model}",
      journal = {\mnras},
         year = 2010,
        month = aug,
       volume = {406},
       number = {2},
        pages = {744-766},
          doi = {10.1111/j.1365-2966.2010.16740.x},
archivePrefix = {arXiv},
       eprint = {0910.3211},
 primaryClass = {astro-ph.GA},
       adsurl = {https://ui.adsabs.harvard.edu/abs/2010MNRAS.406..744C}
}

@ARTICLE{Blanton2009,
       author = {{Blanton}, Michael R. and {Moustakas}, John},
        title = "{Physical Properties and Environments of Nearby Galaxies}",
      journal = {\araa},
         year = 2009,
        month = sep,
       volume = {47},
       number = {1},
        pages = {159-210},
          doi = {10.1146/annurev-astro-082708-101734},
archivePrefix = {arXiv},
       eprint = {0908.3017},
 primaryClass = {astro-ph.GA},
       adsurl = {https://ui.adsabs.harvard.edu/abs/2009ARA&A..47..159B}
}

@ARTICLE{Cortese2021,
       author = {{Cortese} and L. and {Catinella} and B. and {Smith}, R.},
        title = "{The Dawes Review 9: The role of cold gas stripping on the star formation quenching of satellite galaxies}",
      journal = {\pasa},
         year = 2021,
        month = aug,
       volume = {38},
          eid = {e035},
        pages = {e035},
          doi = {10.1017/pasa.2021.18},
archivePrefix = {arXiv},
       eprint = {2104.02193},
 primaryClass = {astro-ph.GA},
       adsurl = {https://ui.adsabs.harvard.edu/abs/2021PASA...38...35C}
}

@ARTICLE{Conroy2013,
       author = {{Conroy}, Charlie},
        title = "{Modeling the Panchromatic Spectral Energy Distributions of Galaxies}",
      journal = {\araa},
         year = 2013,
        month = aug,
       volume = {51},
       number = {1},
        pages = {393-455},
          doi = {10.1146/annurev-astro-082812-141017},
archivePrefix = {arXiv},
       eprint = {1301.7095},
 primaryClass = {astro-ph.CO},
       adsurl = {https://ui.adsabs.harvard.edu/abs/2013ARA&A..51..393C}
}

@ARTICLE{Koprowski2024,
       author = {{Koprowski}, M.~P. and {Wijesekera}, J.~V. and {Dunlop}, J.~S. and {McLeod}, D.~J. and {Micha{\l}owski}, M.~J. and {Lisiecki} and K. and {McLure}, R.~J.},
        title = "{Charting the main sequence of star-forming galaxies out to redshifts $z \lesssim 5.7$}",
      journal = {\aap},
         year = 2024,
        month = nov,
       volume = {691},
          eid = {A164},
        pages = {A164},
          doi = {10.1051/0004-6361/202449948},
archivePrefix = {arXiv},
       eprint = {2403.06575},
 primaryClass = {astro-ph.GA},
       adsurl = {https://ui.adsabs.harvard.edu/abs/2024A&A...691A.164K}
}

@ARTICLE{Baldwin1981,
       author = {{Baldwin}, J.~A. and {Phillips}, M.~M. and {Terlevich}, R.},
        title = "{Classification parameters for the emission-line spectra of extragalactic objects.}",
      journal = {\pasp},
         year = 1981,
        month = feb,
       volume = {93},
        pages = {5-19},
          doi = {10.1086/130766},
       adsurl = {https://ui.adsabs.harvard.edu/abs/1981PASP...93....5B}
}

@ARTICLE{Kennicutt1998b,
       author = {{Kennicutt}, Jr., Robert C.},
        title = "{The Global Schmidt Law in Star-forming Galaxies}",
      journal = {\apj},
         year = 1998,
        month = may,
       volume = {498},
       number = {2},
        pages = {541-552},
          doi = {10.1086/305588},
archivePrefix = {arXiv},
       eprint = {astro-ph/9712213},
 primaryClass = {astro-ph},
       adsurl = {https://ui.adsabs.harvard.edu/abs/1998ApJ...498..541K}
}

@ARTICLE{Meurer1999,
       author = {{Meurer}, Gerhardt R. and {Heckman}, Timothy M. and {Calzetti}, Daniela},
        title = "{Dust Absorption and the Ultraviolet Luminosity Density at z \raisebox{-0.5ex}\textasciitilde 3 as Calibrated by Local Starburst Galaxies}",
      journal = {\apj},
         year = 1999,
        month = aug,
       volume = {521},
       number = {1},
        pages = {64-80},
          doi = {10.1086/307523},
archivePrefix = {arXiv},
       eprint = {astro-ph/9903054},
 primaryClass = {astro-ph},
       adsurl = {https://ui.adsabs.harvard.edu/abs/1999ApJ...521...64M}
}

@ARTICLE{Hao2011,
       author = {{Hao}, Cai-Na and {Kennicutt}, Robert C. and {Johnson}, Benjamin D. and {Calzetti}, Daniela and {Dale}, Daniel A. and {Moustakas}, John},
        title = "{Dust-corrected Star Formation Rates of Galaxies. II. Combinations of Ultraviolet and Infrared Tracers}",
      journal = {\apj},
         year = 2011,
        month = nov,
       volume = {741},
       number = {2},
          eid = {124},
        pages = {124},
          doi = {10.1088/0004-637X/741/2/124},
archivePrefix = {arXiv},
       eprint = {1108.2837},
 primaryClass = {astro-ph.CO},
       adsurl = {https://ui.adsabs.harvard.edu/abs/2011ApJ...741..124H}
}

@ARTICLE{Maraston2005,
       author = {{Maraston}, Claudia},
        title = "{Evolutionary population synthesis: models, analysis of the ingredients and application to high-z galaxies}",
      journal = {\mnras},
         year = 2005,
        month = sep,
       volume = {362},
       number = {3},
        pages = {799-825},
          doi = {10.1111/j.1365-2966.2005.09270.x},
archivePrefix = {arXiv},
       eprint = {astro-ph/0410207},
 primaryClass = {astro-ph},
       adsurl = {https://ui.adsabs.harvard.edu/abs/2005MNRAS.362..799M}
}

@ARTICLE{Calzetti2000,
       author = {{Calzetti}, Daniela and {Armus}, Lee and {Bohlin}, Ralph C. and {Kinney}, Anne L. and {Koornneef}, Jan and {Storchi-Bergmann}, Thaisa},
        title = "{The Dust Content and Opacity of Actively Star-forming Galaxies}",
      journal = {\apj},
         year = 2000,
        month = apr,
       volume = {533},
       number = {2},
        pages = {682-695},
          doi = {10.1086/308692},
archivePrefix = {arXiv},
       eprint = {astro-ph/9911459},
 primaryClass = {astro-ph},
       adsurl = {https://ui.adsabs.harvard.edu/abs/2000ApJ...533..682C}
}

@ARTICLE{Cardelli1989,
       author = {{Cardelli}, Jason A. and {Clayton}, Geoffrey C. and {Mathis}, John S.},
        title = "{The Relationship between Infrared, Optical, and Ultraviolet Extinction}",
      journal = {\apj},
         year = 1989,
        month = oct,
       volume = {345},
        pages = {245},
          doi = {10.1086/167900},
       adsurl = {https://ui.adsabs.harvard.edu/abs/1989ApJ...345..245C}
}

@ARTICLE{Murphy2011,
       author = {{Murphy}, E.~J. and {Condon}, J.~J. and {Schinnerer} and E. and {Kennicutt}, R.~C. and {Calzetti} and D. and {Armus} and L. and {Helou} and G. and {Turner}, J.~L. and {Aniano} and G. and {Beir{\~a}o} and P. and {Bolatto}, A.~D. and {Brandl}, B.~R. and {Croxall}, K.~V. and {Dale}, D.~A. and {Donovan Meyer}, J.~L. and {Draine}, B.~T. and {Engelbracht} and C. and {Hunt}, L.~K. and {Hao} and C. -N. and {Koda} and J. and {Roussel} and H. and {Skibba} and R. and {Smith} and J. -D.~T.},
        title = "{Calibrating Extinction-free Star Formation Rate Diagnostics with 33 GHz Free-free Emission in NGC 6946}",
      journal = {\apj},
         year = 2011,
        month = aug,
       volume = {737},
       number = {2},
          eid = {67},
        pages = {67},
          doi = {10.1088/0004-637X/737/2/67},
archivePrefix = {arXiv},
       eprint = {1105.4877},
 primaryClass = {astro-ph.CO},
       adsurl = {https://ui.adsabs.harvard.edu/abs/2011ApJ...737...67M}
}

@ARTICLE{Condon1992,
       author = {{Condon}, J.~J.},
        title = "{Radio emission from normal galaxies.}",
      journal = {\araa},
         year = 1992,
        month = jan,
       volume = {30},
        pages = {575-611},
          doi = {10.1146/annurev.aa.30.090192.003043},
       adsurl = {https://ui.adsabs.harvard.edu/abs/1992ARA&A..30..575C}
}

@ARTICLE{Calzetti2007,
       author = {{Calzetti} and D. and {Kennicutt}, R.~C. and {Engelbracht}, C.~W. and {Leitherer} and C. and {Draine}, B.~T. and {Kewley} and L. and {Moustakas} and J. and {Sosey} and M. and {Dale}, D.~A. and {Gordon}, K.~D. and {Helou}, G.~X. and {Hollenbach}, D.~J. and {Armus} and L. and {Bendo} and G. and {Bot} and C. and {Buckalew} and B. and {Jarrett} and T. and {Li} and A. and {Meyer} and M. and {Murphy}, E.~J. and {Prescott} and M. and {Regan}, M.~W. and {Rieke}, G.~H. and {Roussel} and H. and {Sheth} and K. and {Smith}, J.~D.~T. and {Thornley}, M.~D. and {Walter}, F.},
        title = "{The Calibration of Mid-Infrared Star Formation Rate Indicators}",
      journal = {\apj},
         year = 2007,
        month = sep,
       volume = {666},
       number = {2},
        pages = {870-895},
          doi = {10.1086/520082},
archivePrefix = {arXiv},
       eprint = {0705.3377},
 primaryClass = {astro-ph},
       adsurl = {https://ui.adsabs.harvard.edu/abs/2007ApJ...666..870C}
}

@ARTICLE{Rieke2009,
       author = {{Rieke}, G.~H. and {Alonso-Herrero} and A. and {Weiner}, B.~J. and {P{\'e}rez-Gonz{\'a}lez}, P.~G. and {Blaylock} and M. and {Donley}, J.~L. and {Marcillac}, D.},
        title = "{Determining Star Formation Rates for Infrared Galaxies}",
      journal = {\apj},
         year = 2009,
        month = feb,
       volume = {692},
       number = {1},
        pages = {556-573},
          doi = {10.1088/0004-637X/692/1/556},
archivePrefix = {arXiv},
       eprint = {0810.4150},
 primaryClass = {astro-ph},
       adsurl = {https://ui.adsabs.harvard.edu/abs/2009ApJ...692..556R}
}

@INCOLLECTION{Roberts1975,
       author = {{Roberts}, Morton S.},
        title = "{Radio Observations of Neutral Hydrogen in Galaxies}",
    booktitle = {Galaxies and the Universe},
         year = 1975,
       editor = {{Sandage}, Allan and {Sandage}, Mary and {Kristian}, Jerome},
        pages = {309},
       adsurl = {https://ui.adsabs.harvard.edu/abs/1975gaun.book..309R}
}

@ARTICLE{Giovanelli1983,
       author = {{Giovanelli} and R. and {Haynes}, M.~P.},
        title = "{The HI extent and deficiency of spiral galaxies in the Virgo cluster.}",
      journal = {\aj},
         year = 1983,
        month = jul,
       volume = {88},
        pages = {881-908},
          doi = {10.1086/113376},
       adsurl = {https://ui.adsabs.harvard.edu/abs/1983AJ.....88..881G}
}

@ARTICLE{Haynes2011,
       author = {{Haynes}, Martha P. and {Giovanelli}, Riccardo and {Martin}, Ann M. and {Hess}, Kelley M. and {Saintonge}, Am{\'e}lie and {Adams}, Elizabeth A.~K. and {Hallenbeck}, Gregory and {Hoffman} and G. Lyle and {Huang}, Shan and {Kent}, Brian R. and {Koopmann}, Rebecca A. and {Papastergis}, Emmanouil and {Stierwalt}, Sabrina and {Balonek}, Thomas J. and {Craig}, David W. and {Higdon}, Sarah J.~U. and {Kornreich}, David A. and {Miller}, Jeffrey R. and {O'Donoghue}, Aileen A. and {Olowin}, Ronald P. and {Rosenberg}, Jessica L. and {Spekkens}, Kristine and {Troischt}, Parker and {Wilcots}, Eric M.},
        title = "{The Arecibo Legacy Fast ALFA Survey: The {\ensuremath{\alpha}}.40 H I Source Catalog, Its Characteristics and Their Impact on the Derivation of the H I Mass Function}",
      journal = {\aj},
         year = 2011,
        month = nov,
       volume = {142},
       number = {5},
          eid = {170},
        pages = {170},
          doi = {10.1088/0004-6256/142/5/170},
archivePrefix = {arXiv},
       eprint = {1109.0027},
 primaryClass = {astro-ph.CO},
       adsurl = {https://ui.adsabs.harvard.edu/abs/2011AJ....142..170H}
}

@ARTICLE{Meyer2017,
       author = {{Meyer}, Martin and {Robotham}, Aaron and {Obreschkow}, Danail and {Westmeier}, Tobias and {Duffy}, Alan R. and {Staveley-Smith}, Lister},
        title = "{Tracing HI Beyond the Local Universe}",
      journal = {\pasa},
         year = 2017,
        month = nov,
       volume = {34},
        pages = {52},
          doi = {10.1017/pasa.2017.31},
archivePrefix = {arXiv},
       eprint = {1705.04210},
 primaryClass = {astro-ph.CO},
       adsurl = {https://ui.adsabs.harvard.edu/abs/2017PASA...34...52M}
}

@ARTICLE{Solomon1997,
       author = {{Solomon}, P.~M. and {Downes} and D. and {Radford}, S.~J.~E. and {Barrett}, J.~W.},
        title = "{The Molecular Interstellar Medium in Ultraluminous Infrared Galaxies}",
      journal = {\apj},
         year = 1997,
        month = mar,
       volume = {478},
       number = {1},
        pages = {144-161},
          doi = {10.1086/303765},
archivePrefix = {arXiv},
       eprint = {astro-ph/9610166},
 primaryClass = {astro-ph},
       adsurl = {https://ui.adsabs.harvard.edu/abs/1997ApJ...478..144S}
}

@ARTICLE{Bolatto2013,
       author = {{Bolatto}, Alberto D. and {Wolfire}, Mark and {Leroy}, Adam K.},
        title = "{The CO-to-H$_{2}$ Conversion Factor}",
      journal = {\araa},
         year = 2013,
        month = aug,
       volume = {51},
       number = {1},
        pages = {207-268},
          doi = {10.1146/annurev-astro-082812-140944},
archivePrefix = {arXiv},
       eprint = {1301.3498},
 primaryClass = {astro-ph.GA},
       adsurl = {https://ui.adsabs.harvard.edu/abs/2013ARA&A..51..207B}
}

@ARTICLE{Tacconi2020,
       author = {{Tacconi}, Linda J. and {Genzel}, Reinhard and {Sternberg}, Amiel},
        title = "{The Evolution of the Star-Forming Interstellar Medium Across Cosmic Time}",
      journal = {\araa},
         year = 2020,
        month = aug,
       volume = {58},
        pages = {157-203},
          doi = {10.1146/annurev-astro-082812-141034},
archivePrefix = {arXiv},
       eprint = {2003.06245},
 primaryClass = {astro-ph.GA},
       adsurl = {https://ui.adsabs.harvard.edu/abs/2020ARA&A..58..157T}
}

@ARTICLE{Genzel2015,
       author = {{Genzel} and R. and {Tacconi}, L.~J. and {Lutz} and D. and {Saintonge} and A. and {Berta} and S. and {Magnelli} and B. and {Combes} and F. and {Garc{\'\i}a-Burillo} and S. and {Neri} and R. and {Bolatto} and A. and {Contini} and T. and {Lilly} and S. and {Boissier} and J. and {Boone} and F. and {Bouch{\'e}} and N. and {Bournaud} and F. and {Burkert} and A. and {Carollo} and M. and {Colina} and L. and {Cooper}, M.~C. and {Cox} and P. and {Feruglio} and C. and {F{\"o}rster Schreiber}, N.~M. and {Freundlich} and J. and {Gracia-Carpio} and J. and {Juneau} and S. and {Kovac} and K. and {Lippa} and M. and {Naab} and T. and {Salome} and P. and {Renzini} and A. and {Sternberg} and A. and {Walter} and F. and {Weiner} and B. and {Weiss} and A. and {Wuyts}, S.},
        title = "{Combined CO and Dust Scaling Relations of Depletion Time and Molecular Gas Fractions with Cosmic Time, Specific Star-formation Rate, and Stellar Mass}",
      journal = {\apj},
         year = 2015,
        month = feb,
       volume = {800},
       number = {1},
          eid = {20},
        pages = {20},
          doi = {10.1088/0004-637X/800/1/20},
archivePrefix = {arXiv},
       eprint = {1409.1171},
 primaryClass = {astro-ph.GA},
       adsurl = {https://ui.adsabs.harvard.edu/abs/2015ApJ...800...20G}
}

@ARTICLE{Genzel2012,
       author = {{Genzel} and R. and {Tacconi}, L.~J. and {Combes} and F. and {Bolatto} and A. and {Neri} and R. and {Sternberg} and A. and {Cooper}, M.~C. and {Bouch{\'e}} and N. and {Bournaud} and F. and {Burkert} and A. and {Comerford} and J. and {Cox} and P. and {Davis} and M. and {F{\"o}rster Schreiber}, N.~M. and {Garcia-Burillo} and S. and {Gracia-Carpio} and J. and {Lutz} and D. and {Naab} and T. and {Newman} and S. and {Saintonge} and A. and {Shapiro} and K. and {Shapley} and A. and {Weiner}, B.},
        title = "{The Metallicity Dependence of the CO {\textrightarrow} H$_{2}$ Conversion Factor in $z \geqslant 1$ Star-forming Galaxies}",
      journal = {\apj},
         year = 2012,
        month = feb,
       volume = {746},
       number = {1},
          eid = {69},
        pages = {69},
          doi = {10.1088/0004-637X/746/1/69},
archivePrefix = {arXiv},
       eprint = {1106.2098},
 primaryClass = {astro-ph.CO},
       adsurl = {https://ui.adsabs.harvard.edu/abs/2012ApJ...746...69G}
}

@ARTICLE{Narayanan2012,
       author = {{Narayanan}, Desika and {Krumholz}, Mark R. and {Ostriker}, Eve C. and {Hernquist}, Lars},
        title = "{A general model for the CO-H$_{2}$ conversion factor in galaxies with applications to the star formation law}",
      journal = {\mnras},
         year = 2012,
        month = apr,
       volume = {421},
       number = {4},
        pages = {3127-3146},
          doi = {10.1111/j.1365-2966.2012.20536.x},
archivePrefix = {arXiv},
       eprint = {1110.3791},
 primaryClass = {astro-ph.GA},
       adsurl = {https://ui.adsabs.harvard.edu/abs/2012MNRAS.421.3127N}
}

@ARTICLE{Leroy2011,
       author = {{Leroy}, Adam K. and {Bolatto}, Alberto and {Gordon}, Karl and {Sandstrom}, Karin and {Gratier}, Pierre and {Rosolowsky}, Erik and {Engelbracht}, Charles W. and {Mizuno}, Norikazu and {Corbelli}, Edvige and {Fukui}, Yasuo and {Kawamura}, Akiko},
        title = "{The CO-to-H$_{2}$ Conversion Factor from Infrared Dust Emission across the Local Group}",
      journal = {\apj},
         year = 2011,
        month = aug,
       volume = {737},
       number = {1},
          eid = {12},
        pages = {12},
          doi = {10.1088/0004-637X/737/1/12},
archivePrefix = {arXiv},
       eprint = {1102.4618},
 primaryClass = {astro-ph.CO},
       adsurl = {https://ui.adsabs.harvard.edu/abs/2011ApJ...737...12L}
}

@ARTICLE{Magdis2012,
       author = {{Magdis}, Georgios E. and {Daddi} and E. and {B{\'e}thermin} and M. and {Sargent} and M. and {Elbaz} and D. and {Pannella} and M. and {Dickinson} and M. and {Dannerbauer} and H. and {da Cunha} and E. and {Walter} and F. and {Rigopoulou} and D. and {Charmandaris} and V. and {Hwang}, H.~S. and {Kartaltepe}, J.},
        title = "{The Evolving Interstellar Medium of Star-forming Galaxies since z = 2 as Probed by Their Infrared Spectral Energy Distributions}",
      journal = {\apj},
         year = 2012,
        month = nov,
       volume = {760},
       number = {1},
          eid = {6},
        pages = {6},
          doi = {10.1088/0004-637X/760/1/6},
archivePrefix = {arXiv},
       eprint = {1210.1035},
 primaryClass = {astro-ph.CO},
       adsurl = {https://ui.adsabs.harvard.edu/abs/2012ApJ...760....6M}
}

@ARTICLE{Scoville2016,
       author = {{Scoville} and N. and {Sheth} and K. and {Aussel} and H. and {Vanden Bout} and P. and {Capak} and P. and {Bongiorno} and A. and {Casey}, C.~M. and {Murchikova} and L. and {Koda} and J. and {{\'A}lvarez-M{\'a}rquez} and J. and {Lee} and N. and {Laigle} and C. and {McCracken}, H.~J. and {Ilbert} and O. and {Pope} and A. and {Sanders} and D. and {Chu} and J. and {Toft} and S. and {Ivison}, R.~J. and {Manohar}, S.},
        title = "{ISM Masses and the Star formation Law at Z = 1 to 6: ALMA Observations of Dust Continuum in 145 Galaxies in the COSMOS Survey Field}",
      journal = {\apj},
         year = 2016,
        month = apr,
       volume = {820},
       number = {2},
          eid = {83},
        pages = {83},
          doi = {10.3847/0004-637X/820/2/83},
archivePrefix = {arXiv},
       eprint = {1511.05149},
 primaryClass = {astro-ph.GA},
       adsurl = {https://ui.adsabs.harvard.edu/abs/2016ApJ...820...83S}
}

@ARTICLE{Concas2019,
       author = {{Concas}, Alice and {Popesso}, Paola},
        title = "{A new empirical method to estimate the molecular gas mass in galaxies}",
      journal = {\mnras},
         year = 2019,
        month = jun,
       volume = {486},
       number = {1},
        pages = {L91-L95},
          doi = {10.1093/mnrasl/slz065},
archivePrefix = {arXiv},
       eprint = {1905.02214},
 primaryClass = {astro-ph.GA},
       adsurl = {https://ui.adsabs.harvard.edu/abs/2019MNRAS.486L..91C}
}

@ARTICLE{Gruver2009,
       author = {{G{\"u}ver}, Tolga and {{\"O}zel}, Feryal},
        title = "{The relation between optical extinction and hydrogen column density in the Galaxy}",
      journal = {\mnras},
         year = 2009,
        month = dec,
       volume = {400},
       number = {4},
        pages = {2050-2053},
          doi = {10.1111/j.1365-2966.2009.15598.x},
archivePrefix = {arXiv},
       eprint = {0903.2057},
 primaryClass = {astro-ph.GA},
       adsurl = {https://ui.adsabs.harvard.edu/abs/2009MNRAS.400.2050G}
}

@ARTICLE{Leja2019,
       author = {{Leja}, Joel and {Carnall}, Adam C. and {Johnson}, Benjamin D. and {Conroy}, Charlie and {Speagle}, Joshua S.},
        title = "{How to Measure Galaxy Star Formation Histories. II. Nonparametric Models}",
      journal = {\apj},
         year = 2019,
        month = may,
       volume = {876},
       number = {1},
          eid = {3},
        pages = {3},
          doi = {10.3847/1538-4357/ab133c},
archivePrefix = {arXiv},
       eprint = {1811.03637},
 primaryClass = {astro-ph.GA},
       adsurl = {https://ui.adsabs.harvard.edu/abs/2019ApJ...876....3L}
}

@ARTICLE{Carnall2019,
       author = {{Carnall}, Adam C. and {Leja}, Joel and {Johnson}, Benjamin D. and {McLure}, Ross J. and {Dunlop}, James S. and {Conroy}, Charlie},
        title = "{How to Measure Galaxy Star Formation Histories. I. Parametric Models}",
      journal = {\apj},
         year = 2019,
        month = mar,
       volume = {873},
       number = {1},
          eid = {44},
        pages = {44},
          doi = {10.3847/1538-4357/ab04a2},
archivePrefix = {arXiv},
       eprint = {1811.03635},
 primaryClass = {astro-ph.GA},
       adsurl = {https://ui.adsabs.harvard.edu/abs/2019ApJ...873...44C}
}

@ARTICLE{Kriek2009,
       author = {{Kriek}, Mariska and {van Dokkum}, Pieter G. and {Labb{\'e}}, Ivo and {Franx}, Marijn and {Illingworth}, Garth D. and {Marchesini}, Danilo and {Quadri}, Ryan F.},
        title = "{An Ultra-Deep Near-Infrared Spectrum of a Compact Quiescent Galaxy at z = 2.2}",
      journal = {\apj},
         year = 2009,
        month = jul,
       volume = {700},
       number = {1},
        pages = {221-231},
          doi = {10.1088/0004-637X/700/1/221},
archivePrefix = {arXiv},
       eprint = {0905.1692},
 primaryClass = {astro-ph.CO},
       adsurl = {https://ui.adsabs.harvard.edu/abs/2009ApJ...700..221K}
}

@ARTICLE{Leja2017,
       author = {{Leja}, Joel and {Johnson}, Benjamin D. and {Conroy}, Charlie and {van Dokkum}, Pieter G. and {Byler}, Nell},
        title = "{Deriving Physical Properties from Broadband Photometry with Prospector: Description of the Model and a Demonstration of its Accuracy Using 129 Galaxies in the Local Universe}",
      journal = {\apj},
         year = 2017,
        month = mar,
       volume = {837},
       number = {2},
          eid = {170},
        pages = {170},
          doi = {10.3847/1538-4357/aa5ffe},
archivePrefix = {arXiv},
       eprint = {1609.09073},
 primaryClass = {astro-ph.GA},
       adsurl = {https://ui.adsabs.harvard.edu/abs/2017ApJ...837..170L}
}

@ARTICLE{Maraston2010,
       author = {{Maraston}, Claudia and {Pforr}, Janine and {Renzini}, Alvio and {Daddi}, Emanuele and {Dickinson}, Mark and {Cimatti}, Andrea and {Tonini}, Chiara},
        title = "{Star formation rates and masses of $z \sim 2$ galaxies from multicolour photometry}",
      journal = {\mnras},
         year = 2010,
        month = sep,
       volume = {407},
       number = {2},
        pages = {830-845},
          doi = {10.1111/j.1365-2966.2010.16973.x},
archivePrefix = {arXiv},
       eprint = {1004.4546},
 primaryClass = {astro-ph.CO},
       adsurl = {https://ui.adsabs.harvard.edu/abs/2010MNRAS.407..830M}
}

@ARTICLE{DaCunha2008,
       author = {{da Cunha}, Elisabete and {Charlot}, St{\'e}phane and {Elbaz}, David},
        title = "{A simple model to interpret the ultraviolet, optical and infrared emission from galaxies}",
      journal = {\mnras},
         year = 2008,
        month = aug,
       volume = {388},
       number = {4},
        pages = {1595-1617},
          doi = {10.1111/j.1365-2966.2008.13535.x},
archivePrefix = {arXiv},
       eprint = {0806.1020},
 primaryClass = {astro-ph},
       adsurl = {https://ui.adsabs.harvard.edu/abs/2008MNRAS.388.1595D}
}

@ARTICLE{Boquien2019,
       author = {{Boquien} and M. and {Burgarella} and D. and {Roehlly} and Y. and {Buat} and V. and {Ciesla} and L. and {Corre} and D. and {Inoue}, A.~K. and {Salas}, H.},
        title = "{CIGALE: a python Code Investigating GALaxy Emission}",
      journal = {\aap},
         year = 2019,
        month = feb,
       volume = {622},
          eid = {A103},
        pages = {A103},
          doi = {10.1051/0004-6361/201834156},
archivePrefix = {arXiv},
       eprint = {1811.03094},
 primaryClass = {astro-ph.GA},
       adsurl = {https://ui.adsabs.harvard.edu/abs/2019A&A...622A.103B}
}

@ARTICLE{Pacifici2016,
       author = {{Pacifici}, Camilla and {Kassin}, Susan A. and {Weiner}, Benjamin J. and {Holden}, Bradford and {Gardner}, Jonathan P. and {Faber}, Sandra M. and {Ferguson}, Henry C. and {Koo}, David C. and {Primack}, Joel R. and {Bell}, Eric F. and {Dekel}, Avishai and {Gawiser}, Eric and {Giavalisco}, Mauro and {Rafelski}, Marc and {Simons}, Raymond C. and {Barro}, Guillermo and {Croton}, Darren J. and {Dav{\'e}}, Romeel and {Fontana}, Adriano and {Grogin}, Norman A. and {Koekemoer}, Anton M. and {Lee}, Seong-Kook and {Salmon}, Brett and {Somerville}, Rachel and {Behroozi}, Peter},
        title = "{The Evolution of Star Formation Histories of Quiescent Galaxies}",
      journal = {\apj},
         year = 2016,
        month = nov,
       volume = {832},
       number = {1},
          eid = {79},
        pages = {79},
          doi = {10.3847/0004-637X/832/1/79},
archivePrefix = {arXiv},
       eprint = {1609.03572},
 primaryClass = {astro-ph.GA},
       adsurl = {https://ui.adsabs.harvard.edu/abs/2016ApJ...832...79P}
}

@ARTICLE{Lee2010,
       author = {{Lee}, Seong-Kook and {Ferguson}, Henry C. and {Somerville}, Rachel S. and {Wiklind}, Tommy and {Giavalisco}, Mauro},
        title = "{The Estimation of Star Formation Rates and Stellar Population Ages of High-redshift Galaxies from Broadband Photometry}",
      journal = {\apj},
         year = 2010,
        month = dec,
       volume = {725},
       number = {2},
        pages = {1644-1651},
          doi = {10.1088/0004-637X/725/2/1644},
archivePrefix = {arXiv},
       eprint = {1010.1966},
 primaryClass = {astro-ph.CO},
       adsurl = {https://ui.adsabs.harvard.edu/abs/2010ApJ...725.1644L}
}

@ARTICLE{Worthey1994,
       author = {{Worthey}, Guy},
        title = "{Comprehensive Stellar Population Models and the Disentanglement of Age and Metallicity Effects}",
      journal = {\apjs},
         year = 1994,
        month = nov,
       volume = {95},
        pages = {107},
          doi = {10.1086/192096},
       adsurl = {https://ui.adsabs.harvard.edu/abs/1994ApJS...95..107W}
}

@ARTICLE{Chevallard2016,
       author = {{Chevallard}, Jacopo and {Charlot}, St{\'e}phane},
        title = "{Modelling and interpreting spectral energy distributions of galaxies with BEAGLE}",
      journal = {\mnras},
         year = 2016,
        month = oct,
       volume = {462},
       number = {2},
        pages = {1415-1443},
          doi = {10.1093/mnras/stw1756},
archivePrefix = {arXiv},
       eprint = {1603.03037},
 primaryClass = {astro-ph.GA},
       adsurl = {https://ui.adsabs.harvard.edu/abs/2016MNRAS.462.1415C}
}

@ARTICLE{Carnall2018,
       author = {{Carnall}, A.~C. and {McLure}, R.~J. and {Dunlop}, J.~S. and {Dav{\'e}}, R.},
        title = "{Inferring the star formation histories of massive quiescent galaxies with BAGPIPES: evidence for multiple quenching mechanisms}",
      journal = {\mnras},
         year = 2018,
        month = nov,
       volume = {480},
       number = {4},
        pages = {4379-4401},
          doi = {10.1093/mnras/sty2169},
archivePrefix = {arXiv},
       eprint = {1712.04452},
 primaryClass = {astro-ph.GA},
       adsurl = {https://ui.adsabs.harvard.edu/abs/2018MNRAS.480.4379C}
}

@ARTICLE{Ilbert2006,
       author = {{Ilbert} and O. and {Arnouts} and S. and {McCracken}, H.~J. and {Bolzonella} and M. and {Bertin} and E. and {Le F{\`e}vre} and O. and {Mellier} and Y. and {Zamorani} and G. and {Pell{\`o}} and R. and {Iovino} and A. and {Tresse} and L. and {Le Brun} and V. and {Bottini} and D. and {Garilli} and B. and {Maccagni} and D. and {Picat}, J.~P. and {Scaramella} and R. and {Scodeggio} and M. and {Vettolani} and G. and {Zanichelli} and A. and {Adami} and C. and {Bardelli} and S. and {Cappi} and A. and {Charlot} and S. and {Ciliegi} and P. and {Contini} and T. and {Cucciati} and O. and {Foucaud} and S. and {Franzetti} and P. and {Gavignaud} and I. and {Guzzo} and L. and {Marano} and B. and {Marinoni} and C. and {Mazure} and A. and {Meneux} and B. and {Merighi} and R. and {Paltani} and S. and {Pollo} and A. and {Pozzetti} and L. and {Radovich} and M. and {Zucca} and E. and {Bondi} and M. and {Bongiorno} and A. and {Busarello} and G. and {de La Torre} and S. and {Gregorini} and L. and {Lamareille} and F. and {Mathez} and G. and {Merluzzi} and P. and {Ripepi} and V. and {Rizzo} and D. and {Vergani}, D.},
        title = "{Accurate photometric redshifts for the CFHT legacy survey calibrated using the VIMOS VLT deep survey}",
      journal = {\aap},
         year = 2006,
        month = oct,
       volume = {457},
       number = {3},
        pages = {841-856},
          doi = {10.1051/0004-6361:20065138},
archivePrefix = {arXiv},
       eprint = {astro-ph/0603217},
 primaryClass = {astro-ph},
       adsurl = {https://ui.adsabs.harvard.edu/abs/2006A&A...457..841I}
}

@ARTICLE{Bolzonella2000,
       author = {{Bolzonella} and M. and {Miralles} and J. -M. and {Pell{\'o}}, R.},
        title = "{Photometric redshifts based on standard SED fitting procedures}",
      journal = {\aap},
         year = 2000,
        month = nov,
       volume = {363},
        pages = {476-492},
          doi = {10.48550/arXiv.astro-ph/0003380},
archivePrefix = {arXiv},
       eprint = {astro-ph/0003380},
 primaryClass = {astro-ph},
       adsurl = {https://ui.adsabs.harvard.edu/abs/2000A&A...363..476B}
}

@ARTICLE{Brammer2008,
       author = {{Brammer}, Gabriel B. and {van Dokkum}, Pieter G. and {Coppi}, Paolo},
        title = "{EAZY: A Fast, Public Photometric Redshift Code}",
      journal = {\apj},
         year = 2008,
        month = oct,
       volume = {686},
       number = {2},
        pages = {1503-1513},
          doi = {10.1086/591786},
archivePrefix = {arXiv},
       eprint = {0807.1533},
 primaryClass = {astro-ph},
       adsurl = {https://ui.adsabs.harvard.edu/abs/2008ApJ...686.1503B}
}

@BOOK{Binney1998,
       author = {{Binney}, James and {Merrifield}, Michael},
        title = "{Galactic Astronomy}",
      address = {Princeton, NJ},
    publisher = {Princeton University Press},
         year = 1998,
       adsurl = {https://ui.adsabs.harvard.edu/abs/1998gaas.book.....B}
}

@ARTICLE{Ellison2020,
       author = {{Ellison}, Sara L. and {Thorp}, Mallory D. and {Lin}, Lihwai and {Pan}, Hsi-An and {Bluck}, Asa F.~L. and {Scudder}, Jillian M. and {Teimoorinia}, Hossen and {S{\'a}nchez}, Sebastian F. and {Sargent}, Mark},
        title = "{The ALMaQUEST survey - III. Scatter in the resolved star-forming main sequence is primarily due to variations in star formation efficiency}",
      journal = {\mnras},
         year = 2020,
        month = mar,
       volume = {493},
       number = {1},
        pages = {L39-L43},
          doi = {10.1093/mnrasl/slz179},
archivePrefix = {arXiv},
       eprint = {1911.11887},
 primaryClass = {astro-ph.GA},
       adsurl = {https://ui.adsabs.harvard.edu/abs/2020MNRAS.493L..39E}
}

@ARTICLE{Ellison2021,
       author = {{Ellison}, Sara L. and {Lin}, Lihwai and {Thorp}, Mallory D. and {Pan}, Hsi-An and {Scudder}, Jillian M. and {S{\'a}nchez}, Sebastian F. and {Bluck}, Asa F.~L. and {Maiolino}, Roberto},
        title = "{The ALMaQUEST Survey - V. The non-universality of kpc-scale star formation relations and the factors that drive them}",
      journal = {\mnras},
         year = 2021,
        month = mar,
       volume = {501},
       number = {4},
        pages = {4777-4797},
          doi = {10.1093/mnras/staa3822},
archivePrefix = {arXiv},
       eprint = {2012.04771},
 primaryClass = {astro-ph.GA},
       adsurl = {https://ui.adsabs.harvard.edu/abs/2021MNRAS.501.4777E}
}

@ARTICLE{Renzini2015,
       author = {{Renzini}, Alvio and {Peng}, Ying-jie},
        title = "{An Objective Definition for the Main Sequence of Star-forming Galaxies}",
      journal = {\apjl},
         year = 2015,
        month = mar,
       volume = {801},
       number = {2},
          eid = {L29},
        pages = {L29},
          doi = {10.1088/2041-8205/801/2/L29},
archivePrefix = {arXiv},
       eprint = {1502.01027},
 primaryClass = {astro-ph.GA},
       adsurl = {https://ui.adsabs.harvard.edu/abs/2015ApJ...801L..29R}
}

@ARTICLE{Dekel2013,
       author = {{Dekel} and A. and {Zolotov} and A. and {Tweed} and D. and {Cacciato} and M. and {Ceverino} and D. and {Primack}, J.~R.},
        title = "{Toy models for galaxy formation versus simulations}",
      journal = {\mnras},
         year = 2013,
        month = oct,
       volume = {435},
       number = {2},
        pages = {999-1019},
          doi = {10.1093/mnras/stt1338},
archivePrefix = {arXiv},
       eprint = {1303.3009},
 primaryClass = {astro-ph.CO},
       adsurl = {https://ui.adsabs.harvard.edu/abs/2013MNRAS.435..999D}
}

@ARTICLE{Neistein2008,
       author = {{Neistein}, Eyal and {Dekel}, Avishai},
        title = "{Merger rates of dark matter haloes}",
      journal = {\mnras},
         year = 2008,
        month = aug,
       volume = {388},
       number = {4},
        pages = {1792-1802},
          doi = {10.1111/j.1365-2966.2008.13525.x},
archivePrefix = {arXiv},
       eprint = {0802.0198},
 primaryClass = {astro-ph},
       adsurl = {https://ui.adsabs.harvard.edu/abs/2008MNRAS.388.1792N}
}

@ARTICLE{McBride2009,
       author = {{McBride}, James and {Fakhouri}, Onsi and {Ma}, Chung-Pei},
        title = "{Mass accretion rates and histories of dark matter haloes}",
      journal = {\mnras},
         year = 2009,
        month = oct,
       volume = {398},
       number = {4},
        pages = {1858-1868},
          doi = {10.1111/j.1365-2966.2009.15329.x},
archivePrefix = {arXiv},
       eprint = {0902.3659},
 primaryClass = {astro-ph.CO},
       adsurl = {https://ui.adsabs.harvard.edu/abs/2009MNRAS.398.1858M}
}

@ARTICLE{Salmon2015,
       author = {{Salmon}, Brett and {Papovich}, Casey and {Finkelstein}, Steven L. and {Tilvi}, Vithal and {Finlator}, Kristian and {Behroozi}, Peter and {Dahlen}, Tomas and {Dav{\'e}}, Romeel and {Dekel}, Avishai and {Dickinson}, Mark and {Ferguson}, Henry C. and {Giavalisco}, Mauro and {Long}, James and {Lu}, Yu and {Mobasher}, Bahram and {Reddy}, Naveen and {Somerville}, Rachel S. and {Wechsler}, Risa H.},
        title = "{The Relation between Star Formation Rate and Stellar Mass for Galaxies at $3.5 \leq z \leq 6.5$ in {CANDELS}}",
      journal = {\apj},
         year = 2015,
        month = feb,
       volume = {799},
       number = {2},
          eid = {183},
        pages = {183},
          doi = {10.1088/0004-637X/799/2/183},
archivePrefix = {arXiv},
       eprint = {1407.6012},
 primaryClass = {astro-ph.GA},
       adsurl = {https://ui.adsabs.harvard.edu/abs/2015ApJ...799..183S}
}

@ARTICLE{Tasca2015,
       author = {{Tasca}, L.~A.~M. and {Le F{\`e}vre} and O. and {Hathi}, N.~P. and {Schaerer} and D. and {Ilbert} and O. and {Zamorani} and G. and {Lemaux}, B.~C. and {Cassata} and P. and {Garilli} and B. and {Le Brun} and V. and {Maccagni} and D. and {Pentericci} and L. and {Thomas} and R. and {Vanzella} and E. and {Zucca} and E. and {Amorin} and R. and {Bardelli} and S. and {Cassar{\`a}}, L.~P. and {Castellano} and M. and {Cimatti} and A. and {Cucciati} and O. and {Durkalec} and A. and {Fontana} and A. and {Giavalisco} and M. and {Grazian} and A. and {Paltani} and S. and {Ribeiro} and B. and {Scodeggio} and M. and {Sommariva} and V. and {Talia} and M. and {Tresse} and L. and {Vergani} and D. and {Capak} and P. and {Charlot} and S. and {Contini} and T. and {de la Torre} and S. and {Dunlop} and J. and {Fotopoulou} and S. and {Koekemoer} and A. and {L{\'o}pez-Sanjuan} and C. and {Mellier} and Y. and {Pforr} and J. and {Salvato} and M. and {Scoville} and N. and {Taniguchi} and Y. and {Wang}, P.~W.},
        title = "{The evolving star formation rate: M$_{{\ensuremath{\star}}}$ relation and sSFR since $z \simeq 5$ from the VUDS spectroscopic survey}",
      journal = {\aap},
         year = 2015,
        month = sep,
       volume = {581},
          eid = {A54},
        pages = {A54},
          doi = {10.1051/0004-6361/201425379},
archivePrefix = {arXiv},
       eprint = {1411.5687},
 primaryClass = {astro-ph.GA},
       adsurl = {https://ui.adsabs.harvard.edu/abs/2015A&A...581A..54T}
}

@ARTICLE{Popesso2019,
       author = {{Popesso} and P. and {Morselli} and L. and {Concas} and A. and {Schreiber} and C. and {Rodighiero} and G. and {Cresci} and G. and {Belli} and S. and {Ilbert} and O. and {Erfanianfar} and G. and {Mancini} and C. and {Inami} and H. and {Dickinson} and M. and {Pannella} and M. and {Elbaz}, D.},
        title = "{The main sequence of star-forming galaxies - II. A non-evolving slope at the high-mass end}",
      journal = {\mnras},
         year = 2019,
        month = dec,
       volume = {490},
       number = {4},
        pages = {5285-5299},
          doi = {10.1093/mnras/stz2635},
archivePrefix = {arXiv},
       eprint = {1909.07760},
 primaryClass = {astro-ph.GA},
       adsurl = {https://ui.adsabs.harvard.edu/abs/2019MNRAS.490.5285P}
}

@ARTICLE{Whitaker2012,
       author = {{Whitaker}, Katherine E. and {van Dokkum}, Pieter G. and {Brammer}, Gabriel and {Franx}, Marijn},
        title = "{The Star Formation Mass Sequence Out to z = 2.5}",
      journal = {\apjl},
         year = 2012,
        month = aug,
       volume = {754},
       number = {2},
          eid = {L29},
        pages = {L29},
          doi = {10.1088/2041-8205/754/2/L29},
archivePrefix = {arXiv},
       eprint = {1205.0547},
 primaryClass = {astro-ph.CO},
       adsurl = {https://ui.adsabs.harvard.edu/abs/2012ApJ...754L..29W}
}

@ARTICLE{Whitaker2014,
       author = {{Whitaker}, Katherine E. and {Franx}, Marijn and {Leja}, Joel and {van Dokkum}, Pieter G. and {Henry}, Alaina and {Skelton}, Rosalind E. and {Fumagalli}, Mattia and {Momcheva}, Ivelina G. and {Brammer}, Gabriel B. and {Labb{\'e}}, Ivo and {Nelson}, Erica J. and {Rigby}, Jane R.},
        title = "{Constraining the Low-mass Slope of the Star Formation Sequence at $0.5 < z < 2.5$}",
      journal = {\apj},
         year = 2014,
        month = nov,
       volume = {795},
       number = {2},
          eid = {104},
        pages = {104},
          doi = {10.1088/0004-637X/795/2/104},
archivePrefix = {arXiv},
       eprint = {1407.1843},
 primaryClass = {astro-ph.GA},
       adsurl = {https://ui.adsabs.harvard.edu/abs/2014ApJ...795..104W}
}

@ARTICLE{Speagle2014,
       author = {{Speagle}, J.~S. and {Steinhardt}, C.~L. and {Capak}, P.~L. and {Silverman}, J.~D.},
        title = "{A Highly Consistent Framework for the Evolution of the Star-Forming ``Main Sequence'' from $z \sim 0$--6}",
      journal = {\apjs},
         year = 2014,
        month = oct,
       volume = {214},
       number = {2},
          eid = {15},
        pages = {15},
          doi = {10.1088/0067-0049/214/2/15},
archivePrefix = {arXiv},
       eprint = {1405.2041},
 primaryClass = {astro-ph.GA},
       adsurl = {https://ui.adsabs.harvard.edu/abs/2014ApJS..214...15S}
}

@ARTICLE{Schreiber2015,
       author = {{Schreiber} and C. and {Pannella} and M. and {Elbaz} and D. and {B{\'e}thermin} and M. and {Inami} and H. and {Dickinson} and M. and {Magnelli} and B. and {Wang} and T. and {Aussel} and H. and {Daddi} and E. and {Juneau} and S. and {Shu} and X. and {Sargent}, M.~T. and {Buat} and V. and {Faber}, S.~M. and {Ferguson}, H.~C. and {Giavalisco} and M. and {Koekemoer}, A.~M. and {Magdis} and G. and {Morrison}, G.~E. and {Papovich} and C. and {Santini} and P. and {Scott}, D.},
        title = "{The Herschel view of the dominant mode of galaxy growth from z = 4 to the present day}",
      journal = {\aap},
         year = 2015,
        month = mar,
       volume = {575},
          eid = {A74},
        pages = {A74},
          doi = {10.1051/0004-6361/201425017},
archivePrefix = {arXiv},
       eprint = {1409.5433},
 primaryClass = {astro-ph.GA},
       adsurl = {https://ui.adsabs.harvard.edu/abs/2015A&A...575A..74S}
}

@ARTICLE{Tomczak2016,
       author = {{Tomczak}, Adam R. and {Quadri}, Ryan F. and {Tran}, Kim-Vy H. and {Labb{\'e}}, Ivo and {Straatman}, Caroline M.~S. and {Papovich}, Casey and {Glazebrook}, Karl and {Allen}, Rebecca and {Brammer}, Gabreil B. and {Cowley}, Michael and {Dickinson}, Mark and {Elbaz}, David and {Inami}, Hanae and {Kacprzak}, Glenn G. and {Morrison}, Glenn E. and {Nanayakkara}, Themiya and {Persson} and S. Eric and {Rees}, Glen A. and {Salmon}, Brett and {Schreiber}, Corentin and {Spitler}, Lee R. and {Whitaker}, Katherine E.},
        title = "{The SFR--$M_*$ Relation and Empirical Star-Formation Histories from {ZFOURGE} at $0.5 < z < 4$}",
      journal = {\apj},
         year = 2016,
        month = feb,
       volume = {817},
       number = {2},
          eid = {118},
        pages = {118},
          doi = {10.3847/0004-637X/817/2/118},
archivePrefix = {arXiv},
       eprint = {1510.06072},
 primaryClass = {astro-ph.GA},
       adsurl = {https://ui.adsabs.harvard.edu/abs/2016ApJ...817..118T}
}

@ARTICLE{Noeske2007a,
       author = {{Noeske}, K.~G. and {Faber}, S.~M. and {Weiner}, B.~J. and {Koo}, D.~C. and {Primack}, J.~R. and {Dekel} and A. and {Papovich} and C. and {Conselice}, C.~J. and {Le Floc'h} and E. and {Rieke}, G.~H. and {Coil}, A.~L. and {Lotz}, J.~M. and {Somerville}, R.~S. and {Bundy}, K.},
        title = "{Star Formation in AEGIS Field Galaxies since z=1.1: Staged Galaxy Formation and a Model of Mass-dependent Gas Exhaustion}",
      journal = {\apjl},
         year = 2007,
        month = may,
       volume = {660},
       number = {1},
        pages = {L47-L50},
          doi = {10.1086/517927},
archivePrefix = {arXiv},
       eprint = {astro-ph/0703056},
 primaryClass = {astro-ph},
       adsurl = {https://ui.adsabs.harvard.edu/abs/2007ApJ...660L..47N}
}

@ARTICLE{Noeske2007b,
       author = {{Noeske}, K.~G. and {Weiner}, B.~J. and {Faber}, S.~M. and {Papovich} and C. and {Koo}, D.~C. and {Somerville}, R.~S. and {Bundy} and K. and {Conselice}, C.~J. and {Newman}, J.~A. and {Schiminovich} and D. and {Le Floc'h} and E. and {Coil}, A.~L. and {Rieke}, G.~H. and {Lotz}, J.~M. and {Primack}, J.~R. and {Barmby} and P. and {Cooper}, M.~C. and {Davis} and M. and {Ellis}, R.~S. and {Fazio}, G.~G. and {Guhathakurta} and P. and {Huang} and J. and {Kassin}, S.~A. and {Martin}, D.~C. and {Phillips}, A.~C. and {Rich}, R.~M. and {Small}, T.~A. and {Willmer}, C.~N.~A. and {Wilson}, G.},
        title = "{Star Formation in AEGIS Field Galaxies since z=1.1: The Dominance of Gradually Declining Star Formation, and the Main Sequence of Star-forming Galaxies}",
      journal = {\apjl},
         year = 2007,
        month = may,
       volume = {660},
       number = {1},
        pages = {L43-L46},
          doi = {10.1086/517926},
archivePrefix = {arXiv},
       eprint = {astro-ph/0701924},
 primaryClass = {astro-ph},
       adsurl = {https://ui.adsabs.harvard.edu/abs/2007ApJ...660L..43N}
}

@ARTICLE{Daddi2007,
       author = {{Daddi} and E. and {Alexander}, D.~M. and {Dickinson} and M. and {Gilli} and R. and {Renzini} and A. and {Elbaz} and D. and {Cimatti} and A. and {Chary} and R. and {Frayer} and D. and {Bauer}, F.~E. and {Brandt}, W.~N. and {Giavalisco} and M. and {Grogin}, N.~A. and {Huynh} and M. and {Kurk} and J. and {Mignoli} and M. and {Morrison} and G. and {Pope} and A. and {Ravindranath}, S.},
        title = "{Multiwavelength Study of Massive Galaxies at z\raisebox{-0.5ex}\textasciitilde2. II. Widespread Compton-thick Active Galactic Nuclei and the Concurrent Growth of Black Holes and Bulges}",
      journal = {\apj},
         year = 2007,
        month = nov,
       volume = {670},
       number = {1},
        pages = {173-189},
          doi = {10.1086/521820},
archivePrefix = {arXiv},
       eprint = {0705.2832},
 primaryClass = {astro-ph},
       adsurl = {https://ui.adsabs.harvard.edu/abs/2007ApJ...670..173D}
}

@ARTICLE{Elbaz2007,
       author = {{Elbaz} and D. and {Daddi} and E. and {Le Borgne} and D. and {Dickinson} and M. and {Alexander}, D.~M. and {Chary} and R. -R. and {Starck} and J. -L. and {Brandt}, W.~N. and {Kitzbichler} and M. and {MacDonald} and E. and {Nonino} and M. and {Popesso} and P. and {Stern} and D. and {Vanzella}, E.},
        title = "{The reversal of the star formation-density relation in the distant universe}",
      journal = {\aap},
         year = 2007,
        month = jun,
       volume = {468},
       number = {1},
        pages = {33-48},
          doi = {10.1051/0004-6361:20077525},
archivePrefix = {arXiv},
       eprint = {astro-ph/0703653},
 primaryClass = {astro-ph},
       adsurl = {https://ui.adsabs.harvard.edu/abs/2007A&A...468...33E}
}

@ARTICLE{Cano-Diaz2016,
       author = {{Cano-D{\'\i}az} and M. and {S{\'a}nchez}, S.~F. and {Zibetti} and S. and {Ascasibar} and Y. and {Bland-Hawthorn} and J. and {Ziegler} and B. and {Gonz{\'a}lez Delgado}, R.~M. and {Walcher}, C.~J. and {Garc{\'\i}a-Benito} and R. and {Mast} and D. and {Mendoza-P{\'e}rez}, M.~A. and {Falc{\'o}n-Barroso} and J. and {Galbany} and L. and {Husemann} and B. and {Kehrig} and C. and {Marino}, R.~A. and {S{\'a}nchez-Bl{\'a}zquez} and P. and {L{\'o}pez-Cob{\'a}} and C. and {L{\'o}pez-S{\'a}nchez}, {\'A}. R. and {Vilchez}, J.~M.},
        title = "{Spatially Resolved Star Formation Main Sequence of Galaxies in the CALIFA Survey}",
      journal = {\apjl},
         year = 2016,
        month = apr,
       volume = {821},
       number = {2},
          eid = {L26},
        pages = {L26},
          doi = {10.3847/2041-8205/821/2/L26},
archivePrefix = {arXiv},
       eprint = {1602.02770},
 primaryClass = {astro-ph.GA},
       adsurl = {https://ui.adsabs.harvard.edu/abs/2016ApJ...821L..26C}
}

@ARTICLE{Hsieh2017,
       author = {{Hsieh}, B.~C. and {Lin}, Lihwai and {Lin}, J.~H. and {Pan}, H.~A. and {Hsu}, C.~H. and {S{\'a}nchez}, S.~F. and {Cano-D{\'\i}az} and M. and {Zhang} and K. and {Yan} and R. and {Barrera-Ballesteros}, J.~K. and {Boquien} and M. and {Riffel} and R. and {Brownstein} and J. and {Cruz-Gonz{\'a}lez} and I. and {Hagen} and A. and {Ibarra} and H. and {Pan} and K. and {Bizyaev} and D. and {Oravetz} and D. and {Simmons}, A.},
        title = "{SDSS-IV MaNGA: Spatially Resolved Star Formation Main Sequence and LI(N)ER Sequence}",
      journal = {\apjl},
         year = 2017,
        month = dec,
       volume = {851},
       number = {2},
          eid = {L24},
        pages = {L24},
          doi = {10.3847/2041-8213/aa9d80},
archivePrefix = {arXiv},
       eprint = {1711.09162},
 primaryClass = {astro-ph.GA},
       adsurl = {https://ui.adsabs.harvard.edu/abs/2017ApJ...851L..24H}
}

@ARTICLE{Medling2018,
       author = {{Medling}, Anne M. and {Cortese}, Luca and {Croom}, Scott M. and {Green}, Andrew W. and {Groves}, Brent and {Hampton}, Elise and {Ho} and I. -Ting and {Davies}, Luke J.~M. and {Kewley}, Lisa J. and {Moffett}, Amanda J. and {Schaefer}, Adam L. and {Taylor}, Edward and {Zafar}, Tayyaba and {Bekki}, Kenji and {Bland-Hawthorn}, Joss and {Bloom}, Jessica V. and {Brough}, Sarah and {Bryant}, Julia J. and {Catinella}, Barbara and {Cecil}, Gerald and {Colless}, Matthew and {Couch}, Warrick J. and {Drinkwater}, Michael J. and {Driver}, Simon P. and {Federrath}, Christoph and {Foster}, Caroline and {Goldstein}, Gregory and {Goodwin}, Michael and {Hopkins}, Andrew and {Lawrence}, J.~S. and {Leslie}, Sarah K. and {Lewis}, Geraint F. and {Lorente}, Nuria P.~F. and {Owers}, Matt S. and {McDermid}, Richard and {Richards}, Samuel N. and {Sharp}, Robert and {Scott}, Nicholas and {Sweet}, Sarah M. and {Taranu}, Dan S. and {Tescari}, Edoardo and {Tonini}, Chiara and {van de Sande}, Jesse and {Walcher} and C. Jakob and {Wright}, Angus},
        title = "{The SAMI Galaxy Survey: spatially resolving the main sequence of star formation}",
      journal = {\mnras},
         year = 2018,
        month = apr,
       volume = {475},
       number = {4},
        pages = {5194-5214},
          doi = {10.1093/mnras/sty127},
archivePrefix = {arXiv},
       eprint = {1801.04283},
 primaryClass = {astro-ph.GA},
       adsurl = {https://ui.adsabs.harvard.edu/abs/2018MNRAS.475.5194M}
}

@ARTICLE{Ilbert2010,
       author = {{Ilbert} and O. and {Salvato} and M. and {Le Floc'h} and E. and {Aussel} and H. and {Capak} and P. and {McCracken}, H.~J. and {Mobasher} and B. and {Kartaltepe} and J. and {Scoville} and N. and {Sanders}, D.~B. and {Arnouts} and S. and {Bundy} and K. and {Cassata} and P. and {Kneib} and J. -P. and {Koekemoer} and A. and {Le F{\`e}vre} and O. and {Lilly} and S. and {Surace} and J. and {Taniguchi} and Y. and {Tasca} and L. and {Thompson} and D. and {Tresse} and L. and {Zamojski} and M. and {Zamorani} and G. and {Zucca}, E.},
        title = "{Galaxy Stellar Mass Assembly Between $0.2 < z < 2$ from the S-COSMOS Survey}",
      journal = {\apj},
         year = 2010,
        month = feb,
       volume = {709},
       number = {2},
        pages = {644-663},
          doi = {10.1088/0004-637X/709/2/644},
archivePrefix = {arXiv},
       eprint = {0903.0102},
 primaryClass = {astro-ph.CO},
       adsurl = {https://ui.adsabs.harvard.edu/abs/2010ApJ...709..644I}
}

@ARTICLE{Behroozi2019,
       author = {{Behroozi}, Peter and {Wechsler}, Risa H. and {Hearin}, Andrew P. and {Conroy}, Charlie},
        title = "{UNIVERSEMACHINE: The correlation between galaxy growth and dark matter halo assembly from z = 0-10}",
      journal = {\mnras},
         year = 2019,
        month = sep,
       volume = {488},
       number = {3},
        pages = {3143-3194},
          doi = {10.1093/mnras/stz1182},
archivePrefix = {arXiv},
       eprint = {1806.07893},
 primaryClass = {astro-ph.GA},
       adsurl = {https://ui.adsabs.harvard.edu/abs/2019MNRAS.488.3143B}
}

@ARTICLE{Rodriguez-Puebla2017,
       author = {{Rodr{\'\i}guez-Puebla}, Aldo and {Primack}, Joel R. and {Avila-Reese}, Vladimir and {Faber}, S.~M.},
        title = "{Constraining the galaxy-halo connection over the last 13.3 Gyr: star formation histories, galaxy mergers and structural properties}",
      journal = {\mnras},
         year = 2017,
        month = sep,
       volume = {470},
       number = {1},
        pages = {651-687},
          doi = {10.1093/mnras/stx1172},
archivePrefix = {arXiv},
       eprint = {1703.04542},
 primaryClass = {astro-ph.GA},
       adsurl = {https://ui.adsabs.harvard.edu/abs/2017MNRAS.470..651R}
}

@ARTICLE{Mandelbaum2016,
       author = {{Mandelbaum}, Rachel and {Wang}, Wenting and {Zu}, Ying and {White}, Simon and {Henriques}, Bruno and {More}, Surhud},
        title = "{Strong bimodality in the host halo mass of central galaxies from galaxy-galaxy lensing}",
      journal = {\mnras},
         year = 2016,
        month = apr,
       volume = {457},
       number = {3},
        pages = {3200-3218},
          doi = {10.1093/mnras/stw188},
archivePrefix = {arXiv},
       eprint = {1509.06762},
 primaryClass = {astro-ph.GA},
       adsurl = {https://ui.adsabs.harvard.edu/abs/2016MNRAS.457.3200M}
}

@ARTICLE{Lee2015,
       author = {{Lee}, Nicholas and {Sanders}, D.~B. and {Casey}, Caitlin M. and {Toft}, Sune and {Scoville}, N.~Z. and {Hung}, Chao-Ling and {Le Floc'h}, Emeric and {Ilbert}, Olivier and {Zahid} and H. Jabran and {Aussel}, Herv{\'e} and {Capak}, Peter and {Kartaltepe}, Jeyhan S. and {Kewley}, Lisa J. and {Li}, Yanxia and {Schawinski}, Kevin and {Sheth}, Kartik and {Xiao}, Quanbao},
        title = "{A Turnover in the Galaxy Main Sequence of Star Formation at $M_{*} \sim 10^{10} M_{\odot}$ for Redshifts $z < 1.3$}",
      journal = {\apj},
         year = 2015,
        month = mar,
       volume = {801},
       number = {2},
          eid = {80},
        pages = {80},
          doi = {10.1088/0004-637X/801/2/80},
archivePrefix = {arXiv},
       eprint = {1501.01080},
 primaryClass = {astro-ph.GA},
       adsurl = {https://ui.adsabs.harvard.edu/abs/2015ApJ...801...80L}
}

@ARTICLE{Abramson2014,
       author = {{Abramson}, Louis E. and {Kelson}, Daniel D. and {Dressler}, Alan and {Poggianti}, Bianca and {Gladders}, Michael D. and {Oemler}, Jr., Augustus and {Vulcani}, Benedetta},
        title = "{The Mass-independence of Specific Star Formation Rates in Galactic Disks}",
      journal = {\apjl},
         year = 2014,
        month = apr,
       volume = {785},
       number = {2},
          eid = {L36},
        pages = {L36},
          doi = {10.1088/2041-8205/785/2/L36},
archivePrefix = {arXiv},
       eprint = {1402.7076},
 primaryClass = {astro-ph.GA},
       adsurl = {https://ui.adsabs.harvard.edu/abs/2014ApJ...785L..36A}
}

@ARTICLE{Guo2015,
       author = {{Guo}, Yicheng and {Ferguson}, Henry C. and {Bell}, Eric F. and {Koo}, David C. and {Conselice}, Christopher J. and {Giavalisco}, Mauro and {Kassin}, Susan and {Lu}, Yu and {Lucas}, Ray and {Mandelker}, Nir and {McIntosh}, Daniel H. and {Primack}, Joel R. and {Ravindranath}, Swara and {Barro}, Guillermo and {Ceverino}, Daniel and {Dekel}, Avishai and {Faber}, Sandra M. and {Fang}, Jerome J. and {Koekemoer}, Anton M. and {Noeske}, Kai and {Rafelski}, Marc and {Straughn}, Amber},
        title = "{Clumpy Galaxies in CANDELS. I. The Definition of UV Clumps and the Fraction of Clumpy Galaxies at $0.5 < z < 3$}",
      journal = {\apj},
         year = 2015,
        month = feb,
       volume = {800},
       number = {1},
          eid = {39},
        pages = {39},
          doi = {10.1088/0004-637X/800/1/39},
archivePrefix = {arXiv},
       eprint = {1410.7398},
 primaryClass = {astro-ph.GA},
       adsurl = {https://ui.adsabs.harvard.edu/abs/2015ApJ...800...39G}
}

@ARTICLE{Leslie2020,
       author = {{Leslie}, Sarah K. and {Schinnerer}, Eva and {Liu}, Daizhong and {Magnelli}, Benjamin and {Algera}, Hiddo and {Karim}, Alexander and {Davidzon}, Iary and {Gozaliasl}, Ghassem and {Jim{\'e}nez-Andrade}, Eric F. and {Lang}, Philipp and {Sargent}, Mark T. and {Novak}, Mladen and {Groves}, Brent and {Smol{\v{c}}i{\'c}}, Vernesa and {Zamorani}, Giovanni and {Vaccari}, Mattia and {Battisti}, Andrew and {Vardoulaki}, Eleni and {Peng}, Yingjie and {Kartaltepe}, Jeyhan},
        title = "{The VLA-COSMOS 3 GHz Large Project: Evolution of Specific Star Formation Rates out to $z \sim 5$}",
      journal = {\apj},
         year = 2020,
        month = aug,
       volume = {899},
       number = {1},
          eid = {58},
        pages = {58},
          doi = {10.3847/1538-4357/aba044},
archivePrefix = {arXiv},
       eprint = {2006.13937},
 primaryClass = {astro-ph.GA},
       adsurl = {https://ui.adsabs.harvard.edu/abs/2020ApJ...899...58L}
}

@ARTICLE{Franx2008,
       author = {{Franx}, Marijn and {van Dokkum}, Pieter G. and {F{\"o}rster Schreiber}, Natascha M. and {Wuyts}, Stijn and {Labb{\'e}}, Ivo and {Toft}, Sune},
        title = "{Structure and Star Formation in Galaxies out to z = 3: Evidence for Surface Density Dependent Evolution and Upsizing}",
      journal = {\apj},
         year = 2008,
        month = dec,
       volume = {688},
       number = {2},
        pages = {770-788},
          doi = {10.1086/592431},
archivePrefix = {arXiv},
       eprint = {0808.2642},
 primaryClass = {astro-ph},
       adsurl = {https://ui.adsabs.harvard.edu/abs/2008ApJ...688..770F}
}

@ARTICLE{Williams2009,
       author = {{Williams}, Rik J. and {Quadri}, Ryan F. and {Franx}, Marijn and {van Dokkum}, Pieter and {Labb{\'e}}, Ivo},
        title = "{Detection of Quiescent Galaxies in a Bicolor Sequence from Z = 0-2}",
      journal = {\apj},
         year = 2009,
        month = feb,
       volume = {691},
       number = {2},
        pages = {1879-1895},
          doi = {10.1088/0004-637X/691/2/1879},
archivePrefix = {arXiv},
       eprint = {0806.0625},
 primaryClass = {astro-ph},
       adsurl = {https://ui.adsabs.harvard.edu/abs/2009ApJ...691.1879W}
}

@ARTICLE{Bell2004,
       author = {{Bell}, Eric F. and {Wolf}, Christian and {Meisenheimer}, Klaus and {Rix}, Hans-Walter and {Borch}, Andrea and {Dye}, Simon and {Kleinheinrich}, Martina and {Wisotzki}, Lutz and {McIntosh}, Daniel H.},
        title = "{Nearly 5000 Distant Early-Type Galaxies in COMBO-17: A Red Sequence and Its Evolution since $z \sim 1$}",
      journal = {\apj},
         year = 2004,
        month = jun,
       volume = {608},
       number = {2},
        pages = {752-767},
          doi = {10.1086/420778},
archivePrefix = {arXiv},
       eprint = {astro-ph/0303394},
 primaryClass = {astro-ph},
       adsurl = {https://ui.adsabs.harvard.edu/abs/2004ApJ...608..752B}
}

@ARTICLE{Whitaker2011,
       author = {{Whitaker}, Katherine E. and {Labb{\'e}}, Ivo and {van Dokkum}, Pieter G. and {Brammer}, Gabriel and {Kriek}, Mariska and {Marchesini}, Danilo and {Quadri}, Ryan F. and {Franx}, Marijn and {Muzzin}, Adam and {Williams}, Rik J. and {Bezanson}, Rachel and {Illingworth}, Garth D. and {Lee}, Kyoung-Soo and {Lundgren}, Britt and {Nelson}, Erica J. and {Rudnick}, Gregory and {Tal}, Tomer and {Wake}, David A.},
        title = "{The NEWFIRM Medium-band Survey: Photometric Catalogs, Redshifts, and the Bimodal Color Distribution of Galaxies out to z \raisebox{-0.5ex}\textasciitilde 3}",
      journal = {\apj},
         year = 2011,
        month = jul,
       volume = {735},
       number = {2},
          eid = {86},
        pages = {86},
          doi = {10.1088/0004-637X/735/2/86},
archivePrefix = {arXiv},
       eprint = {1105.4609},
 primaryClass = {astro-ph.CO},
       adsurl = {https://ui.adsabs.harvard.edu/abs/2011ApJ...735...86W}
}

@ARTICLE{Baldry2004,
       author = {{Baldry}, Ivan K. and {Glazebrook}, Karl and {Brinkmann}, Jon and {Ivezi{\'c}}, {\v{Z}}eljko and {Lupton}, Robert H. and {Nichol}, Robert C. and {Szalay}, Alexander S.},
        title = "{Quantifying the Bimodal Color-Magnitude Distribution of Galaxies}",
      journal = {\apj},
         year = 2004,
        month = jan,
       volume = {600},
       number = {2},
        pages = {681-694},
          doi = {10.1086/380092},
archivePrefix = {arXiv},
       eprint = {astro-ph/0309710},
 primaryClass = {astro-ph},
       adsurl = {https://ui.adsabs.harvard.edu/abs/2004ApJ...600..681B}
}

@ARTICLE{Kodama1997,
       author = {{Kodama} and T. and {Arimoto}, N.},
        title = "{Origin of the colour-magnitude relation of elliptical galaxies.}",
      journal = {\aap},
         year = 1997,
        month = apr,
       volume = {320},
        pages = {41-53},
          doi = {10.48550/arXiv.astro-ph/9609160},
archivePrefix = {arXiv},
       eprint = {astro-ph/9609160},
 primaryClass = {astro-ph},
       adsurl = {https://ui.adsabs.harvard.edu/abs/1997A&A...320...41K}
}

@ARTICLE{MacArthur2004,
       author = {{MacArthur}, Lauren A. and {Courteau}, St{\'e}phane and {Bell}, Eric and {Holtzman}, Jon A.},
        title = "{Structure of Disk-dominated Galaxies. II. Color Gradients and Stellar Population Models}",
      journal = {\apjs},
         year = 2004,
        month = jun,
       volume = {152},
       number = {2},
        pages = {175-199},
          doi = {10.1086/383525},
archivePrefix = {arXiv},
       eprint = {astro-ph/0401437},
 primaryClass = {astro-ph},
       adsurl = {https://ui.adsabs.harvard.edu/abs/2004ApJS..152..175M}
}

@ARTICLE{Gallazzi2005,
       author = {{Gallazzi}, Anna and {Charlot}, St{\'e}phane and {Brinchmann}, Jarle and {White}, Simon D.~M. and {Tremonti}, Christy A.},
        title = "{The ages and metallicities of galaxies in the local universe}",
      journal = {\mnras},
         year = 2005,
        month = sep,
       volume = {362},
       number = {1},
        pages = {41-58},
          doi = {10.1111/j.1365-2966.2005.09321.x},
archivePrefix = {arXiv},
       eprint = {astro-ph/0506539},
 primaryClass = {astro-ph},
       adsurl = {https://ui.adsabs.harvard.edu/abs/2005MNRAS.362...41G}
}

@ARTICLE{Kauffmann2003b,
       author = {{Kauffmann}, Guinevere and {Heckman}, Timothy M. and {White}, Simon D.~M. and {Charlot}, St{\'e}phane and {Tremonti}, Christy and {Peng}, Eric W. and {Seibert}, Mark and {Brinkmann}, Jon and {Nichol}, Robert C. and {SubbaRao}, Mark and {York}, Don},
        title = "{The dependence of star formation history and internal structure on stellar mass for {}10$^{5}$ low-redshift galaxies}",
      journal = {\mnras},
         year = 2003,
        month = may,
       volume = {341},
       number = {1},
        pages = {54-69},
          doi = {10.1046/j.1365-8711.2003.06292.x},
archivePrefix = {arXiv},
       eprint = {astro-ph/0205070},
 primaryClass = {astro-ph},
       adsurl = {https://ui.adsabs.harvard.edu/abs/2003MNRAS.341...54K}
}

@ARTICLE{Balogh1999,
       author = {{Balogh}, Michael L. and {Morris}, Simon L. and {Yee}, H.~K.~C. and {Carlberg}, R.~G. and {Ellingson}, Erica},
        title = "{Differential Galaxy Evolution in Cluster and Field Galaxies at $z \sim 0.3$}",
      journal = {\apj},
         year = 1999,
        month = dec,
       volume = {527},
       number = {1},
        pages = {54-79},
          doi = {10.1086/308056},
archivePrefix = {arXiv},
       eprint = {astro-ph/9906470},
 primaryClass = {astro-ph},
       adsurl = {https://ui.adsabs.harvard.edu/abs/1999ApJ...527...54B}
}

@ARTICLE{Bruzual1983,
       author = {{Bruzual A.}, G.},
        title = "{Spectral evolution of galaxies. I. Early-type systems.}",
      journal = {\apj},
         year = 1983,
        month = oct,
       volume = {273},
        pages = {105-127},
          doi = {10.1086/161352},
       adsurl = {https://ui.adsabs.harvard.edu/abs/1983ApJ...273..105B}
}

@ARTICLE{Dressler1983,
       author = {{Dressler} and A. and {Gunn}, J.~E.},
        title = "{Spectroscopy of galaxies in distant clusters. II. The population of the 3C 295 cluster.}",
      journal = {\apj},
         year = 1983,
        month = jul,
       volume = {270},
        pages = {7-19},
          doi = {10.1086/161093},
       adsurl = {https://ui.adsabs.harvard.edu/abs/1983ApJ...270....7D}
}

@ARTICLE{CidFernandes2011,
       author = {{Cid Fernandes} and R. and {Stasi{\'n}ska} and G. and {Mateus} and A. and {Vale Asari}, N.},
        title = "{A comprehensive classification of galaxies in the Sloan Digital Sky Survey: how to tell true from fake AGN?}",
      journal = {\mnras},
         year = 2011,
        month = may,
       volume = {413},
       number = {3},
        pages = {1687-1699},
          doi = {10.1111/j.1365-2966.2011.18244.x},
archivePrefix = {arXiv},
       eprint = {1012.4426},
 primaryClass = {astro-ph.CO},
       adsurl = {https://ui.adsabs.harvard.edu/abs/2011MNRAS.413.1687C}
}

@ARTICLE{Gavazzi2002,
       author = {{Gavazzi} and G. and {Boselli} and A. and {Pedotti} and P. and {Gallazzi} and A. and {Carrasco}, L.},
        title = "{H{\ensuremath{\alpha}} surface photometry of galaxies in the Virgo cluster. IV. The current star formation in nearby clusters of galaxies}",
      journal = {\aap},
         year = 2002,
        month = dec,
       volume = {396},
        pages = {449-461},
          doi = {10.1051/0004-6361:20021403},
archivePrefix = {arXiv},
       eprint = {astro-ph/0209616},
 primaryClass = {astro-ph},
       adsurl = {https://ui.adsabs.harvard.edu/abs/2002A&A...396..449G}
}

@ARTICLE{Salim2007,
       author = {{Salim}, Samir and {Rich} and R. Michael and {Charlot}, St{\'e}phane and {Brinchmann}, Jarle and {Johnson}, Benjamin D. and {Schiminovich}, David and {Seibert}, Mark and {Mallery}, Ryan and {Heckman}, Timothy M. and {Forster}, Karl and {Friedman}, Peter G. and {Martin} and D. Christopher and {Morrissey}, Patrick and {Neff}, Susan G. and {Small}, Todd and {Wyder}, Ted K. and {Bianchi}, Luciana and {Donas}, Jos{\'e} and {Lee}, Young-Wook and {Madore}, Barry F. and {Milliard}, Bruno and {Szalay}, Alex S. and {Welsh}, Barry Y. and {Yi}, Sukyoung K.},
        title = "{UV Star Formation Rates in the Local Universe}",
      journal = {\apjs},
         year = 2007,
        month = dec,
       volume = {173},
       number = {2},
        pages = {267-292},
          doi = {10.1086/519218},
archivePrefix = {arXiv},
       eprint = {0704.3611},
 primaryClass = {astro-ph},
       adsurl = {https://ui.adsabs.harvard.edu/abs/2007ApJS..173..267S}
}

@ARTICLE{Martin2007,
       author = {{Martin} and D. Christopher and {Wyder}, Ted K. and {Schiminovich}, David and {Barlow}, Tom A. and {Forster}, Karl and {Friedman}, Peter G. and {Morrissey}, Patrick and {Neff}, Susan G. and {Seibert}, Mark and {Small}, Todd and {Welsh}, Barry Y. and {Bianchi}, Luciana and {Donas}, Jos{\'e} and {Heckman}, Timothy M. and {Lee}, Young-Wook and {Madore}, Barry F. and {Milliard}, Bruno and {Rich} and R. Michael and {Szalay}, Alex S. and {Yi}, Sukyoung K.},
        title = "{The UV-Optical Galaxy Color-Magnitude Diagram. III. Constraints on Evolution from the Blue to the Red Sequence}",
      journal = {\apjs},
         year = 2007,
        month = dec,
       volume = {173},
       number = {2},
        pages = {342-356},
          doi = {10.1086/516639},
archivePrefix = {arXiv},
       eprint = {astro-ph/0703281},
 primaryClass = {astro-ph},
       adsurl = {https://ui.adsabs.harvard.edu/abs/2007ApJS..173..342M}
}

@ARTICLE{Wyder2007,
       author = {{Wyder}, Ted K. and {Martin} and D. Christopher and {Schiminovich}, David and {Seibert}, Mark and {Budav{\'a}ri}, Tam{\'a}s and {Treyer}, Marie A. and {Barlow}, Tom A. and {Forster}, Karl and {Friedman}, Peter G. and {Morrissey}, Patrick and {Neff}, Susan G. and {Small}, Todd and {Bianchi}, Luciana and {Donas}, Jos{\'e} and {Heckman}, Timothy M. and {Lee}, Young-Wook and {Madore}, Barry F. and {Milliard}, Bruno and {Rich} and R. Michael and {Szalay}, Alex S. and {Welsh}, Barry Y. and {Yi}, Sukyoung K.},
        title = "{The UV-Optical Galaxy Color-Magnitude Diagram. I. Basic Properties}",
      journal = {\apjs},
         year = 2007,
        month = dec,
       volume = {173},
       number = {2},
        pages = {293-314},
          doi = {10.1086/521402},
archivePrefix = {arXiv},
       eprint = {0706.3938},
 primaryClass = {astro-ph},
       adsurl = {https://ui.adsabs.harvard.edu/abs/2007ApJS..173..293W}
}

@ARTICLE{Cappellari2011,
       author = {{Cappellari}, Michele and {Emsellem}, Eric and {Krajnovi{\'c}}, Davor and {McDermid}, Richard M. and {Scott}, Nicholas and {Verdoes Kleijn}, G.~A. and {Young}, Lisa M. and {Alatalo}, Katherine and {Bacon} and R. and {Blitz}, Leo and {Bois}, Maxime and {Bournaud}, Fr{\'e}d{\'e}ric and {Bureau} and M. and {Davies}, Roger L. and {Davis}, Timothy A. and {de Zeeuw}, P.~T. and {Duc}, Pierre-Alain and {Khochfar}, Sadegh and {Kuntschner}, Harald and {Lablanche}, Pierre-Yves and {Morganti}, Raffaella and {Naab}, Thorsten and {Oosterloo}, Tom and {Sarzi}, Marc and {Serra}, Paolo and {Weijmans}, Anne-Marie},
        title = "{The ATLAS$^{3D}$ project - I. A volume-limited sample of 260 nearby early-type galaxies: science goals and selection criteria}",
      journal = {\mnras},
         year = 2011,
        month = may,
       volume = {413},
       number = {2},
        pages = {813-836},
          doi = {10.1111/j.1365-2966.2010.18174.x},
archivePrefix = {arXiv},
       eprint = {1012.1551},
 primaryClass = {astro-ph.CO},
       adsurl = {https://ui.adsabs.harvard.edu/abs/2011MNRAS.413..813C}
}

@ARTICLE{Cappellari2013,
       author = {{Cappellari}, Michele and {Scott}, Nicholas and {Alatalo}, Katherine and {Blitz}, Leo and {Bois}, Maxime and {Bournaud}, Fr{\'e}d{\'e}ric and {Bureau} and M. and {Crocker}, Alison F. and {Davies}, Roger L. and {Davis}, Timothy A. and {de Zeeuw}, P.~T. and {Duc}, Pierre-Alain and {Emsellem}, Eric and {Khochfar}, Sadegh and {Krajnovi{\'c}}, Davor and {Kuntschner}, Harald and {McDermid}, Richard M. and {Morganti}, Raffaella and {Naab}, Thorsten and {Oosterloo}, Tom and {Sarzi}, Marc and {Serra}, Paolo and {Weijmans}, Anne-Marie and {Young}, Lisa M.},
        title = "{The ATLAS$^{3D}$ project - XV. Benchmark for early-type galaxies scaling relations from 260 dynamical models: mass-to-light ratio, dark matter, Fundamental Plane and Mass Plane}",
      journal = {\mnras},
         year = 2013,
        month = jul,
       volume = {432},
       number = {3},
        pages = {1709-1741},
          doi = {10.1093/mnras/stt562},
archivePrefix = {arXiv},
       eprint = {1208.3522},
 primaryClass = {astro-ph.CO},
       adsurl = {https://ui.adsabs.harvard.edu/abs/2013MNRAS.432.1709C}
}

@ARTICLE{Cappellari2016,
       author = {{Cappellari}, Michele},
        title = "{Structure and Kinematics of Early-Type Galaxies from Integral Field Spectroscopy}",
      journal = {\araa},
         year = 2016,
        month = sep,
       volume = {54},
        pages = {597-665},
          doi = {10.1146/annurev-astro-082214-122432},
archivePrefix = {arXiv},
       eprint = {1602.04267},
 primaryClass = {astro-ph.GA},
       adsurl = {https://ui.adsabs.harvard.edu/abs/2016ARA&A..54..597C}
}

@ARTICLE{Cappellari2008,
       author = {{Cappellari}, Michele},
        title = "{Measuring the inclination and mass-to-light ratio of axisymmetric galaxies via anisotropic Jeans models of stellar kinematics}",
      journal = {\mnras},
         year = 2008,
        month = oct,
       volume = {390},
       number = {1},
        pages = {71-86},
          doi = {10.1111/j.1365-2966.2008.13754.x},
archivePrefix = {arXiv},
       eprint = {0806.0042},
 primaryClass = {astro-ph},
       adsurl = {https://ui.adsabs.harvard.edu/abs/2008MNRAS.390...71C}
}

@ARTICLE{DiTeodoro2015,
       author = {{Di Teodoro}, E.~M. and {Fraternali}, F.},
        title = "{$^{3D}$ BAROLO: a new 3D algorithm to derive rotation curves of galaxies}",
      journal = {\mnras},
         year = 2015,
        month = aug,
       volume = {451},
       number = {3},
        pages = {3021-3033},
          doi = {10.1093/mnras/stv1213},
archivePrefix = {arXiv},
       eprint = {1505.07834},
 primaryClass = {astro-ph.GA},
       adsurl = {https://ui.adsabs.harvard.edu/abs/2015MNRAS.451.3021D}
}

@ARTICLE{begeman1989,
       author = {{Begeman}, K.~G.},
        title = "{HI rotation curves of spiral galaxies. I. NGC 3198.}",
      journal = {\aap},
         year = 1989,
        month = oct,
       volume = {223},
        pages = {47-60},
       adsurl = {https://ui.adsabs.harvard.edu/abs/1989A&A...223...47B}
}

@ARTICLE{Graham2018,
       author = {{Graham}, Mark T. and {Cappellari}, Michele and {Li}, Hongyu and {Mao}, Shude and {Bershady}, Matthew A. and {Bizyaev}, Dmitry and {Brinkmann}, Jonathan and {Brownstein}, Joel R. and {Bundy}, Kevin and {Drory}, Niv and {Law}, David R. and {Pan}, Kaike and {Thomas}, Daniel and {Wake}, David A. and {Weijmans}, Anne-Marie and {Westfall}, Kyle B. and {Yan}, Renbin},
        title = "{SDSS-IV MaNGA: stellar angular momentum of about 2300 galaxies: unveiling the bimodality of massive galaxy properties}",
      journal = {\mnras},
         year = 2018,
        month = jul,
       volume = {477},
       number = {4},
        pages = {4711-4737},
          doi = {10.1093/mnras/sty504},
archivePrefix = {arXiv},
       eprint = {1802.08213},
 primaryClass = {astro-ph.GA},
       adsurl = {https://ui.adsabs.harvard.edu/abs/2018MNRAS.477.4711G}
}

@ARTICLE{deBlok2008,
       author = {{de Blok}, W.~J.~G. and {Walter} and F. and {Brinks} and E. and {Trachternach} and C. and {Oh} and S. -H. and {Kennicutt}, Jr., R.~C.},
        title = "{High-Resolution Rotation Curves and Galaxy Mass Models from THINGS}",
      journal = {\aj},
         year = 2008,
        month = dec,
       volume = {136},
       number = {6},
        pages = {2648-2719},
          doi = {10.1088/0004-6256/136/6/2648},
archivePrefix = {arXiv},
       eprint = {0810.2100},
 primaryClass = {astro-ph},
       adsurl = {https://ui.adsabs.harvard.edu/abs/2008AJ....136.2648D}
}

@ARTICLE{Fall2018,
       author = {{Fall} and S. Michael and {Romanowsky}, Aaron J.},
        title = "{Angular Momentum and Galaxy Formation Revisited: Scaling Relations for Disks and Bulges}",
      journal = {\apj},
         year = 2018,
        month = dec,
       volume = {868},
       number = {2},
          eid = {133},
        pages = {133},
          doi = {10.3847/1538-4357/aaeb27},
archivePrefix = {arXiv},
       eprint = {1808.02525},
 primaryClass = {astro-ph.GA},
       adsurl = {https://ui.adsabs.harvard.edu/abs/2018ApJ...868..133F}
}

@ARTICLE{Emsellem2007,
       author = {{Emsellem}, Eric and {Cappellari}, Michele and {Krajnovi{\'c}}, Davor and {van de Ven}, Glenn and {Bacon} and R. and {Bureau} and M. and {Davies}, Roger L. and {de Zeeuw}, P.~T. and {Falc{\'o}n-Barroso}, Jes{\'u}s and {Kuntschner}, Harald and {McDermid}, Richard and {Peletier}, Reynier F. and {Sarzi}, Marc},
        title = "{The SAURON project - IX. A kinematic classification for early-type galaxies}",
      journal = {\mnras},
         year = 2007,
        month = aug,
       volume = {379},
       number = {2},
        pages = {401-417},
          doi = {10.1111/j.1365-2966.2007.11752.x},
archivePrefix = {arXiv},
       eprint = {astro-ph/0703531},
 primaryClass = {astro-ph},
       adsurl = {https://ui.adsabs.harvard.edu/abs/2007MNRAS.379..401E}
}

\end{document}